\documentclass[iop]{emulateapj}
\usepackage{graphicx,natbib}

\begin{document}

\shorttitle{AN ATLAS OF GALAXY SEDS}
\shortauthors{Brown et al.}

\date{\today}
\title{AN ATLAS OF GALAXY SPECTRAL ENERGY DISTRIBUTIONS FROM THE ULTRAVIOLET TO THE MID-INFRARED}
\author{
Michael J. I. Brown\altaffilmark{1,2,3},
John Moustakas\altaffilmark{4},
J.-D. T. Smith\altaffilmark{5},
Elisabete da Cunha\altaffilmark{6},
T. H. Jarrett\altaffilmark{7},
Masatoshi Imanishi\altaffilmark{8,9,10},
Lee Armus\altaffilmark{11},
Bernhard R. Brandl\altaffilmark{12},
J. E. G. Peek\altaffilmark{13,14}}
\altaffiltext{1}{School of Physics, Monash University, Clayton, Victoria 3800, Australia}
\altaffiltext{2}{Monash Centre for Astrophysics, Monash University, Clayton, Victoria, 3800, Australia}
\altaffiltext{3}{ARC Future Fellow}
\altaffiltext{4}{Department of Physics and Astronomy, Siena College, 515 Loudon Road, Loudonville, NY 12211, USA}
\altaffiltext{5}{Department of Physics and Astronomy, University of Toledo, Ritter Obs., MS \#113, Toledo, OH 43606, USA}
\altaffiltext{6}{Max-Planck-Institut fŸr Astronomie, Kšnigstuhl 17, D-69117 Heidelberg, Germany}
\altaffiltext{7}{Department of Astronomy, University of Cape Town, Private Bag X3, Rondebosch 7701, South Africa}
\altaffiltext{8}{Subaru Telescope, 650 North A'ohoku Place, Hilo, HI 96720, USA}
\altaffiltext{9}{Department of Astronomy, School of Science, Graduate University for Advanced Studies (SOKENDAI), Mitaka, Tokyo 181-8588, Japan}
\altaffiltext{10}{National Astronomical Observatory of Japan, 2-21-1 Osawa, Mitaka, Tokyo 181-8588, Japan}
\altaffiltext{11}{Spitzer Science Center, California Institute of Technology, Pasadena, CA, USA}
\altaffiltext{12}{Leiden Observatory, Leiden University, P.O. Box 9513, 2300 RA Leiden, The Netherlands}
\altaffiltext{13}{Department of Astronomy, Columbia University, New York, NY, USA}
\altaffiltext{14}{Hubble Fellow}

\begin{abstract}
We present an atlas of 129 spectral energy distributions for nearby galaxies, with wavelength coverage spanning from the UV to the mid-infrared. Our atlas spans a broad range of galaxy types, including ellipticals, spirals, merging galaxies, blue compact dwarfs and luminous infrared galaxies. We have combined ground-based optical drift-scan spectrophotometry with infrared spectroscopy from {\it Spitzer} and {\it Akari}, with gaps in spectral coverage being filled using MAGPHYS spectral energy distribution models. The spectroscopy and models were normalized, constrained and verified with matched-aperture photometry measured from {\it Swift}, GALEX, SDSS, 2MASS, {\it Spitzer} and WISE images. The availability of 26 photometric bands allowed us to identify and mitigate systematic errors present in the data. Comparison of our spectral energy distributions with other template libraries and the observed colors of galaxies indicates that we have smaller systematic errors than existing atlases, while spanning a broader range of galaxy types. Relative to the prior literature, our atlas will provide improved K-corrections, photometric redshifts and star-formation rate calibrations.
\end{abstract}
\keywords{atlases --- galaxies: general --- galaxies: photometry ---  techniques: spectroscopic} 

\maketitle

\section{Introduction}

Templates and models of galaxy spectral energy distributions (SEDs) are often essential for deriving the physical properties of galaxies from observables. K-corrections, where rest-frame colors and luminosities are derived from observed galaxy photometry, rely on galaxy SED templates and models \citep[e.g.,][]{oke68,pen76,col80}. Photometric redshifts rely on accurate models of the relationship between observed galaxy photometry and redshift. While this relationship can be modeled empirically for some galaxies using polynomial fits and neural networks \citep[e.g.,][]{con95,fir03}, models and templates are required for photometric redshifts of faint and high redshift galaxies.

The most extensively used galaxy SED template libraries are those of \cite{col80} and \cite{kin96}. These SEDs have proved exceptionally useful over several decades, but have understandable limitations. The wavelength coverage of \cite{col80} and \cite{kin96} templates is limited to the ultraviolet and optical, so they are often extended into the infrared  with stellar population synthesis models \citep[e.g.,][]{bc03}. The \cite{col80} and \cite{kin96} spectra are often of galaxy nuclei, which may not be representative of integrated galaxy spectra. These templates also predate the availability of precisely-calibrated wide-field imaging, which can be used normalize and verify SEDs. Without such imaging, there is the potential for systematic error, especially when combining spectra from different instruments using different extraction apertures.

With the advent of imaging and spectroscopy from Infrared Space Observatory \citep[ISO,][]{kes96}, {\it Spitzer} Space Telescope \citep{wer04},  {\it Akari} \citep[Astro-F,][]{mur07} and Wide-field Infrared Space Explorer \citep[WISE;][]{wri10}, it became both important and possible to extend galaxy SED templates into the mid-infrared \citep[e.g.,][]{cha01,pol07,rie09}. These templates include emission from dust, silicate absorption and features attributed to polycyclic aromatic hydrocarbons (PAHs), which dominate the mid-infrared spectra of star forming galaxies. However, large aperture spectrophotometry over a broad wavelength range remained unavailable (or unutilized), and consequently many of these templates rely upon model galaxy spectra rather than observed galaxy spectra.

Galaxy SEDs can be modeled using models of stellar populations, nebular emission lines, dust obscuration and dust emission \citep[e.g.,][]{tin68, sil98, charlot01, bc03, mar05, dac08, veg08,pac12}. The models have improved considerably over the decades, with advances in relevant theory and stellar libraries, combined with better constraints from observed galaxy spectra and photometry. When provided with sufficiently precise photometry, galaxy SED models do a remarkably good job of reproducing galaxy spectra (see \S\ref{sec:sed}). 

Unfortunately, SED models with a large number of free parameters cannot be expected to reproduce the SED of a galaxy with a complex star formation history when only limited photometry is available \citep[e.g.,][]{pfo12}. Similarly, SED models are not always the best means of predicting the observed colors of galaxies as a function of redshift, as models can generate unrealistic spectra. These issues also apply (albeit to a lesser extent) to modeling SEDs with combinations of empirical component spectra \citep[e.g.][]{bla07,ass08}. This being the case, there is still a role for galaxy SED templates derived from real galaxies.

\begin{figure*}
\vspace*{3cm}
\plotone{f1a.ps}
\plottwo{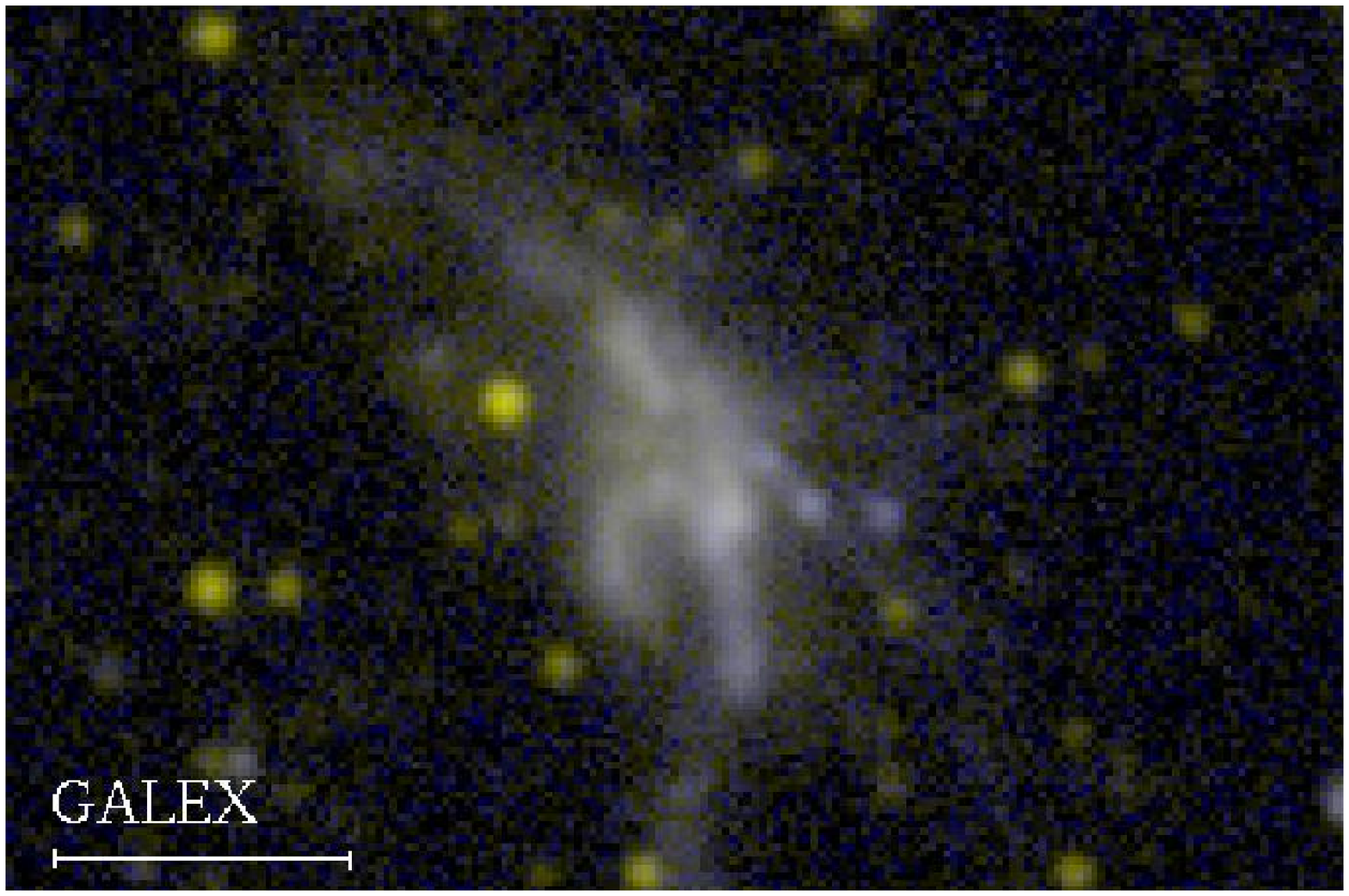}{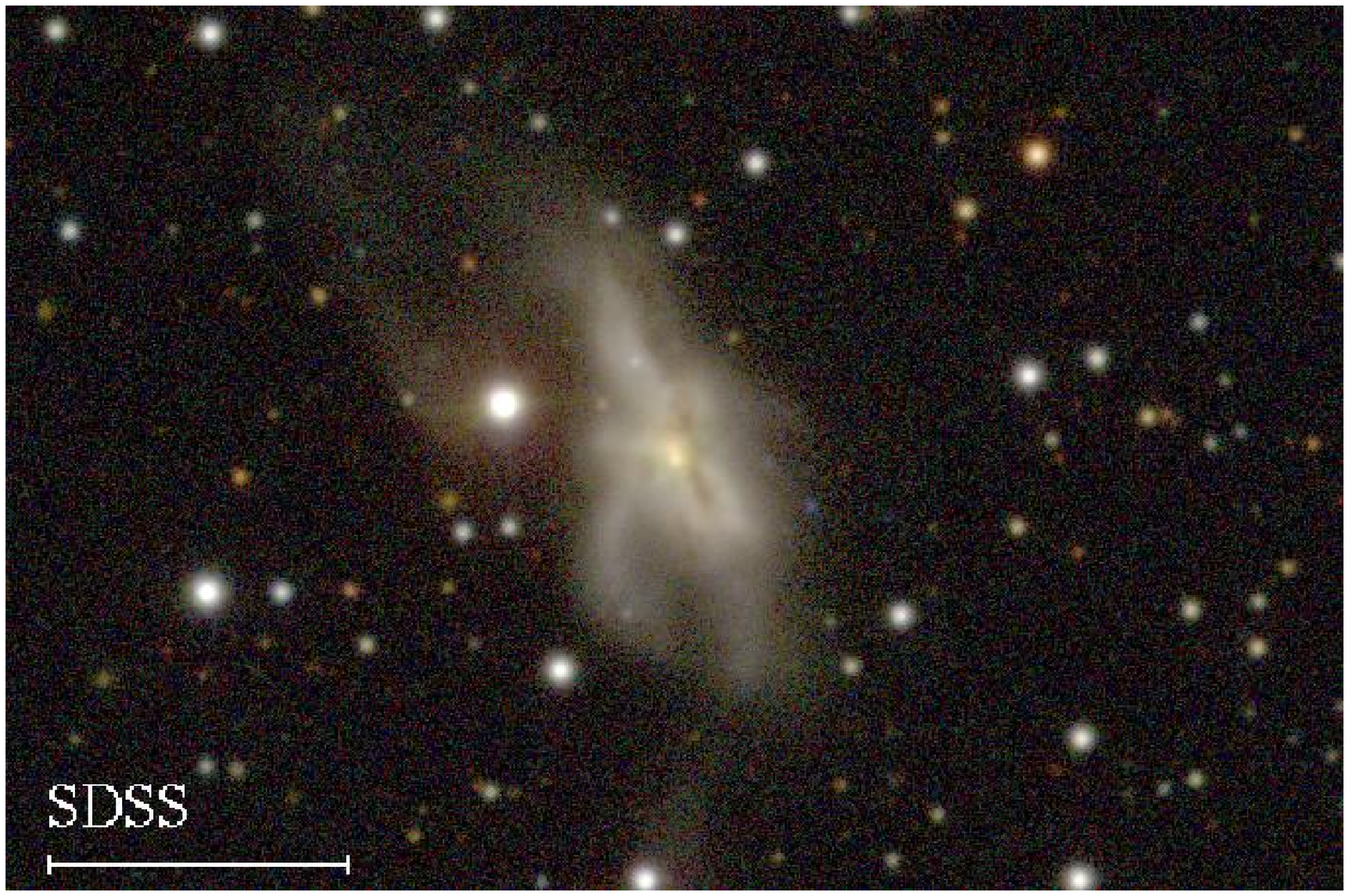}
\plottwo{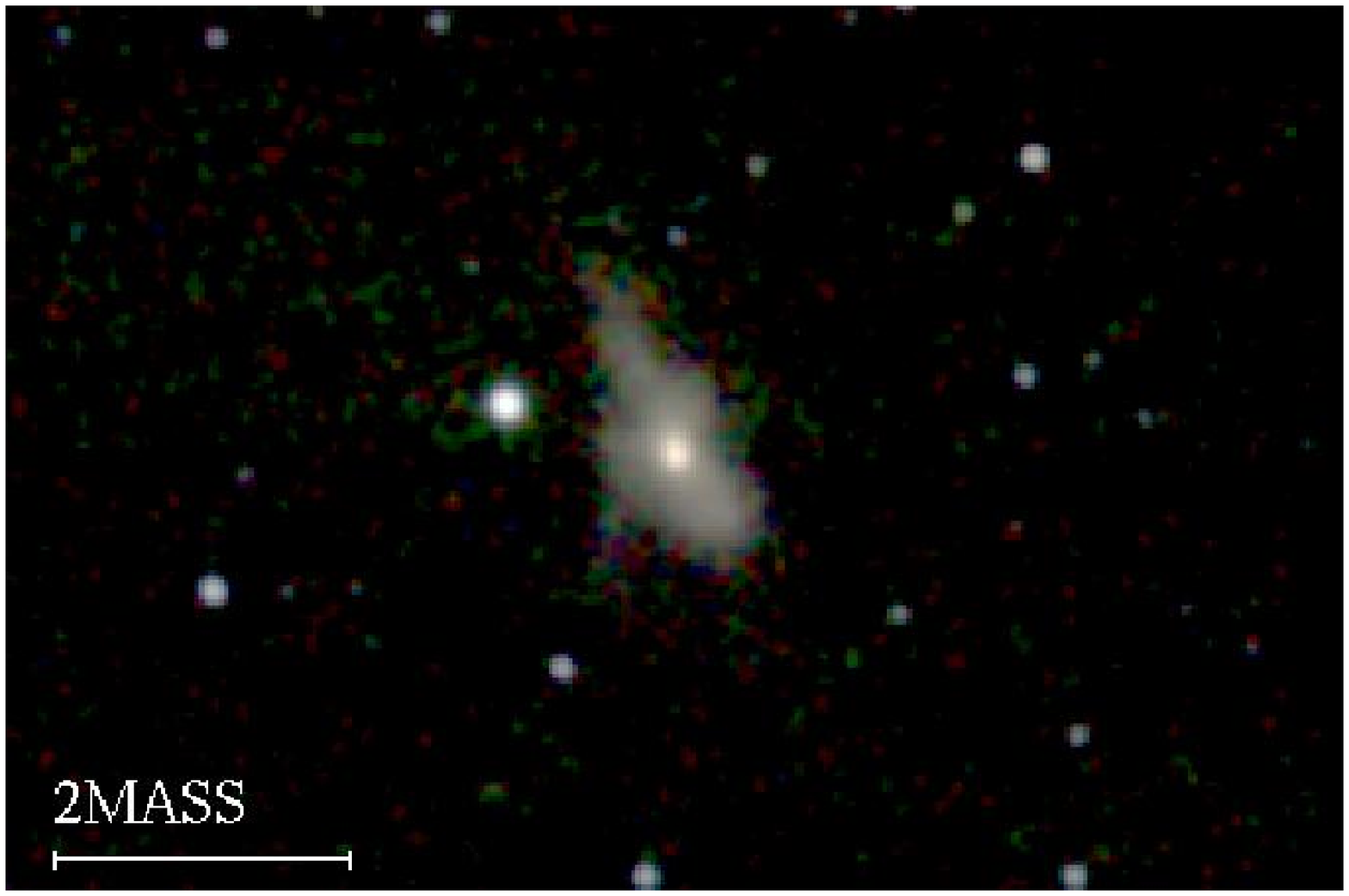}{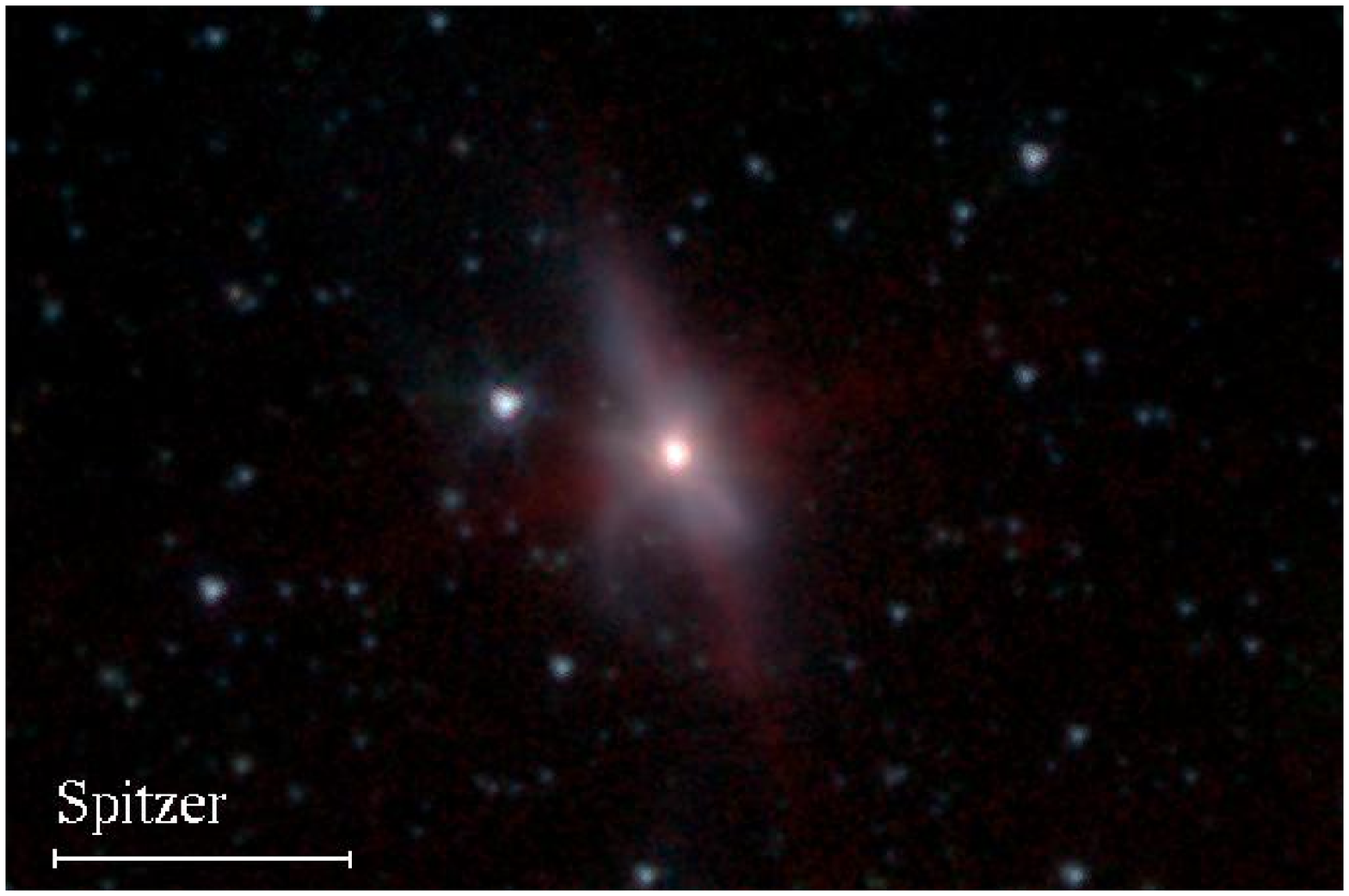}
\caption{The UV to mid-infrared SED of NGC~6240 (top panel), along with some of the GALEX, SDSS, 2MASS and {\it Spitzer} images that were used to constrain and verify the SED. The horizontal bar denotes an angular scale of $1^\prime$. In the top panel, the observed and model spectra are shown in black and grey respectively, while the photometry used to constrain and verify the spectra is shown with red dots.}
\label{fig:pretty}
\end{figure*}

To address this need, we present an atlas of 129 galaxy SEDs with wavelength coverage spanning from the ultraviolet to the mid-infrared. Our atlas spans a broad range of absolute magnitudes ($-14.4>M_g>-22.3$), colors ($0.1<u-g<1.9$) and galaxy types, including ellipticals, spirals, merging galaxies, blue compact dwarfs and luminous infrared galaxies. Our templates make use of optical drift-scan spectroscopy from ground-based telescopes and mid-infrared spectroscopy from {\it Spitzer} and {\it Akari}. The galaxy spectra were scaled and verified using matched-aperture photometry in 26 passbands. In Figure~\ref{fig:pretty} we present one of the SEDs along with color composite images of the GALEX, SDSS and {\it Spitzer} imaging used to constrain and verify the SED.

The structure of our paper is as follows. In \S\ref{sec:data} we describe the galaxy sample, spectra and photometry (all photometry presented in this paper uses the AB~magnitude system). In \S\ref{sec:sed} we describe how we compiled the SEDs, including filling gaps in spectral coverage with models and power-laws. We compare our spectra against previous template libraries and the observed galaxy colors in \S\ref{sec:comp}, and summarize our major conclusions in \S\ref{sec:summary}. In Appendix~\ref{sec:subsample} we introduce an illustrative subsample of 18 galaxies, and in Appendix~\ref{sec:presentation} we present SEDs and optical images for the entire sample. The SEDs, along with images used to constrain and verify the SEDs, are available via http://dx.doi.org/10.5072/03/529D3551F0117.

\section{Data}
\label{sec:data}

Our sample consists of $z<0.05$ galaxies with archival drift-scan optical spectroscopy, all of which have well calibrated multi-wavelength imaging spanning from the UV to the mid-IR. Almost all galaxies in our sample have spectra from {\it Spitzer} Space Telescope's Infrared Spectrograph \citep[IRS; ][]{hou04}, with the exceptions being elliptical galaxies with (approximately) Rayleigh-Jeans spectra in the mid-infrared. The bulk of the optical spectra are taken from \cite{mou06} and \cite{mou10}, who used the Bok 2.3-m and CITO 1.5-m telescopes, with some additional spectra of elliptical galaxies taken from \cite{ken92} and \cite{gav04}. {\it Akari} infrared camera \citep[IRC;][]{ona07} spectroscopy is used when it captures a significant fraction of the total galaxy flux and has high signal-to-noise. Accurate matched-aperture photometry from the UV to the mid-infrared was measured with archival imaging from the Galaxy Evolution Explorer \citep[GALEX;][]{mor07},   {\it Swift} UV/optical monitor telescope \citep[UVOT;][]{rom05}, Sloan Digital Sky Survey III \citep[SDSS III;][]{sdss3}, Two Micron All Sky Survey \citep[2MASS;][]{skr06}, {\it Spitzer} Space Telescope \citep{faz04,rie04} and  Wide-field Infrared Space Explorer \citep[WISE;][]{wri10}. 

Sample selection was set by the availability of suitable spectra and images, which fortunately results in a very diverse (but not necessarily representative) sample of 129 nearby galaxies. Our requirement for multi-wavelength images and spectroscopy results in our sample (largely) being a subset of other surveys of the nearby Universe. For example, our sample size is smaller than that of \citet{mou06}, who presented optical-drift scan spectroscopy of 417 galaxies, and our sample size is smaller than the 202 LIRGs and ULIRGs studied by the Great Observatories All-sky LIRG Survey \citep[GOALS; ][]{arm09}. Readers may find that the utility of our SED atlas is increased by combining it with detailed studies of nearby galaxies (many of which use data that we have now incorporated into this paper), including measurements of star formation rates \citep[e.g., ][]{mou06a,ken09}, fitting or modeling the mid-infrared spectra \citep[e.g., ][]{smi07,ina13,sti13} and estimates of the contribution of AGNs to mid-infrared emission \citep{pet11}. In Table~\ref{table:summary} we summarize the basic properties of our sample, including names, coordinates, morphologies and redshifts.

\subsection{Photometry}

Throughout this paper we use AB apparent magnitudes that are defined by 
\begin{equation}
m = -2.5 {\rm log} \left[ \left( \int R(\nu) \frac{f_\nu(\nu)}{h\nu} d\nu \right) \times \left( \int R(\nu) \frac{g_\nu(\nu)}{h\nu} d\nu \right)^{-1}  \right]
\label{eq:mag}
\end{equation}
\citep[e.g.,][]{hog02} where $f_\nu$ is the SED of the source, $g_\nu$ is the SED of a source that is 3631~Jy at all wavelengths (i.e., $m=0$), $R_\nu$ is the filter response function (defined as electrons per incident photon) and $h\nu$ is the energy of a photon with frequency $\nu$.  Please note that these magnitudes are based on photon counts rather than fluxes, and do not simply correspond to monochromatic flux densities at the effective wavelengths of the relevant filters. 

The 2MASS and MIPS filter curves (available from their respective archives) are defined as electrons per $f_\lambda$ (then renormalised to peak at $1.00$), so we modified these curves by dividing the filter curves by $\lambda$. The MIPS zero-points provided with the archival imaging are defined using a flux density at a specific wavelength and a spectrum where the flux density varies with wavelength (i.e., a Rayleigh-Jeans spectrum). For these data, we applied corrections to the zero-points so the resulting photometry is defined by Equation~\ref{eq:mag}. Photometry originally defined using the Vega magnitude system (e.g., 2MASS) has been shifted into the AB magnitude system\footnote{We assume the AB magnitudes of Vega are $J=0.89$, $H=1.37$ and $K_S=1.84$}. The SDSS calibration is not quite on the AB magnitude system, so we added corrections of $-0.04$ and $+0.02$ magnitudes to the $u$ and $z$ photometry respectively to shift it back to the AB system\footnote{http://www.sdss3.org/dr8/algorithms/fluxcal.php\#SDSStoAB}. 

Foreground dust extinction was modeled using the Planck dust extinction maps \citep{pla11,pla13}. We use the \citet{fit99} model of dust extinction as a function of wavelength, with a modified parameterization for $c_4$ \citep{pee2013b},
\begin{equation}
c_4=4.64-11.64 \times E(B-V).
\end{equation}
This accounts for the enhanced UV extinction observed at high Galactic latitudes \citep{pee2013} relative to the expectation of the \cite{fit99} extinction curve. In this paper we largely use photometry corrected for foreground dust extinction, but in Table~\ref{table:photometry} the photometry is presented without dust extinction corrections (as the corrections are model dependent).

Accurate matched-aperture photometry was essential for normalizing and verifying our spectra, and for constraining models used to fill gaps in our spectral coverage. Where possible, we made use of archival imaging that is readily available online. Ultraviolet imaging was taken from GALEX Release 6 (GR6) and the {\it Swift} UVOT \citep{rom05}, with the latter often requiring stacking of individual exposures using Swarp \citep{ber02}. While some galaxies have ultraviolet imaging from the XMM-Newton optical monitor \citep{mas01}, these images often suffer from scattered light and were not used in the analyses presented here. The only sample galaxy without GALEX or {\it Swift} imaging is UGCA~410 (a blue compact dwarf with weak PAH emission) where we used the International Ultraviolet Explorer fluxes of \citet{kin93}.

All galaxies in the sample have optical imaging from SDSS III \citep{sdss3}, near-infrared imaging from 2MASS \citep{skr06} and mid-infrared imaging from WISE. When galaxies overlapped the SDSS III or 2MASS image boundaries, we used mosaic images from the NASA-Sloan Atlas \citep{bla11}, the 2MASS Large Galaxy Atlas \citep[LGA;][]{jar03} or generated new Swarp mosaics \citep{ber02}. For the WISE bands, we used release version 4.1 atlas images from the WISE All Sky Data Release. When available, we utilized {\it Spitzer} Infrared Camera \citep[IRAC; ][]{faz04}, IRS and Multiband Imaging Photometry for SIRTF \citep[MIPS; ][]{rie04} mosaics available from the {\it Spitzer} Infrared Nearby Galaxies Survey \citep[SINGS;][]{ken03} or the {\it Spitzer} Heritage Archive. For a small number of galaxies we recombined individual exposures (to better remove cosmic rays) or fitted the background when it varied strongly as a function of position. In some instances where multiple images were available for the same galaxy in the same band (e.g., GALEX $NUV$ images from two surveys), we measured photometry using both images to check consistency.

We defined our photometric aperture to (where possible) approximate the rectangular apertures used by \cite{ken92}, \cite{gav04}, \cite{mou06} and \cite{mou10} for the optical drift scan spectrophotometry. In some instances the photometric aperture was reduced in size to mitigate contamination from neighboring sources or improve signal-to-noise at some wavelengths (particularly the ultraviolet and mid-infrared). The photometric apertures used for each of the galaxies are defined in Table~\ref{table:input}. Bright stars are manually flagged, masked and not included in the aperture photometry presented in this paper. As the masked regions are very small relative to the large apertures used to measure galaxy photometry, we did not attempt to model the galaxy light within the masked regions. 

The background of the images was estimated using the median of the pixel values that are outside of the aperture.  The background in the 2MASS and {\it Spitzer} images can vary as a function of position, so we visually inspected all of the images, and for those images with significant background variations we subtracted off a model background determined with IRAF's imsurfit task. Many galaxies in our sample are bright so photon counting noise is likely to be small relative to systematic errors. For the brightest galaxies in our sample, we assumed the uncertainties had a floor of 0.05~mag in the SDSS and 2MASS bands, and we assumed the uncertainties in the other bands had a floor of 0.10~mag. For the fainter galaxies in our sample (i.e., $m_{FUV}>18$, $m_{K_S}>13$), we estimated the uncertainties and corrected for background errors  by measuring fluxes at 24 locations (offset from the galaxy position). The distribution of the fluxes in these apertures was used to determine the uncertainties while the median of the fluxes was used to correct for background errors.  For a small number of star forming galaxies, the 2MASS photometry has very large uncertainties so we used the near-infrared photometry of \cite{eng08} and \cite{vad05} (when available) for constraining and verifying the spectra. 

\subsubsection{Zeropoint and PSF Corrections}

Large aperture photometry will overestimate fluxes if the zero-points have been determined with small apertures that don't capture all the light from relevant standard stars. The {\it Spitzer} IRAC imaging was calibrated using $24^{\prime\prime}$ diameter aperture photometry (without corrections for flux beyond the aperture) and IRAC large aperture photometry suffers from scattered light within the aperture, which can increase measured galaxy fluxes by 10\% or more \citep{faz04}. To mitigate these issues, we used the IRAC point spread function to determine the fraction of light beyond the aperture and we used the corrections described in the IRAC Instrument Handbook\footnote{http://irsa.ipac.caltech.edu/data/SPITZER/docs/irac/iracinstrumenthandbook/} to model scattered light as a function of aperture area. {\it Swift} UVOT photometry was calibrated using $10^{\prime\prime}$ diameter aperture photometry (without corrections for flux beyond the aperture) and we correct for this using approximations to the curves of growth presented in \cite{bre10}. {\it Spitzer} IRS Peak-up blue and red channel imaging is calibrated using 12 and 13 pixel radius apertures (respectively), and we made small corrections for this using the Peak-up Imager PSFs. The WISE all sky release zeropoints were determined by fitting PSF models to the inner region of the PSF, and consequently there are small zero-point errors for large aperture photometry. To correct for this, we added 0.03, 0.04, 0.03 and -0.03 to the measured WISE W1, W2, W3 and W4 magnitudes respectively \citep{jar12}.  

For galaxies with small angular sizes, we are effectively measuring total magnitudes and the smallest apertures approach the size of the GALEX, IRS, MIPS and WISE point spread functions. For the UV and mid-infrared photometry we applied an approximate correction for flux beyond the aperture using radially averaged models of the point spread function. For very small apertures the WISE W4 point spread function results in PSF corrections of 50\% or more, so for isolated galaxies measured with small spectroscopic extraction apertures (less than $30^{\prime\prime}$ on either side) we increased the W4 aperture size to at least $30^{\prime\prime}$ on a side. 

\subsubsection{Photon Coincidence Loss Corrections}

\begin{figure*}
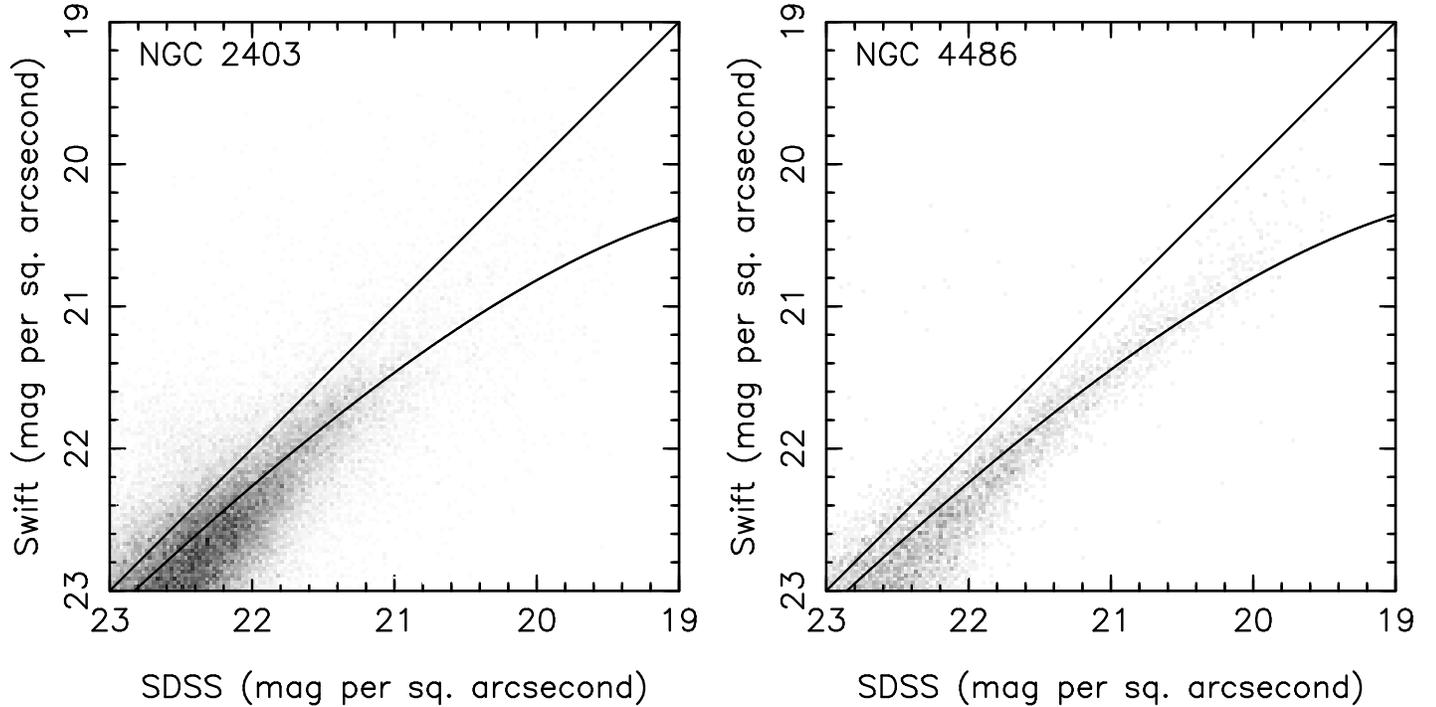

\resizebox{7.5in}{!}{\includegraphics{f2a.ps}\includegraphics{f2b.ps}}
\caption{Comparison of {\it Swift} $U$-band and SDSS $u$-band photometry for NGC~2403 (left) and NGC~4486 (right) prior to corrections for photon coincidence losses. Surface brightness estimates were determined with single pixels (for the same right ascension and declination) in both the {\it Swift} and SDSS images, and $\simeq 0.1$ magnitude corrections were applied to compensate for the difference between the {\it Swift} $U$-band and SDSS $u$-band filters. For extended sources, coincidence losses are significant, even at relatively low surface brightnesses. In both panels the $1:1$ line is shown along with the coincidence loss correction curve defined by  Equation~\ref{eq:ccor}.
\vspace*{0.5cm}
}
\label{fig:coin}
\end{figure*}

The {\it Swift} UVOT is a micro-channel plate intensified CCD and suffers from significant photon coincidence losses \citep{rom05}.  These coincidence losses are well defined for bright point sources but have not been previously determined for extended sources. To measure the photon coincidence losses for extended sources we compared {\it Swift} UVOT $U$-band and SDSS $u$-band surface brightness measurements for bright galaxies. Figure~\ref{fig:coin} compares {\it Swift} UVOT and SDSS measurements of the surface brightness for NGC~2403 and NGC~4486, and the impact of photon coincidence losses can be clearly seen. The losses approach $50\%$ for a surface brightness of $U\sim 20$ per square arcsecond,  corresponding to a photon count rate of 0.5 photons per second per square arcsecond. The photon coincidence losses occur at relatively faint magnitudes due to the large ($4^{\prime\prime} \times 4^{\prime\prime}$)  physical pixel scale of the UVOT detector and the use of 5 pixels when centroiding photon splashes. 

We applied an approximate correction to the photon counts given by
\begin{equation}
C_{cor} = \frac{ -{\rm ln} (1- 80 \times C_{raw}\times ft -2[80 \times C_{raw}\times ft]^2) } { 80\times ft(1-df)}
\label{eq:ccor}
\end{equation}
where $C_{raw}$ is the counts per second per square arcsecond, $ft$ is the frame time (0.011088~s) and df is the deadtime fraction (0.0155844). As this is an approximation, we measured {\it Swift} photometry using images with and without the photon coincidence loss corrections applied, and generally only include measurements when the correction to the photometry is less than 0.3 magnitudes. (To increase the number of blue compact dwarfs in our sample, we include {\it Swift} photometry for  Mrk~1450 with corrections up to 0.5 magnitudes.) A consequence of the coincidence loss corrections and the 0.3 magnitude criterion is much of the {\it Swift} optical photometry (including all the $B$-band) is not included in the final sample. 

\subsubsection{WISE W4 Filter Curve Correction}

\begin{figure}
\resizebox{3.25in}{!}{\includegraphics{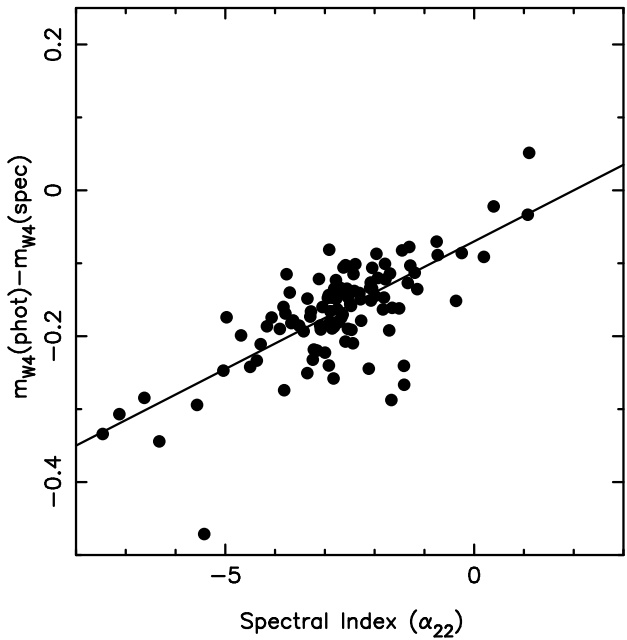}}
\caption{The difference between measured and synthesized W4 magnitudes for galaxies drawn from our sample, plotted as a function of $\sim 22~{\rm \mu m}$ spectral index. The pre-launch WISE W4 filter curves does not  match the on-sky performance \citep{wri10,jar13}, so the measured W4 magnitudes are systematically too bright for galaxies with spectra that differ significantly from the Rayleigh-Jeans approximation. 
\vspace*{0.5cm}
}
\label{fig:w4}
\end{figure}

The pre-launch WISE W4 ($22~{\rm \mu m}$) filter response curve (RSR) does not match the on-sky performance of the WISE W4 measurements \citep{wri10}. 
The WISE photometric measurements are calibrated using a network of A stars and K-M giants \citep{jar11}, with typical spectral indices ($\alpha$) between 1 and 2, where the spectral index is defined by $f_\nu \propto \nu^\alpha$.  Hence, by definition, the photometry is well behaved for objects with Rayleigh-Jeans SEDs, including early-type galaxies. Star-forming galaxies and AGNs with SEDs that rise towards longer wavelengths have WISE W4 magnitudes that are at least 10\% brighter than what is expected from {\it Spitzer} IRS spectrophotometry and {\it Spitzer} $24~{\rm \mu m}$ photometry \citep{wri10, jar13}. As we show in Figure~\ref{fig:w4}, we have been able to measure the anomaly with greater fidelity with our sample, and measure it as a function of $\sim 22~{\rm \mu m}$ spectral index. We find the residual is well approximated by 
\begin{equation}
\Delta m_{W4}  = 0.035 \times (\alpha_{22} - 2 )  
\end{equation}
with an RMS of 0.05 magnitudes. We present our original and corrected WISE W4 photometry in Table~\ref{table:photometry}, with the latter being denoted by ${\rm W4^{\prime}}$.

\subsubsection{The Observed Colors of Nearby Galaxies}
\label{sec:colors}

\begin{figure}
\resizebox{3.25in}{!}{\includegraphics{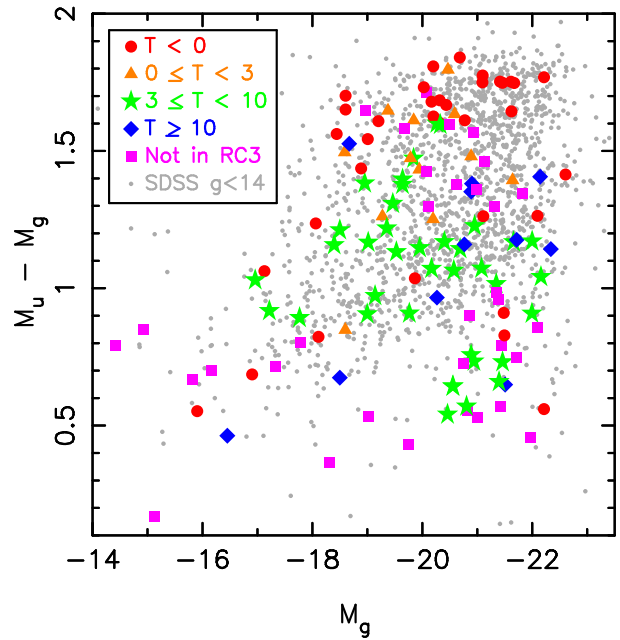}}
\caption{The optical color-magnitude diagram for galaxies in our sample along with $g<14$ galaxies from the NASA-Sloan Atlas. Symbols are a function of RC3 morphological T-Type \citep{rc3}, and sample galaxies that are not in the RC3 have almost exclusively irregular and peculiar morphologies. Our sample spans the range of observed colors for $z\sim 0$ galaxies and a broad range of absolute magnitudes.}
\label{fig:colmag}
\end{figure}

The photometry of the sample galaxies is presented in full in Table~\ref{table:photometry} and the optical color-magnitude diagram for the sample is plotted in Figure~\ref{fig:colmag}.  The symbols in Figure~\ref{fig:colmag} are a function of RC3 morphological T-Type \citep{rc3}, and sample galaxies that are not in the RC3 have almost exclusively irregular and peculiar morphologies (as shown in  Table~\ref{table:summary}). For comparison, in Figure~\ref{fig:colmag} we also plot the colors and magnitudes of $g<14$ galaxies drawn from the NASA-Sloan Atlas \citep{bla11} . Our sample spans the observed range of galaxy colors  ($0.1<u-g<1.9$), a broad range of absolute magnitude ($-14.4>M_g>-22.3$) and a broad range of galaxy types. While our sample spans a broad range of galaxy properties, some galaxy types are not included in the sample (e.g., ultra-compact dwarf galaxies) while other galaxy types (such as LIRGs) are over-represented. We caution that while galaxy morphological types are correlated with galaxy spectra, there are outliers including elliptical galaxies with star formation (e.g., NGC~855) and disk galaxies with low star formation rates (e.g, NGC~4450).

\begin{figure*}
\plottwo{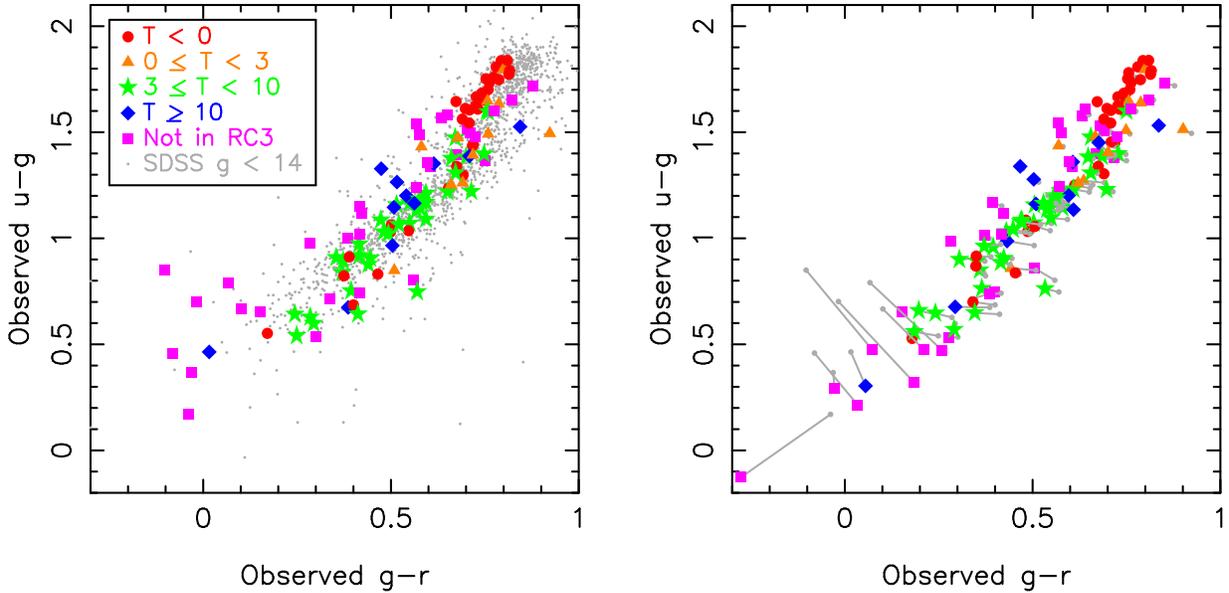}{f5b.ps}
\caption{The $u-g$ and $g-r$ colors of galaxies from our sample (colored symbols), along with $g<14$ galaxies from the NASA-Sloan Atlas (grey points). The left panel shows the observed colors of galaxies while the right panel has had the contribution of nebular emission lines subtracted (\S\ref{sec:sed}). Nebular emission lines can contribute significantly to the broadband photometry of blue compact dwarf galaxies, that fall at the bottom-left of the diagram. Once the contribution of nebular emission lines is subtracted (right panel), the galaxy locus becomes significantly narrower. The loci of our galaxies and the NASA-Sloan atlas galaxies are slightly offset, in part due to the use of aperture and model photometry. Our sample largely spans the range of observed colors for nearby galaxies. 
\vspace*{0.5cm}}
\label{fig:ugr}
\end{figure*}

In Figure~\ref{fig:ugr} we plot the optical colors of sample galaxies and NASA-Sloan Atlas galaxies. The majority of galaxies fall along a relatively narrow locus running diagonally across the diagram. The bluest galaxies have strong emission lines that contribute significantly to the observed $ugr$ photometry, and these galaxies fan across the lower left of the diagram. Many of the galaxies falling to the right of the galaxy locus are heavily obscured (e.g., NGC~660) while those falling to the left have strong Balmer breaks (e.g., NGC~5992).

\begin{figure}
\resizebox{3.25in}{!}{\includegraphics{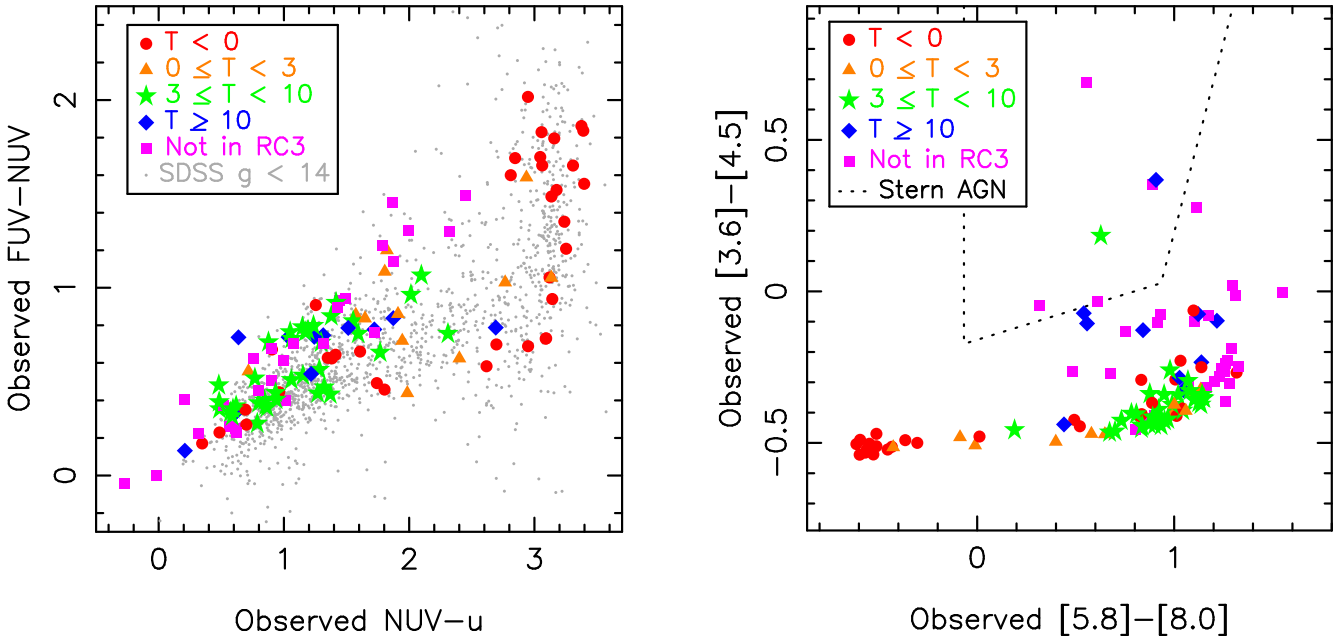}}
\caption{The ultraviolet colors of sample galaxies (colored symbols) and $g<14$ NASA-Sloan Atlas galaxies (grey dots). Early-type galaxies with $NUV-u\sim 3.1$ have a broad range of far ultraviolet colors. The distribution of galaxy colors tightens as one moves towards low metallicity star forming galaxies at the bottom-left of the plot.
\vspace*{0.5cm}}
\label{fig:uv}
\end{figure}

\begin{figure}
\resizebox{3.25in}{!}{\includegraphics{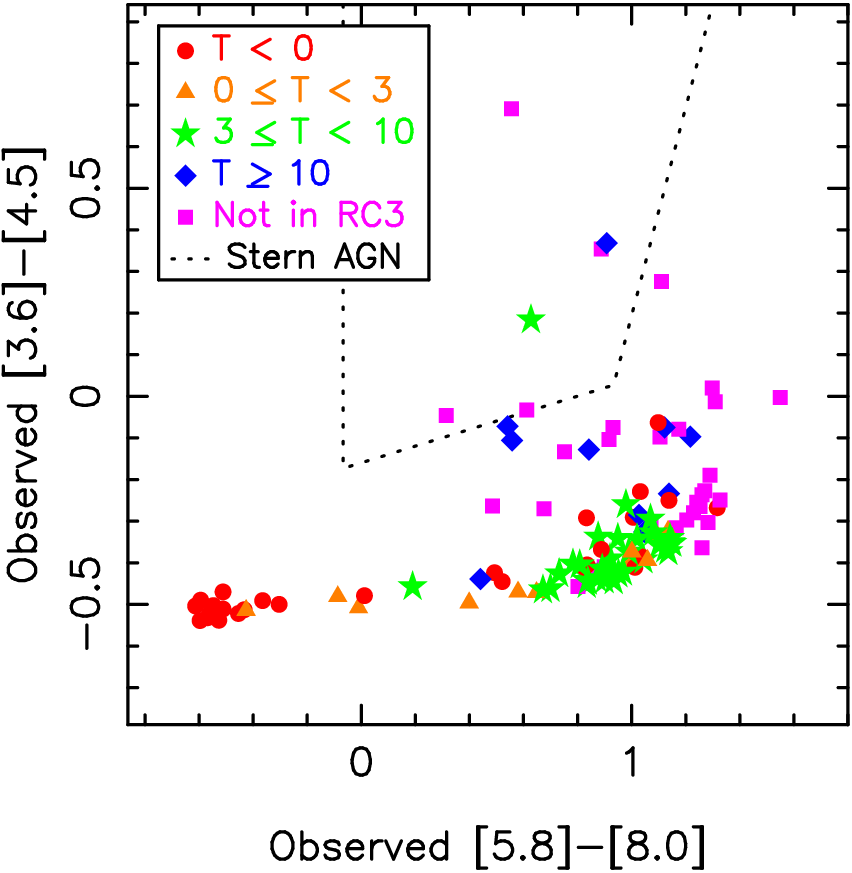}}
\caption{{\it Spitzer} IRAC colors for sample galaxies (not all sample galaxies have photometry in all four IRAC bands), along with the \citet{ste05} AGN selection criterion (dotted line). Galaxies with SEDs dominated by the Rayleigh-Jeans tail of stellar spectra fall at the bottom-left of the diagram, while galaxies with strong PAH emission fall to the right. Galaxies that are above the locus include AGNs (e.g., NGC~1275), galaxies with ice absorption at $\sim 3~{\rm \mu m}$ (e.g., NGC~2623, IC~4553) and galaxies with hot dust emission (e.g., UGCA~166).
\vspace*{0.5cm}}
\label{fig:irac}
\end{figure}

The diversity of the sample is illustrated by Figures~\ref{fig:uv} and ~\ref{fig:irac}, where we plot the ultraviolet and infrared colors of sample galaxies.  Elliptical galaxies have a broad range of ultraviolet continuum slopes, and this has been known for several decades \citep[e.g.,][]{cod79,don87,don07}. The distribution of ultraviolet colors tightens as one moves towards the left of  Figure~\ref{fig:uv}, with blue dwarf galaxies having a relatively narrow range of ultraviolet colors. 

In Figure~\ref{fig:irac} we plot the infrared colors of the sample galaxies. Galaxies with a Rayleigh-Jeans spectrum at $\sim 4~{\rm \mu m}$ fall along the bottom of the Figure, with galaxies with weak and strong PAH emission at $\sim 6~ {\rm \mu m}$ falling on the left and right of the plot respectively. Galaxies that lie above the locus in Figure~\ref{fig:irac} include active galactic nuclei (AGNs), galaxies with ice absorption at $\simeq 3.1~{\rm \mu m}$ and galaxies with hot dust emission. We caution that our large sample does not capture the full diversity of local galaxy populations, as some populations are not included (e.g., ultra compact dwarfs, low surface brightness galaxies, quasars) and it doesn't capture the full diversity of other populations \citep[e.g., LIRGs, ][]{how10}. However, our galaxy sample does capture much of the diversity of local galaxy populations (as discussed in \S\ref{sec:comp}) and illustrates why a relatively small number of galaxy templates can struggle to model the full diversity of local galaxies.

\subsection{Optical Spectra}

The bulk of the optical spectra are from \cite{mou06} and \cite{mou10}, who targeted star-forming galaxies and SINGS galaxies respectively, using both the Bok 2.3~m telescope on Kitt Peak and the CTIO 1.5~m. These spectra have previously been used by \citet{ken09} to calibrate star formation rate indicators and by \citet{mou10} to measure the gas-phase oxygen abundances of local galaxies. The spectral resolution is $R\sim 650$ and the wavelength coverage spans from 3650 to $6900~{\rm \AA}$. As the telluric lines near $6870~{\rm \AA}$ can produce artifacts in the spectra, we truncated the optical spectra at $6830~{\rm \AA}$ when producing the galaxy SEDs. Observations were obtained by scanning a long-slit perpendicularly back and forth across each galaxy to obtain a luminosity weighted integrated spectrum. In \cite{mou10} the region used for extracting the integrated spectrum was similar to the region used by \cite{smi07} to obtain 15--38~${\rm \mu m}$ spectra of SINGS galaxies. Relative spectrophotometric accuracy is estimated to be better than 5\% across the optical wavelength range.

Elliptical galaxies have a broad range of ultraviolet continuum slopes \citep[e.g.,][]{cod79,don87,don07}, and this is not well sampled by the galaxies from \cite{mou10} that also have SDSS III imaging. To span this region of color-space, we supplemented the sample with drift-scan spectrophotometry from \cite{ken92} and \cite{gav04}.  \cite{ken92} also used the Bok~2.3-m telescope while \cite{gav04} used the 1.93-m telescope of the Observatoire de Haute Provence (OHP), the ESO 3.6-m telescope, the Loiano 1.52-m telescope and the San Pedro Martir (SPM) 2.1-m telescope. We caution that the spectra from \citet{ken92} and \citet{gav04} have spectrophotometric accuracy on the order of $\sim 15\%$, which is poorer than the accuracy of the \citet{mou06} and \cite{mou10} spectra. 

\subsection{Spitzer Infrared Spectra}

Almost all of the galaxies in our sample have low resolution 5--38$~{\rm \mu m}$ spectra from the {\it Spitzer} Infrared Spectrograph (IRS). Many of the spectra were obtained for SINGS \citep{ken03,smi07} and Great Observatories All-Sky LIRG Survey \citep[GOALS; ][]{arm09,u12,ina13,sti13}, supplemented by other programs targeting specific galaxy types such as starbursts, low metallicity galaxies and ellipticals \citep[e.g.,][]{bra06,wu06}. For relatively compact galaxies, ``stare'' observations include a significant fraction of the total flux. When using stare observations we used version five of the Cornell Atlas of Spitzer IRS Sources \citep{leb11}. Tapered extractions were used for most galaxies, but for a small number of compact galaxies PSF weighted extractions were used as these had lower random errors than tapered extractions. 

Whenever they were available, we made use of Spitzer IRS spectral maps. For SINGS galaxies spectral maps are available from \cite{smi07} while for the remaining galaxies we used CUBISM \citep{smi07b} to produce new spectral cubes and maps using data downloaded from the Spitzer Heritage Archive. For both the SINGS and new spectral cubes, we extracted spectra using an aperture matched to the optical spectrum or (when this was not possible) we used the largest aperture possible. Extraction apertures and position angles are summarized in Table~\ref{table:input}.

We applied corrections to the IRS spectra so the spectra were continuous and in good agreement with the photometry\footnote{As NGC~1068 is saturated in the {\it Spitzer} images, we used the corrected {\it Spitzer} photometry of \citet{how07}.}. The IRS low resolution spectra come from four distinct orders; SL2 (5.25--7.6$~{\rm \mu m}$),  SL1 (7.5--14$~{\rm \mu m}$), LL2 (14.5--20.75$~{\rm \mu m}$) and LL1 (20--38.5$~{\rm \mu m}$). To merge the spectra together, we fitted straight lines to the continua near the overlap regions and then used these fits to determine scalings. To join the SL1 and LL2 spectra, we fitted from $13.65~{\rm \mu m}$ to $14.30~{\rm \mu m}$ and from $14.0~{\rm \mu m}$ to $15.25~{\rm \mu m}$ respectively. To join the LL2 and LL1 spectra we fitted from $19.70~{\rm \mu m}$ to $20.50~{\rm \mu m}$ and from $20.6~{\rm \mu m}$ to $22.0~{\rm \mu m}$ respective. As poor fits can happen, particularly when dealing with very noisy data, fits were visually inspected and in a small minority of cases manual scalings were applied when merging IRS spectra. 

The IRS spectra were scaled to agree with the {\it Spitzer} $24~{\rm \mu m}$ and WISE ${\rm W4^\prime}$ photometry. Once this was done, some IRS spectra were significantly offset from the {\it Spitzer} $8~{\rm \mu m}$ and WISE $W3$ photometry, in part because the SL1 and SL2 extraction apertures were often smaller than the LL1 and LL2 extraction apertures. We corrected the spectra for this residual, by assuming the residual magnitude difference between the spectra and the photometry varied linearly as a function of wavelength (with the slope being chosen to best match the photometry).  This correction is overly simple, but has relatively little impact on the resulting spectra as the (additional) corrections at $8~{\rm \mu m}$ are on the order of $30\%$ for IRS stare observations and often less than $10\%$ for IRS spectral maps. Galaxies with stare mode spectra were excluded from the sample if the IRS spectrum captured less than $\simeq 25\%$ of the $8~{\rm \mu m}$ flux (some exceptions being made to increase the diversity of the sample). A list of the multiplicative scaling factors used for the {\it Spitzer} IRS spectra as a function of wavelength is provided in Table~\ref{table:specscale}. We caution that for certain science applications, some of our IRS spectra should not be used if the multiplicative scalings are large or vary significantly with wavelength.

\subsection{Akari Infrared Spectra}

Approximately half of the galaxies in the sample have 2--5~${\rm \mu m}$ spectra from the near-infrared channel {\it Akari's} Infrared Camera \citep[IRC;][]{ohy07,ona07}. These spectra are particularly useful for AGNs and LIRGs, where the $\sim 4~\rm {\mu m}$ continuum can include significant non-stellar emission (e.g., NGC~1275, NGC~3690) and the $3.1~{\rm \mu m}$ ice feature is sufficiently deep to impact broad-band photometry (e.g., NGC~2623, IC~4553). We exclusively used {\it Akari} grism spectra, and many of the spectra are drawn from \cite{ima10}\footnote{The galaxies where we used the \citet{ima10} {\it Akari} spectra are
CGCG~049-057,
CGCG~436-030,
CGCG~453-062,
IC~860,
IC~883 (UGC~8387),
IC~4553 (Arp~220),
IC~5298,
Mrk~331,
Mrk~1490,
NGC~1614,
NGC~2388,
NGC~5104,
NGC~6090,
NGC~7674,
UGC~8335 SE (Mrk~273),
UGC~9618 (VV~340),
UGC~9618~N, 
and UGC~12150} 
 while the remainder were downloaded from JAXA's DAta Archives and Transmission System (DARTS) and reduced with the IRC Spectroscopic Toolkit\footnote{http://www.ir.isas.jaxa.jp/ASTRO-F/Observation/DataReduction/IRC/} \citep{ohy07}.  IRC frames were dark-subtracted, linearity-corrected, flat-field corrected, corrected for positional drift and combined using the IRC Spectroscopic Toolkit as described in \citet{ohy07}. Target galaxies were usually detected automatically within the slit by the toolkit, but in some instances target positions were identified manually. Whenever possible, we used an extraction aperture 20~pixels wide, corresponding to $29.2^{\prime\prime}$, but for some galaxies with low signal-to-noise spectra we had to reduce the width of the extraction aperture. For NGC~7714 the initial {\it Akari} reduced spectrum had a systematic error that we removed by remeasuring the background on either side of the extraction aperture.  Extraction apertures, slit sizes and slit position angles are detailed in Table~\ref{table:input}. For approximately 40\% of the galaxies that had low signal-to-noise grim spectra, we applied a 3-pixel boxcar smoothing to the data, corresponding to $0.03~{\rm \mu m}$.

\section{Assembling SEDs Spanning from the UV to the mid-IR}
\label{sec:sed}

Galaxy SEDs were produced by normalizing the observed spectra with our photometry and filling the gaps with spectral coverage with power-law fits or (more frequently) model spectra produced using the Multi-wavelength Analysis of Galaxy Physical Properties code \citep[MAGPHYS; ][]{dac08}. MAGPHYS includes stellar population synthesis models \citep{bc03} combined with a self-consistent model of dust emission, dust absorption and PAH emission. 

\begin{figure}[hbt]
\resizebox{3.25in}{!}{\includegraphics{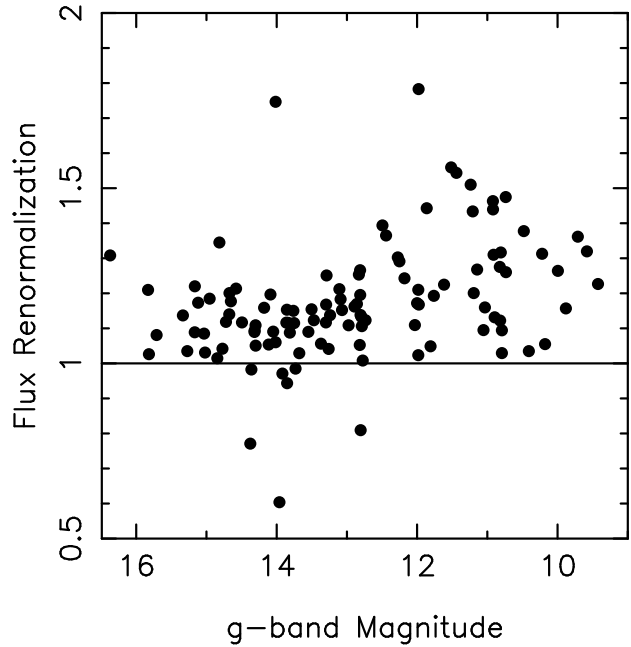}}
\caption{The multiplicative factors used to rescale the optical spectra to match the $g$-band photometry. The optical spectra from \citet{gav04} are not absolute flux calibrated so they are not included in this plot. There is a systematic offset of approximately 10\%, with larger offsets seen for brighter and more extended galaxies. For large galaxies, the background may be over-subtracted due to the background estimate including some galaxy light. Star formation rate indicators calibrated with these spectra (without renormalization) will underestimate the true star formation rate by 10\% or more.
\vspace*{0.5cm}}
\label{fig:rescale}
\end{figure}

The multiplicative factors used to renormalize the optical spectra are plotted in Figure~\ref{fig:rescale} and listed in Table~\ref{table:specscale}. Assuming our SDSS g-band photometry is valid, the measured optical spectra typically underestimate the fluxes of the true optical spectra by approximately 10\%, with larger offsets seen for brighter (and more extended) galaxies. As these underestimates also apply to the nebular emission lines, star formation rate indicators calibrated with the spectra \citep[e.g.,][]{ken09} may systematically underestimate the true star formation rates. A possible cause for this systematic is sky background subtraction using a slit of finite length ($\sim 3.^{\prime}$3 for the Bok 2.3~m telescope), which may be contaminated by galaxy light for the very largest galaxies in the sample (e.g., NGC~4569). A recalibration of star formation rate indicators (e.g., mid-infrared luminosities between 10 and 24 microns) using our SEDs is beyond the scope of this paper and will be the subject of a future work.

MAGPHYS does not include nebular emission lines, so we produced synthetic optical $ugr$ photometry with the nebular emission lines subtracted, using the emission line fluxes from \cite{mou06} and \cite{mou10}. In cases where ${\rm [Ne~III]~3869}$~\AA~ was not measured, we used the relation of \cite{per07} to approximate the ${\rm [Ne~III]~3869}$~\AA~  flux using the measured ${\rm [O~III]~5007}$~\AA~ flux. We show the resulting optical $ugr$ colour-colour diagram in the right panel of Figure~\ref{fig:ugr}. It is immediately clear from this diagram that the galaxy locus is approximately a straight line once the contribution of emission lines has been removed. Very blue star forming galaxies that initially have comparable optical colors can move to different regions of the color-color diagram, in part due to differing contributions from ${\rm [O~III]~5007}$~\AA~ and ${\rm H\alpha}$. Consequently, it can be very difficult to determine some properties of emission line galaxies (e.g., stellar masses) when only broadband photometry is available \citep[e.g.,][]{ate11,sch09}. 

\begin{figure*}
\plottwo{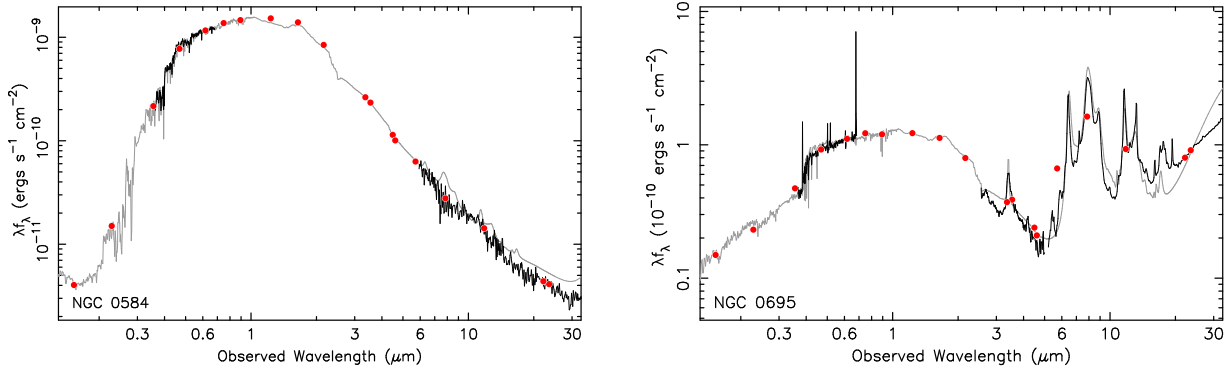}{f9b.ps}
\caption{NGC~584 and NGC~694 observed spectra (black), best-fit MAGPHYS models (grey) and photometry (red dots). While the MAGPHYS models do approximate the observed spectra of these galaxies, they do show significant differences that must be mitigated to produce smooth continuous spectra. The MAGPHYS models can disagree with the observed mid-infrared spectra of galaxies, as our photometry does not accurately constrain the dust temperatures of our galaxies. 
\vspace*{0.5cm}}
\label{fig:magphys}
\end{figure*}

As we illustrate in Figure~\ref{fig:magphys}, the MAGPHYS  and observed spectra can approximate each other remarkably well. However, the MAGPHYS and observed spectra do differ from each other, and simply switching from one to the other can produce false spectral breaks. To produce smooth and continuous spectra, we rescaled the MAGPHYS SEDs near the optical, {\it Akari} and {\it Spitzer} spectra so the models joined the observed spectra smoothly. For example, for each galaxy near $6800~{\rm \AA}$ we determined a scaling factor by comparing the MAGPHYS and optical spectrum fluxes over a wavelength range corresponding to the last $500~{\rm \AA}$ of the optical spectrum. We then multiplied the MAGPHYS fluxes by this scaling factor at $\sim 6800~{\rm \AA}$, and used progressively smaller scaling factors until the MAGPHYS fluxes are unchanged at $10,700~{\rm \AA}$. For galaxies with both {\it Akari} and {\it Spitzer} spectra, we used linear interpolation to fill the small gap between the {\it Akari} and {\it Spitzer} spectra. For  UGCA~410, MAGPHYS did not approximate the spectra near $6~{\rm \mu m}$ so we interpolated between $4.5~{\rm \mu m}$ and the {\it Spitzer} spectra using a double-power law. 

For a small number of early-type galaxies we do not have {\it Spitzer} IRS spectra and we use power-laws to model the spectra of these galaxies at long wavelengths. The $\sim 20~{\rm \mu m}$ spectral index is left as a free parameter, as some galaxies do not precisely match a Rayleigh-Jeans spectrum. There are also some galaxies (e.g., AGNs) where the photometry is better matched (over a limited wavelength range) by a power-law than a MAGPHYS model (e.g., UGC 5101). For these galaxies we replaced the MAGPHYS model with a power-law over the relevant wavelength range. Some galaxies have significant AGN emission in the mid-IR which cannot be approximated by MAGPHYS (NGC~1068, NGC~1275, NGC~6240), so the mid-infrared photometry was excluded when fitting MAGPHYS models to the photometry of these galaxies. Some galaxies with significant AGN emission in the near-IR cannot be approximated by  MAGPHYS models or power-laws, and these had to be rejected from the sample (e.g., Mrk~231).

MAGPHYS struggled to fit the ultraviolet photometry of some LIRGs and AGN host galaxies, which is not unexpected given MAGPHYS does not include AGN emission and does not model very complex dust geometries. We applied two corrections to the spectra of  CGCG~436-030, IC~860, IC~883, IC~5298, II~Zw~96, III~Zw~035, IRAS~08572+3915, IRAS~17208-0014, Mrk~331, NGC~2623, NGC~3690, NGC~5256, NGC~6240, NGC~7674, UGC~5101, UGC~8335~S, UGC~8696 and UGC~9168~N so they were forced to agree with the GALEX $FUV$ and $NUV$ photometry. The two corrections vary linearly with wavelength, with one correction forced to equal zero beyond 2315~\AA~and the other forced to equal zero beyond 3561~\AA. The corrections can be as large as $\sim 60\%$ near the GALEX $FUV$ band, but smaller values are more typical. While the approximations are crude, the resulting spectra should be better than the prior literature (as we discuss in \S\ref{sec:comp}) and will produce improved broad-band colors of LIRGs and AGN host galaxies.

We used the resulting SEDs to determine synthetic photometry, and we compared this with the observed photometry used to normalize and verify the spectra. Any residuals larger than $0.2~{\rm mag}$ are flagged and visually inspected. In some cases outliers were removed by improved masking of foreground stars or improved background subtraction. For some galaxies the signal-to-noise of the optical and {\it Spitzer} spectra was low and these were rejected from the sample. For a handful of galaxies discrepancies were evident in multiple bands and the source of the error could not be identified nor remedied, so the relevant galaxy was rejected from the sample. 





The residuals of the observed photometry and photometry synthesized from the spectra are summarized in Table~\ref{table:res}. As some bands were used to normalize the spectra (i.e, $g$-band), we caution that some bands have small residuals by construction. The median and standard deviation of the offsets between the photometry and spectra are less than 10\% for almost all of the bands. 

\begin{figure}[hbt]
\resizebox{3.25in}{!}{\includegraphics{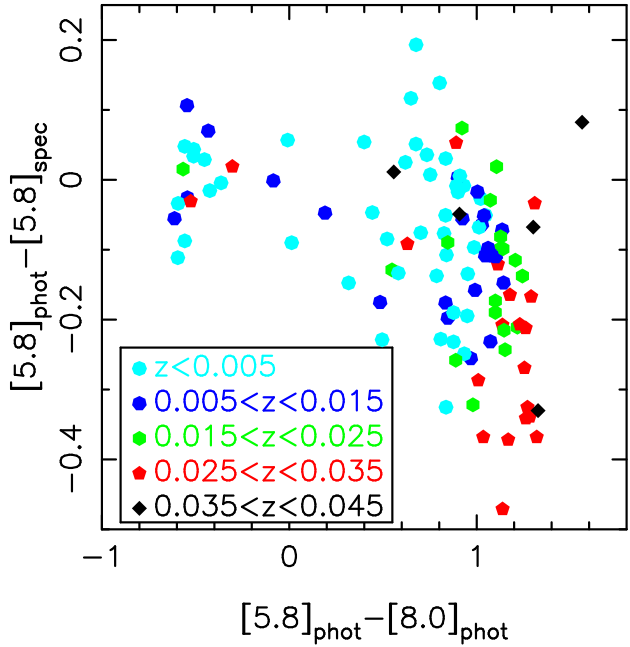}}
\caption{The offsets between the observed and synthesized $5.8~{\rm \mu m}$ photometry of galaxies. The $5.8~{\rm \mu m}$ photometry of star forming galaxies is systematically brighter than the spectroscopy, but we have not (unambiguously) identified the cause of this offset.}
\label{fig:irac3}
\end{figure}

After applying the correction for the W4 filter curve error, the most significant residual is found in $5.8~{\rm \mu m}$ IRAC Channel-3 photometry. In Figure~\ref{fig:irac3} we plot the residuals as a function of galaxy color, with different symbols denoting different redshift ranges. While galaxies with Rayleigh-Jeans spectra show relatively modest errors, the photometry of star forming galaxies is systematically brighter than the spectroscopy. As the presence of the residual depends on SED shape and is not seen in the other IRAC bands, it unlikely to result from errors in the scattered light corrections we have previously applied to the photometry. The $6.2~{\rm \mu m}$ PAH feature is near the red end of the $5.8~{\rm \mu m}$ filter curve, and should start moving out of the band towards the upper end of the redshift range. The systematic offset is removed if we shift the $5.8~{\rm \mu m}$ filter curve redward by 1.5\%, although doing this may not be justified.  The blue end of the IRS spectroscopy also overlaps the $5.8~{\rm \mu m}$ filter curve, and the spectra can have poor signal-to-noise shortward of the $6.2~{\rm \mu m}$ PAH feature. One could apply corrections to the IRS spectroscopy of star forming galaxies, but doing so produces spectra that poorly match the AKARI spectra at shorter wavelengths. We do not apply any corrections to the photometry or spectra $5.8~{\rm \mu m}$, but readers should be cautious of the spectra and photometry near this wavelength.

\begin{figure*}[hbt]
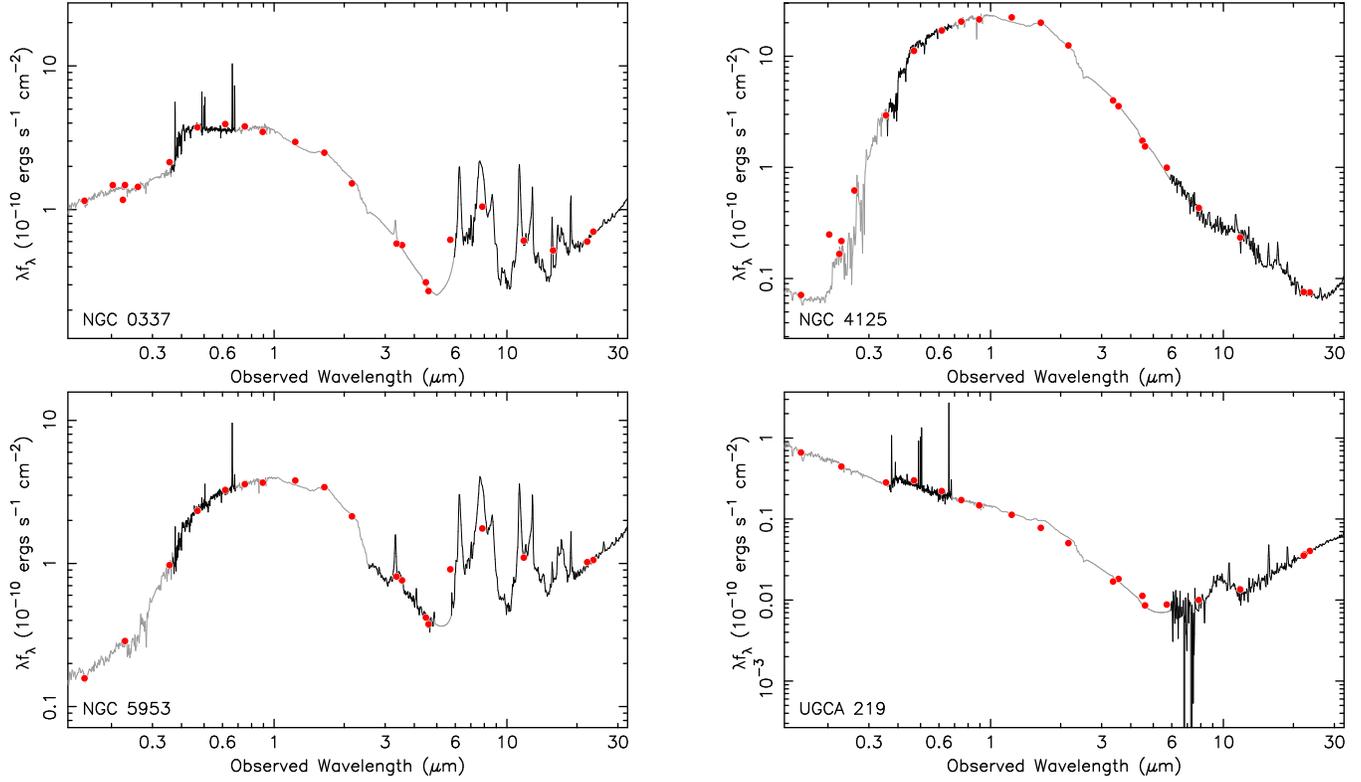

\resizebox{3.25in}{!}{\includegraphics{f11a.ps}}\resizebox{0.5in}{!}{~}\resizebox{3.25in}{!}{\includegraphics{f11b.ps}}
\resizebox{3.25in}{!}{\includegraphics{f11c.ps}}\resizebox{0.5in}{!}{~}\resizebox{3.25in}{!}{\includegraphics{f11d.ps}}
\caption{The spectra of NGC~337, NGC~4125, NGC~5953 and UGCA~219. Observed spectra are shown in black, MAGPHYS models are shown in grey, and photometry used to constrain and verify the spectra are shown with red dots. These galaxies are drawn from the illustrative subset of the sample described in Appendix~\ref{sec:subsample}}
\label{fig:example}
\end{figure*}

In Figure~\ref{fig:example} we provide plots for four of the galaxies drawn from the illustrative subsample described in Appendix~\ref{sec:subsample}. Spectra and images of all the galaxies in the sample are provided in Appendix~\ref{sec:presentation}. Figure~\ref{fig:example} shows good agreement between the photometry and spectroscopy, although we caution that some of this agreement is by construction (\S~\ref{sec:comp} compares the spectra with an independent set of photometry). Figure~\ref{fig:example} also illustrates the diversity of the sample, including a passive early-type galaxy (NGC~4125), a merging galaxy (NGC~5953) and a low metallicity dwarf galaxy (UGCA~219).

\section{Comparison with the Prior Literature}
\label{sec:comp}

Below we directly compare our SEDs (of particular galaxies) with those from the prior literature, and compare how well different template libraries reproduce the observed colors of galaxies. Systematic differences between our SEDs and those of previous template libraries are inevitable due to differences in methodology and data, with much of the data presented here not being available a decade ago. A key difference is the size of the aperture used to measure photometry and extract spectra, with some of the prior literature being restricted to galaxy nuclei \citep[e.g.,][]{kin96}, which may not be representative of entire galaxies. We have deliberately produced spectra for individual galaxies, whereas the prior literature sometimes merged spectra from different galaxies to produce spectra spanning broad wavelength ranges. That said, these limitations and sources of error were well understood and discussed by the relevant authors, who constructed the best SEDs possible with the data then available. 

\begin{figure}
\resizebox{3.25in}{!}{\includegraphics{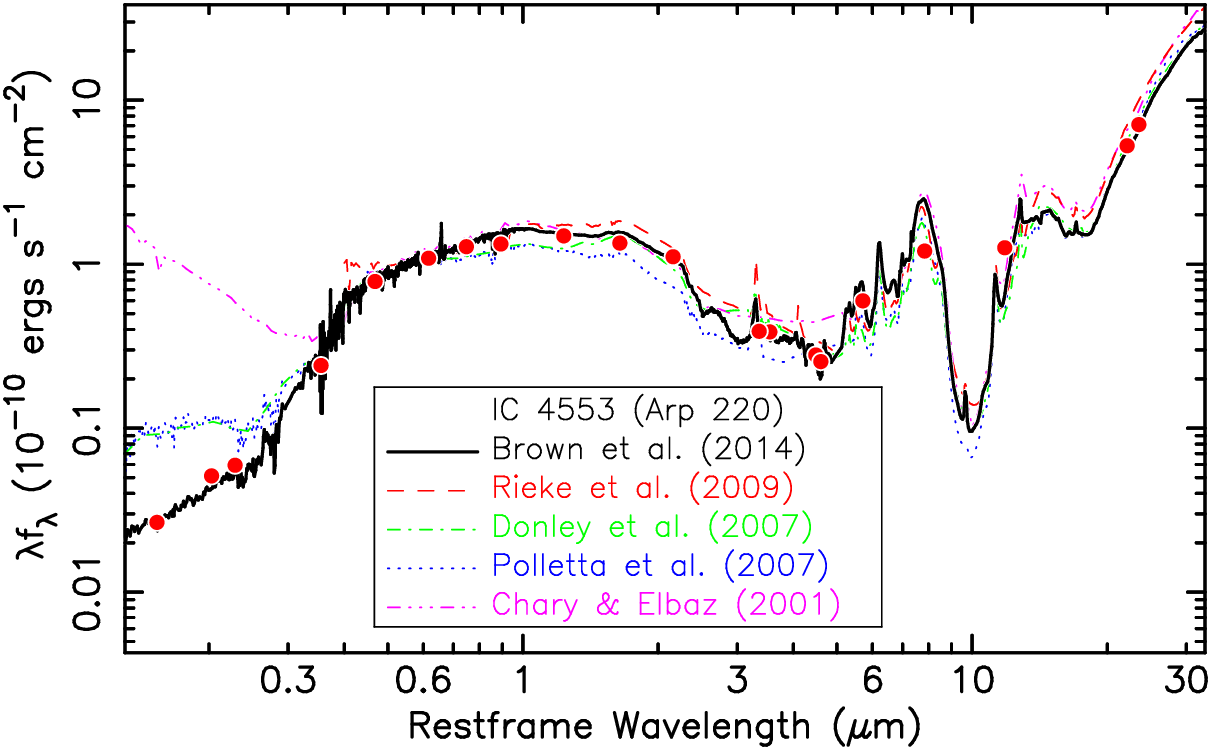}}
\caption{Our spectrum of IC 4553 (Arp~220) shown with IC 4553 template spectra taken from \citet{cha01}, \citet{donley07}, \citet{pol07} and \citet{rie09}. To aid comparisons, the spectra have been normalized at $6000$~\AA~ and our photometry is plotted with red dots. As noted by \citet{cha01}, some LIRG and ULIRG SEDs from the literature were not constrained in the ultraviolet and consequently we see large differences between the templates at $2000$~\AA. The templates show significant differences across the entire wavelength range, and these differences could impact interpretations of galaxy properties derived from these templates.}
\label{fig:arp220}
\end{figure}

The biggest differences between our templates and the prior literature should be for LIRGs and ULIRGs, as these are difficult to model with stellar population synthesis models and are relatively faint in the ultraviolet. In Figure~\ref{fig:arp220} we plot our spectrum of IC~4553 (Arp 220), along with other IC~4553 templates from the recent literature \citep{cha01,donley07,pol07,rie09}. While IC~4553 templates are frequently used to model the spectra of ULIRGs, the properties of this galaxy are unusual \citep[e.g.,][]{soi84} and ULIRG properties do display considerable diversity \citep[e.g.,][]{bra06,how10,u12}. 

It is immediately obvious from Figure~\ref{fig:arp220} that the ultraviolet SEDs vary enormously, with the prior literature overestimating the $\sim 2000$~\AA~ flux of IC~4553 by over 100\%. As with this work, the IC~4553 templates from the prior literature use combinations of spectra and model fits to photometry, using the data available at the time of publication. While it was reasonable for the prior literature to extend their IC~4553 SEDs into the ultraviolet (e.g., to approximate the observed optical colors of high redshift galaxies), previous IC~4553 SEDs had few (if any) constraints in ultraviolet \citep[as noted by ][]{cha01}, and consequently they systematically overestimated the ultraviolet flux of IC~4553. However, we caution that even with GALEX and {\it Swift}, we are constraining the ultraviolet SED of IC~4553 with just three photometric data points. 

\begin{figure}
\resizebox{3.0in}{!}{\includegraphics{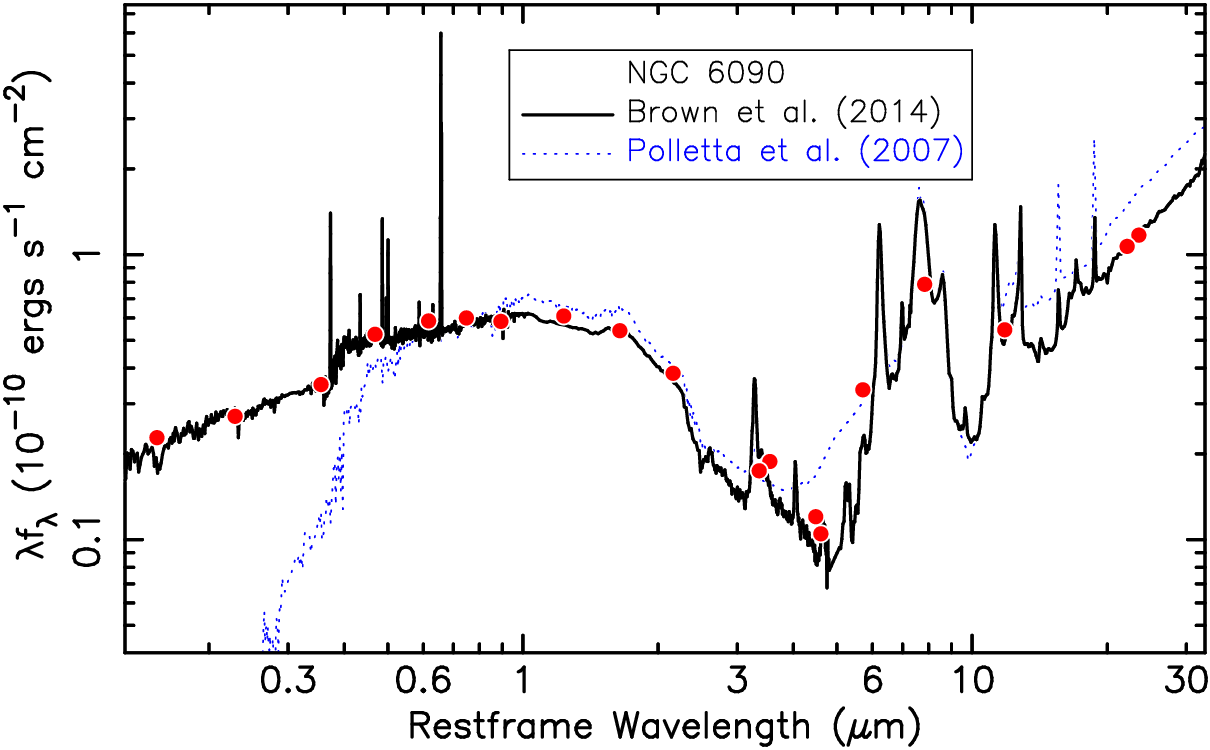}}
\resizebox{3.0in}{!}{\includegraphics{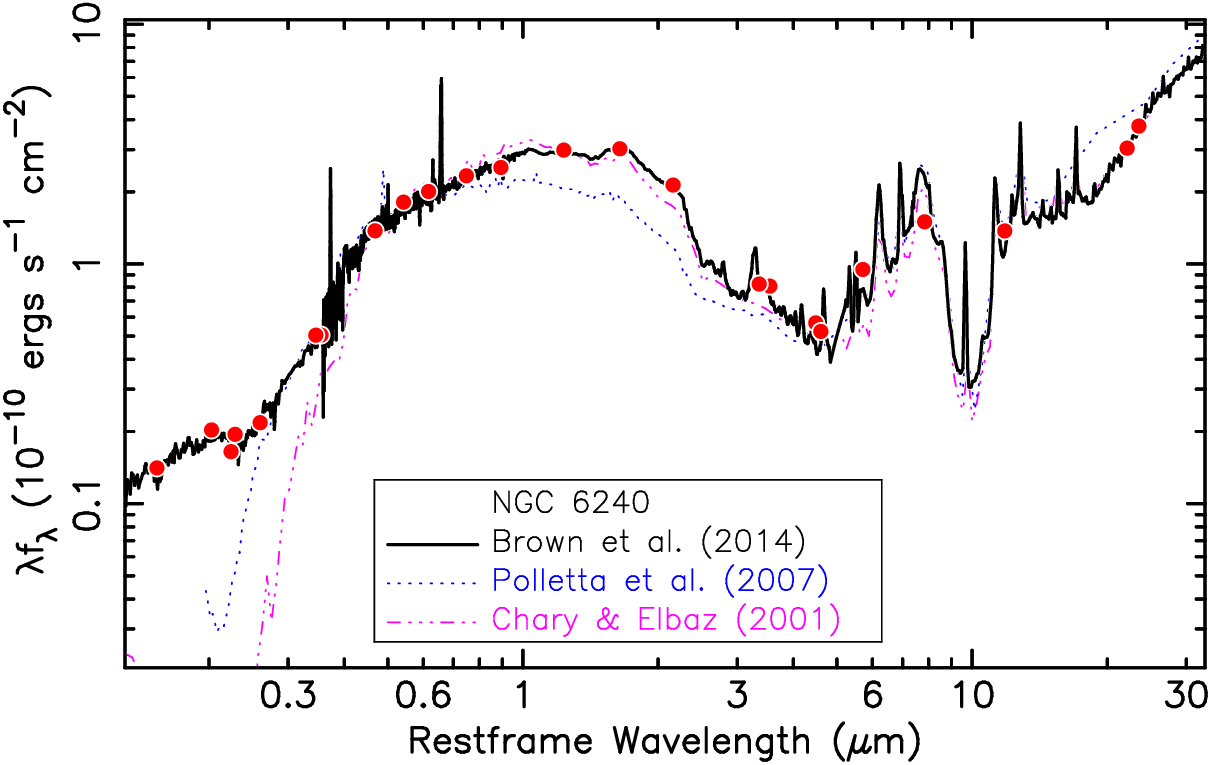}}
\caption{Our templates of NGC~6090 and NGC~6240 shown with the \citet{pol07} templates for the same galaxies. The \citet{pol07} templates for NGC~6090 and NGC~6240 use SEDs and photometry at wavelengths below $2~{\rm \mu m}$ that predate the advent of GALEX and SDSS imaging, and consequently we see significant errors at these wavelengths. The availability of well-calibrated imaging and matched-aperture photometry across a broad wavelength range (including redundant data) greatly simplifies the construction and validation of galaxy SEDs.}
\label{fig:polcomp}
\end{figure}

As can be seen in Figure~\ref{fig:arp220}, the shapes of the various IC~4553 (Arp 220) templates vary in the optical and mid-infrared too, with our spectrum featuring a relatively red $u-r$ color, strong PAH emission and $3.1~{\rm \mu m}$ ice absorption feature. As all of the recent IC~4553 templates make use of (or have been updated with) {\it Spitzer} IRS spectra, the broad agreement beyond $5~{\rm \mu m}$ is not surprising. As IC~4553 template spectra are often used to interpret the observations of high redshift galaxies, it is plausible that errors in template spectra have led to erroneous conclusions. As we illustrate in Figure~\ref{fig:polcomp}, the issues seen in IC~4553 templates are also seen in other LIRG SEDs. However, as we discuss below, there is better agreement between different templates libraries for galaxy types with little dust emission and no AGN content (e.g., ellipticals). 

To compare our ensemble of templates with the prior literature, we generated synthetic colors and compared them to observed galaxy colors. We compare the performance of our templates with those from \citet{col80}, \citet{kin96} and \citet{pol07}, which have been used by a large variety of K-correction and photometric redshift codes over the past two decades \citep[e.g.,][]{fer99,ben00,bol00,ilb09}. For this comparison we used optical and mid-infrared colors of $z\sim 0.3$ galaxies in the Bo\"otes field \citep{jan99,ash09}. By changing redshift we avoid comparing our SEDs with photometry that has the same rest-frame wavelengths as the photometry used to constrain and verify the SEDs. The Bo\"otes field optical imaging, taken from the NOAO Deep Wide-Field Survey, also uses a different set of optical filters than the SDSS. The spectroscopic redshifts used for this comparison are taken from $I<20.4$ AGN and Galaxies Evolution Survey \citep[AGES; ][]{koc12}. The Bo\"otes photometric catalogues are described in \citet{bro08} and we use a magnitude dependent aperture size to capture the vast majority of the galaxy light. 

\begin{figure*}
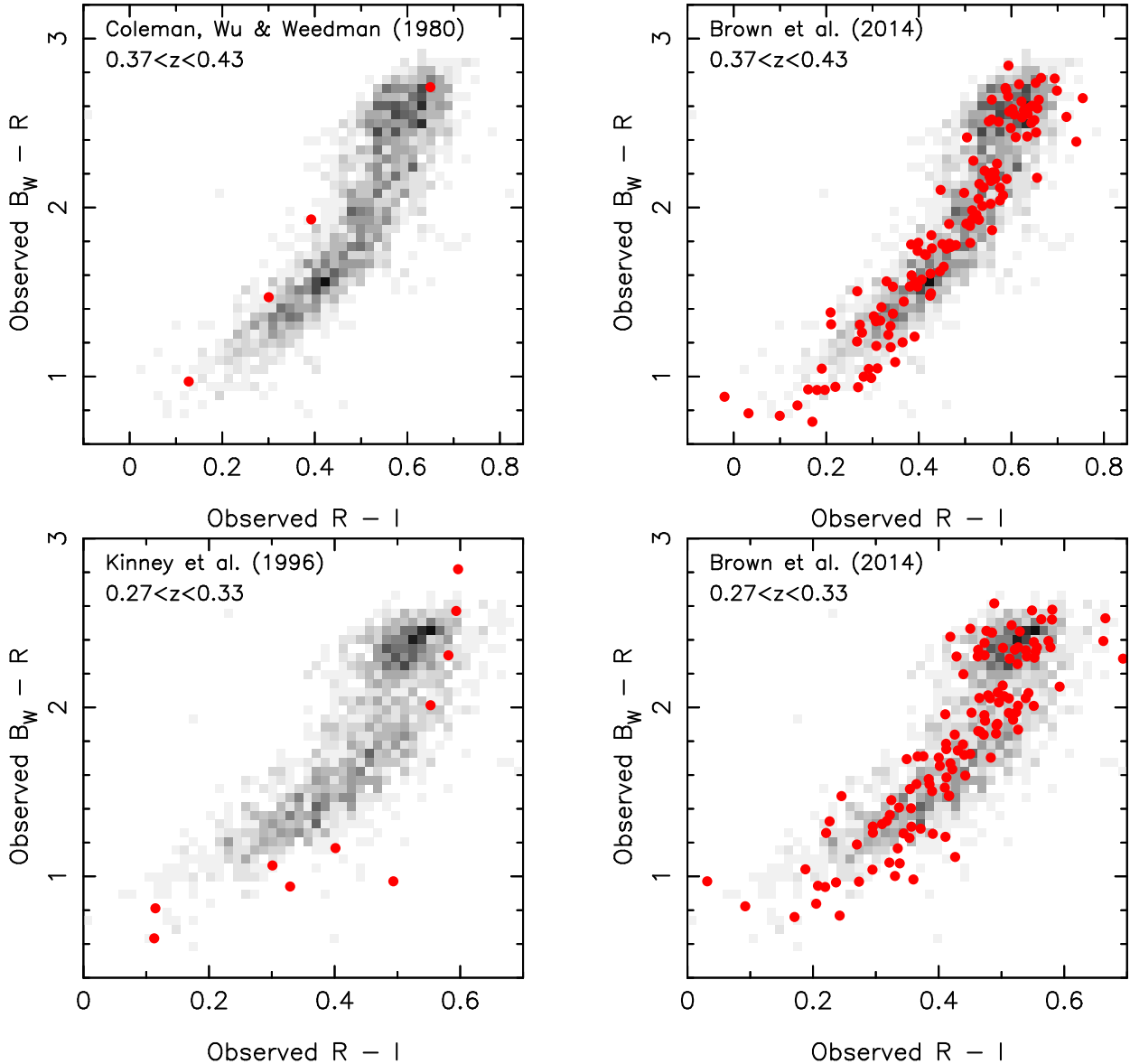

\resizebox{3.0in}{!}{\includegraphics{f14a.ps}}\resizebox{0.5in}{!}{~}\resizebox{3.0in}{!}{\includegraphics{f14b.ps}}
\resizebox{3.0in}{!}{\includegraphics{f14c.ps}}\resizebox{0.5in}{!}{~}\resizebox{3.0in}{!}{\includegraphics{f14d.ps}}
\caption{A comparison of the \citet{col80} and \citet{kin96} templates (left panels) and our templates (right panels) with the observed optical colors of $z=0.4$ and $z=0.3$ galaxies in the Bo\"otes field (greyscale). Both the \citet{col80} and \citet{kin96} templates are slightly offset from the observed galaxy locus, whereas our templates follow the observed galaxy locus.}
\label{fig:cww}
\end{figure*}

Figure~\ref{fig:cww} compares our templates, the \citet{col80} templates and the \citet{kin96} templates with the observed optical colors of $z=0.3$ and $z=0.4$ galaxies in Bo\"otes. Both the \citet{col80} and \citet{kin96} templates are systematically offset from the observed galaxy locus, and in some cases the offset corresponds to several tenths of a magnitude. Our templates broadly follow the observed galaxy loci  in Figure~\ref{fig:cww}. Some templates do fall off the galaxy locus, including blue emission lines galaxies (which are generally fainter than $I=20.4$ at these redshifts) and LIRGs (that are relatively rare at $z\sim 0.3$). Some regions of the color-space that are populated by \citet{kin96} templates are not populated by our templates. Given our large sample size and accurate photometry, we suspect this is due to errors in the \citet{kin96} templates rather than incompleteness in our template library. 

\begin{figure*}
\plottwo{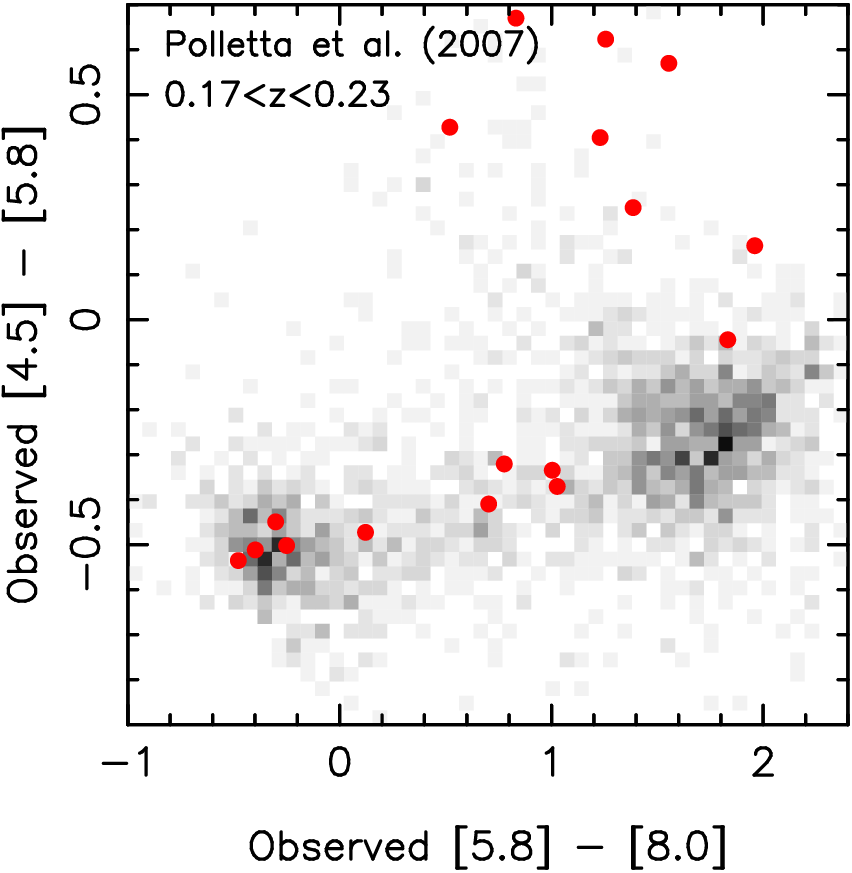}{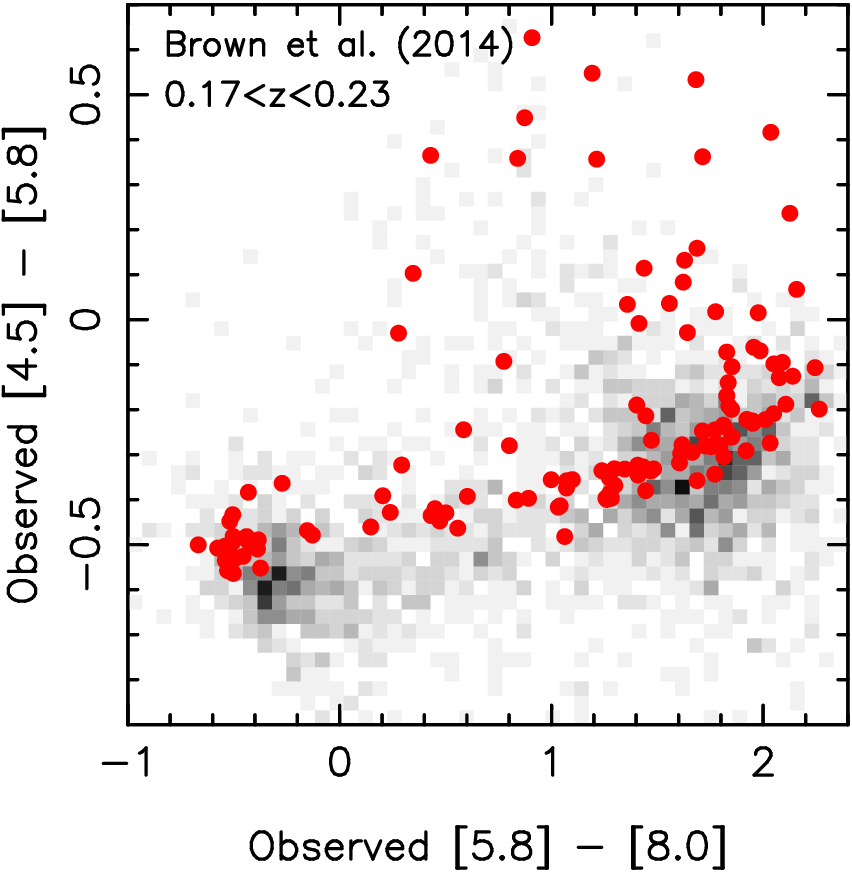}
\caption{A comparison of the \cite{pol07} templates (left panel) and our templates (right panel) with the observed mid-infared colors of $z=0.2$ galaxies in the Bo\"otes field (greyscale). The two loci in each panel correspond to galaxies with Rayleigh-Jeans spectra (left) and significant PAH emission (right). The \cite{pol07} templates do not populate the locus of star-forming galaxies, whereas this region of the color-color diagram is well populated with our templates. It should be noted that our templates span a very broad range of galaxy types, and are not intended to mimic the frequency with which particular galaxies are observed.}
\label{fig:pol}
\end{figure*}

In Figure~\ref{fig:pol} we compare our templates and the \citet{pol07} templates with the observed mid-infrared colors of $z\sim 0.2$ galaxies in Bo\"otes. Galaxies with Rayleigh-Jeans SEDs and strong PAH emission form two distinct loci on the left and right of the plots. The PAH locus is not populated by the \citet{pol07} templates, while multiple templates from our library populate the PAH locus. Our ability to populate the color-color diagram with templates is a consequence of the input photometry and spectroscopy, combined with our large sample size. However, even with 129 templates some regions of the color-color diagram are sparsely populated, including galaxies with mid-infrared colors that don't quite match Rayleigh-Jeans SEDs (e.g., galaxies similar to NGC~4450 and NGC~4725).

\section{Summary}
\label{sec:summary}

We have constructed an atlas of 129  SEDs for nearby galaxies. The atlas combines optical, {\it Spitzer} and {\it Akari} spectra with MAGPHYS models, all of which have been normalized, constrained and verified with matched-aperture photometry. The redundancy of the 26 bands of matched-aperture photometry allows us to identify and mitigate systematic errors known to be present in the imaging data, including scattered light in IRAC images, coincidence losses in the {\it Swift} UVOT imaging and errors in the pre-launch WISE W4 filter curve. Comparison of the photometry and spectroscopy reveals that spectra used to calibrate star formation rate indicators systematically underestimates ${\rm H\alpha}$ fluxes by 10\% or more. The SEDs are typically in good agreement with the photometry, with residuals being less than 10\% in most cases. The 129 SEDs and associated images are available via http://dx.doi.org/10.5072/03/529D3551F0117.

Our atlas of SEDs spans a broad range of galaxy absolute magnitudes ($-14.4>M_g>-22.3$), colors ($0.1<u-g<1.9$) and types, including ellipticals, spirals, merging galaxies, blue compact dwarfs and luminous infrared galaxies. Within individual object classes we see considerable diversity in our SEDs. For example, there is the well known diversity of ultraviolet SEDs for elliptical galaxies, and some elliptical galaxies have significant star formation and dust emission in the mid-infrared. Multi-wavelength photometry reveals the diversity of galaxy properties and shows that this diversity cannot be adequately modeled with a small number of galaxy templates.

Our SEDs differ significantly from those in the prior literature, particularly for LIRGs. Relative to the prior literature, our SEDs provide a better match to the observed optical and mid-infrared colors of $z<0.5$ galaxies. Our atlas will provide improved K-corrections, photometric redshifts and local analogues of distant galaxies than existing template libraries. However, significant improvements can still be made including extending the atlas to longer wavelengths and filling gaps in the spectral coverage in the ultraviolet and near-infrared.

\begin{acknowledgments}

We wish to thank Vianney Lebouteiller for responding to our queries about and making improvements to the Cornell Atlas of Spitzer/IRS Sources \citep[CASSIS][]{leb11}. We wish to thank Michelle Cluver for using and providing feedback on a preliminary version of the templates \citep{clu13}. We also thank Richard Beare, Mark Brodwin, Vassilis Charmandaris, Simon Driver, Peppo Gavazzi, Will Hartley,  Hanae Inami, Rob Kennicutt, and Lauranne Lanz for their assistance and feedback as the spectral templates were being developed. Michael Brown acknowledges financial support from The Australian Research Council (FT100100280), the Monash Research Accelerator Program (MRA) and the International Science Linkages Program. Part of the research for this paper was conducted during a visit to the Aspen Center for Physics in June 2012. Masa Imanishi is supported by Grants-in-Aid for Scientific Research (no. 22012006).

{\it Swift} UVOT was designed and built in collaboration between MSSL, PSU, SwRI, Swales Aerospace and GSFC, and was launched by NASA. GALEX is a NASA Small Explorer, launched in 2003 April. We gratefully acknowledge NASA's support for construction, operation and science analysis for the GALEX mission, developed in cooperation with the Centre National d'Etudes Spatiales of France and the Korean Ministry of Science and Technology.

Funding for SDSS-III has been provided by the Alfred P. Sloan Foundation, the Participating Institutions, the National Science Foundation, and the U.S. Department of Energy Office of Science. The SDSS-III web site is http://www.sdss3.org/. SDSS-III is managed by the Astrophysical Research Consortium for the Participating Institutions of the SDSS-III Collaboration including the University of Arizona, the Brazilian Participation Group, Brookhaven National Laboratory, University of Cambridge, University of Florida, the French Participation Group, the German Participation Group, the Instituto de Astrofisica de Canarias, the Michigan State/Notre Dame/JINA Participation Group, Johns Hopkins University, Lawrence Berkeley National Laboratory, Max Planck Institute for Astrophysics, New Mexico State University, New York University, Ohio State University, Pennsylvania State University, University of Portsmouth, Princeton University, the Spanish Participation Group, University of Tokyo, University of Utah, Vanderbilt University, University of Virginia, University of Washington, and Yale University. The NASA-Sloan Atlas was created by Michael Blanton, with extensive help and testing from Eyal Kazin, Guangtun Zhu, Adrian Price-Whelan, John Moustakas, Demitri Muna, Renbin Yan and Benjamin Weaver. Funding for the NASA-Sloan Atlas has been provided by the NASA Astrophysics Data Analysis Program (08-ADP08-0072) and the NSF (AST-1211644). 

This research is based in part on observations taken with telescopes of the National Optical Astronomy Observatory, which is operated by the Association of Universities for Research in Astronomy (AURA) under cooperative agreement with the National Science Foundation. This publication makes use of data products from the Two Micron All Sky Survey, which is a joint project of the University of Massachusetts and the Infrared Processing and Analysis Center/California Institute of Technology, funded by the National Aeronautics and Space Administration and the National Science Foundation. This research is based in part on observations with AKARI, a JAXA project with the participation of ESA. 

This work is based in part on observations made with the {\it Spitzer} Space Telescope, obtained from the NASA/ IPAC Infrared Science Archive, both of which are operated by the Jet Propulsion Laboratory, California Institute of Technology under a contract with the National Aeronautics and Space Administration. The Cornell Atlas of Spitzer/IRS Sources (CASSIS) is a product of the Infrared Science Center at Cornell University, supported by NASA and JPL. This publication makes use of data products from the Wide-field Infrared Survey Explorer, which is a joint project of the University of California, Los Angeles, and the Jet Propulsion Laboratory/California Institute of Technology, funded by the National Aeronautics and Space Administration. 

{\it Facilities:} \facility{Akari (IRC)}, \facility{Bok (Boller \& Chivens spectrograph)}, \facility{CTIO:1.5m (R-C spectrograph)}, \facility{CITO:2MASS}, \facility{FLWO:2MASS}, \facility{GALEX}, \facility{Mayall (MOSAIC-1 wide-field camera)}, \facility{MMT (Hectospec)}, \facility{Sloan}, \facility{Spitzer (IRAC, IRS, MIPS)}, \facility{Swift}, \facility{WISE}

\end{acknowledgments}

\clearpage

\LongTables
\begin{deluxetable*}{ccccccccccc}
\tablecolumns{11}
\tabletypesize{\tiny}
\tablecaption{Basic properties of the sample (complete table is available online)\label{table:summary}}
\tablehead{
  \colhead{Name}                               &
  \colhead{J2000 Coordinates}       &
  \colhead{Morphology}                    &
  \colhead{T-type}                    &
   \colhead{BPT}        &
  \colhead{$E(B-V)$}                          &
  \colhead{$g$}                                   &
  \colhead{$u-g$}                               &
  \colhead{$M_g$}                             &
  \colhead{$cz$}                                 &
  \colhead{$D_L$} \\
  \colhead{}                   &
  \colhead{}                   &
  \colhead{}                   &
  \colhead{}                   &
  \colhead{Class}                   &
  \colhead{(mag)}         &
  \colhead{(mag)}         &
  \colhead{(mag)}         &
  \colhead{(mag)}         &
  \colhead{(${\rm km~s^{-1}}$)}         &
  \colhead{(Mpc)}	}
\startdata               
 Arp 256 N       & 00 18 50.2  -10 21 44  &    SB(s)c       &   4.5 &  SF     &   0.03 &  14.32 &   0.84 &  -20.97 &   8178 &  114.3 \\ 
 Arp 256 S       & 00 18 50.9  -10 22 39  &    SB(s)b       &   3.0 &  SF     &   0.03 &  14.36 &   0.75 &  -20.91 &   8118 &  113.4 \\ 
 NGC 0337        & 00 59 50.0  -07 34 35  &    SBd          &   7.0 &  SF     &   0.10 &  11.98 &   0.91 &  -19.77 &   1647 &   22.4 \\ 
 CGCG 436-030    & 01 20 02.6  +14 21 43  &    Pec          &   -   &  SF     &   0.06 &  14.58 &   1.00 &  -20.99 &   9347 &  129.9 \\ 
 NGC 0474        & 01 20 06.7  +03 24 56  &    (R)SA(s)0    &  -2.0 &  -      &   0.04 &  12.28 &   1.61 &  -20.30 &   2336 &   32.8 \\ 
 NGC 0520        & 01 24 35.6  +03 47 33  &    Pec          &  99.0 &  SF/AGN &   0.04 &  11.98 &   1.35 &  -20.90 &   2276 &   37.7 \\ 
      NGC 0584   & 01 31 20.7  -06 52 05  &      E4         &  -5.0 &  -      &   0.04 &  11.19 &   1.68 &  -20.32 &   1802 &   20.1 \\ 
      NGC 0628   & 01 36 41.8  +15 47 01  &      SAc        &   5.0 &  -      &   0.10 &  10.92 &   1.16 &  -18.40 &    657 &    7.3 \\
\enddata
\end{deluxetable*}

\begin{deluxetable*}{cccccccccc}
\tablecolumns{10}
\tabletypesize{\tiny}
\tablecaption{Photometric and Spectroscopic Apertures (complete table is available online)\label{table:input}}
\tablehead{
  \colhead{}                              &
  \multicolumn{2}{c}{Photometric Aperture}   &
  \colhead{Optical}                   &
  \multicolumn{2}{c}{Akari Aperture}      &
  \multicolumn{2}{c}{{\it Spitzer} IRS SL Aperture}            &
  \multicolumn{2}{c}{{\it Spitzer} IRS LL Aperture}            \\
  \cline{2-3}\cline{5-6}\cline{7-8}\cline{9-10}
\colhead{Name}      &
\colhead{Size}      &
\colhead{P.A.}      &
\colhead{ Spectrum }       &
\colhead{Size}       &
\colhead{P.A.}       &
\colhead{Size}      &
\colhead{P.A.}      &
\colhead{Size}      &
\colhead{P.A.}      \\
\colhead{}      &
\colhead{(arcsec.)}      &
\colhead{(deg.)}      &
\colhead{   }       &
\colhead{(arcsec.)}      &
\colhead{(deg.)}      &
\colhead{(arcsec.)}      &
\colhead{(deg.)}      &
\colhead{(arcsec.)}    &
\colhead{(deg.)}      \\
}
\startdata               
  Arp 256 N        & $  40 \times  60 $  & $  90 $  & M06  & -                    & -   & $ 33.3 \times 18.5 $ & 154 & $ 40.6 \times 40.6 $ &  71 \\ 
 Arp 256 S        & $  40 \times  40 $  & $  90 $  & M06  & $ 29.2 \times 60.0 $ & 246 & $ 25.9 \times 18.5 $ & 153 & $ 61.0 \times 25.4 $ &  69 \\ 
 NGC 0337         & $  95 \times  55 $  & $  70 $  & M10  & -                    & -   & $ 53.7 \times 31.5 $ & 149 & $ 121.9 \times 71.1 $ &  65 \\ 
 CGCG 436-030     & $  35 \times  40 $  & $  90 $  & M06  & $ 7.3 \times 60.0 $ & 248 & $ 3.6^{\prime\prime}$ Slit   & -28 & $ 10.6^{\prime\prime}$ Slit   &  67 \\ 
 NGC 0474         & $  60 \times 150 $  & $  90 $  & M06  & -                    & -   & -                    & -   & -                    & -   \\ 
 NGC 0520         & $ 140 \times 100 $  & $  90 $  & M06  & $ 29.2 \times 60.0 $ & 248 & $ 75.8 \times 57.4 $ & -27 & $ 106.7 \times 96.5 $ & -111 \\ 
      NGC 0584    & $ 138 \times  55 $  & $  70 $  & M10  & -                    & -   & $ 27.8 \times 18.5 $ & 148 & $ 40.6 \times 35.6 $ &  64 \\ 
      NGC 0628    & $ 346 \times  55 $  & $  70 $  & M10  & -                    & -   & $ 48.1 \times 27.7 $ & 152 & $ 330.2 \times 50.8 $ &  68 \\
\enddata
\end{deluxetable*}

\setlength{\tabcolsep}{0.02in}

\begin{deluxetable*}{ccccccccccc}
\tablecolumns{10}
\tabletypesize{\tiny}
\tablecaption{AB magnitudes for the sample without dust extinction corrections (complete table is available online)\label{table:photometry}}
\tablehead{
  \colhead{Name}    &
  \colhead{FUV}       &
  \colhead{UVW2}       &
   \colhead{UVM2}       &
 \colhead{NUV}       &
  \colhead{UVW1}       &
  \colhead{U}       &
  \colhead{u}       &
  \colhead{g}       &
  \colhead{V}      \\
  \colhead{ }       &
  \colhead{r}       &
  \colhead{i}       &
  \colhead{z}       &
  \colhead{J}       &
  \colhead{H}       &
  \colhead{$K_S$}       &
  \colhead{W1}       &
  \colhead{[3.6]}       &
  \colhead{[4.5]}      \\
  \colhead{ }             &
  \colhead{W2}        &
  \colhead{[5.8]}        &
  \colhead{[8.0]}         &
 \colhead{W3}          &
 \colhead{PUI Blue}         &
 \colhead{W4}         &
 \colhead{W4$^\prime$}        &
 \colhead{PUI Red}    &
  \colhead{[24]}           }
\startdata               
 Arp 256 N        & $ 16.26 \pm 0.10 $  & -                     & -                     & $ 15.91 \pm 0.10 $  & -                     & -                     & $ 15.31 \pm 0.05 $  & $ 14.43 \pm 0.05 $  & -                    \\ 
 & $ 14.03 \pm 0.05 $  & $ 13.77 \pm 0.05 $  & $ 13.62 \pm 0.05 $  & $ 13.33 \pm 0.05 $  & $ 13.28 \pm 0.05 $  & $ 13.34 \pm 0.05 $  & $ 13.83 \pm 0.10 $  & $ 13.85 \pm 0.10 $  & $ 14.22 \pm 0.10 $ \\ 
 & $ 14.34 \pm 0.10 $  & $ 13.12 \pm 0.10 $  & $ 11.98 \pm 0.10 $  & $ 12.06 \pm 0.10 $  & -                     & $ 10.99 \pm 0.10 $  & $ 11.25 \pm 0.10 $  & -                     & $ 11.23 \pm 0.10 $ \\ 
 Arp 256 S        & $ 16.37 \pm 0.10 $  & -                     & -                     & $ 16.00 \pm 0.10 $  & -                     & -                     & $ 15.26 \pm 0.05 $  & $ 14.48 \pm 0.05 $  & -                    \\ 
 & $ 14.05 \pm 0.05 $  & $ 13.80 \pm 0.05 $  & $ 13.58 \pm 0.05 $  & $ 13.25 \pm 0.05 $  & $ 13.03 \pm 0.05 $  & $ 13.15 \pm 0.05 $  & $ 13.28 \pm 0.10 $  & $ 13.16 \pm 0.10 $  & $ 13.40 \pm 0.10 $ \\ 
 & $ 13.49 \pm 0.10 $  & $ 11.92 \pm 0.10 $  & -                     & $ 10.53 \pm 0.10 $  & -                     & $  8.95 \pm 0.10 $  & $  9.12 \pm 0.10 $  & -                     & $  8.90 \pm 0.10 $ \\ 
 NGC 0337         & $ 15.32 \pm 0.10 $  & $ 14.77 \pm 0.10 $  & $ 14.95 \pm 0.10 $  & $ 14.63 \pm 0.10 $  & $ 14.31 \pm 0.10 $  & -                     & $ 13.37 \pm 0.05 $  & $ 12.36 \pm 0.05 $  & -                    \\ 
 & $ 11.88 \pm 0.05 $  & $ 11.65 \pm 0.05 $  & $ 11.51 \pm 0.05 $  & $ 11.27 \pm 0.05 $  & $ 11.12 \pm 0.05 $  & $ 11.34 \pm 0.05 $  & $ 11.89 \pm 0.10 $  & $ 11.85 \pm 0.10 $  & $ 12.24 \pm 0.10 $ \\
\enddata
\end{deluxetable*}

\begin{deluxetable}{ccccccccccc}
\tablecolumns{6}
\tabletypesize{\tiny}
\tablecaption{Residuals between photometry and spectra.\label{table:res}}
\tablehead{
  \colhead{Filter}    &
  \colhead{$\lambda_{eff}$} &
  \colhead{FWHM bandwidth} &
  \multicolumn{2}{c}{$m_{spec}-m_{phot}$} & 
  \colhead{Number of Galaxies} \\
  \colhead{}    &
  \colhead{(${\rm \mu m}$)} &
  \colhead{(${\rm \mu m}$)} &
  \colhead{Median}       &
  \colhead{Std. Dev.}       &
   \colhead{} \\
   }
\startdata               
 $ FUV   $  & 0.1531 & 0.0227 & 0.00 & 0.07 & 126 \\
 $ UVW2  $  & 0.2026 & 0.0557 & 0.05 & 0.06 & 30 \\
 $ UVM2  $  & 0.2238 & 0.0511 & -0.03 & 0.10 & 31 \\
 $ NUV   $  & 0.2286 & 0.0795 & -0.01 & 0.07 & 126 \\
 $ UVW1  $  & 0.2598 & 0.0681 & -0.03 & 0.08 & 26 \\
 $ U     $  & 0.3459 & 0.0557 & 0.08 & 0.14 & 5 \\
 $ u     $  & 0.3551 & 0.0582 & -0.01 & 0.06 & 129 \\
 $ g     $  & 0.4681 & 0.1262 & 0.00 & 0.00 & 129 \\
\enddata
\end{deluxetable}

\begin{deluxetable}{ccccccccccc}
\tablecolumns{6}
\tabletypesize{\tiny}
\tablecaption{Factors used to rescale spectra to match photometry(complete table is available online)\label{table:specscale}}
\tablehead{
  \colhead{Name}                                &
  \colhead{Optical}                               &
  \colhead{Akari}                                  &
 \multicolumn{3}{c}{{\it Spitzer} IRS}   \\ 
  \colhead{} &
  \colhead{} &
  \colhead{} &
  \colhead{$8~{\rm \mu m}$}   &
  \colhead{$12~{\rm \mu m}$} &
  \colhead{$20~{\rm \mu m}$} 
}
\startdata
 Arp 256 N        & 1.09  & -     & 2.42  & 2.03  & 0.97 \\ 
 Arp 256 S        & 0.98  & 0.98  & 1.35  & 1.35  & 1.29 \\ 
 NGC 0337         & 1.02  & -     & 1.03  & 1.07  & 1.06 \\ 
 CGCG 436-030     & 1.21  & 0.94  & 1.17  & 1.11  & 0.99 \\ 
 NGC 0474         & 1.30  & -     & -     & -     & -    \\ 
 NGC 0520         & 1.21  & 1.84  & 1.06  & 1.11  & 1.10 \\ 
      NGC 0584    & 1.20  & -     & 2.22  & 2.12  & 1.38 \\ 
      NGC 0628    & 1.44  & -     & 4.15  & 4.90  & 1.27 \\
\enddata
\end{deluxetable}
 



\appendix

\section{An Illustrative Subsample}
\label{sec:subsample}

\begin{figure}[hbt]
\plotone{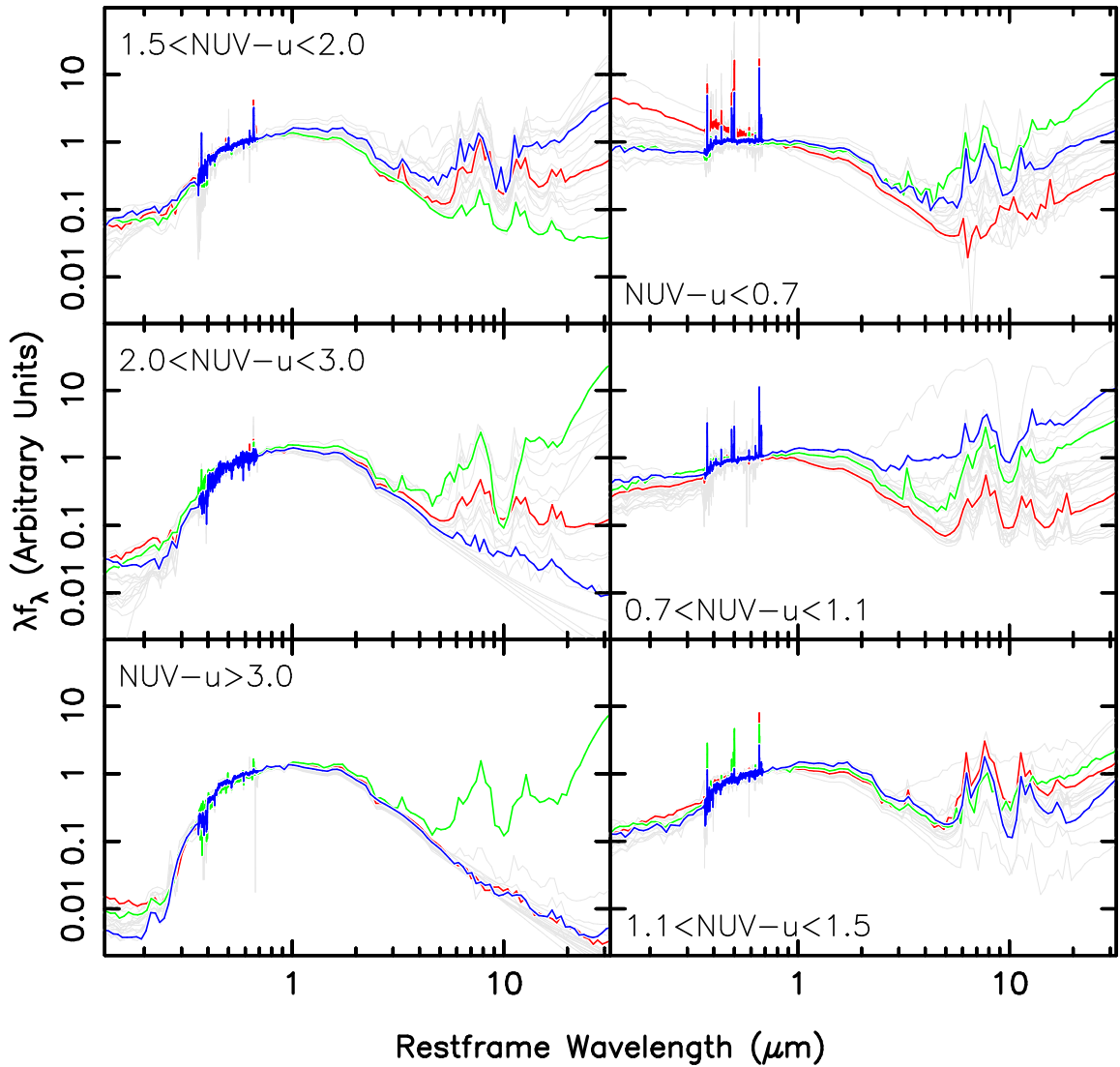}
\caption{The spectra of the 18 subsample galaxies. Each panel shows a different range of $NUV-u$ color, and for comparison the spectra of main sample galaxies are shown in grey. The subsample spans much of the color range of the full sample, and includes a diversity of galaxy types, including ellipticals, spirals, blue compact dwarfs, mergers, LIRGs and starbursts.}
\label{fig:subsetspec}
\end{figure}

\begin{figure}[hbt]
\plottwo{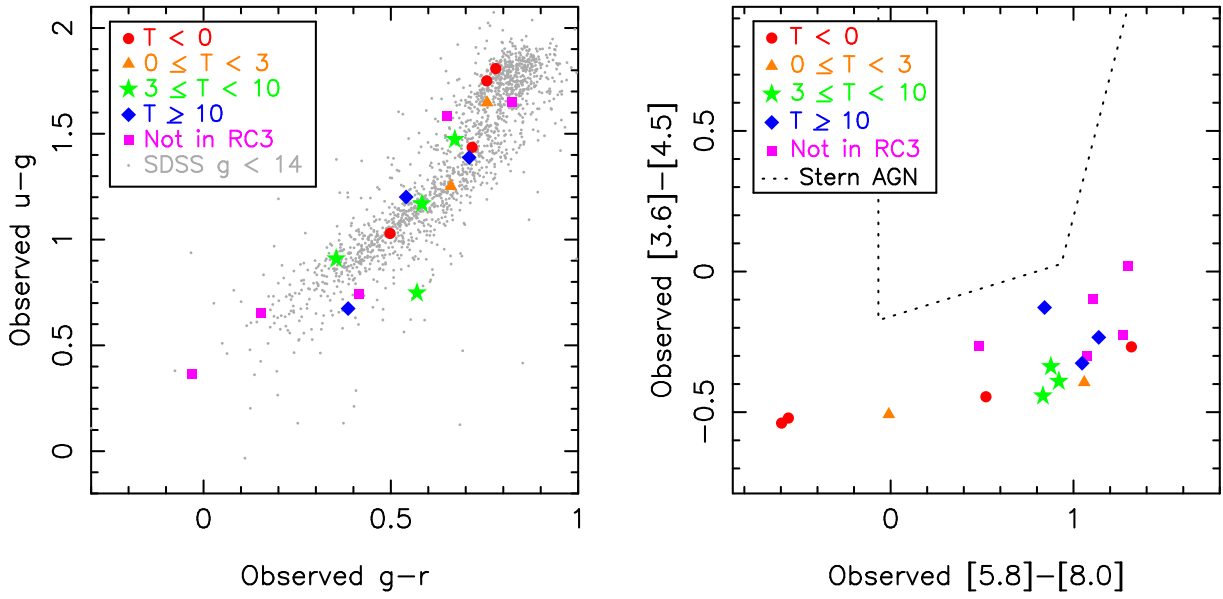}{f17b.ps}
\caption{The observed optical and {\it Spitzer} IRAC colors of the 18 subsample galaxies listed in Table~\ref{table:subset}. The subsample is intended to span the observed range of galaxy types and colors seen in the nearby Universe, and is not intended to be representative of nearby galaxy populations. 
\vspace*{0.5cm}}
\label{fig:subcol}
\end{figure}

In Figure~\ref{fig:subsetspec} and Table~\ref{table:subset} we present a illustrative subsample of 18 galaxy spectra. We present spectra of individual galaxies rather than averaging spectra, as averaging can produce spectra that do not match those of real galaxies. As we are providing spectra of individual galaxies, it is also possible for others to add spectra to these templates as new data becomes available. The subsample was selected to span a broad range of $NUV-u$ color, and for each $NUV-u$ color bin shown in Figure~\ref{fig:subsetspec} we chose 3 galaxies which have differing mid-infrared SEDs. Where possible, we selected galaxies that had {\it Akari} spectra, particularly when galaxies had strong mid-infrared emission and were thus likely to have strong PAH emission and the $3.1~{\rm \mu m}$ ice absorption feature. The observed colors of the subsample galaxies are plotted in Figure~\ref{fig:subcol}. We caution that our subsample doesn't include some of the most extreme galaxies from the overall sample (e.g., IRAS~08572+3915, UGC~5101) and none of the subsample galaxies meet the \cite{ste05} AGN selection criterion. With some exceptions, including LIRGs, the optical colors of the subsample galaxies lie close to the locus of NASA-Sloan Atlas galaxies.

\begin{deluxetable}{cccccccccccc}
\tablecolumns{11}
\tabletypesize{\tiny}
\tablecaption{Basic properties of the subsample\label{table:subset}}
\tablehead{
  \colhead{Name}                               &
  \colhead{Morphology}                      &
  \colhead{T-type}                               &
  \colhead{$g$}                                   &
  \colhead{$M_{NUV}-M_u$}               &
  \colhead{$M_{[8.0]}-M_{[24]}$}           &
  \colhead{Notes}}
\startdata
 NGC 0337         &    SBd          &   7.0 &  12.36 &   0.96 &  0.79 & \\ 
 NGC 0695         &    S0           &  -2.0 &  13.90 &   1.17 &  0.72 & Interacting galaxy\\ 
 NGC 3079         &    SB(s)c       &   7.0 &  11.07 &   1.31 &  0.27 & Seyfert 2\\ 
      Mrk 33      &      Im pec     &  10.0 &  13.36 &   0.59 &  2.09 & Wolf-Rayet Galaxy\\ 
 UGCA 219         &    Sc           &   -   &  14.78 &   0.01 &  2.76 & Blue Compact Dwarf\\ 
 NGC 3521         &    SABbc        &   4.0 &  10.38 &   2.02 &  0.05 & \\ 
 NGC 3690         &    Pec          &   9.0 &  12.07 &   0.72 &  2.19 & Galaxy merger and Wolf-Rayet Galaxy\\ 
      NGC 4125    &      E6 pec     &  -5.0 &  10.89 &   3.44 & -0.70 & Elliptical galaxy without UV upturn\\ 
 NGC 4138         &    SA(r)0       &  -1.0 &  11.86 &   1.83 &  0.04 & \\ 
      NGC 4552    &      E          &  -5.0 &  10.94 &   3.01 & -0.74 & Elliptical galaxy with UV upturn\\ 
      NGC 4725    &      SABab pec  &   2.0 &  11.08 &   2.52 & -0.20 & Seyfert 2\\ 
 NGC 5256         &    Pec          &  99.0 &  13.82 &   1.35 &  2.14 & Galaxy merger\\ 
 CGCG 049-057     &    Irr          &   -   &  15.16 &   3.17 &  2.32 & LIRG\\ 
 NGC 5953         &    Sa           &   1.0 &  12.65 &   1.89 &  0.67 & \\ 
 IC 4553          &    Pec          &   -   &  13.85 &   2.35 &  3.27 & ULIRG Arp 220\\ 
 NGC 6090         &    Pec          &   -   &  14.20 &   0.82 &  1.76 & Galaxy merger and LIRG\\ 
 NGC 6240         &    Pec          &  90.0 &  13.45 &   1.69 &  2.34 & \\ 
 II Zw 096        &    Pec          &   -   &  14.30 &   0.49 &  2.91 & Galaxy merger and starburst\\ 

\hline
\enddata
\end{deluxetable}

\section{Presentation of the Spectra}
\label{sec:presentation}

\begin{figure}[hbt]
\plottwo{f18_001a.ps}{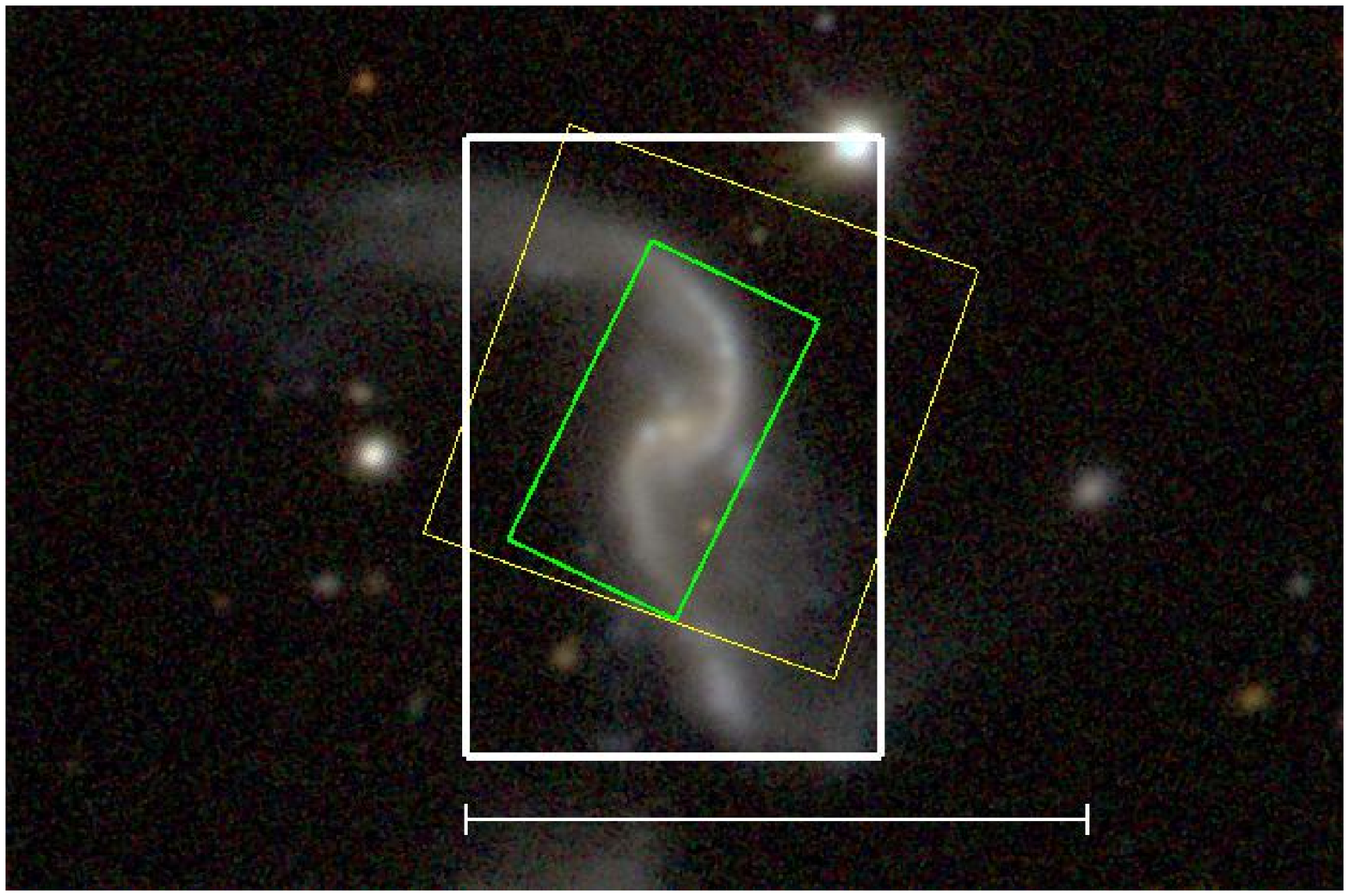}\\
\plottwo{f18_002a.ps}{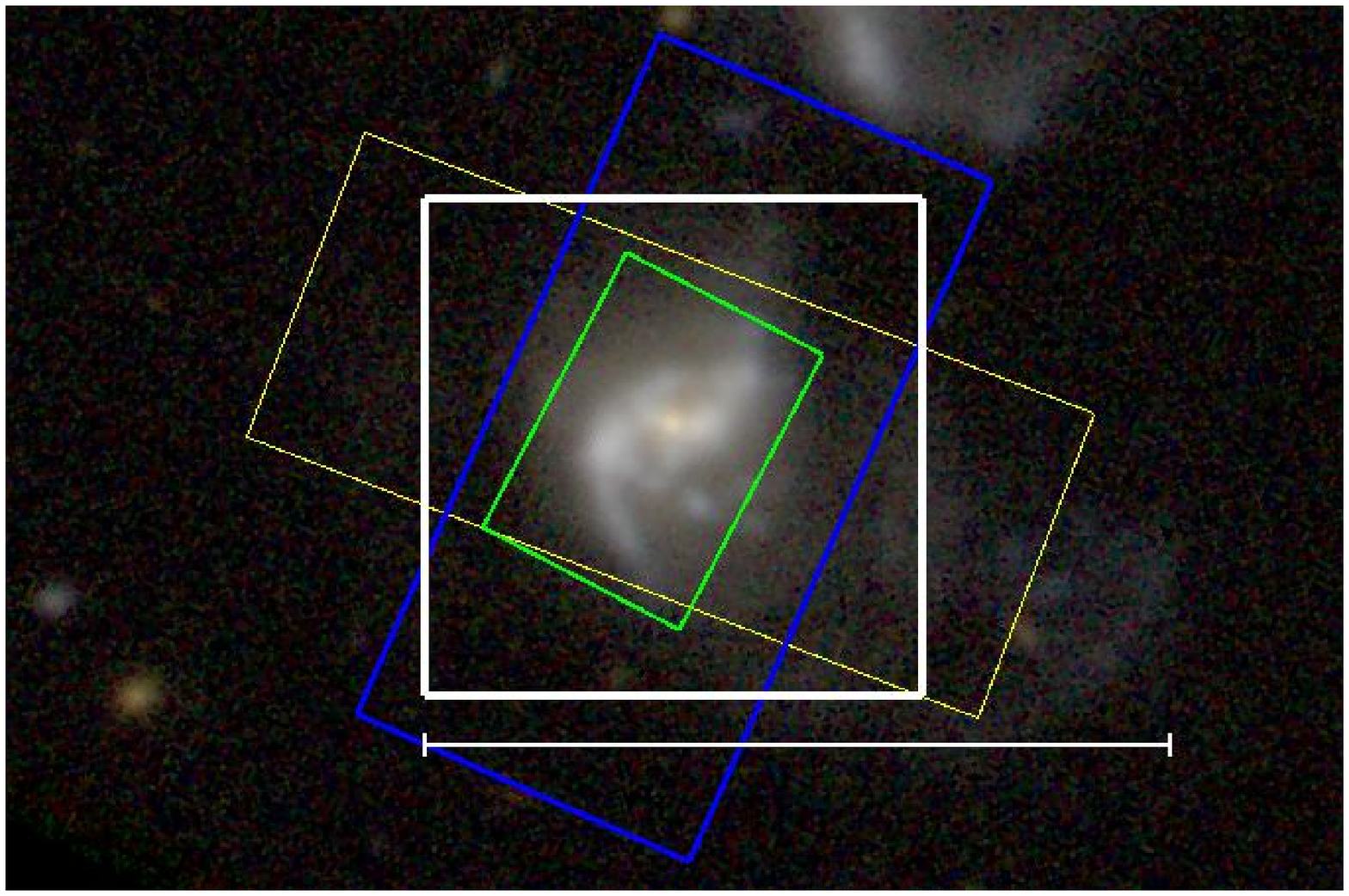}\\
\plottwo{f18_003a.ps}{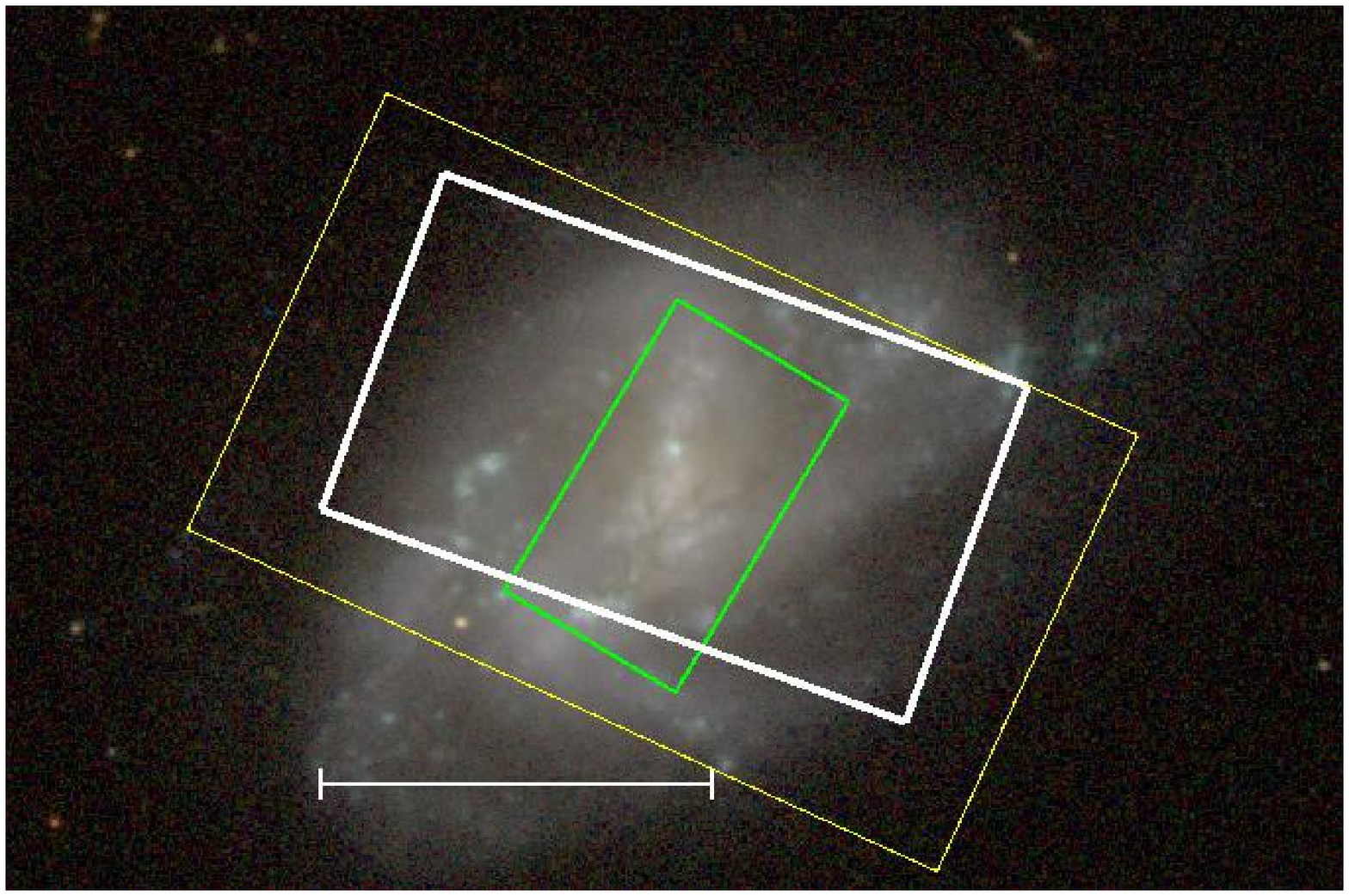}\\
\plottwo{f18_004a.ps}{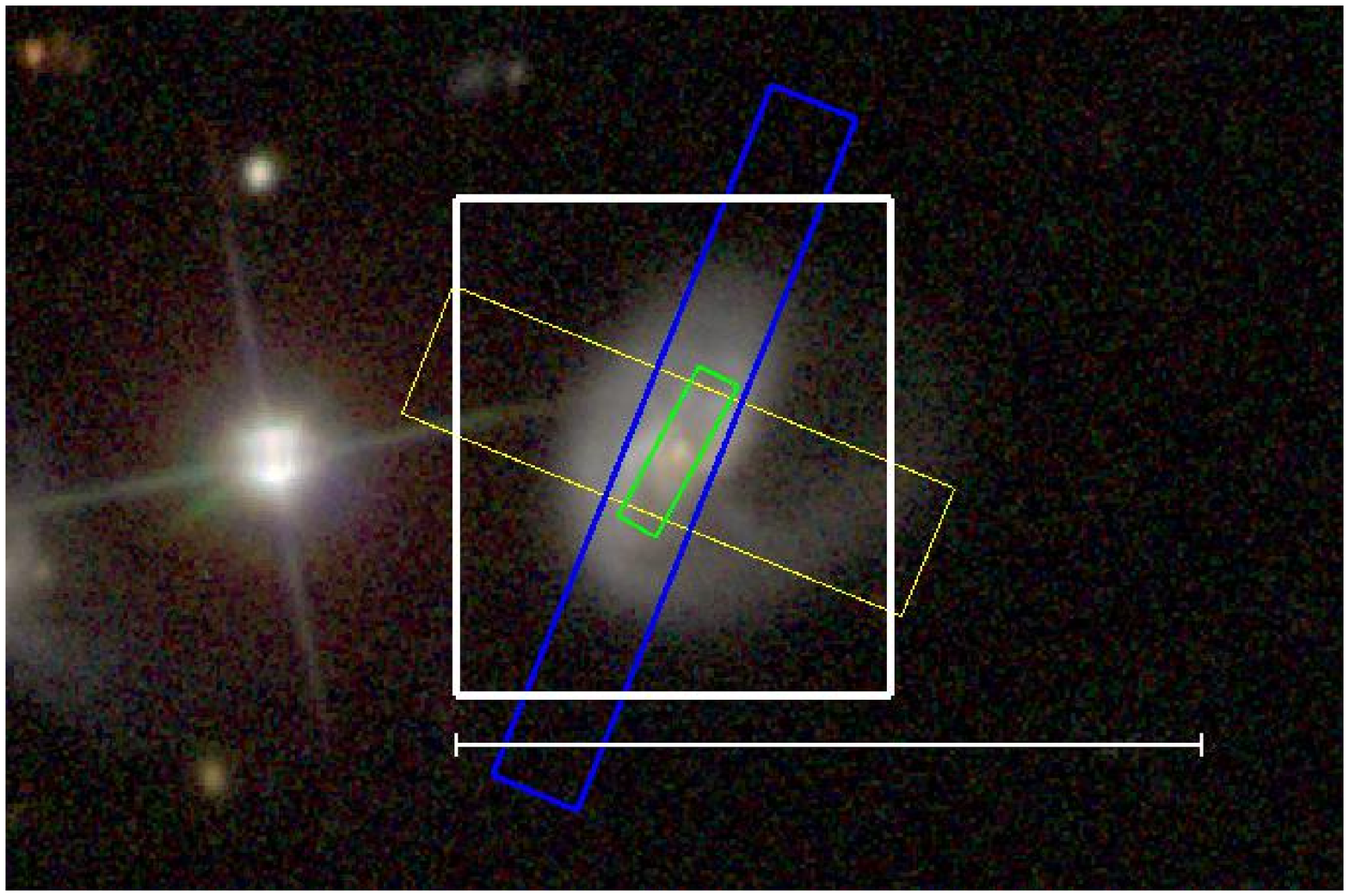}\\
\caption{Galaxy SEDs from the UV to the mid-IR. In the left-panel the observed and model spectra are shown in black and grey respectively, while the photometry used to constrain and verify the spectra is shown with red dots. In the right panel we plot the photometric aperture (thick white rectangle), the {\it Akari} extraction aperture (blue rectangle), the {\it Spitzer} SL extraction aperture (green rectangle) and the {\it Spitzer} LL extraction aperture (yellow rectangle). For galaxies with {\it Spitzer} stare mode spectra, we show a region corresponding to a quarter of the slit length. For scale, the horizontal bar denotes $1^{\prime}$. [{\it See the electronic edition of the Supplement for the complete Figure.}]}
\label{fig:allspec}
\end{figure}

In Figure~\ref{fig:allspec} we present the spectra and $gri$ images of the sample galaxies (the full figure is available online). Overlaid on the $gri$ images are the photometric apertures and the apertures used to extract the {\it Akari} and {\it Spitzer} spectra (defined in Table~\ref{table:input}). When {\it Spitzer} stare mode spectra were used, the  length of the slit used for the extraction varies with wavelength, so we plot a quarter of the slit width in these instances. The plots of the individual galaxy spectra include the dust corrected photometry used to constrain and verify the spectroscopy (also see Table~\ref{table:photometry}), including corrections for {\it Swift} UVOT coincidence losses and errors in the WISE W4 filter curve. 

\clearpage

\begin{figure}[hbt]
\figurenum{\ref{fig:allspec} continued}
\plottwo{f18_005a.ps}{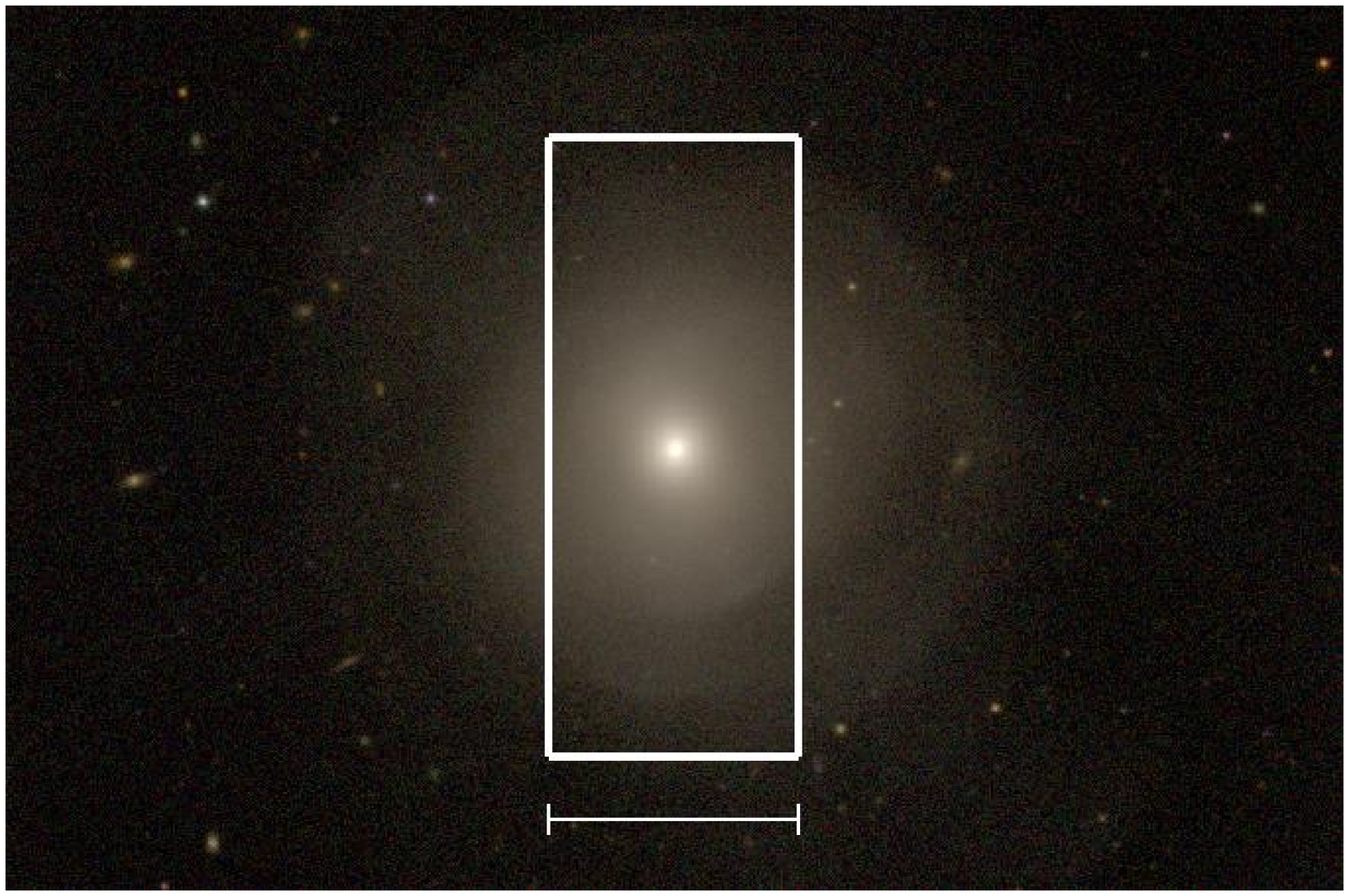}\\
\plottwo{f18_006a.ps}{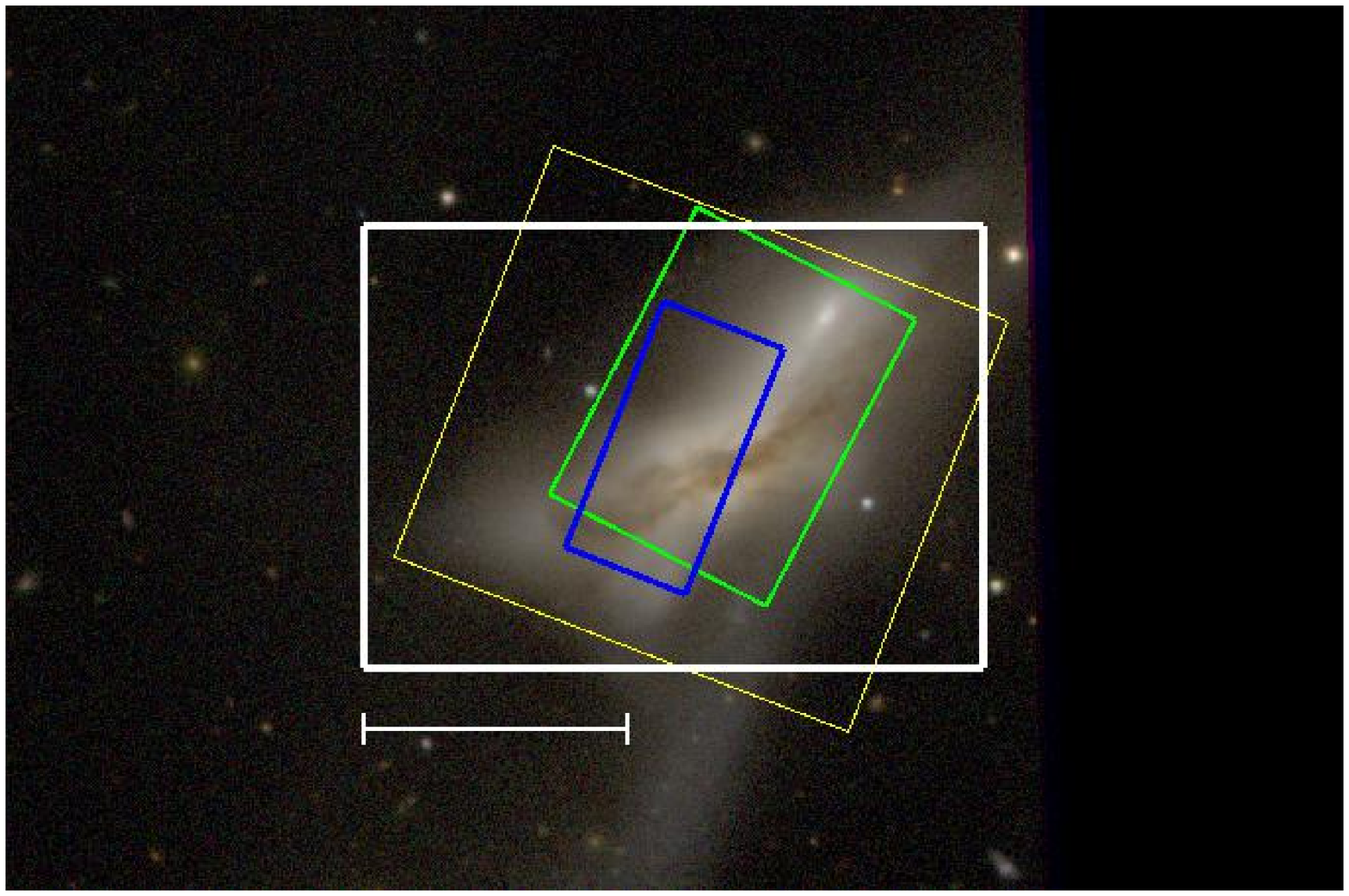}\\
\plottwo{f18_007a.ps}{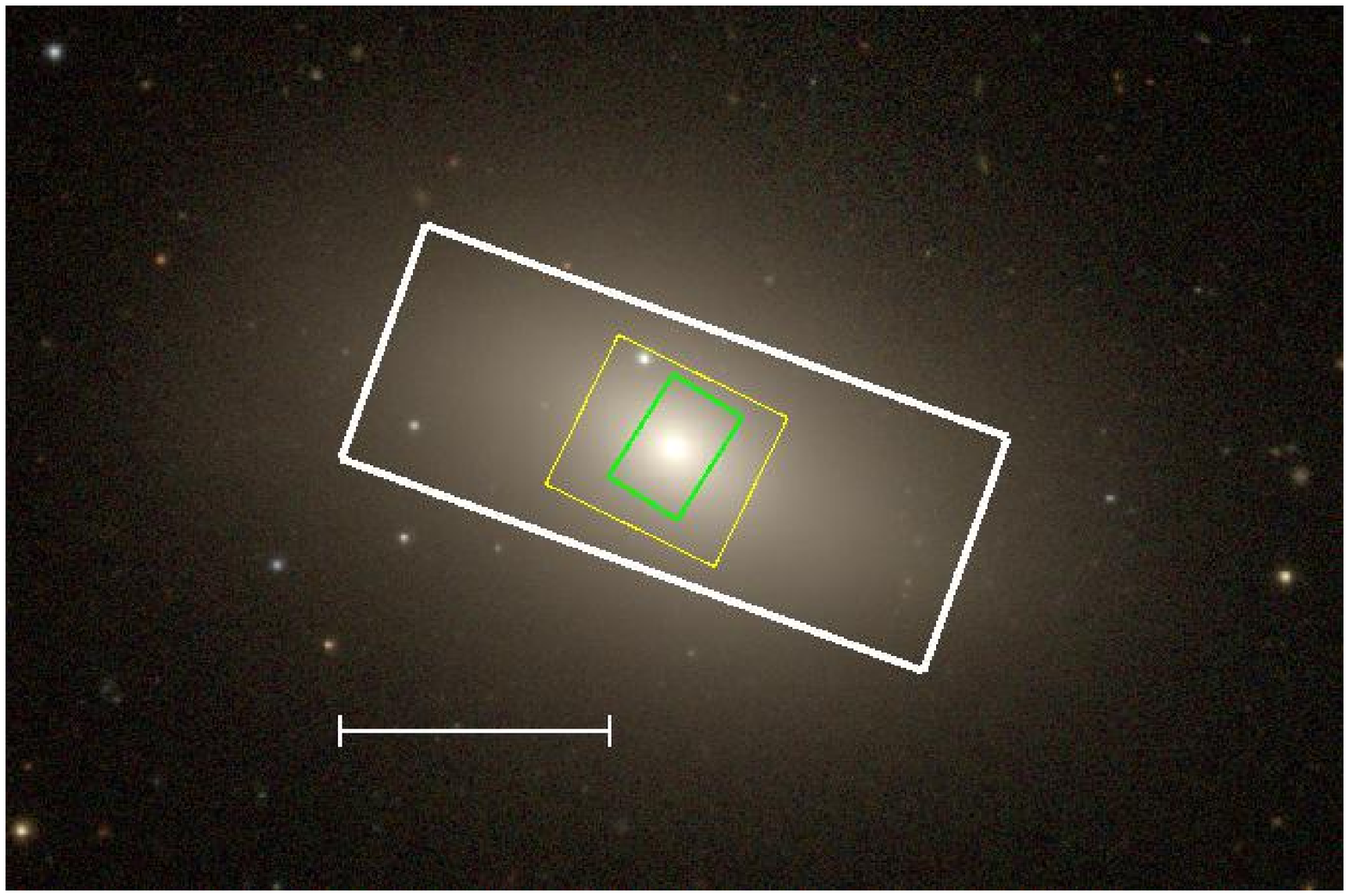}\\
\plottwo{f18_008a.ps}{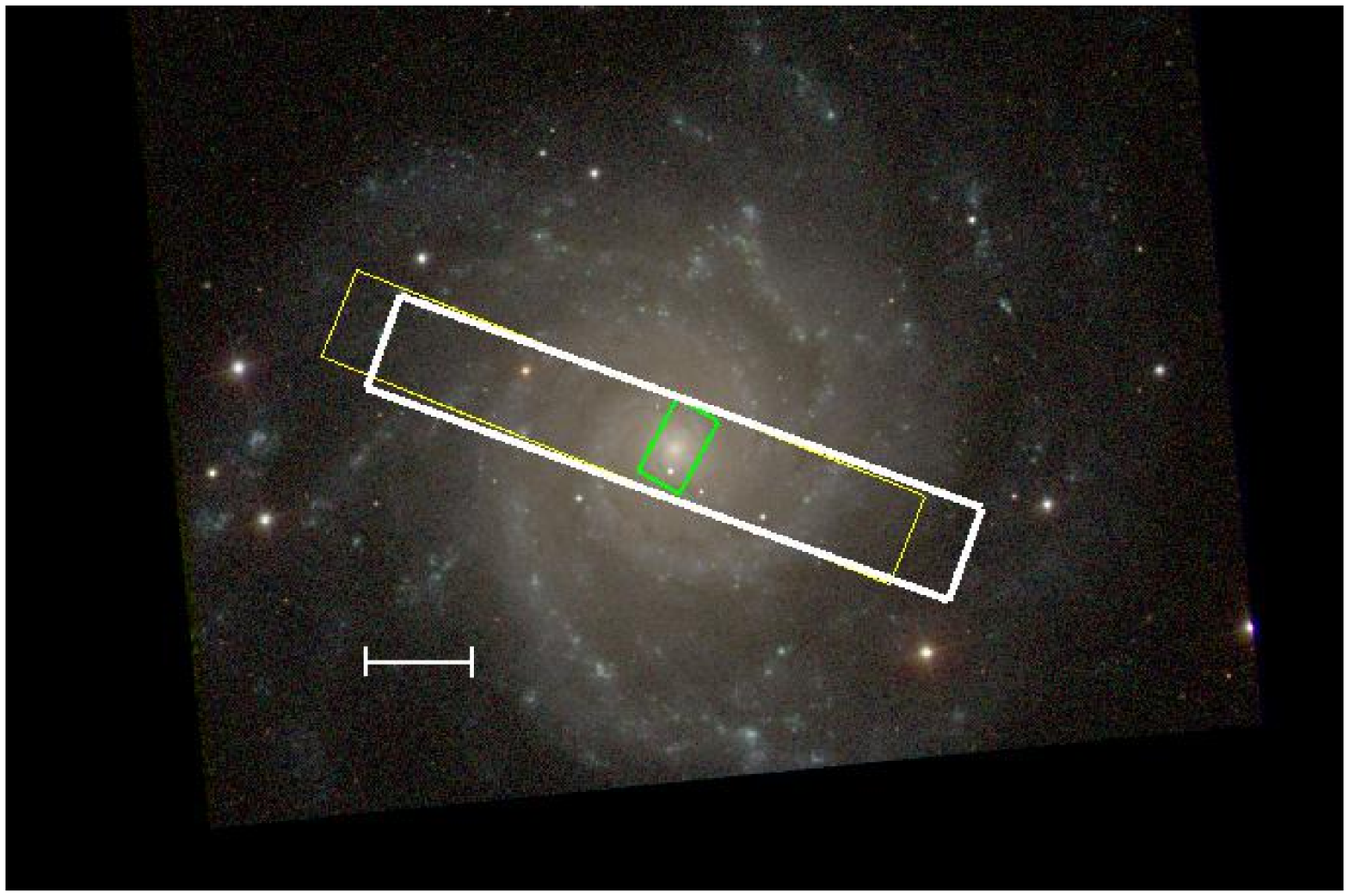}\\
\caption{Galaxy SEDs from the UV to the mid-IR. In the left-panel the observed and model spectra are shown in black and grey respectively, while the photometry used to constrain and verify the spectra is shown with red dots. In the right panel we plot the photometric aperture (thick white rectangle), the {\it Akari} extraction aperture (blue rectangle), the {\it Spitzer} SL extraction aperture (green rectangle) and the {\it Spitzer} LL extraction aperture (yellow rectangle). For galaxies with {\it Spitzer} stare mode spectra, we show a region corresponding to a quarter of the slit length. For scale, the horizontal bar denotes $1^{\prime}$. [{\it See the electronic edition of the Supplement for the complete Figure.}]}
\end{figure}

\clearpage

\begin{figure}[hbt]
\figurenum{\ref{fig:allspec} continued}
\plottwo{f18_009a.ps}{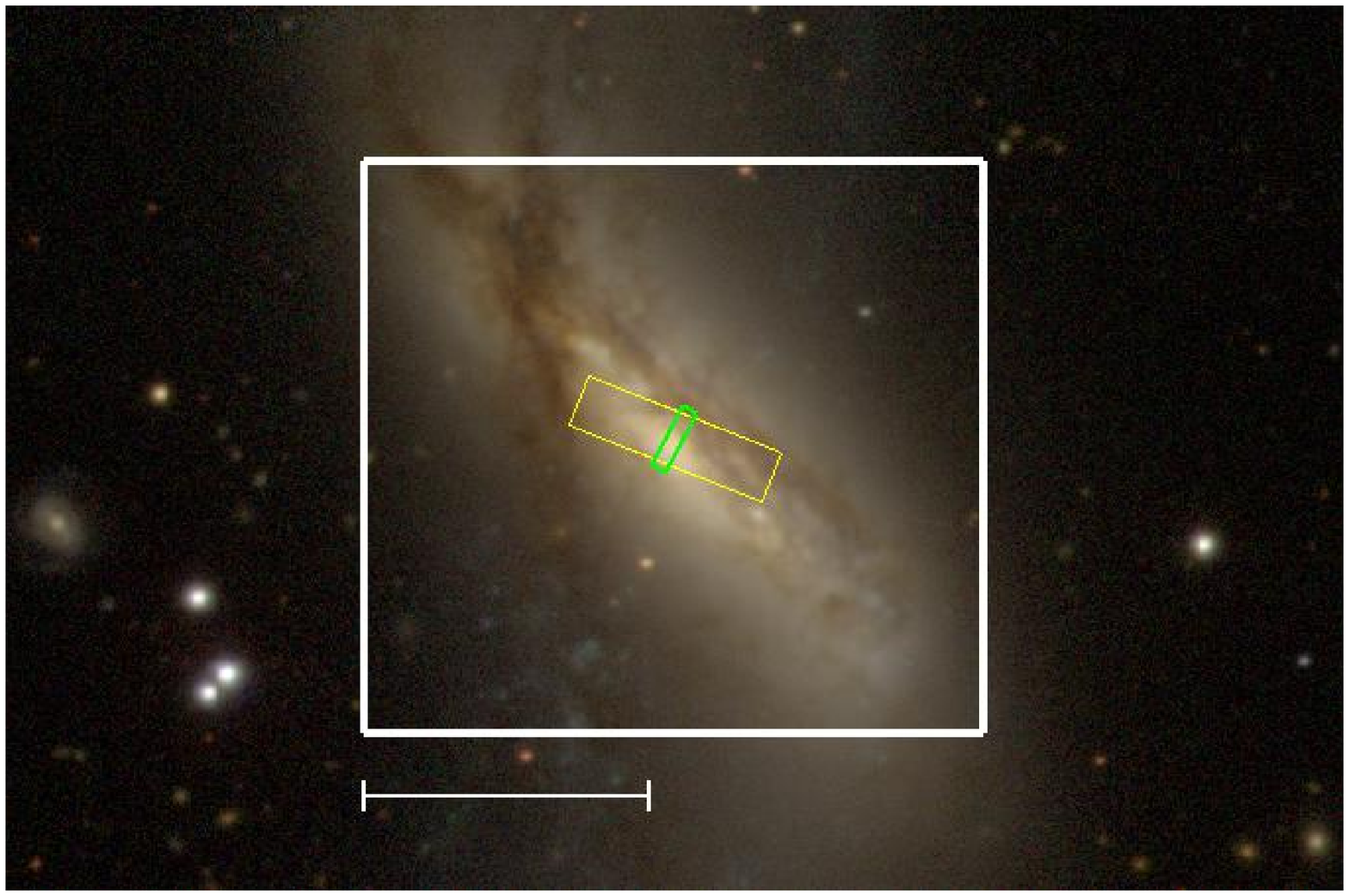}\\
\plottwo{f18_010a.ps}{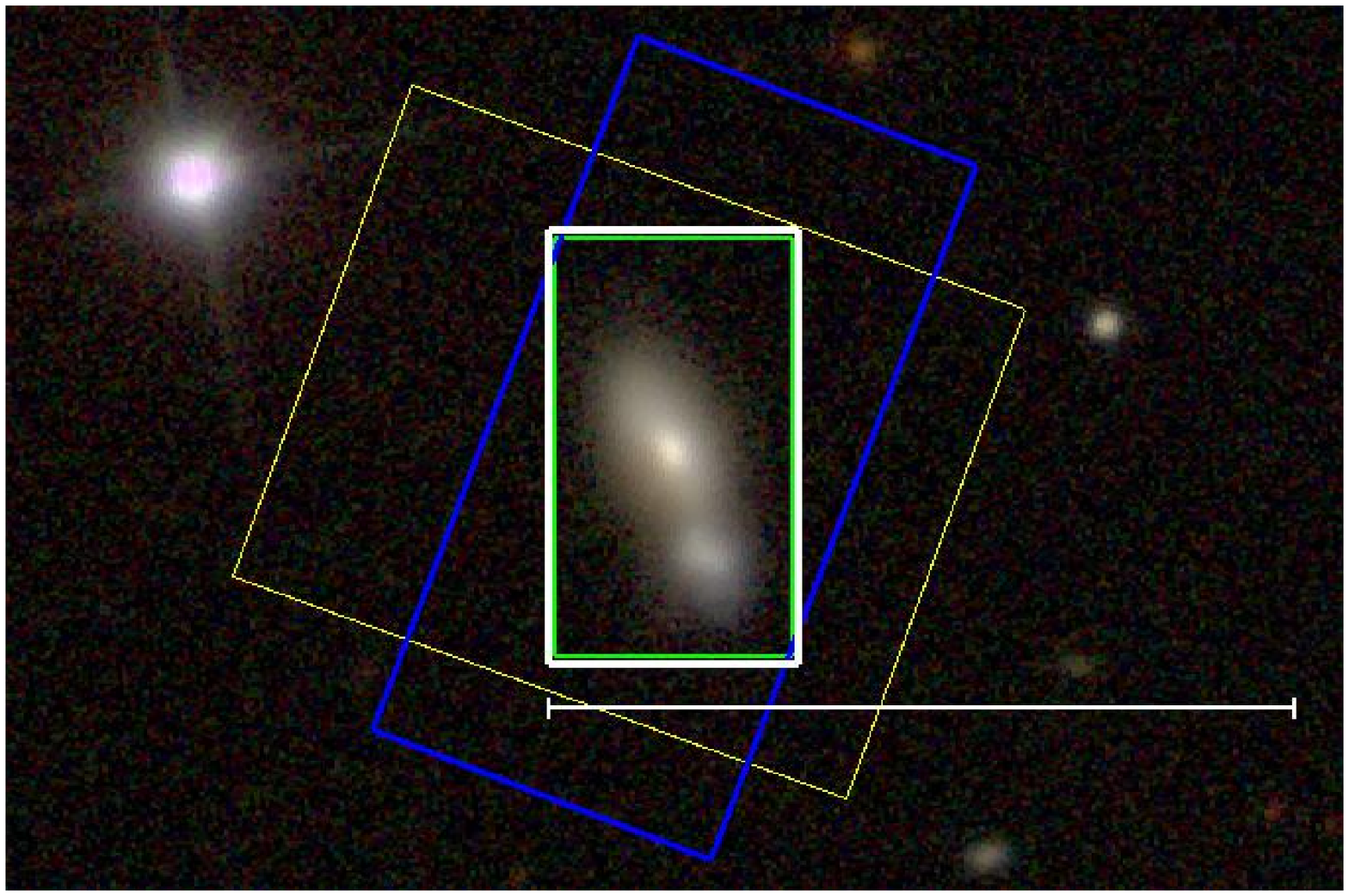}\\
\plottwo{f18_011a.ps}{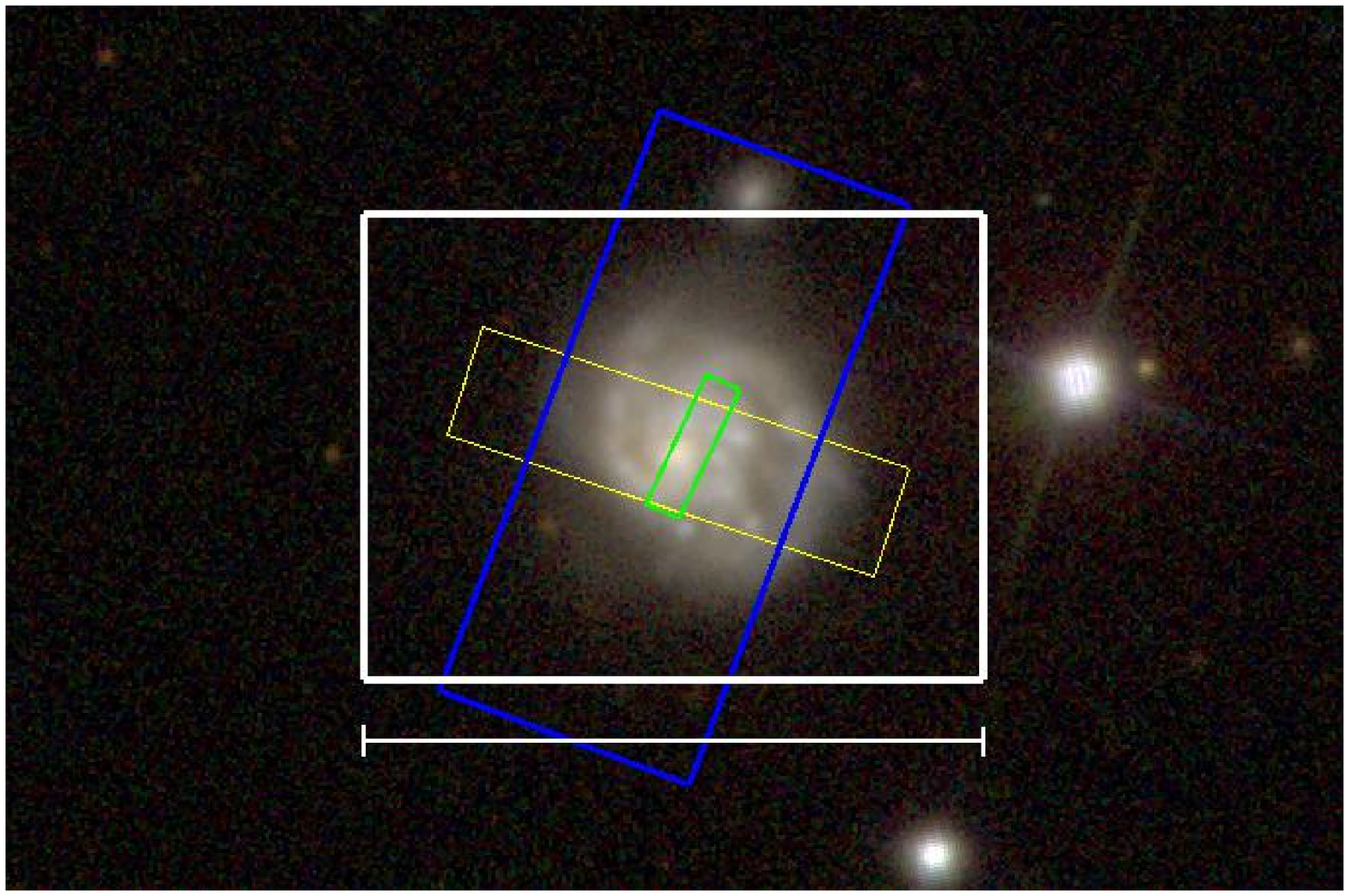}\\
\plottwo{f18_012a.ps}{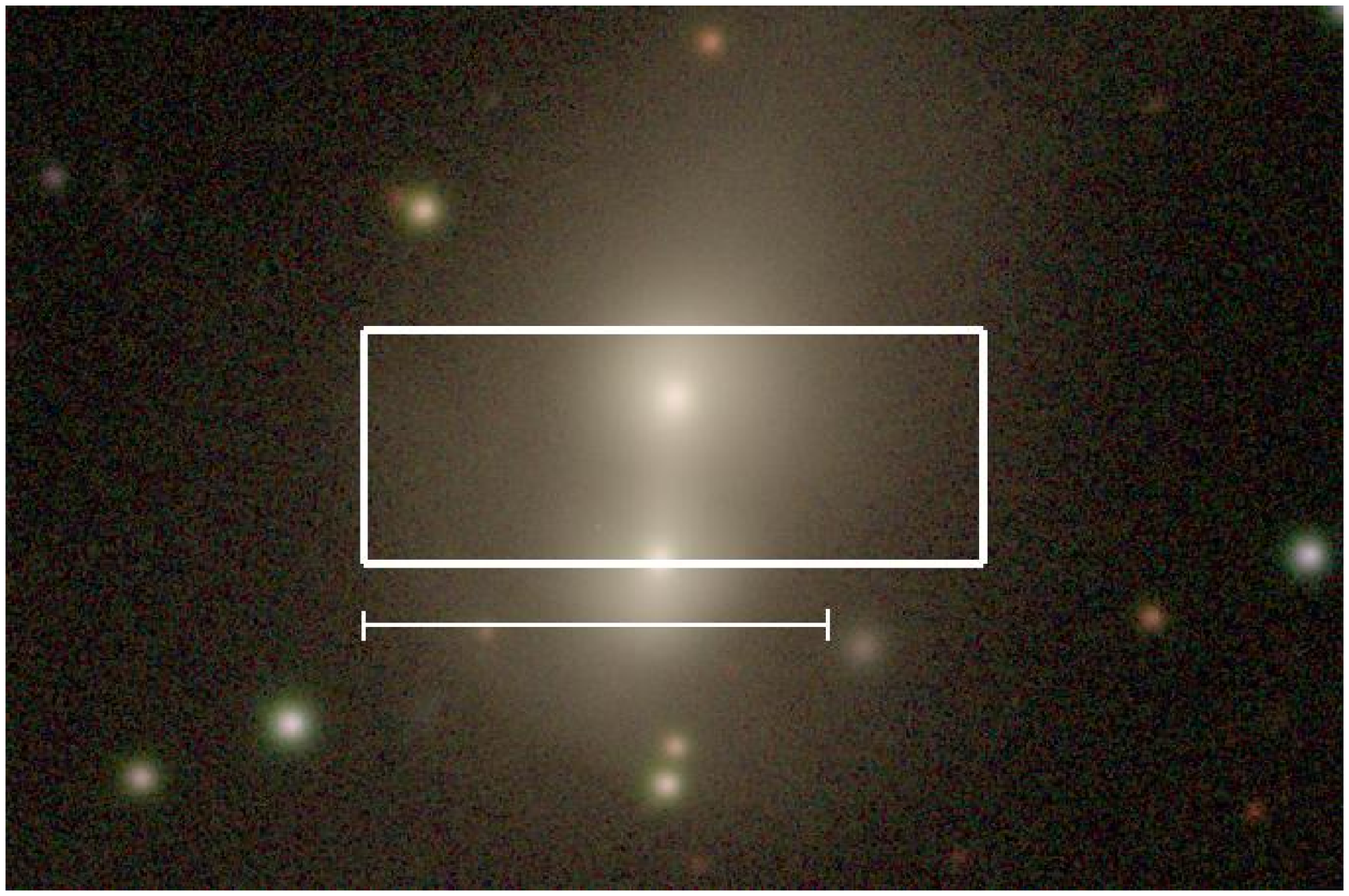}\\
\caption{Galaxy SEDs from the UV to the mid-IR. In the left-panel the observed and model spectra are shown in black and grey respectively, while the photometry used to constrain and verify the spectra is shown with red dots. In the right panel we plot the photometric aperture (thick white rectangle), the {\it Akari} extraction aperture (blue rectangle), the {\it Spitzer} SL extraction aperture (green rectangle) and the {\it Spitzer} LL extraction aperture (yellow rectangle). For galaxies with {\it Spitzer} stare mode spectra, we show a region corresponding to a quarter of the slit length. For scale, the horizontal bar denotes $1^{\prime}$. [{\it See the electronic edition of the Supplement for the complete Figure.}]}
\end{figure}

\clearpage

\begin{figure}[hbt]
\figurenum{\ref{fig:allspec} continued}
\plottwo{f18_013a.ps}{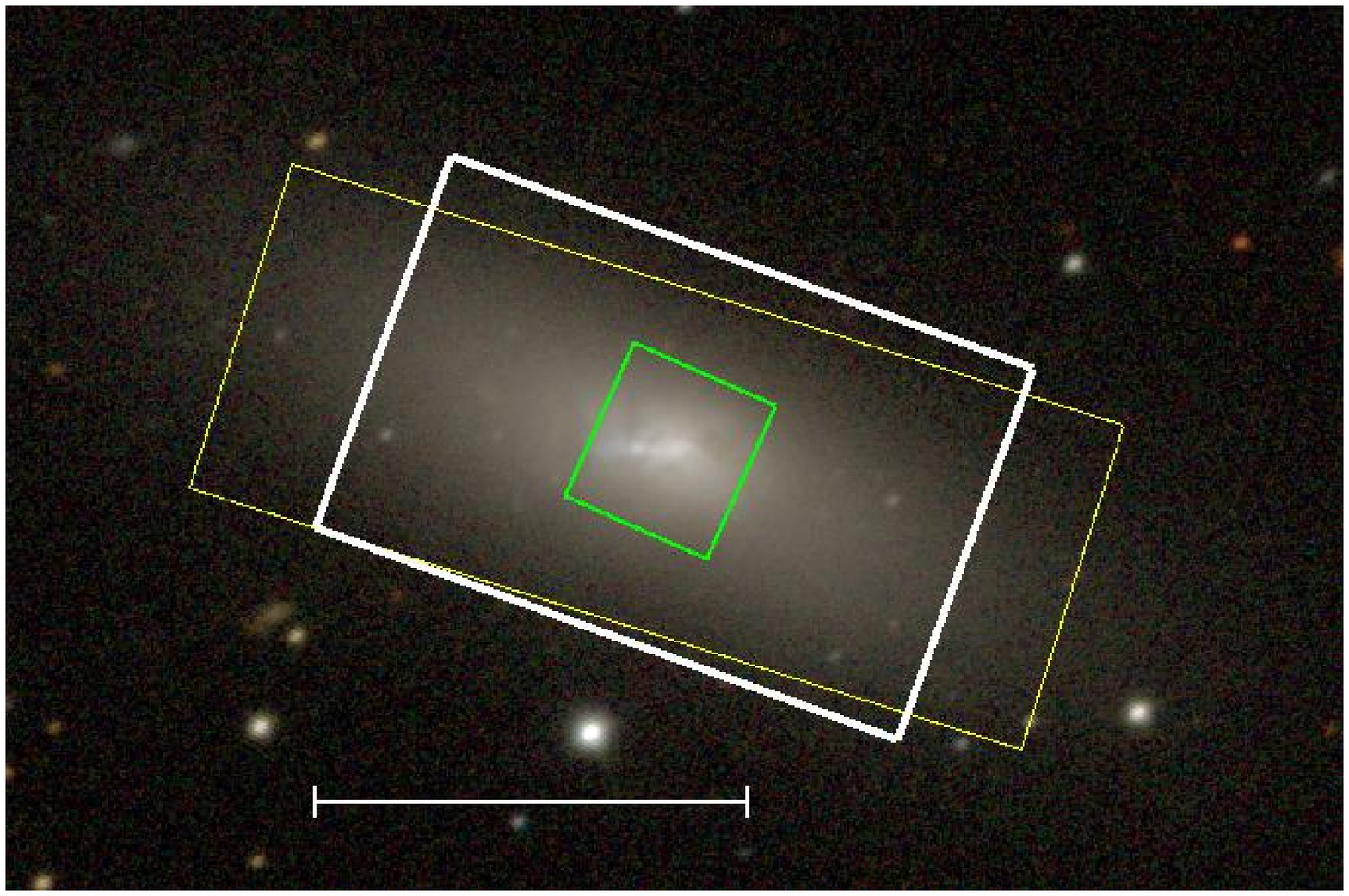}\\
\plottwo{f18_014a.ps}{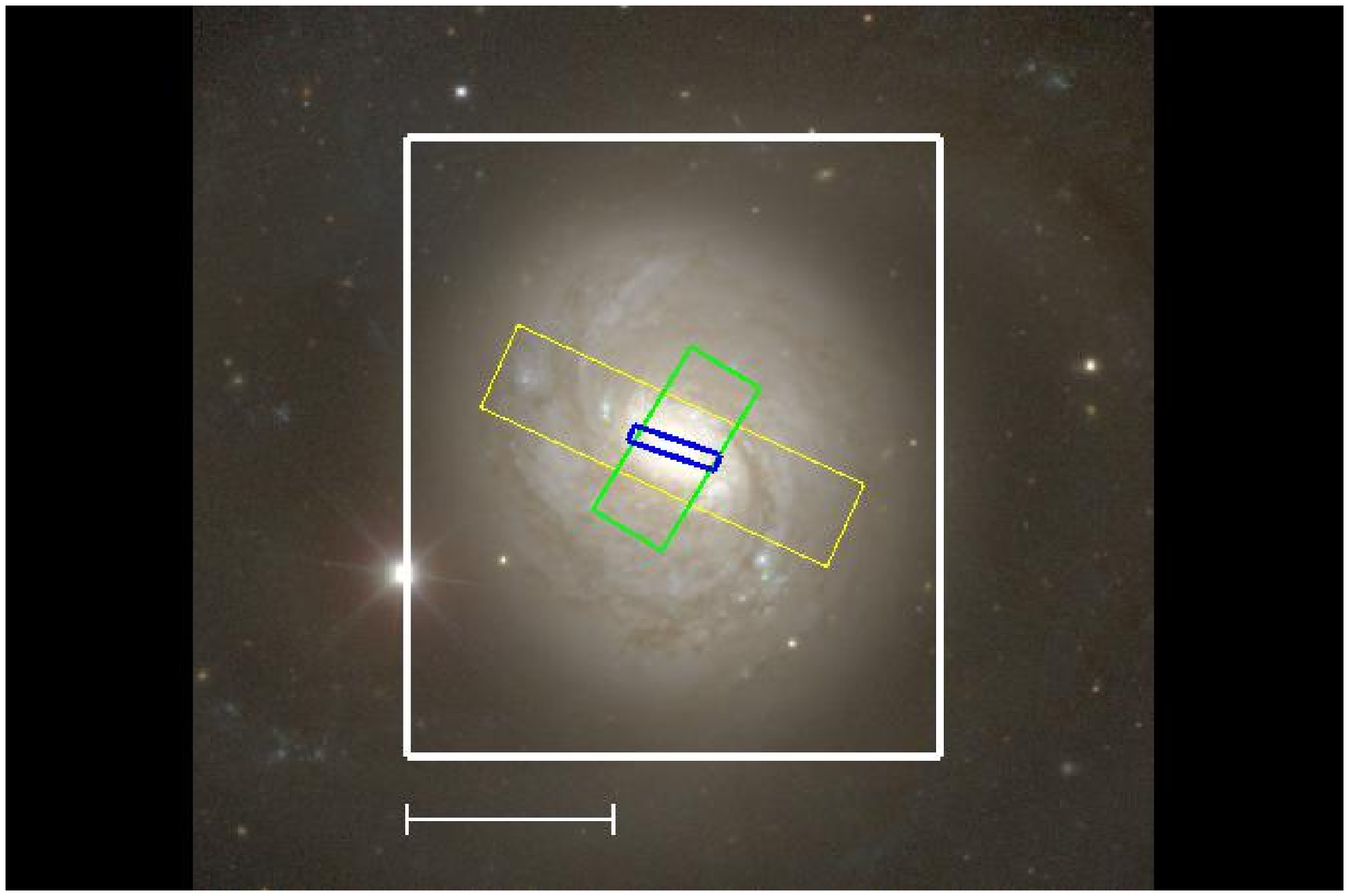}\\
\plottwo{f18_015a.ps}{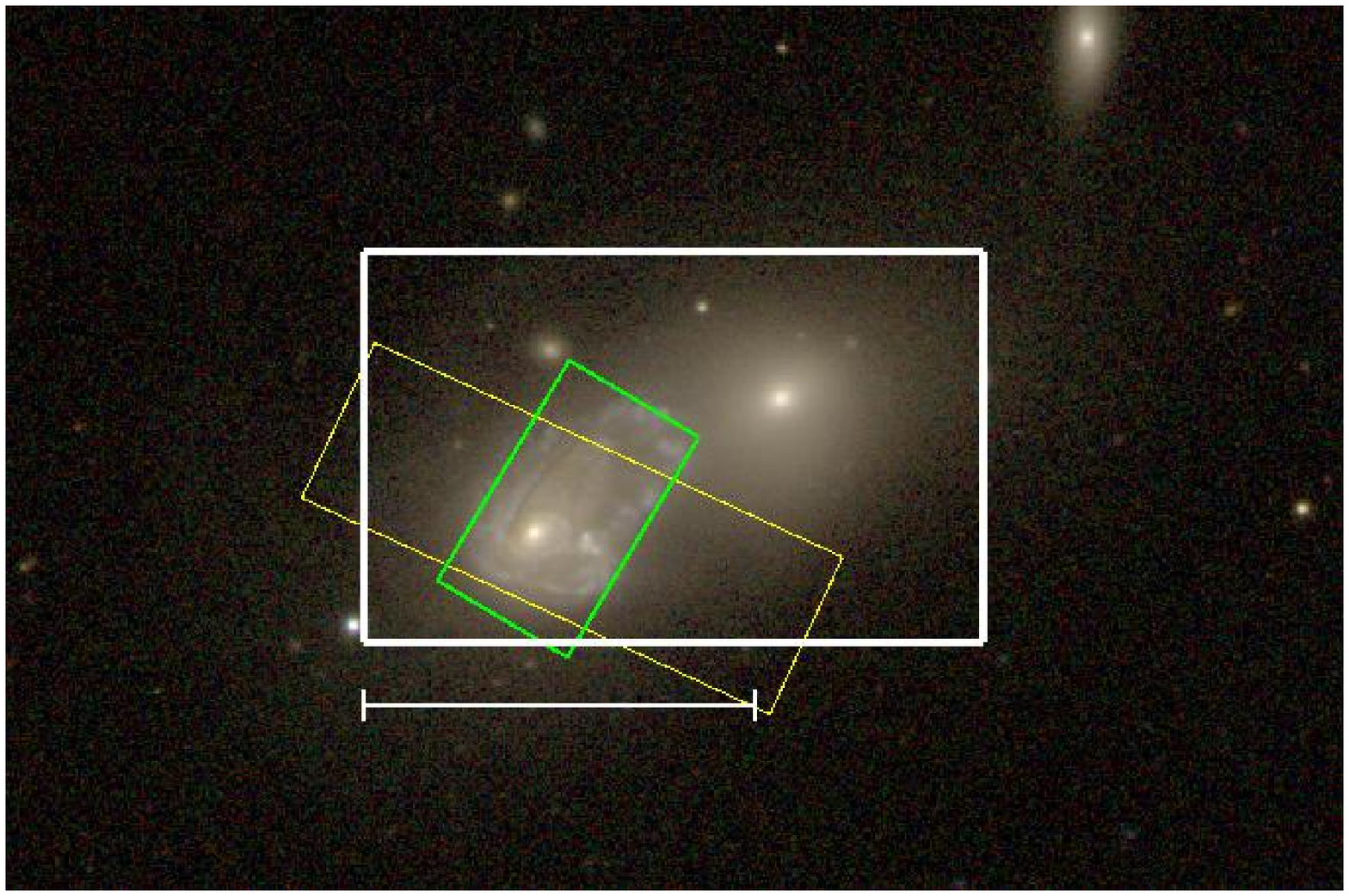}\\
\plottwo{f18_016a.ps}{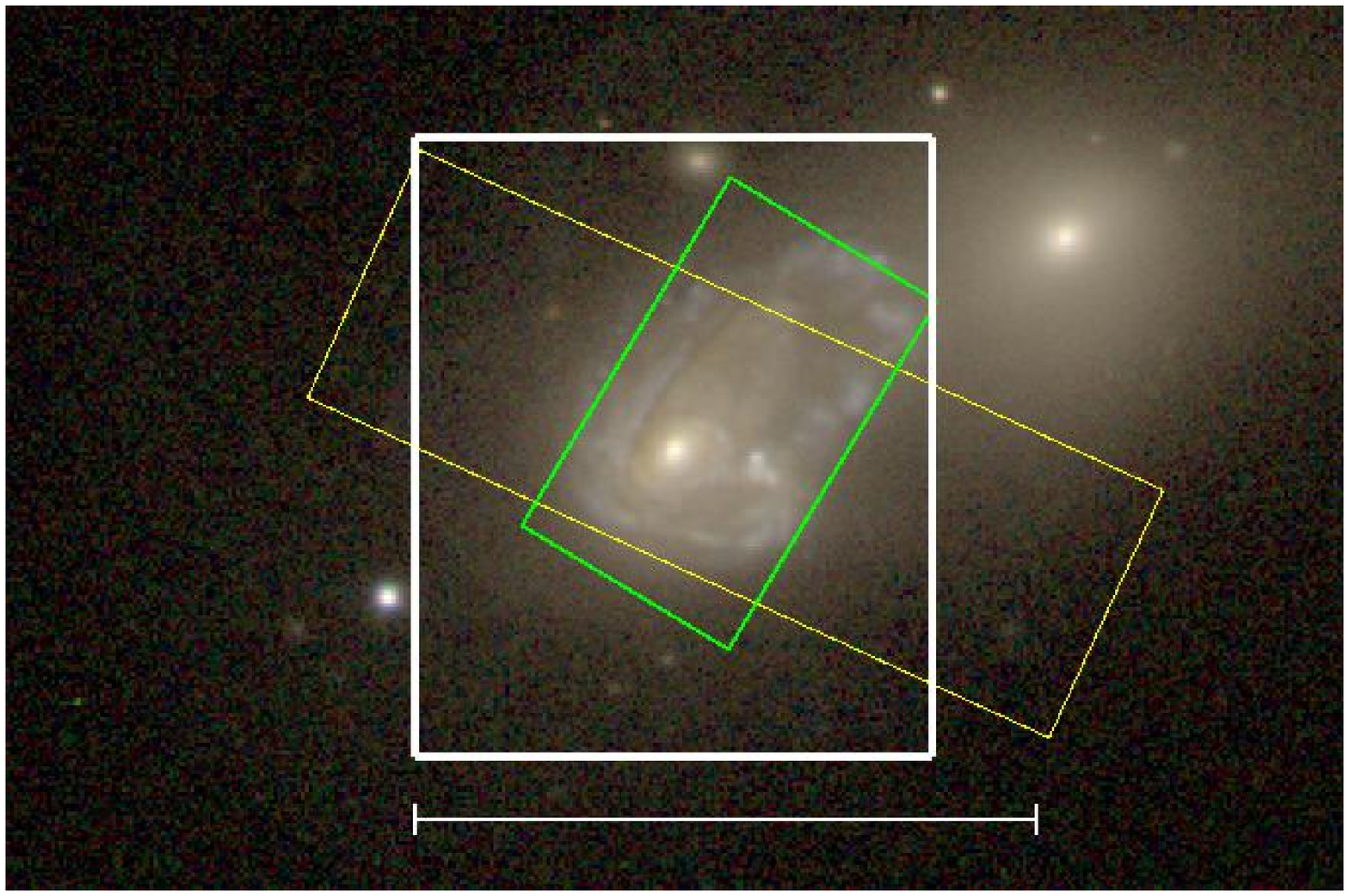}\\
\caption{Galaxy SEDs from the UV to the mid-IR. In the left-panel the observed and model spectra are shown in black and grey respectively, while the photometry used to constrain and verify the spectra is shown with red dots. In the right panel we plot the photometric aperture (thick white rectangle), the {\it Akari} extraction aperture (blue rectangle), the {\it Spitzer} SL extraction aperture (green rectangle) and the {\it Spitzer} LL extraction aperture (yellow rectangle). For galaxies with {\it Spitzer} stare mode spectra, we show a region corresponding to a quarter of the slit length. For scale, the horizontal bar denotes $1^{\prime}$. [{\it See the electronic edition of the Supplement for the complete Figure.}]}
\end{figure}

\clearpage

\begin{figure}[hbt]
\figurenum{\ref{fig:allspec} continued}
\plottwo{f18_017a.ps}{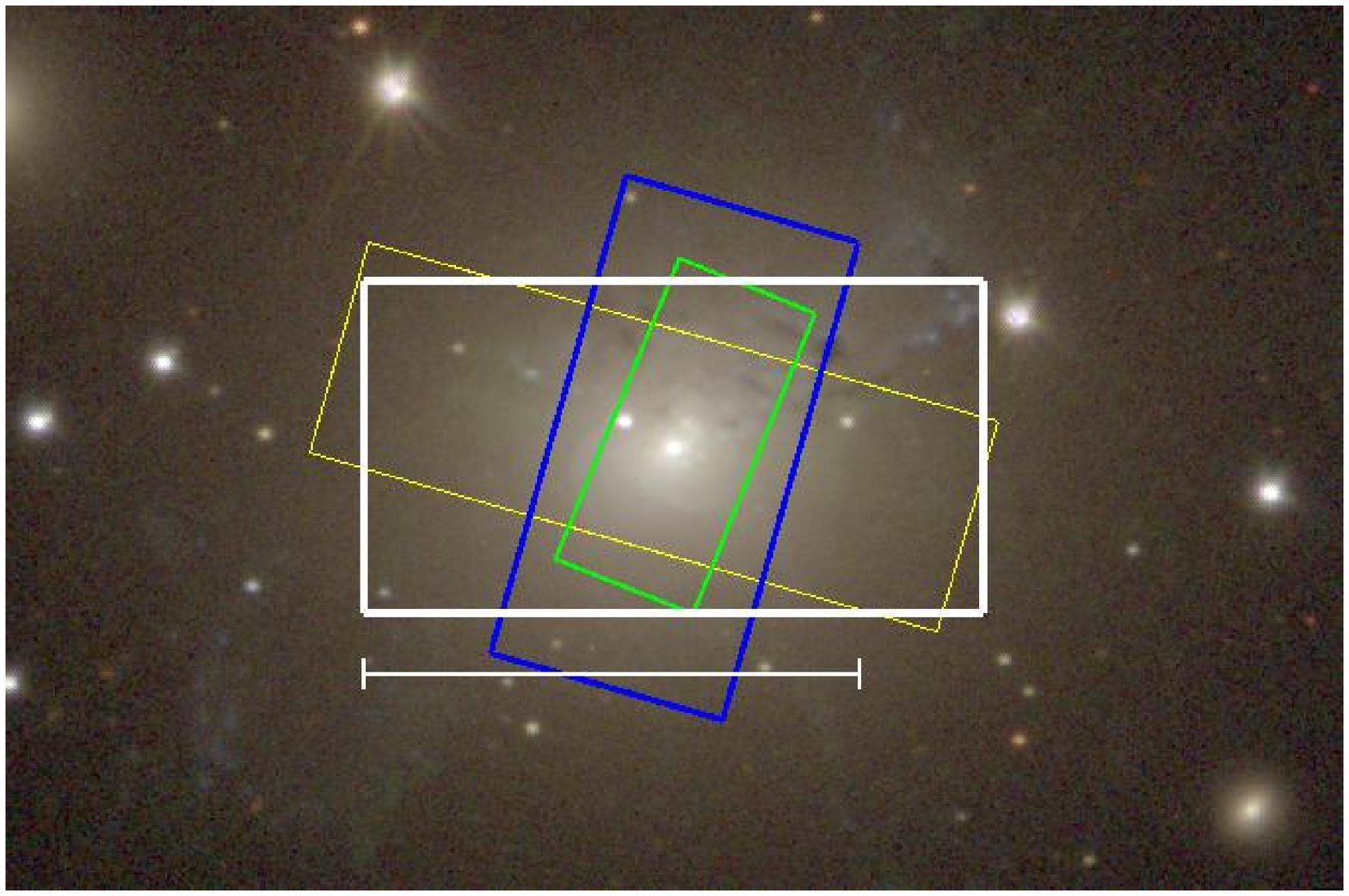}\\
\plottwo{f18_018a.ps}{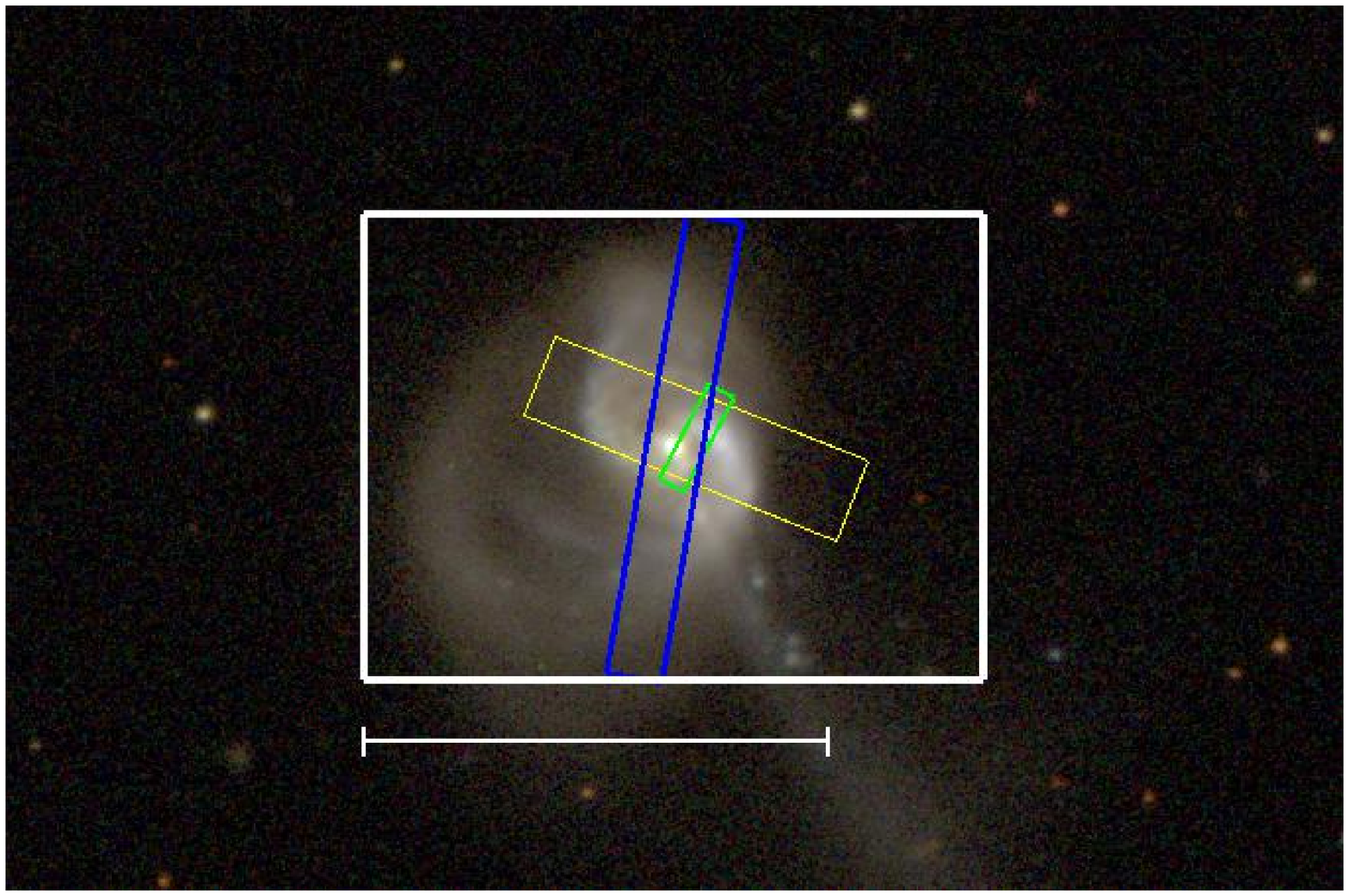}\\
\plottwo{f18_019a.ps}{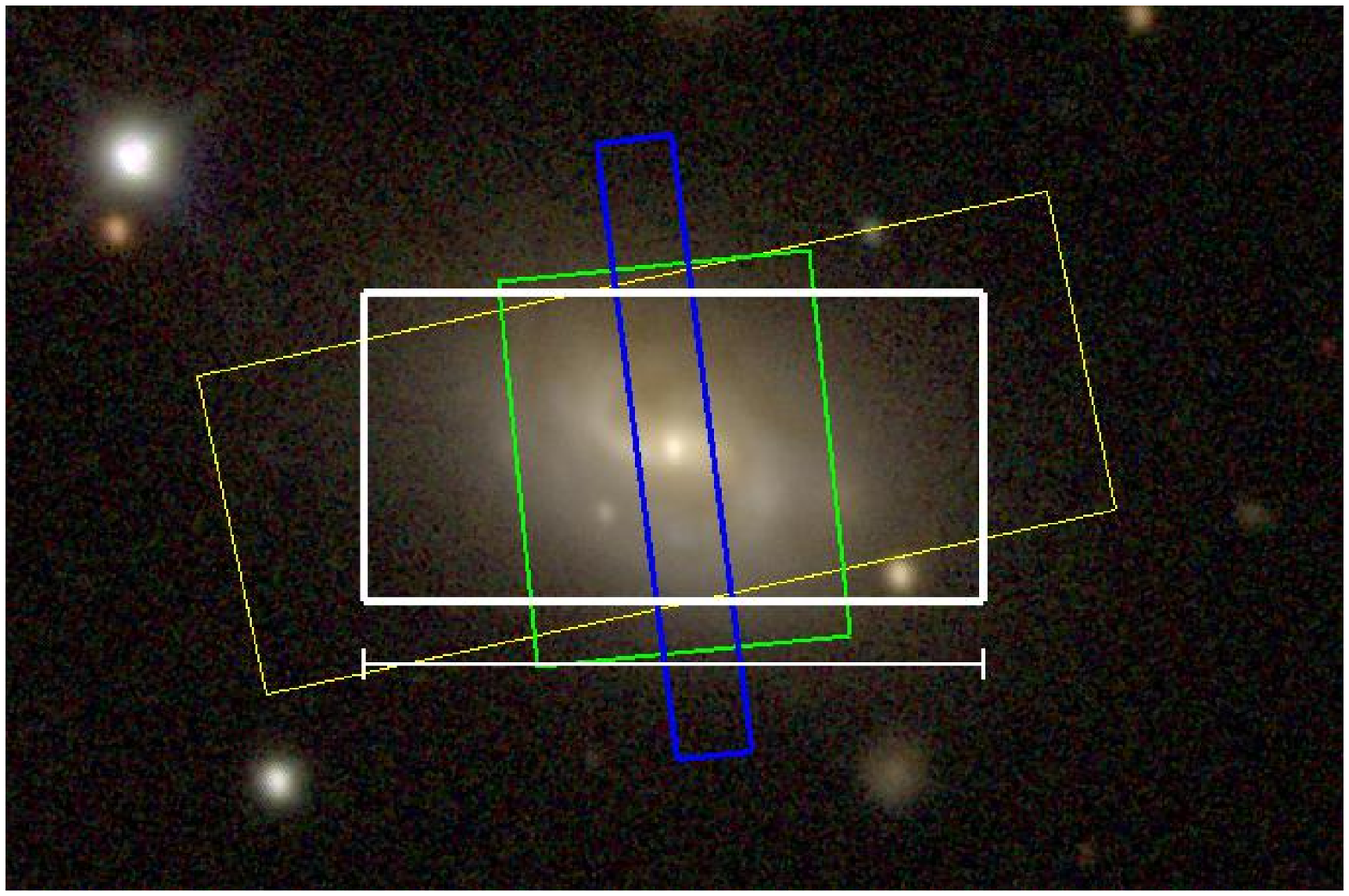}\\
\plottwo{f18_020a.ps}{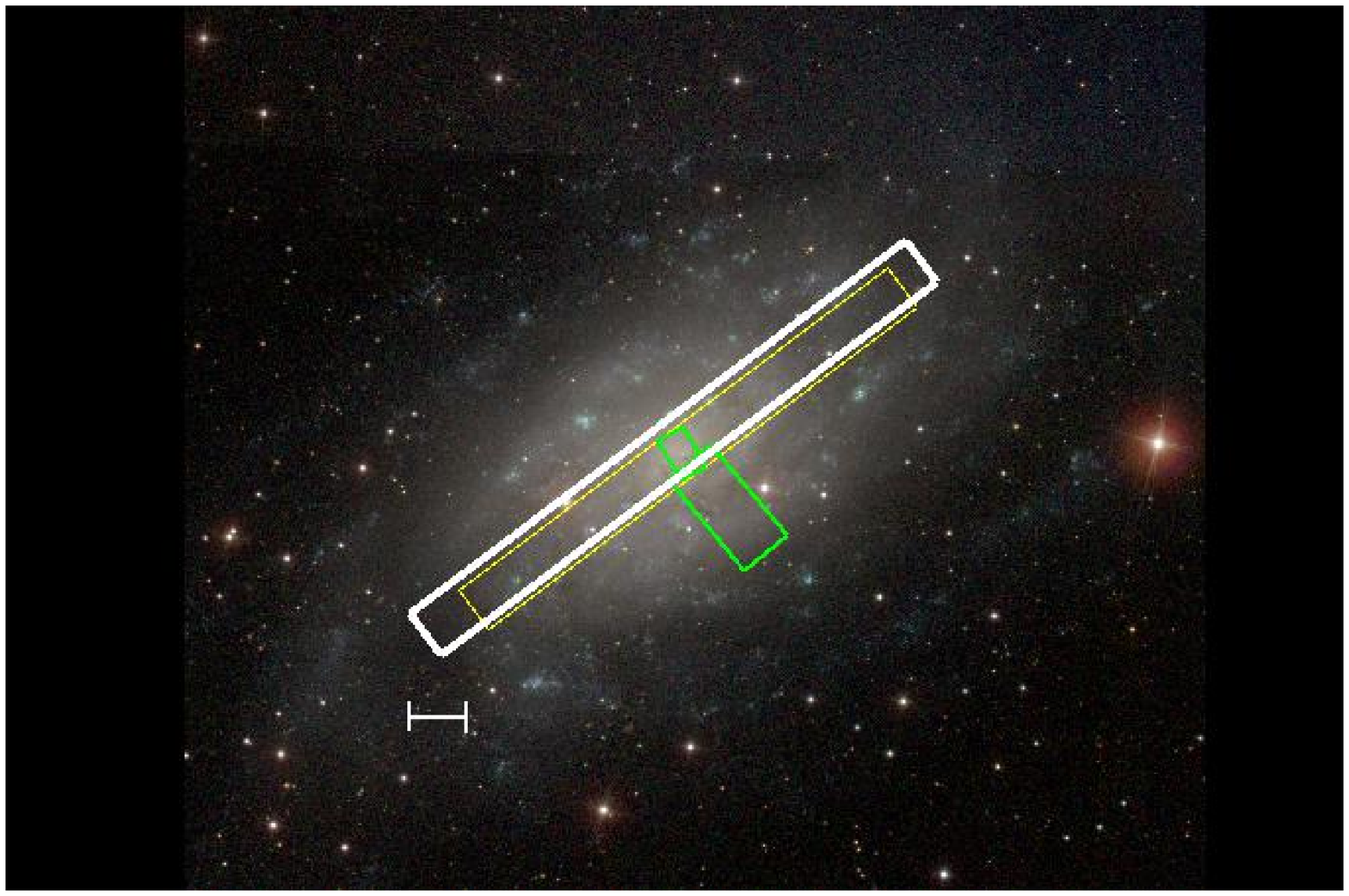}\\
\caption{Galaxy SEDs from the UV to the mid-IR. In the left-panel the observed and model spectra are shown in black and grey respectively, while the photometry used to constrain and verify the spectra is shown with red dots. In the right panel we plot the photometric aperture (thick white rectangle), the {\it Akari} extraction aperture (blue rectangle), the {\it Spitzer} SL extraction aperture (green rectangle) and the {\it Spitzer} LL extraction aperture (yellow rectangle). For galaxies with {\it Spitzer} stare mode spectra, we show a region corresponding to a quarter of the slit length. For scale, the horizontal bar denotes $1^{\prime}$. [{\it See the electronic edition of the Supplement for the complete Figure.}]}
\end{figure}

\clearpage

\begin{figure}[hbt]
\figurenum{\ref{fig:allspec} continued}
\plottwo{f18_021a.ps}{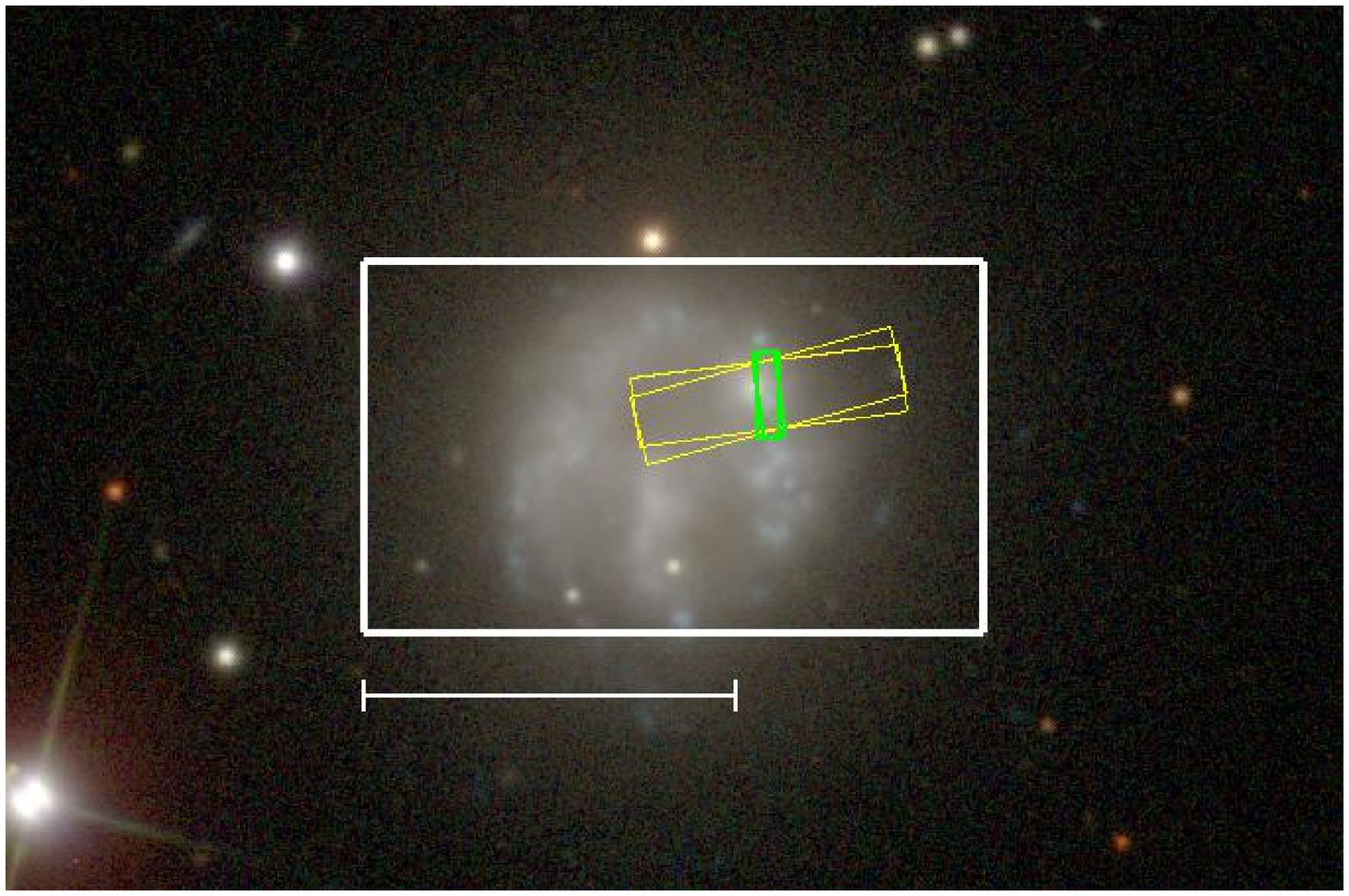}\\
\plottwo{f18_022a.ps}{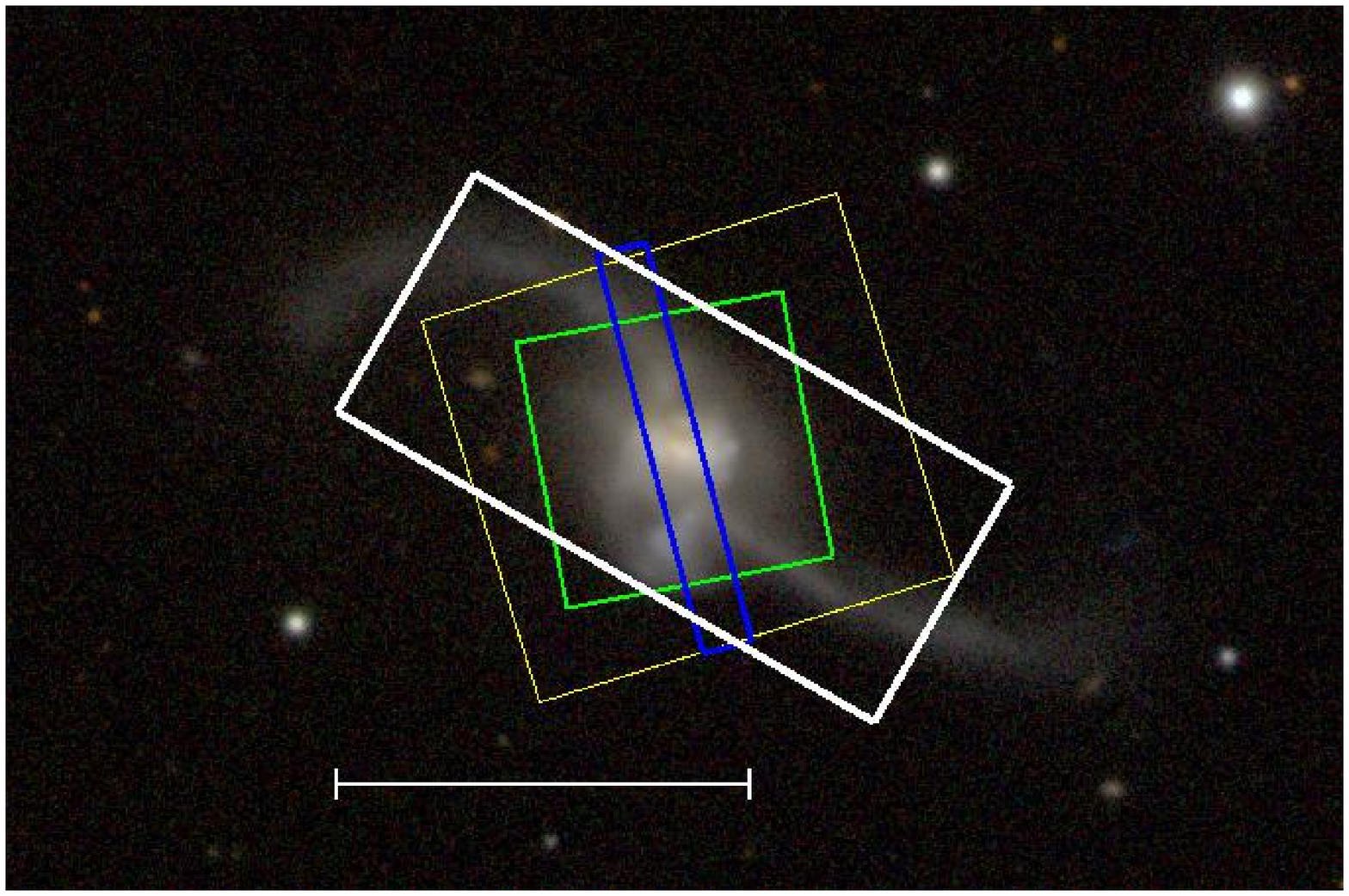}\\
\plottwo{f18_023a.ps}{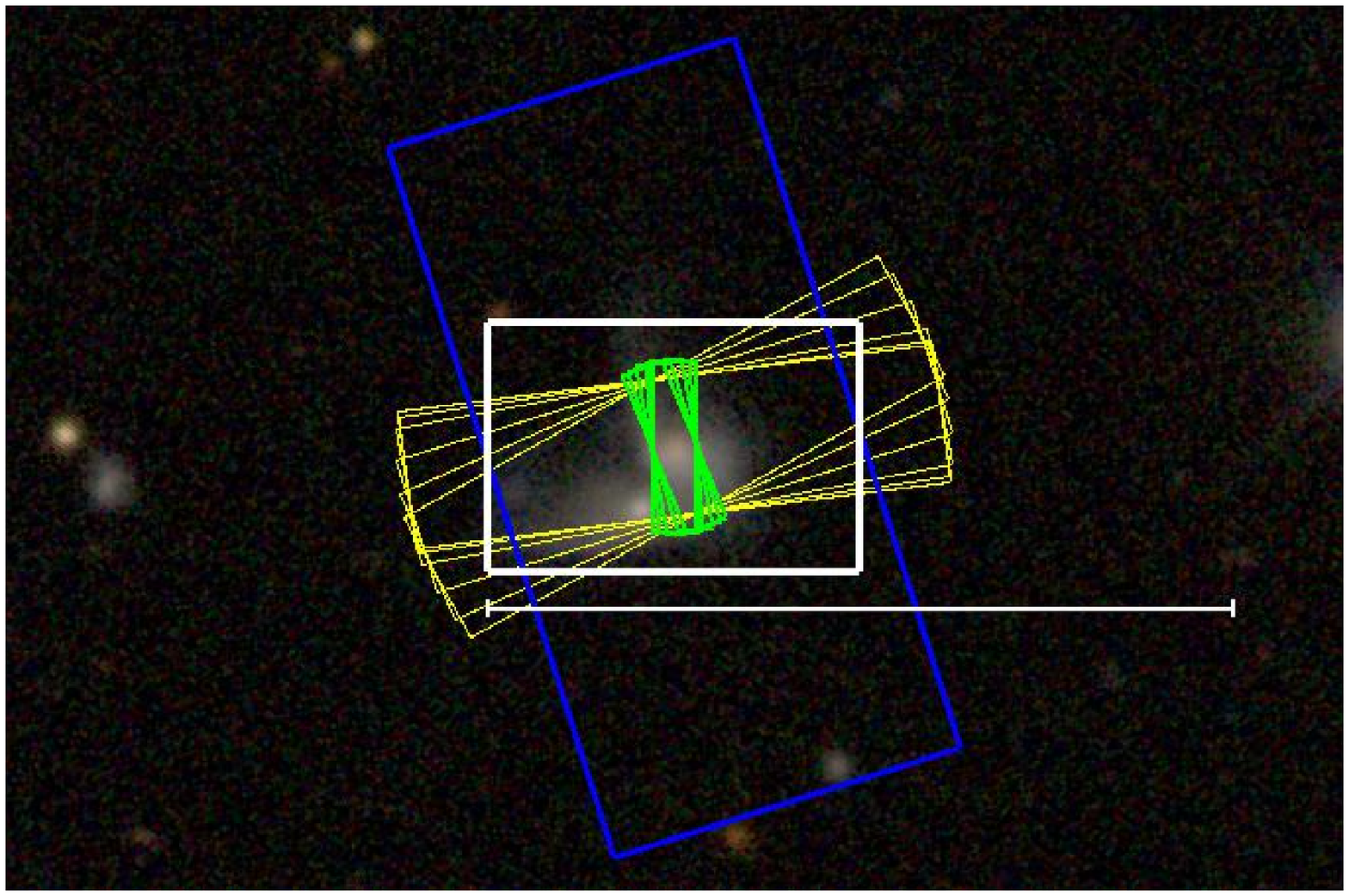}\\
\plottwo{f18_024a.ps}{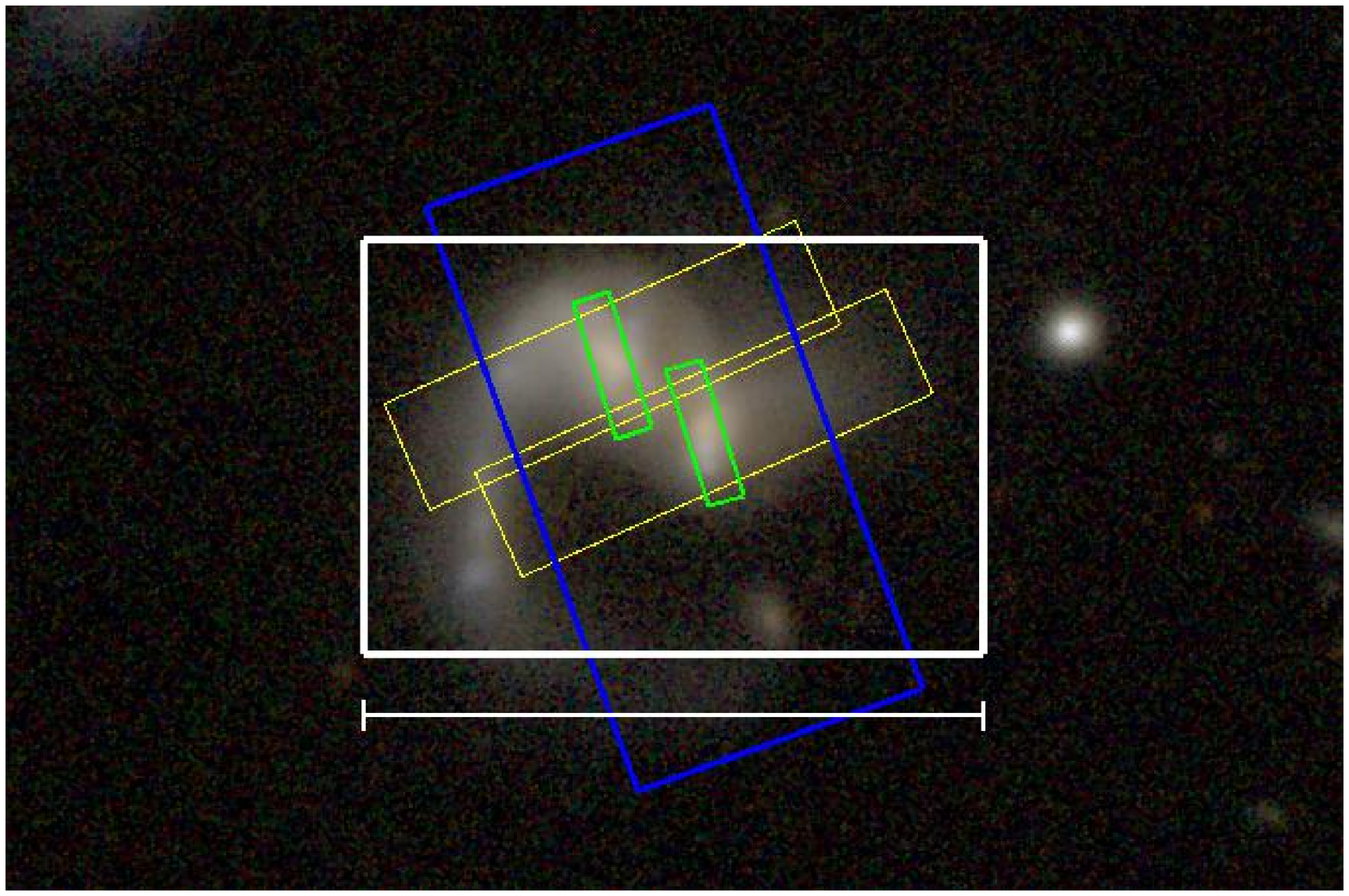}\\
\caption{Galaxy SEDs from the UV to the mid-IR. In the left-panel the observed and model spectra are shown in black and grey respectively, while the photometry used to constrain and verify the spectra is shown with red dots. In the right panel we plot the photometric aperture (thick white rectangle), the {\it Akari} extraction aperture (blue rectangle), the {\it Spitzer} SL extraction aperture (green rectangle) and the {\it Spitzer} LL extraction aperture (yellow rectangle). For galaxies with {\it Spitzer} stare mode spectra, we show a region corresponding to a quarter of the slit length. For scale, the horizontal bar denotes $1^{\prime}$. [{\it See the electronic edition of the Supplement for the complete Figure.}]}
\end{figure}

\clearpage

\begin{figure}[hbt]
\figurenum{\ref{fig:allspec} continued}
\plottwo{f18_025a.ps}{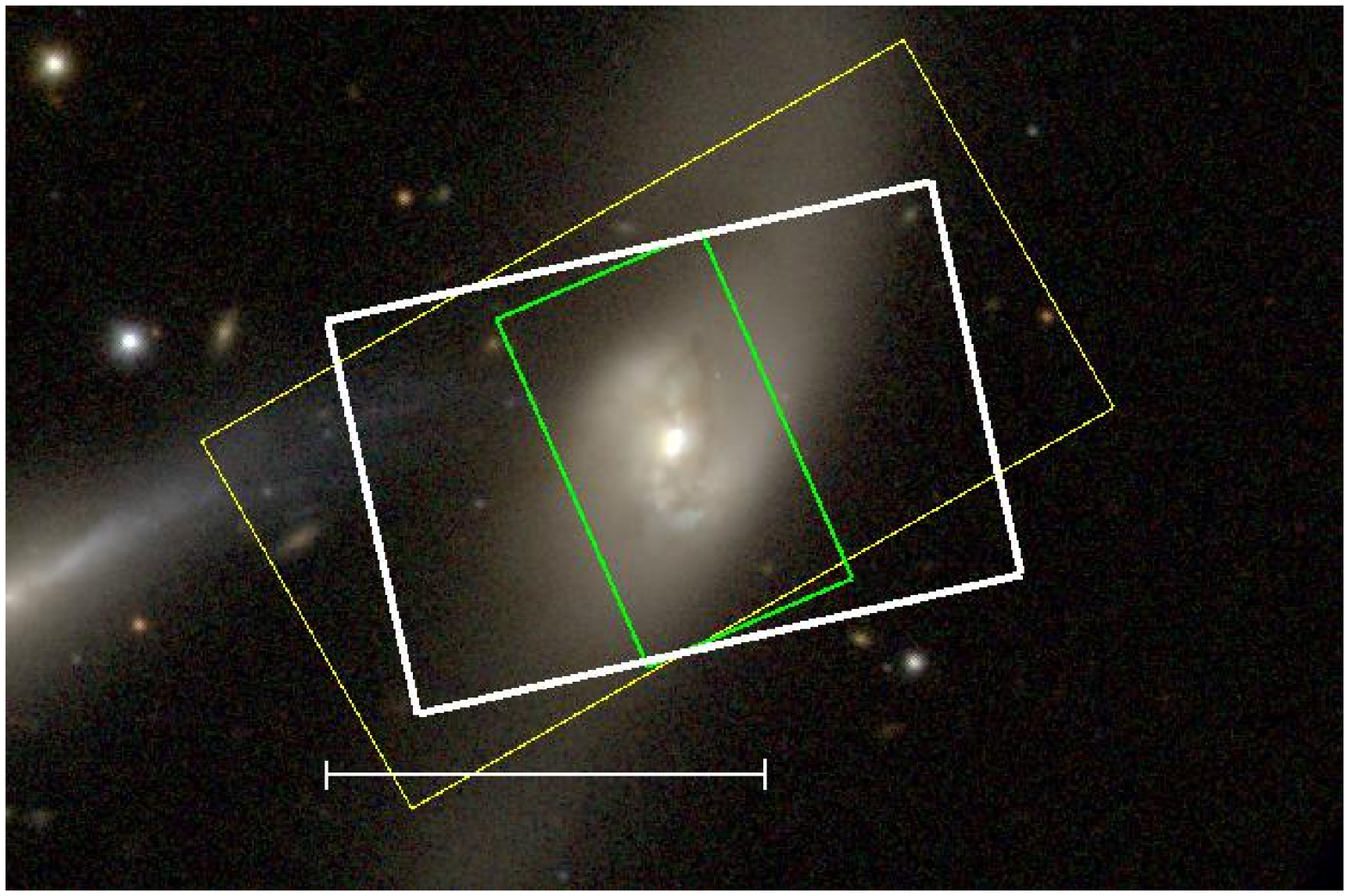}\\
\plottwo{f18_026a.ps}{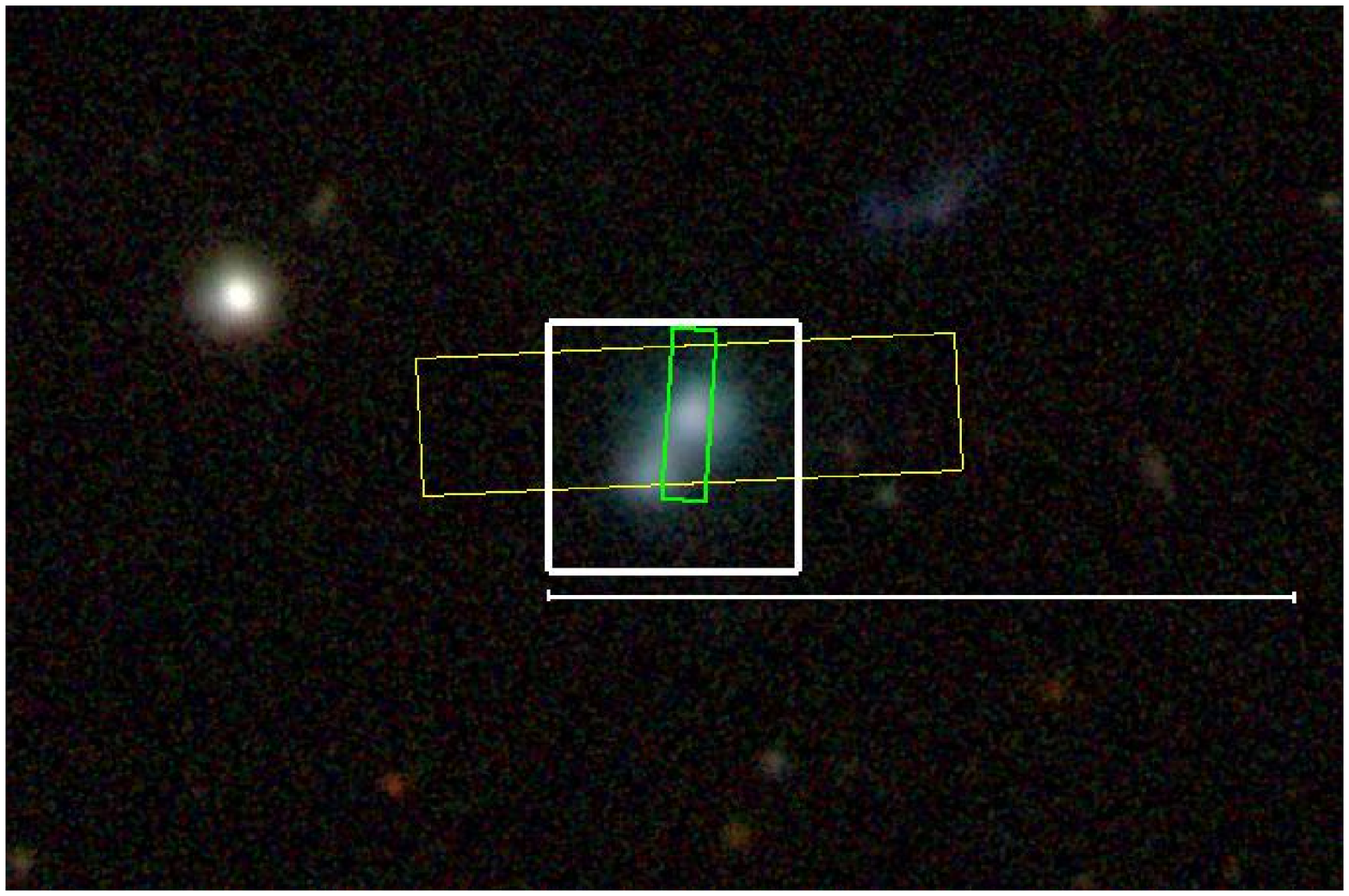}\\
\plottwo{f18_027a.ps}{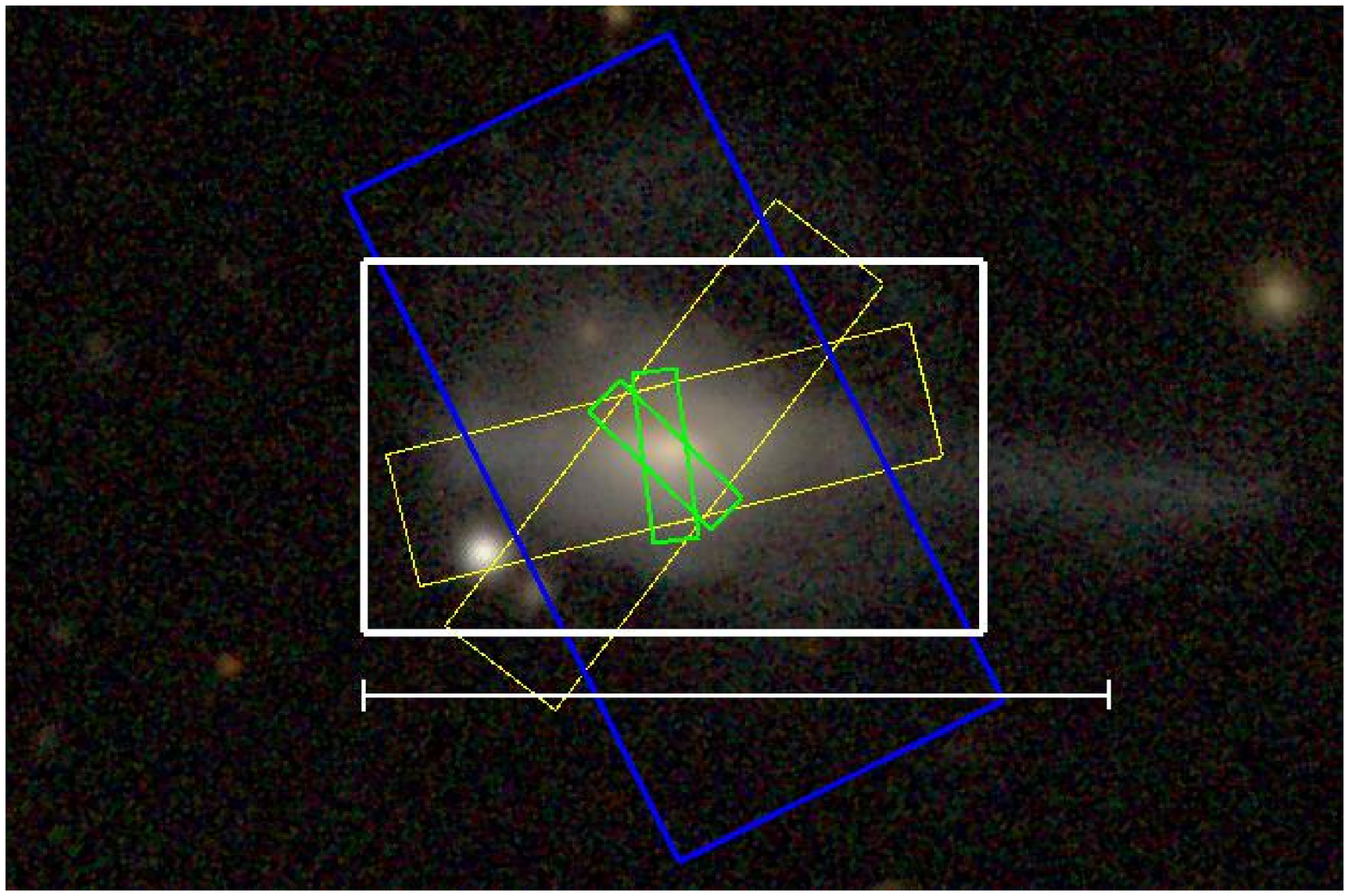}\\
\plottwo{f18_028a.ps}{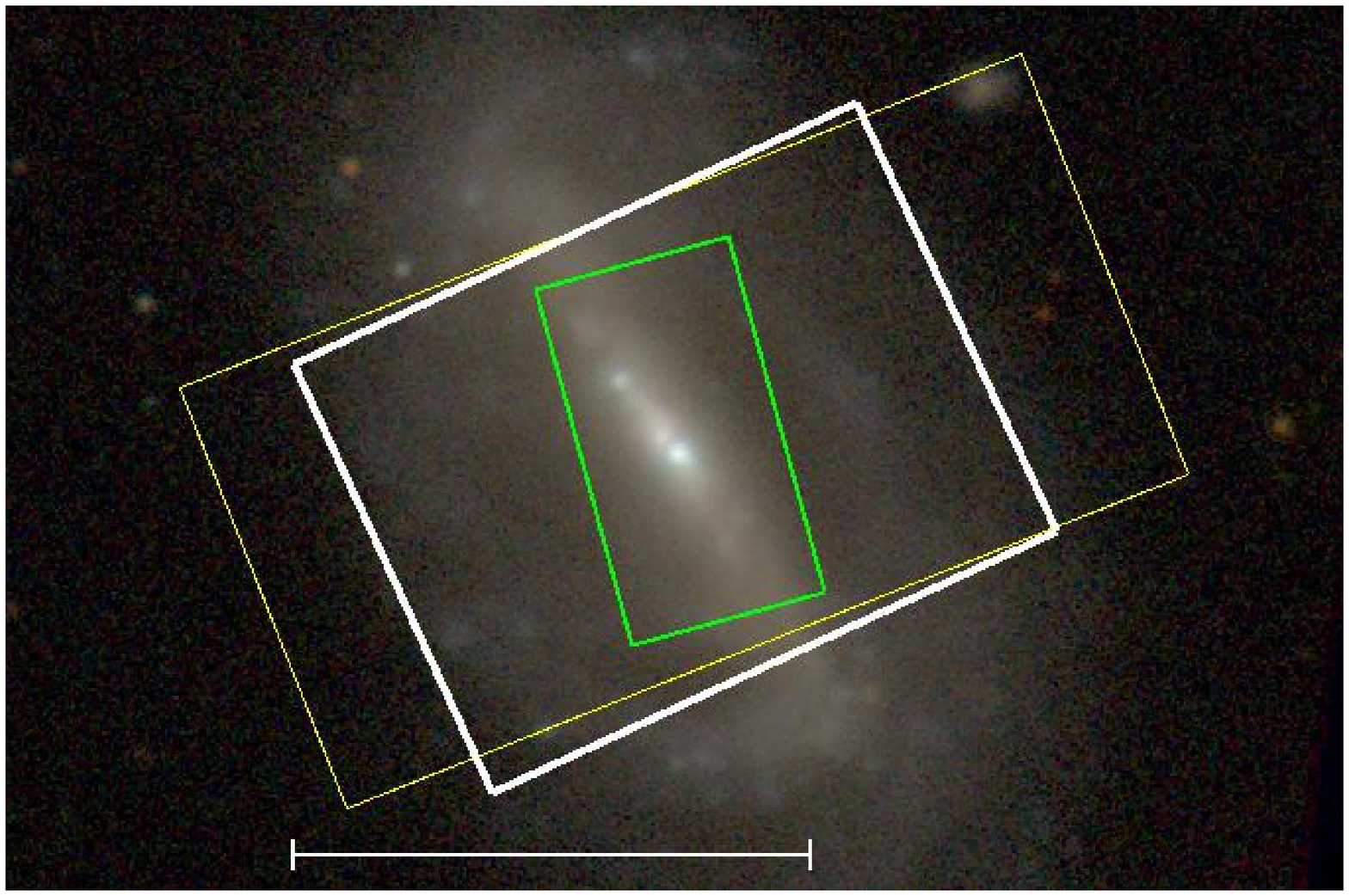}\\
\caption{Galaxy SEDs from the UV to the mid-IR. In the left-panel the observed and model spectra are shown in black and grey respectively, while the photometry used to constrain and verify the spectra is shown with red dots. In the right panel we plot the photometric aperture (thick white rectangle), the {\it Akari} extraction aperture (blue rectangle), the {\it Spitzer} SL extraction aperture (green rectangle) and the {\it Spitzer} LL extraction aperture (yellow rectangle). For galaxies with {\it Spitzer} stare mode spectra, we show a region corresponding to a quarter of the slit length. For scale, the horizontal bar denotes $1^{\prime}$. [{\it See the electronic edition of the Supplement for the complete Figure.}]}
\end{figure}

\clearpage

\begin{figure}[hbt]
\figurenum{\ref{fig:allspec} continued}
\plottwo{f18_029a.ps}{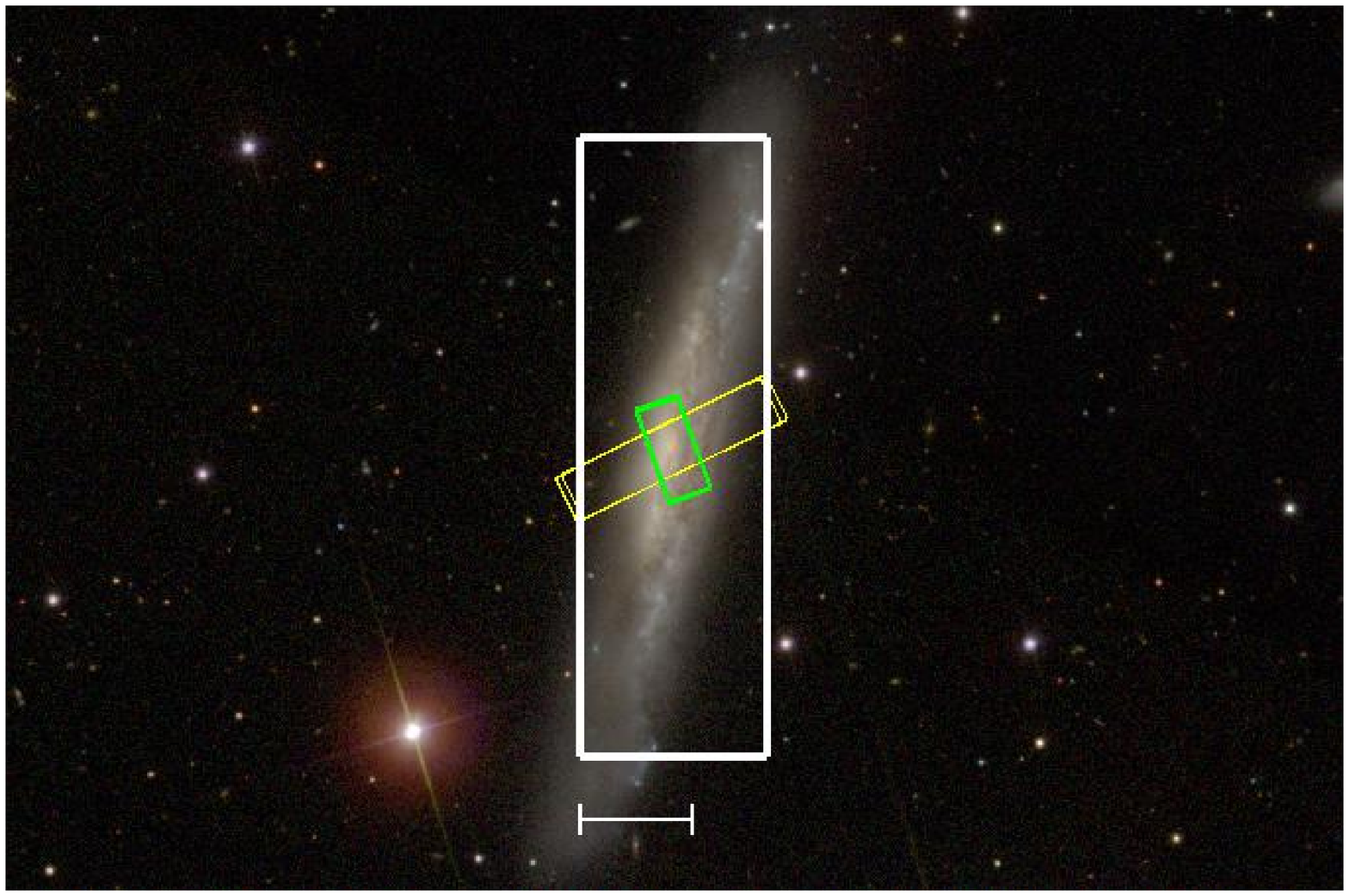}\\
\plottwo{f18_030a.ps}{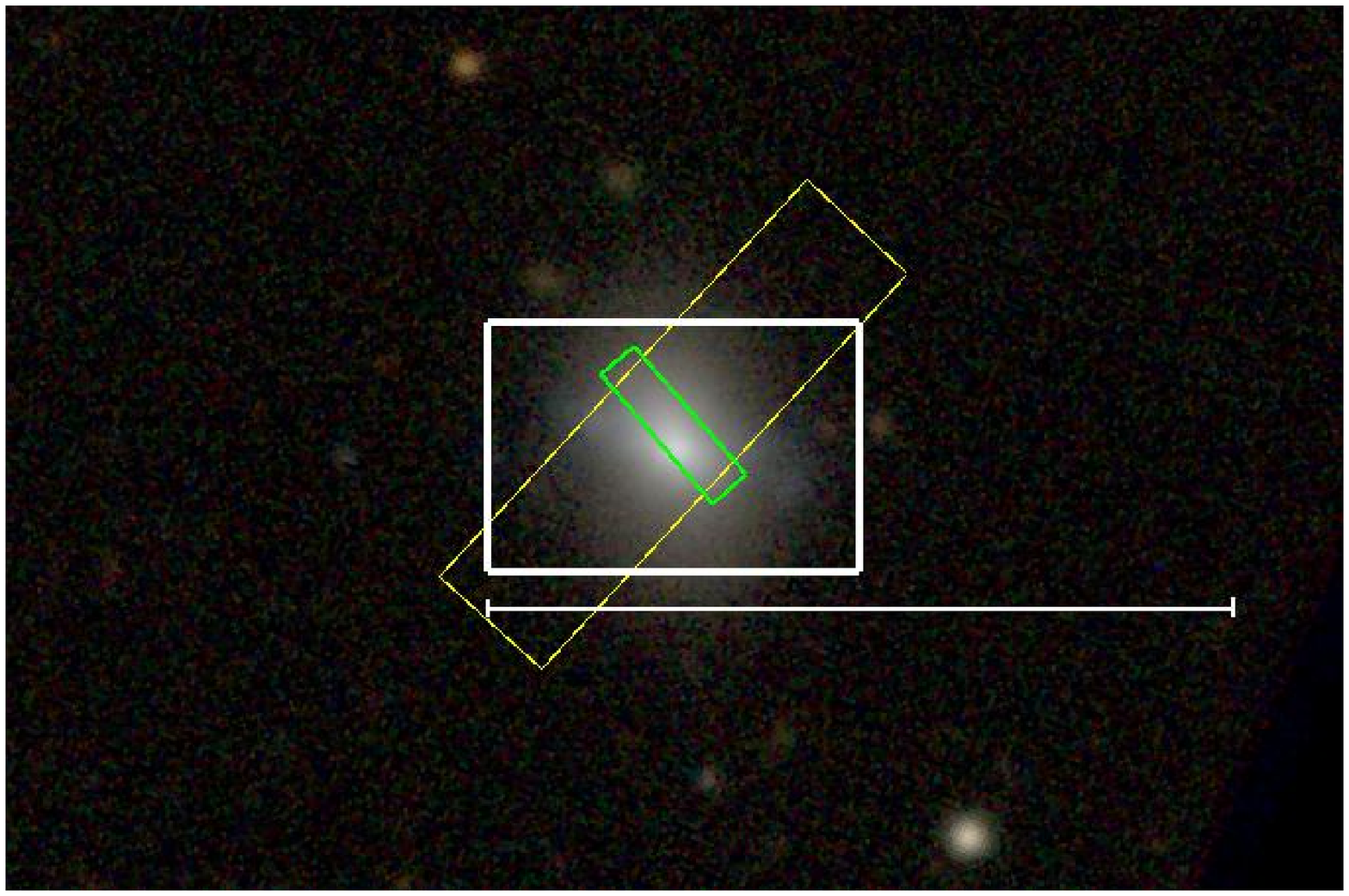}\\
\plottwo{f18_031a.ps}{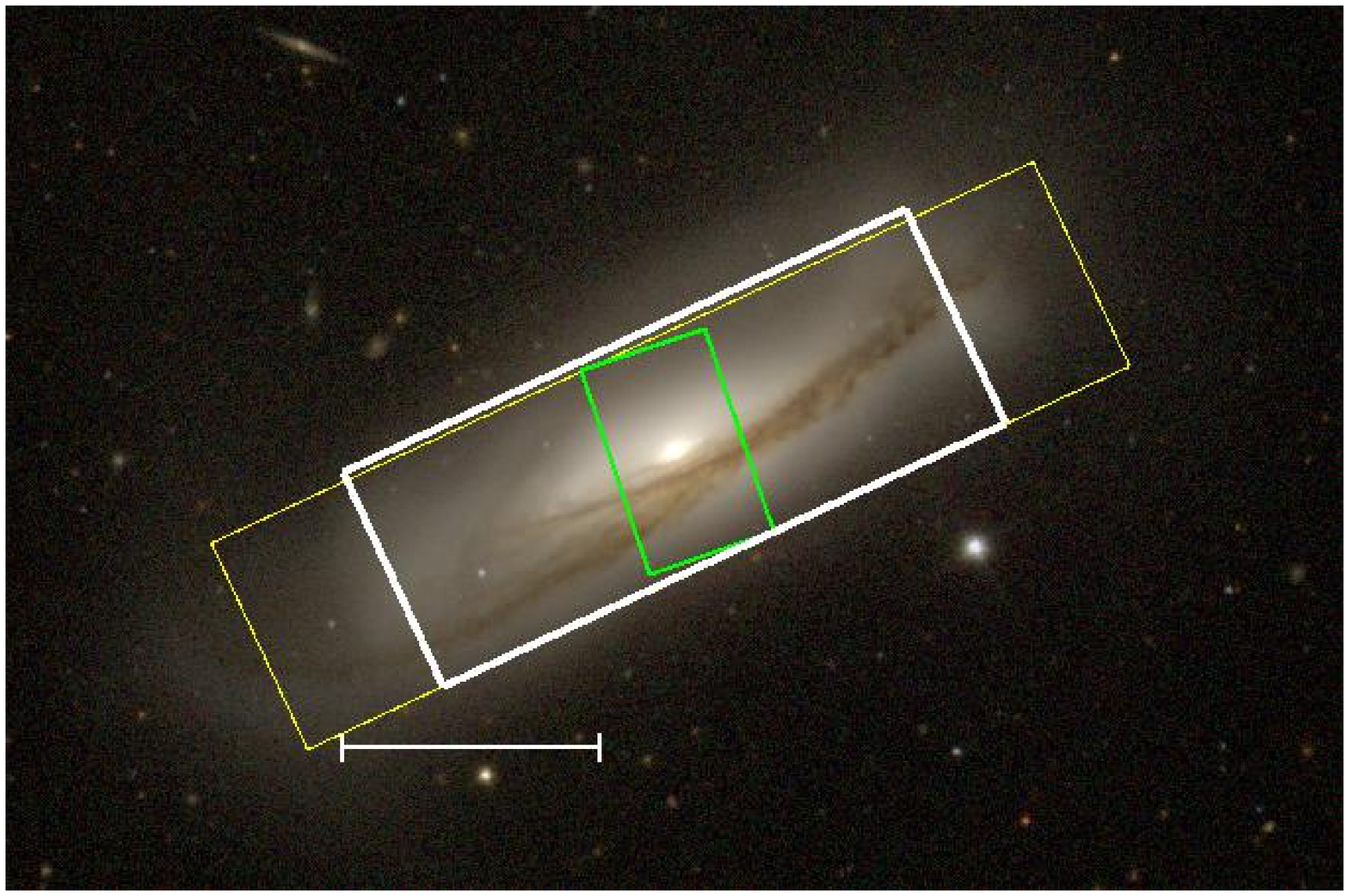}\\
\plottwo{f18_032a.ps}{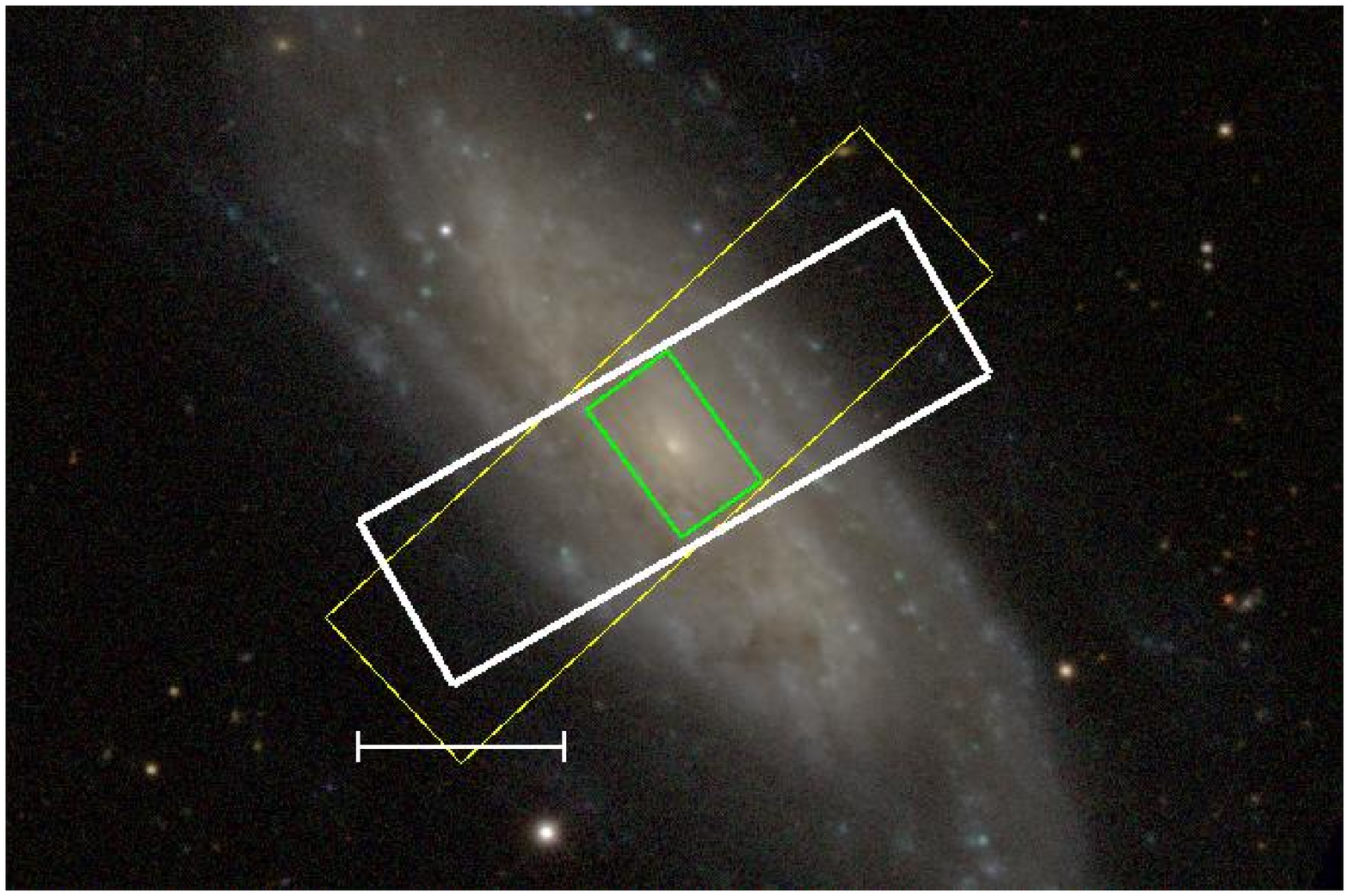}\\
\caption{Galaxy SEDs from the UV to the mid-IR. In the left-panel the observed and model spectra are shown in black and grey respectively, while the photometry used to constrain and verify the spectra is shown with red dots. In the right panel we plot the photometric aperture (thick white rectangle), the {\it Akari} extraction aperture (blue rectangle), the {\it Spitzer} SL extraction aperture (green rectangle) and the {\it Spitzer} LL extraction aperture (yellow rectangle). For galaxies with {\it Spitzer} stare mode spectra, we show a region corresponding to a quarter of the slit length. For scale, the horizontal bar denotes $1^{\prime}$. [{\it See the electronic edition of the Supplement for the complete Figure.}]}
\end{figure}

\clearpage

\begin{figure}[hbt]
\figurenum{\ref{fig:allspec} continued}
\plottwo{f18_033a.ps}{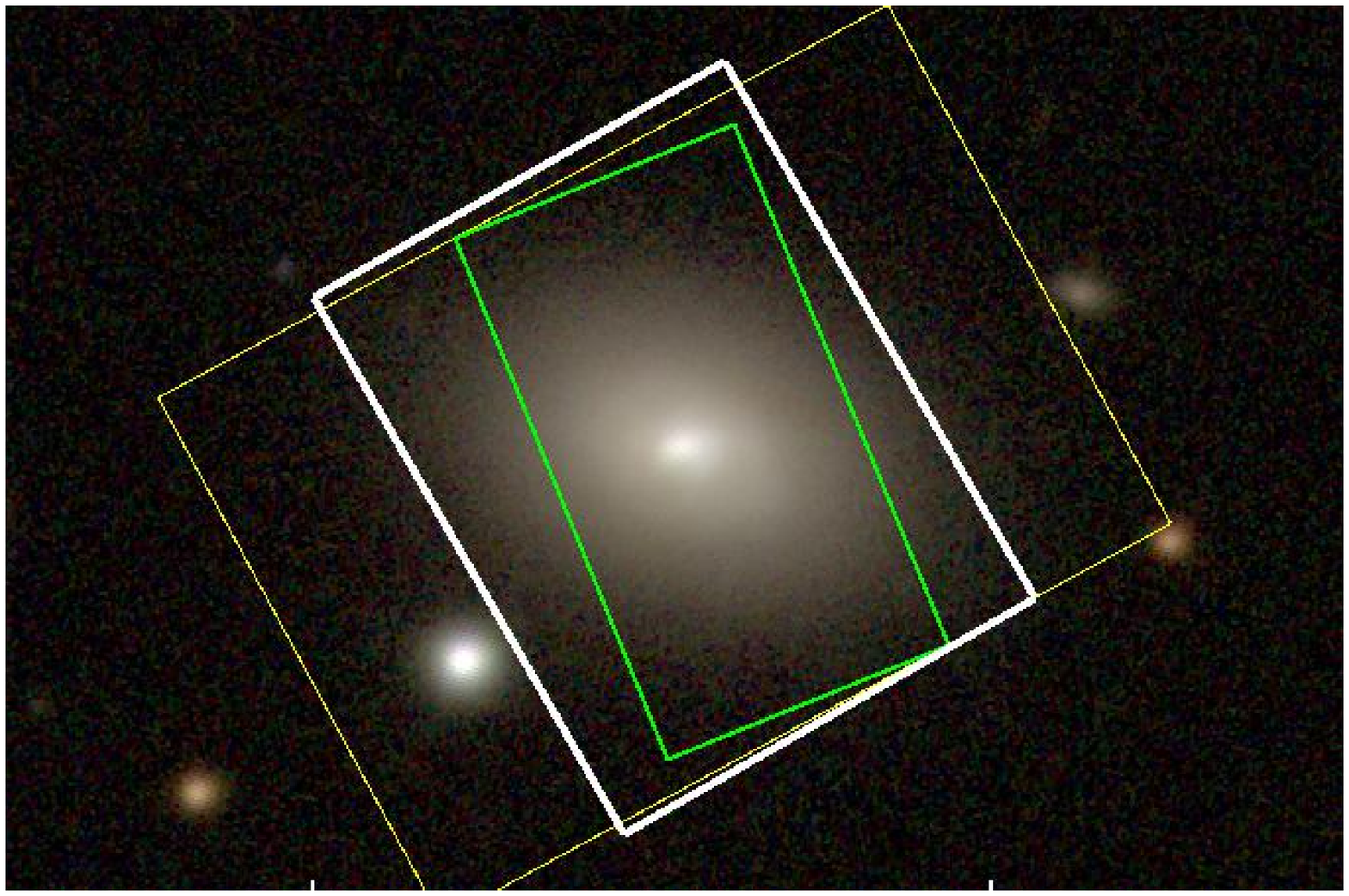}\\
\plottwo{f18_034a.ps}{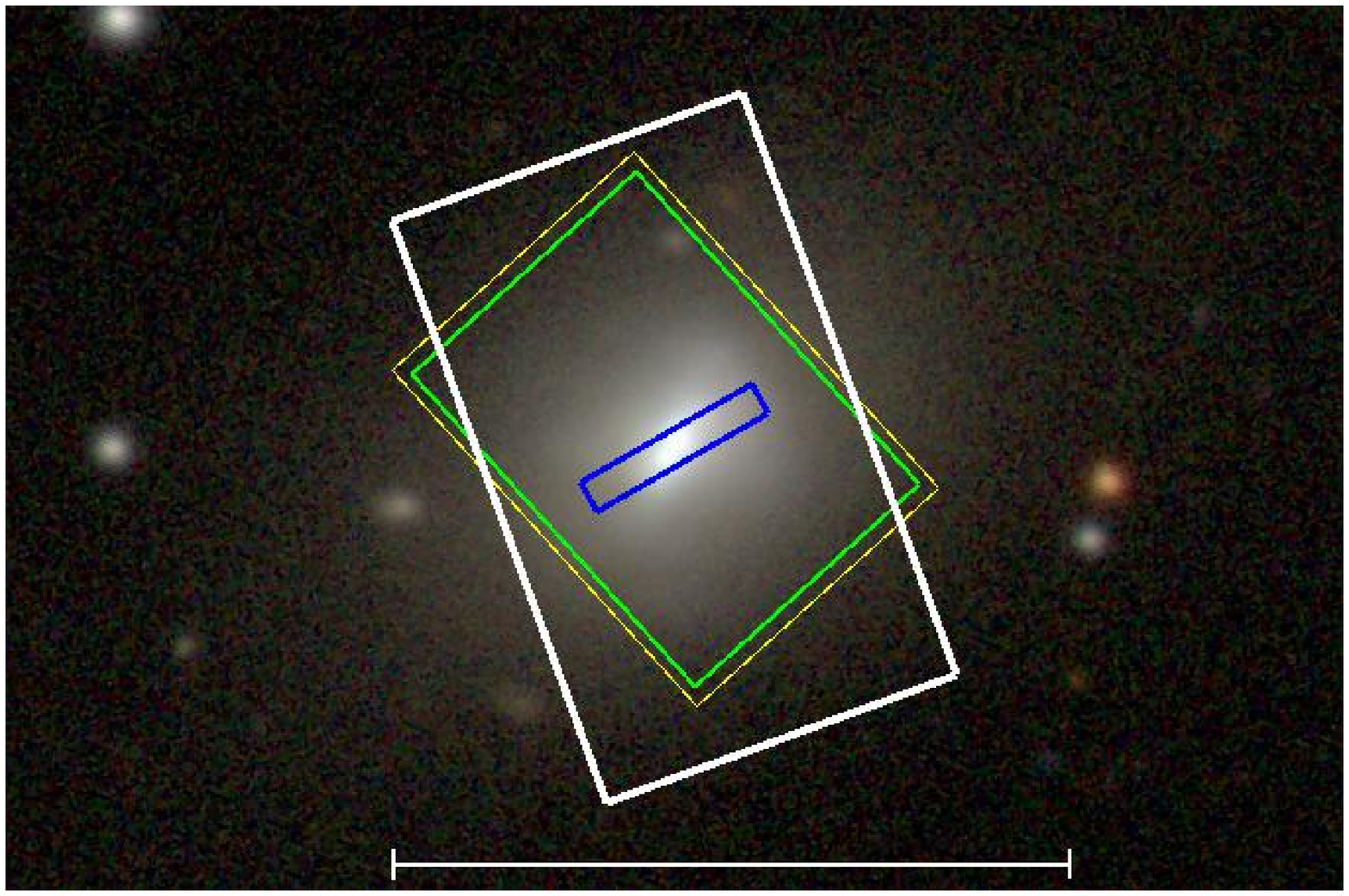}\\
\plottwo{f18_035a.ps}{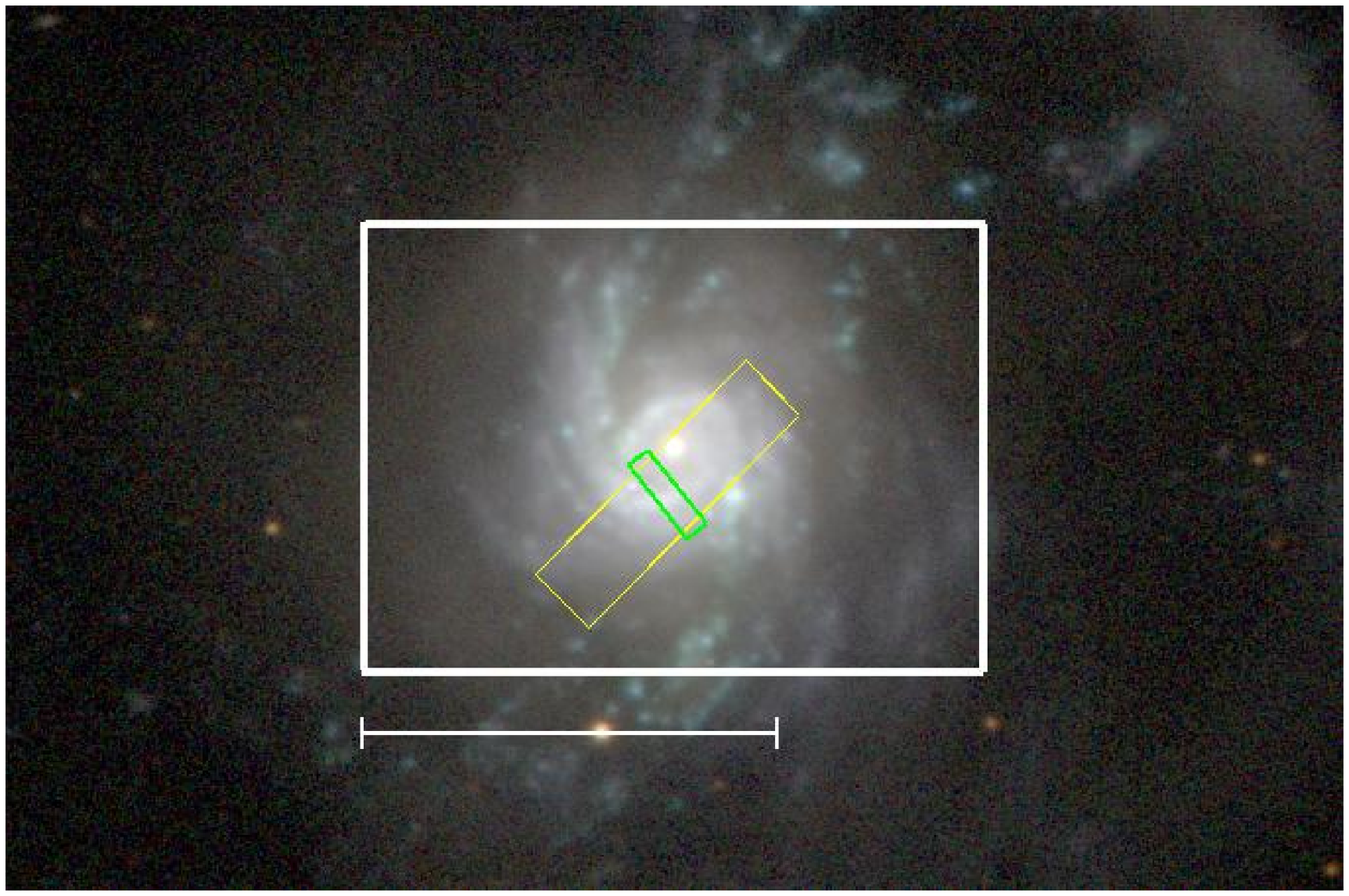}\\
\plottwo{f18_036a.ps}{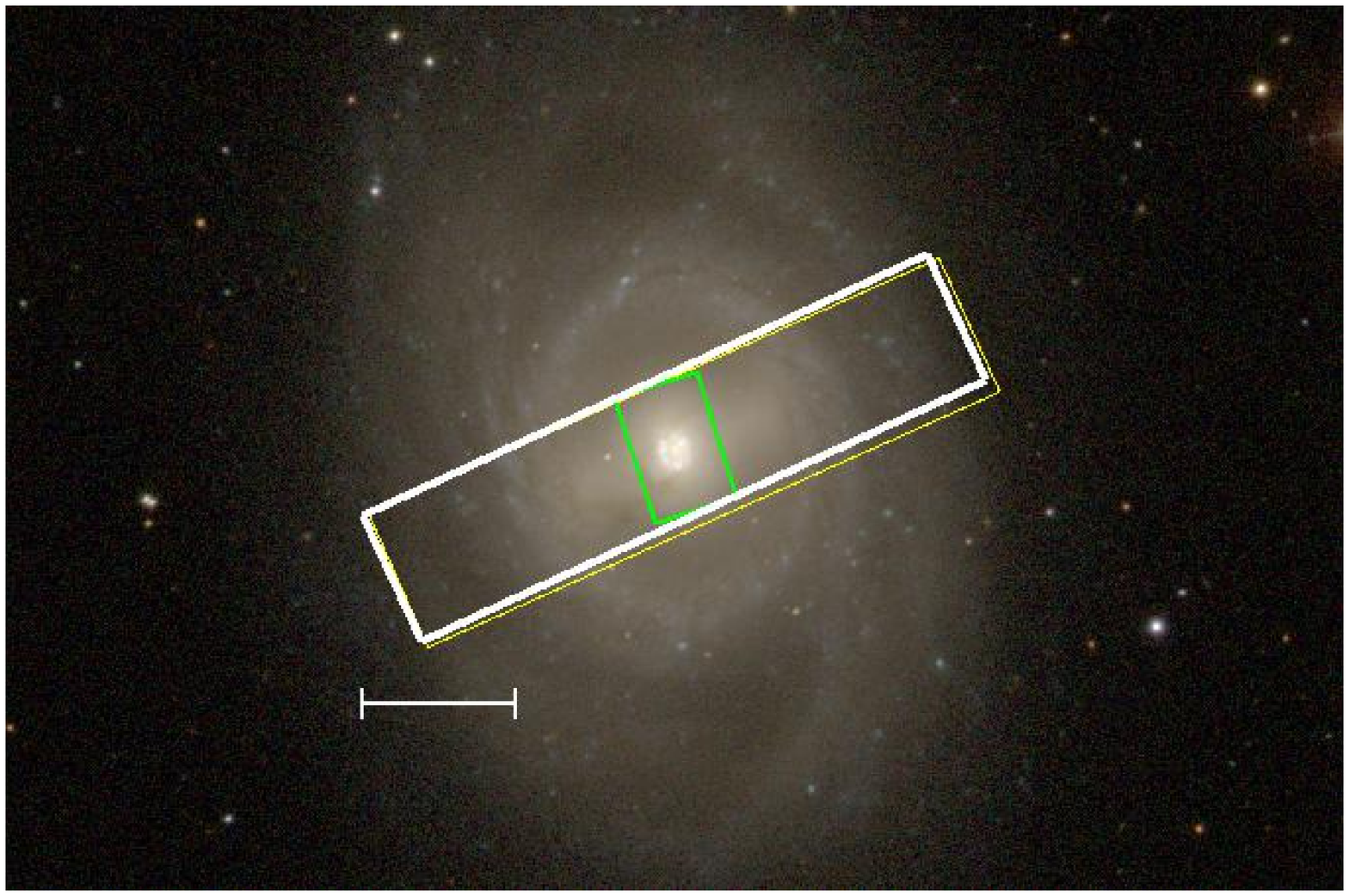}\\
\caption{Galaxy SEDs from the UV to the mid-IR. In the left-panel the observed and model spectra are shown in black and grey respectively, while the photometry used to constrain and verify the spectra is shown with red dots. In the right panel we plot the photometric aperture (thick white rectangle), the {\it Akari} extraction aperture (blue rectangle), the {\it Spitzer} SL extraction aperture (green rectangle) and the {\it Spitzer} LL extraction aperture (yellow rectangle). For galaxies with {\it Spitzer} stare mode spectra, we show a region corresponding to a quarter of the slit length. For scale, the horizontal bar denotes $1^{\prime}$. [{\it See the electronic edition of the Supplement for the complete Figure.}]}
\end{figure}

\clearpage

\begin{figure}[hbt]
\figurenum{\ref{fig:allspec} continued}
\plottwo{f18_037a.ps}{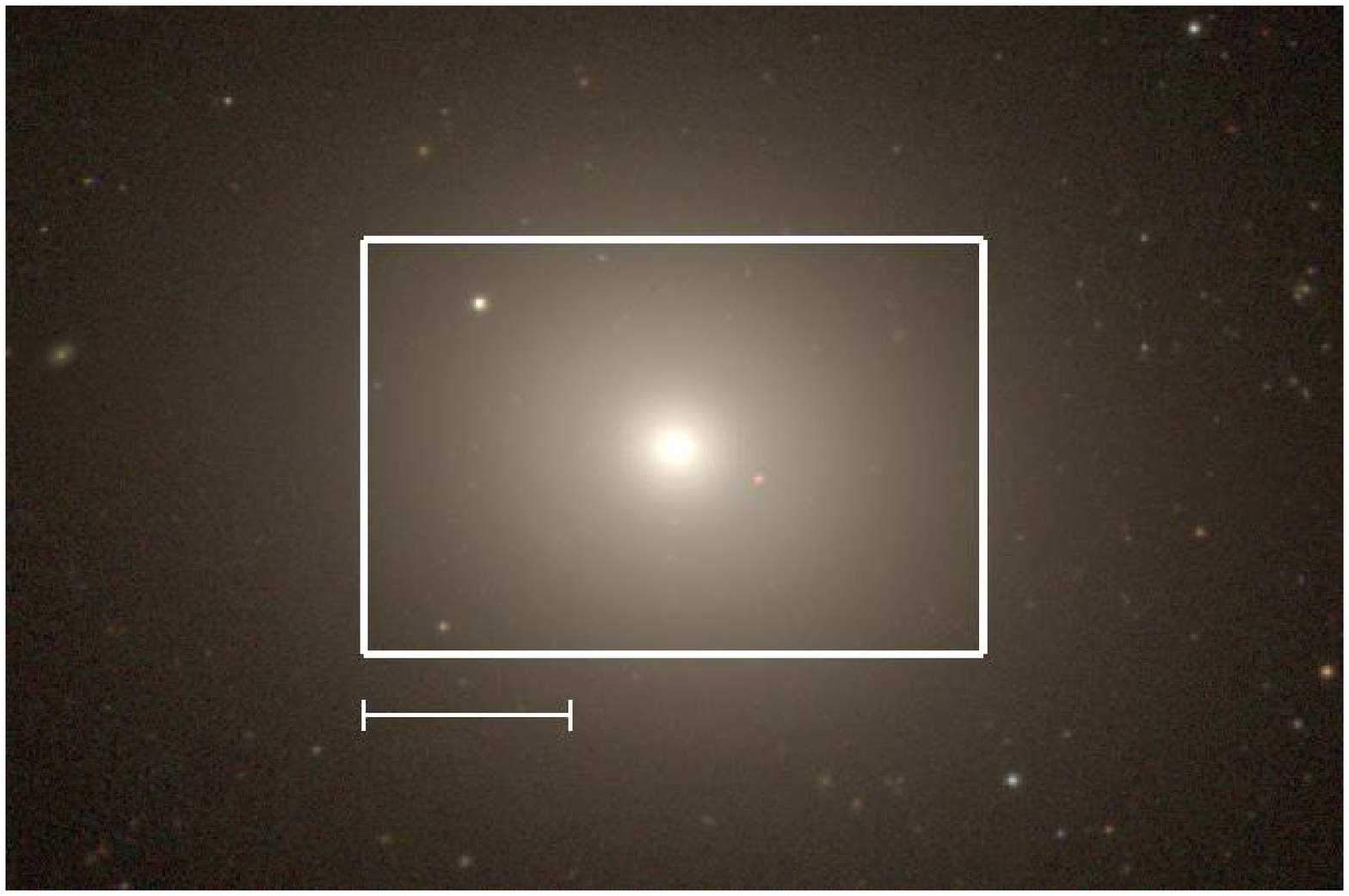}\\
\plottwo{f18_038a.ps}{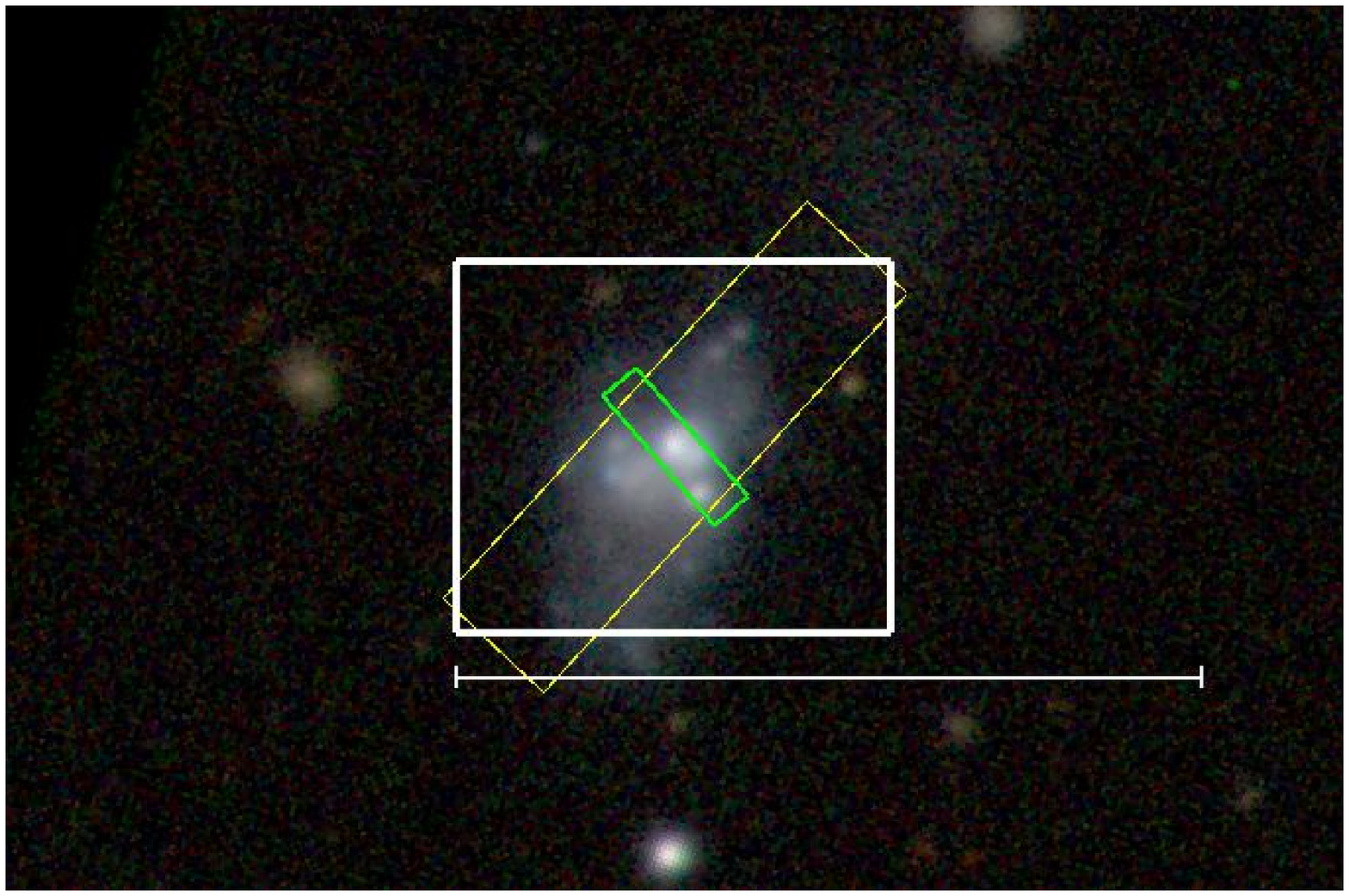}\\
\plottwo{f18_039a.ps}{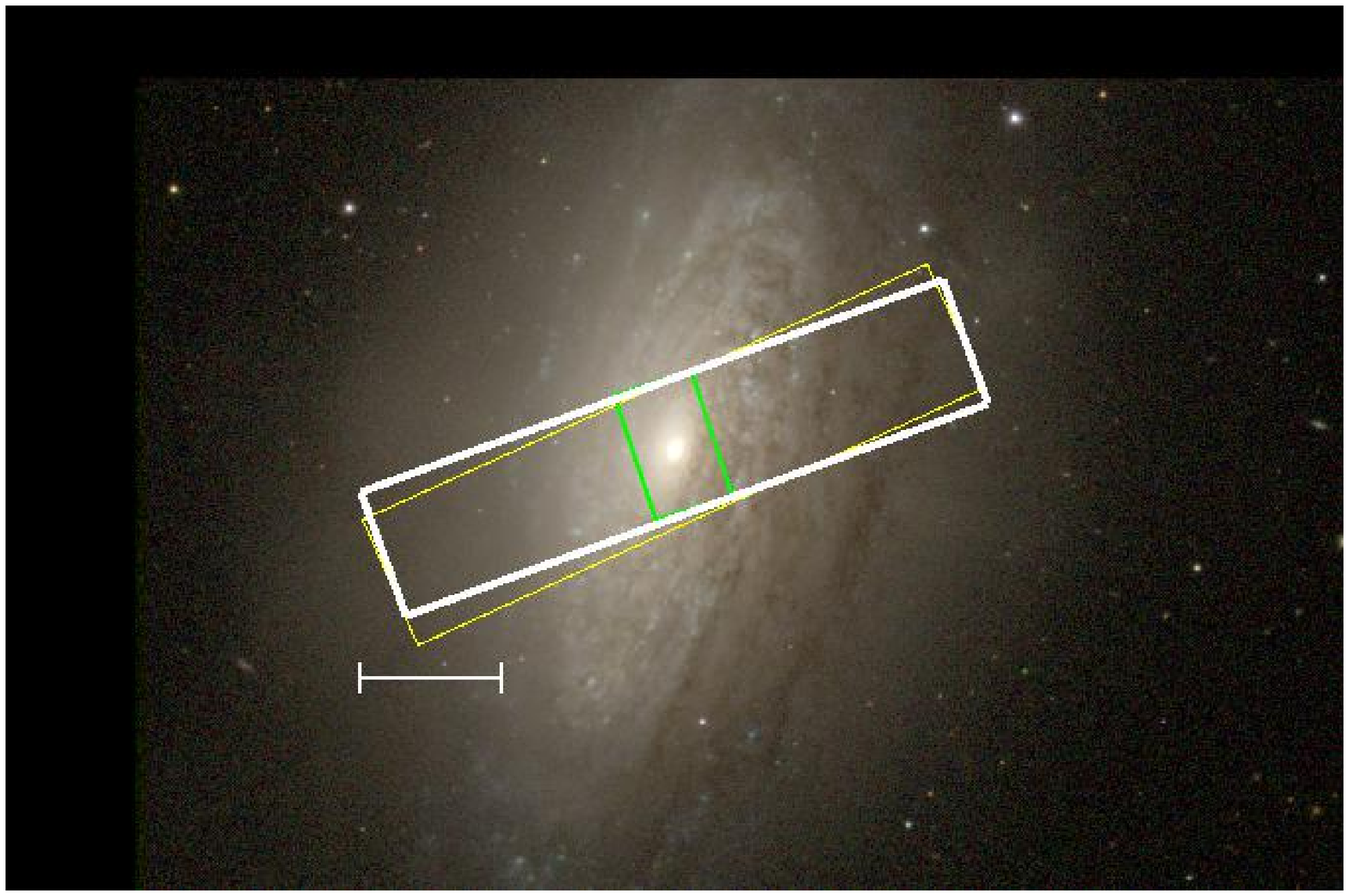}\\
\plottwo{f18_040a.ps}{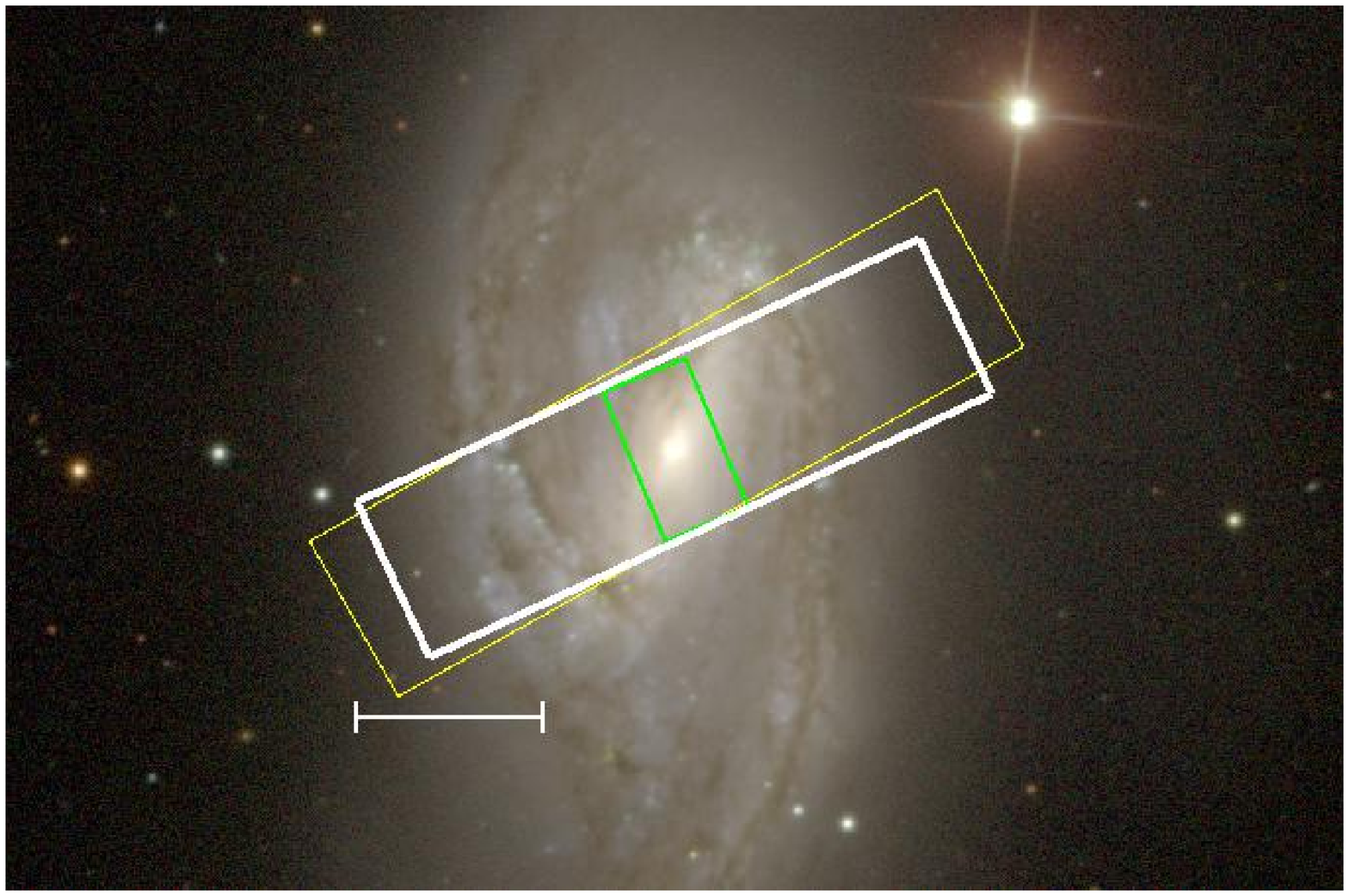}\\
\caption{Galaxy SEDs from the UV to the mid-IR. In the left-panel the observed and model spectra are shown in black and grey respectively, while the photometry used to constrain and verify the spectra is shown with red dots. In the right panel we plot the photometric aperture (thick white rectangle), the {\it Akari} extraction aperture (blue rectangle), the {\it Spitzer} SL extraction aperture (green rectangle) and the {\it Spitzer} LL extraction aperture (yellow rectangle). For galaxies with {\it Spitzer} stare mode spectra, we show a region corresponding to a quarter of the slit length. For scale, the horizontal bar denotes $1^{\prime}$. [{\it See the electronic edition of the Supplement for the complete Figure.}]}
\end{figure}

\clearpage

\begin{figure}[hbt]
\figurenum{\ref{fig:allspec} continued}
\plottwo{f18_041a.ps}{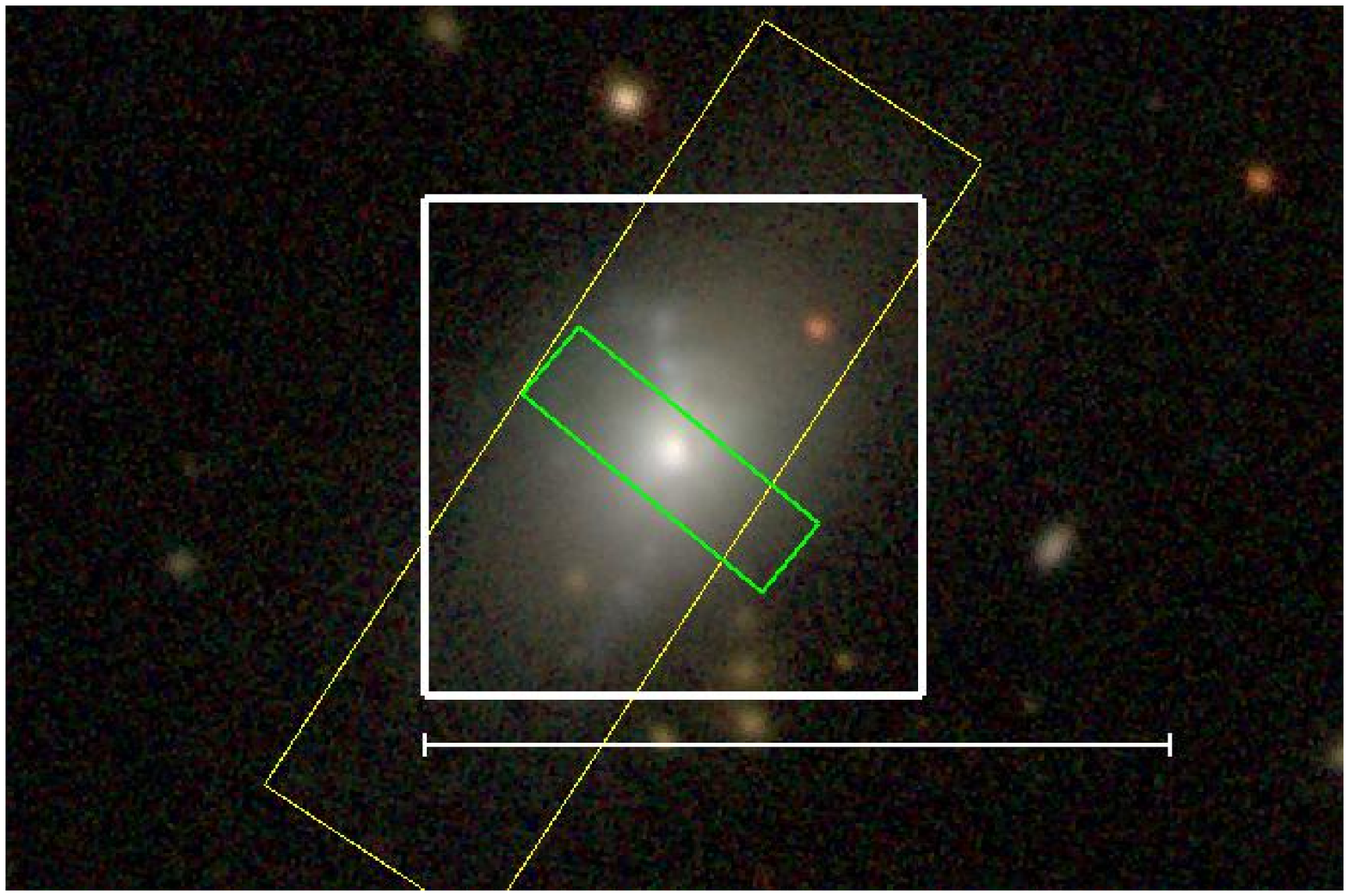}\\
\plottwo{f18_042a.ps}{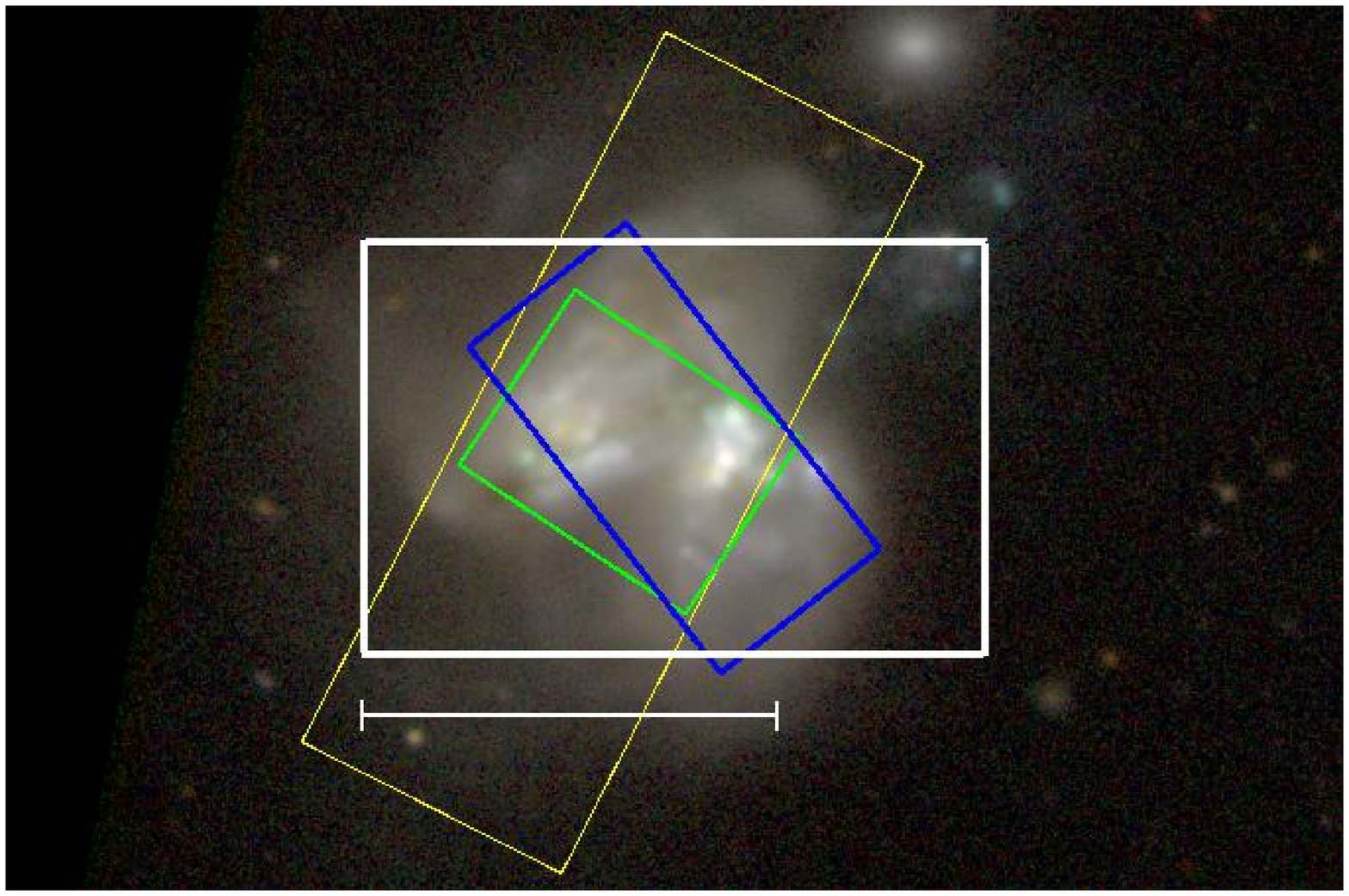}\\
\plottwo{f18_043a.ps}{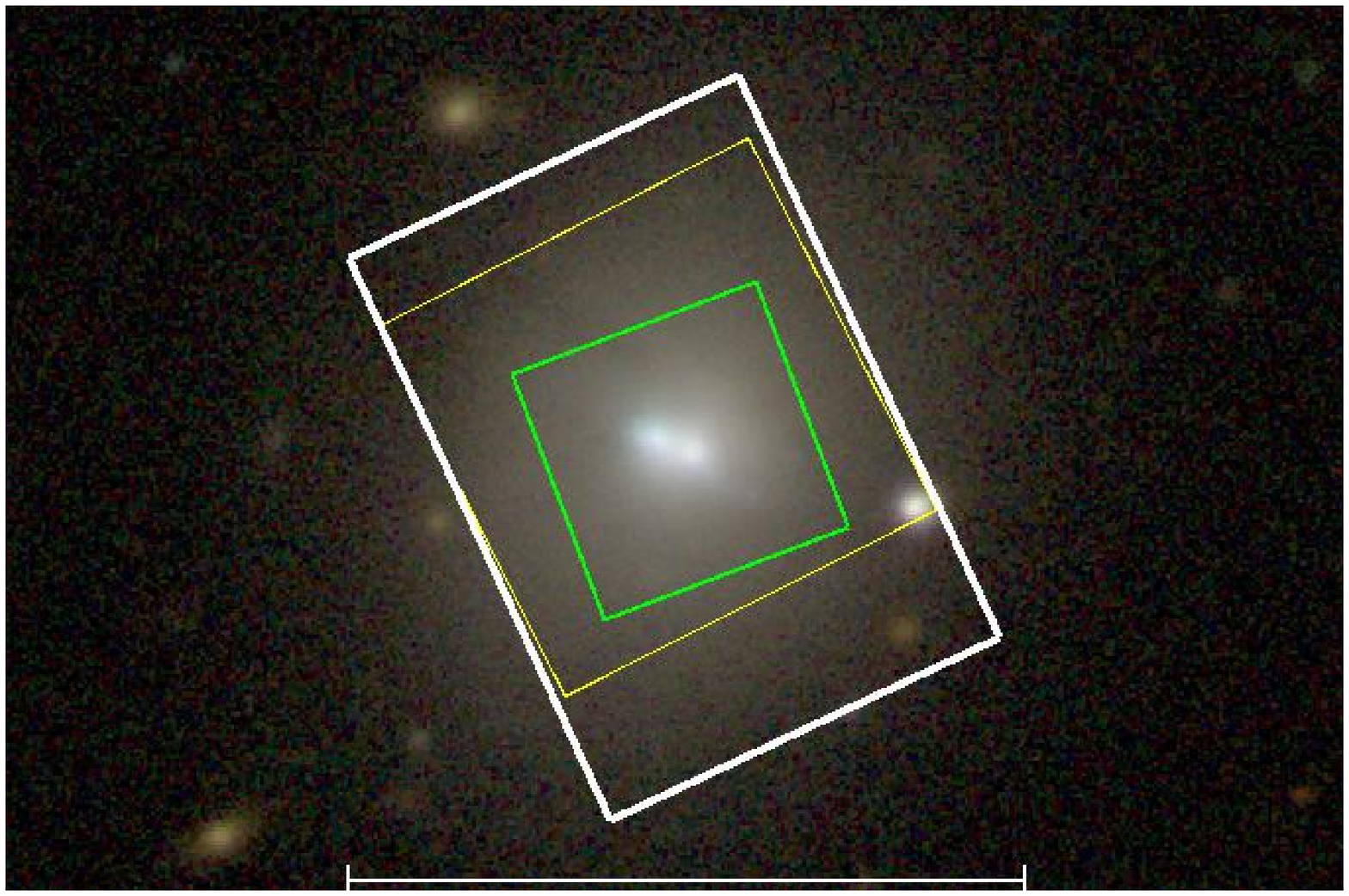}\\
\plottwo{f18_044a.ps}{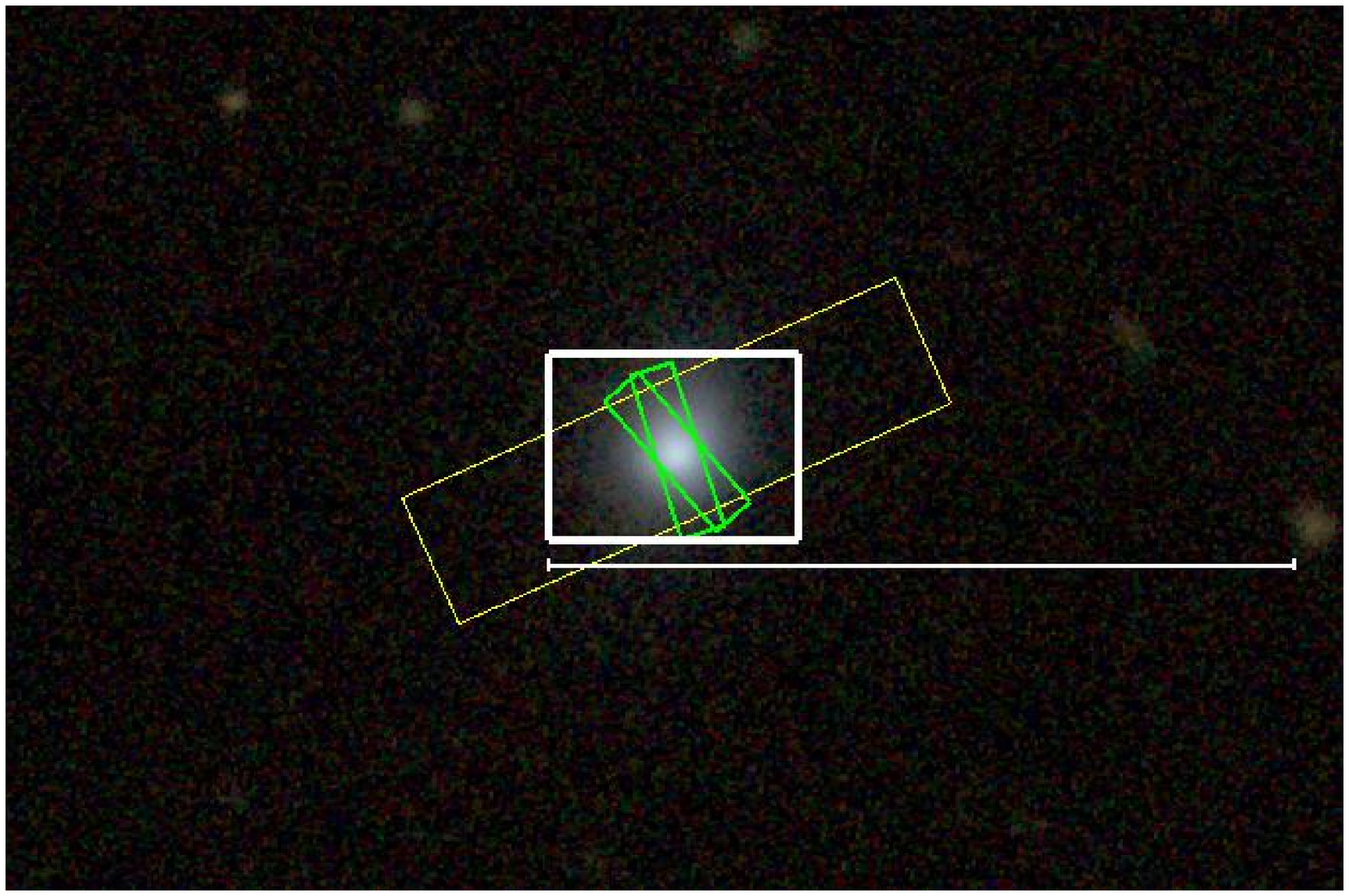}\\
\caption{Galaxy SEDs from the UV to the mid-IR. In the left-panel the observed and model spectra are shown in black and grey respectively, while the photometry used to constrain and verify the spectra is shown with red dots. In the right panel we plot the photometric aperture (thick white rectangle), the {\it Akari} extraction aperture (blue rectangle), the {\it Spitzer} SL extraction aperture (green rectangle) and the {\it Spitzer} LL extraction aperture (yellow rectangle). For galaxies with {\it Spitzer} stare mode spectra, we show a region corresponding to a quarter of the slit length. For scale, the horizontal bar denotes $1^{\prime}$. [{\it See the electronic edition of the Supplement for the complete Figure.}]}
\end{figure}

\clearpage

\begin{figure}[hbt]
\figurenum{\ref{fig:allspec} continued}
\plottwo{f18_045a.ps}{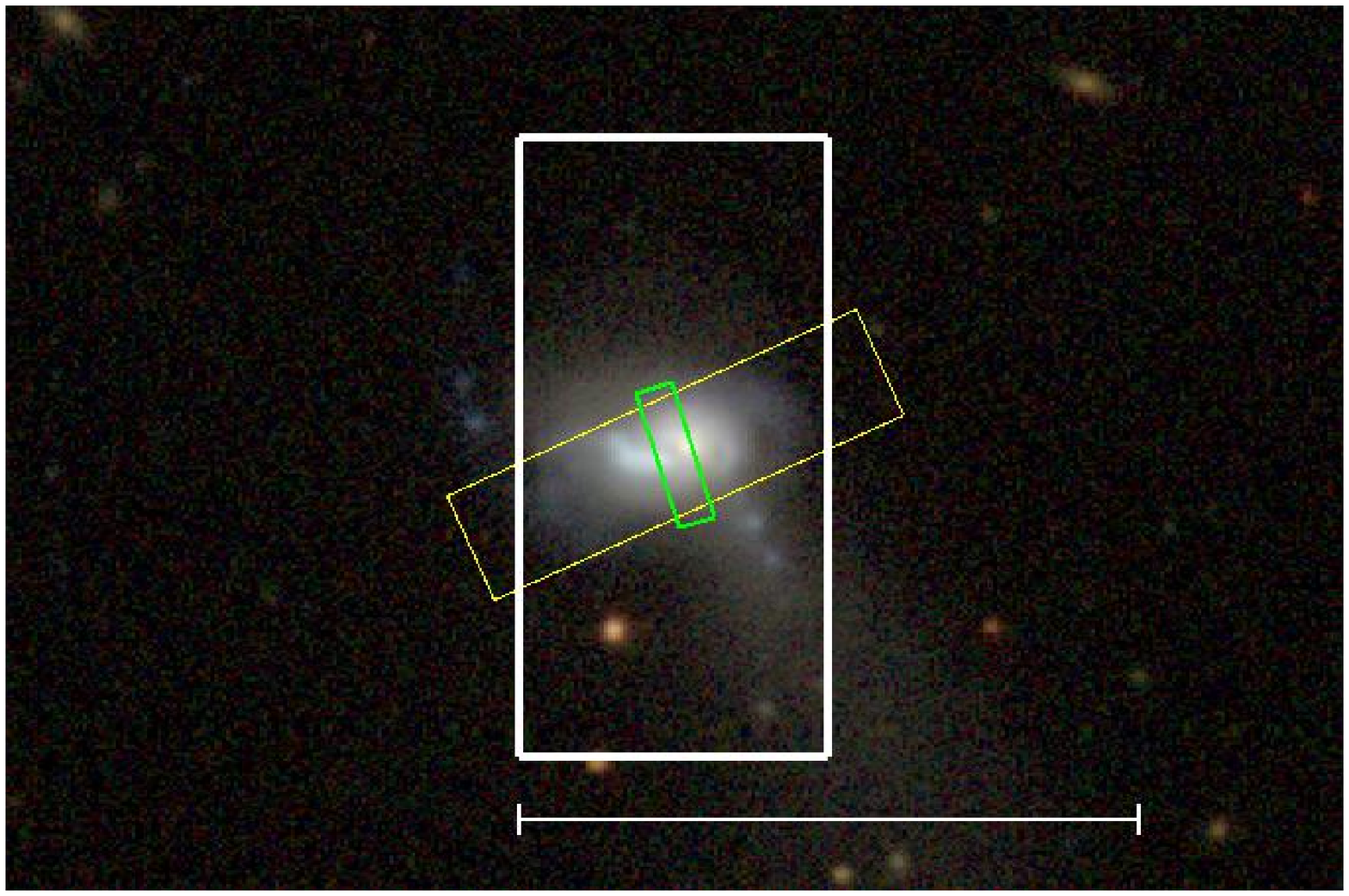}\\
\plottwo{f18_046a.ps}{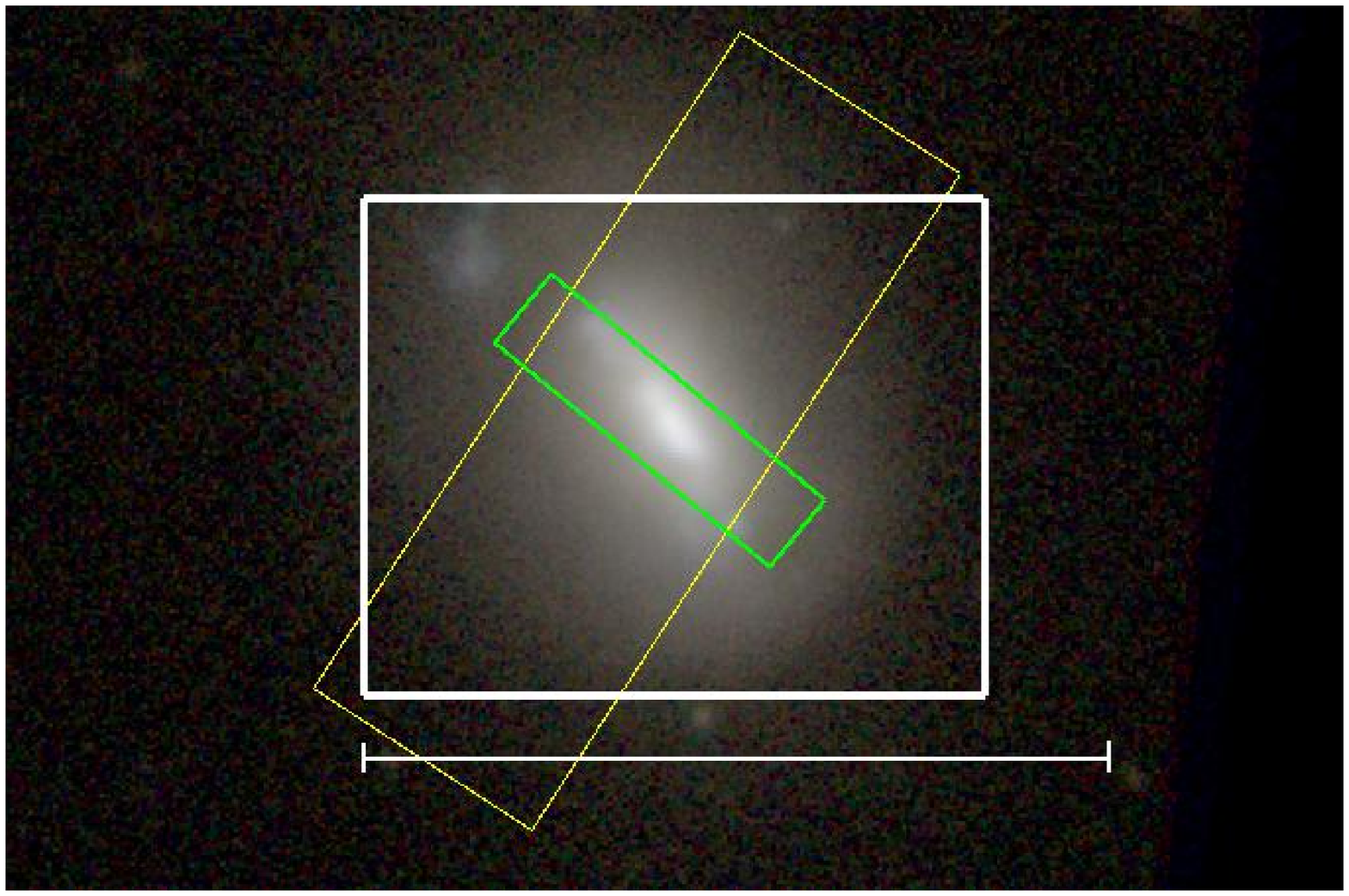}\\
\plottwo{f18_047a.ps}{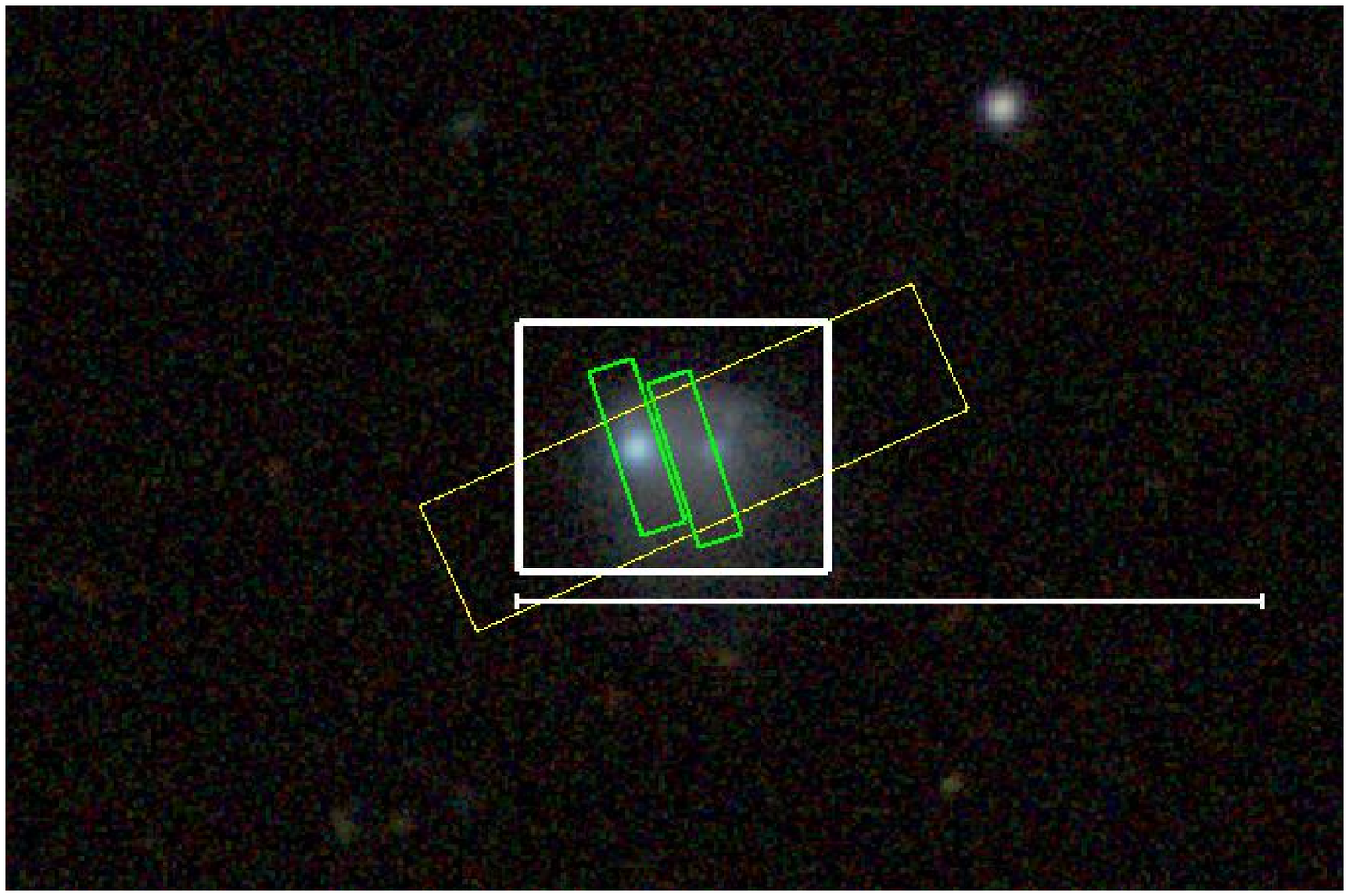}\\
\plottwo{f18_048a.ps}{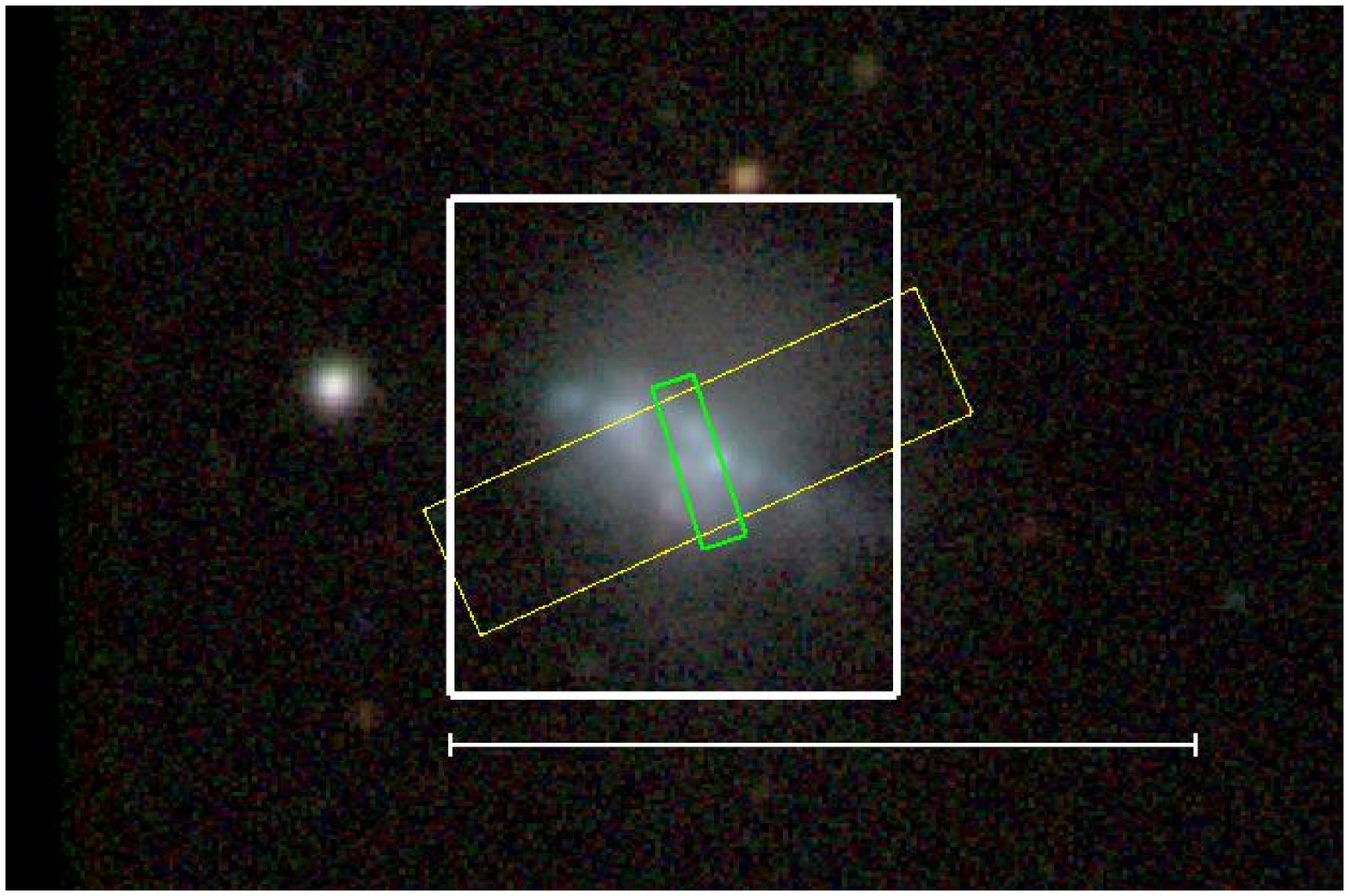}\\
\caption{Galaxy SEDs from the UV to the mid-IR. In the left-panel the observed and model spectra are shown in black and grey respectively, while the photometry used to constrain and verify the spectra is shown with red dots. In the right panel we plot the photometric aperture (thick white rectangle), the {\it Akari} extraction aperture (blue rectangle), the {\it Spitzer} SL extraction aperture (green rectangle) and the {\it Spitzer} LL extraction aperture (yellow rectangle). For galaxies with {\it Spitzer} stare mode spectra, we show a region corresponding to a quarter of the slit length. For scale, the horizontal bar denotes $1^{\prime}$. [{\it See the electronic edition of the Supplement for the complete Figure.}]}
\end{figure}

\clearpage

\begin{figure}[hbt]
\figurenum{\ref{fig:allspec} continued}
\plottwo{f18_049a.ps}{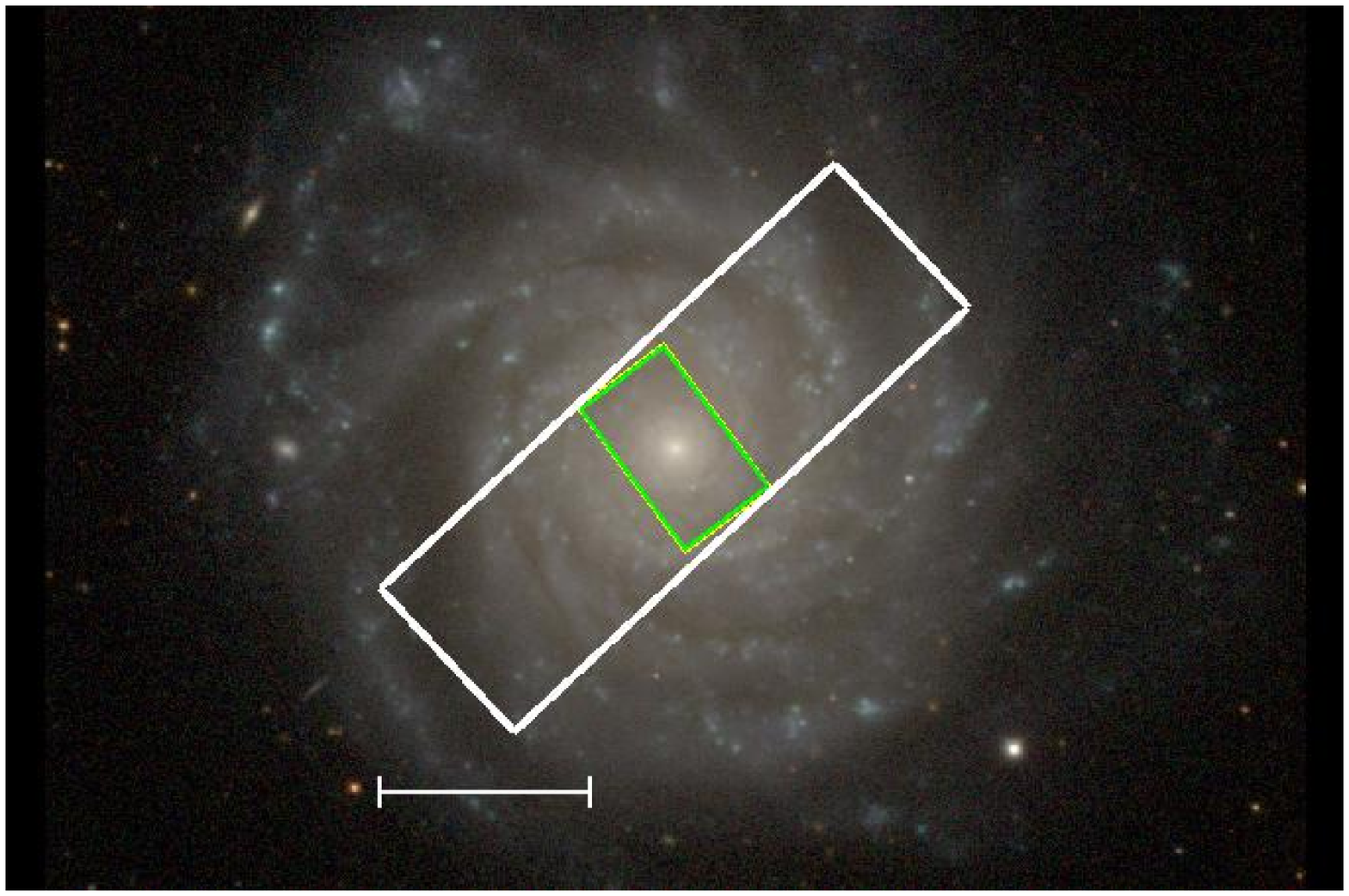}\\
\plottwo{f18_050a.ps}{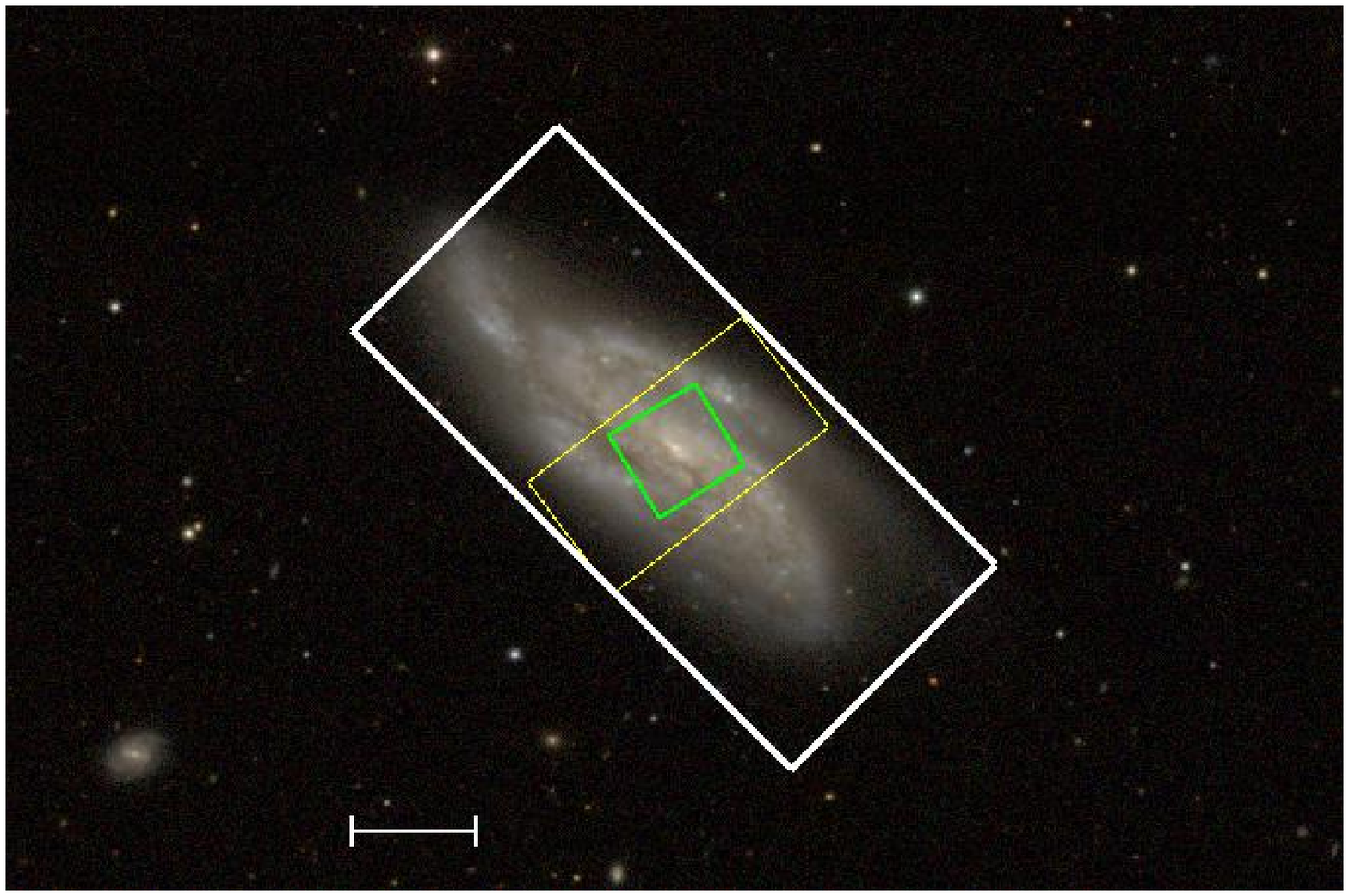}\\
\plottwo{f18_051a.ps}{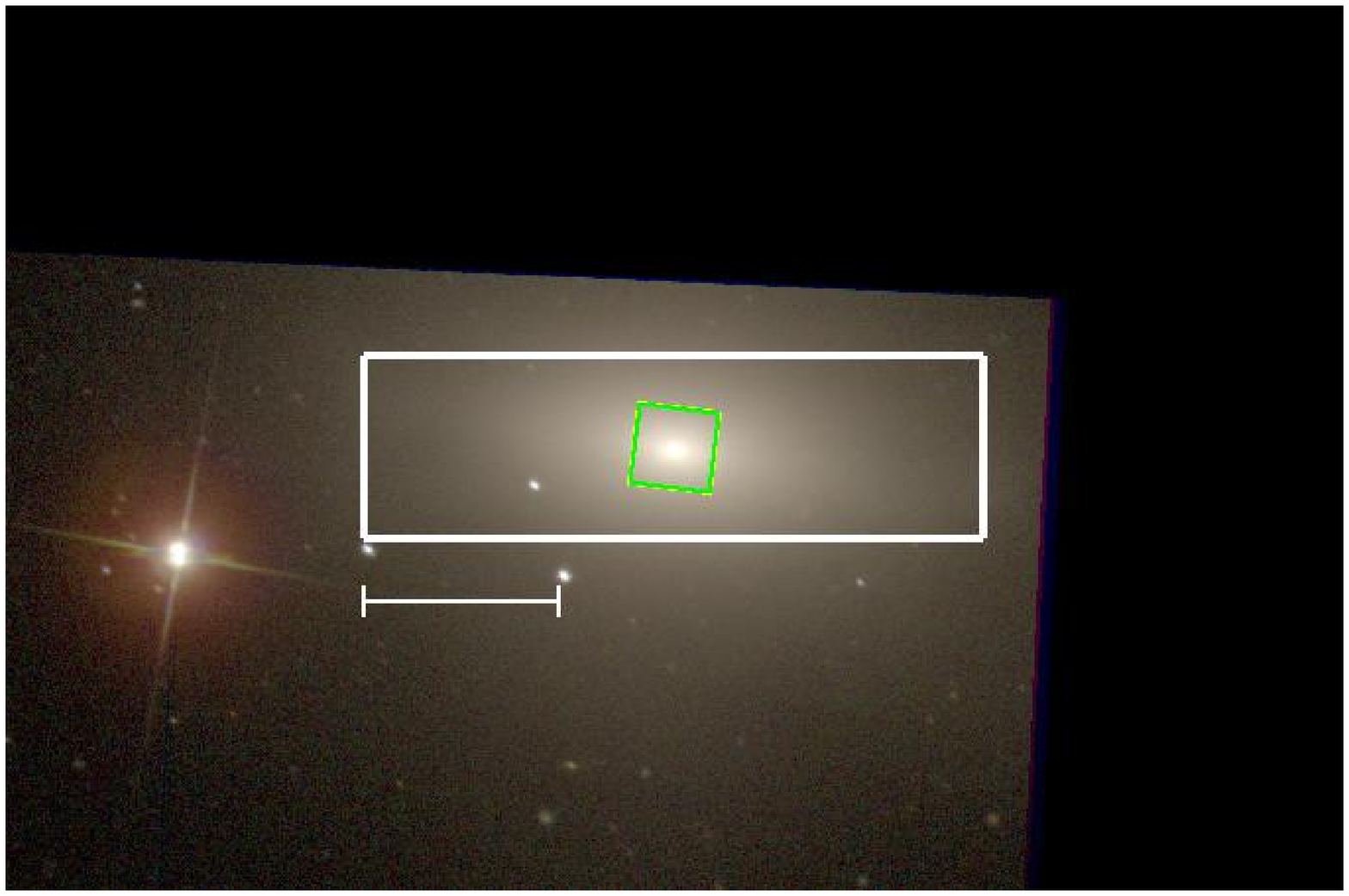}\\
\plottwo{f18_052a.ps}{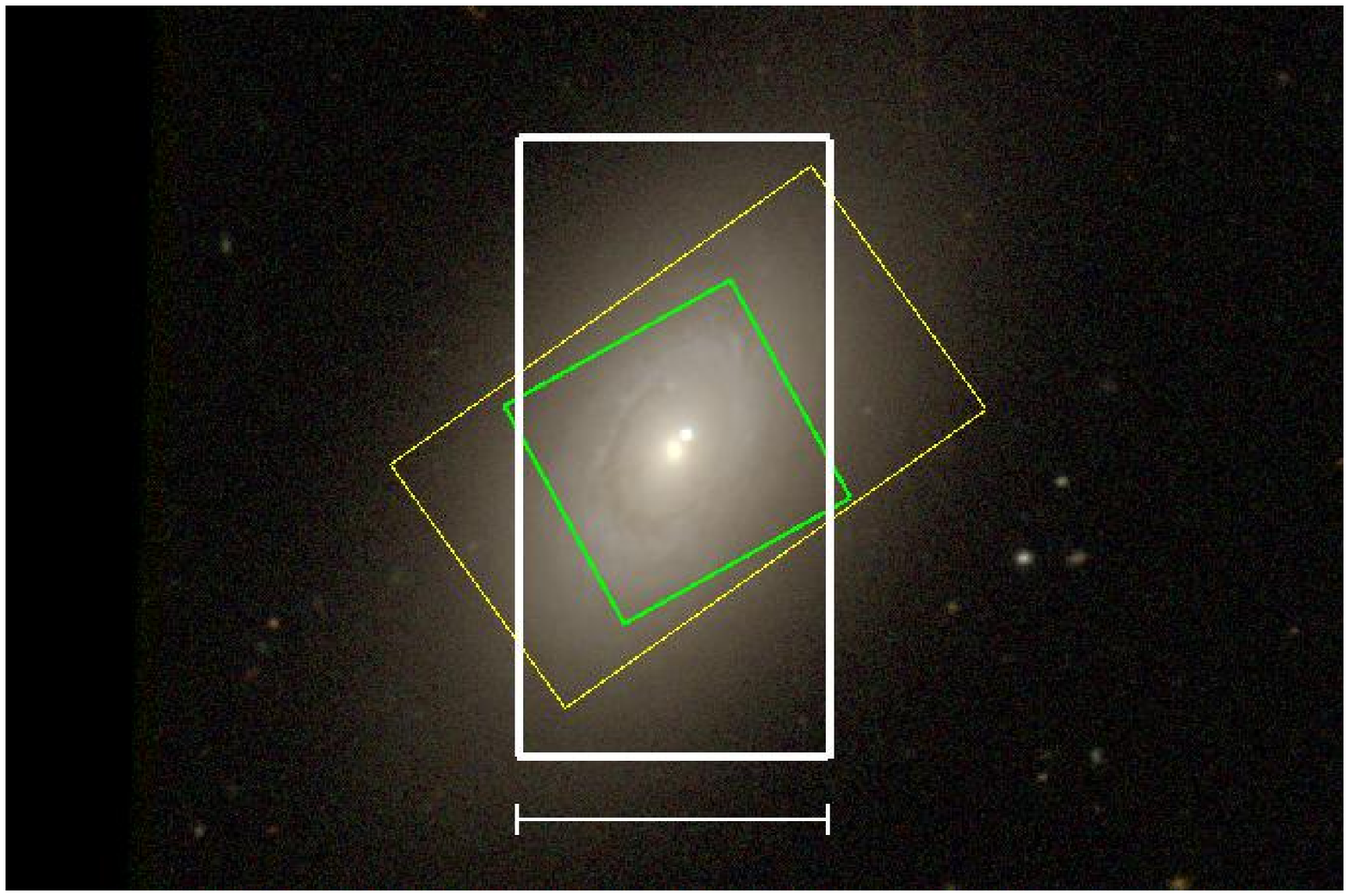}\\
\caption{Galaxy SEDs from the UV to the mid-IR. In the left-panel the observed and model spectra are shown in black and grey respectively, while the photometry used to constrain and verify the spectra is shown with red dots. In the right panel we plot the photometric aperture (thick white rectangle), the {\it Akari} extraction aperture (blue rectangle), the {\it Spitzer} SL extraction aperture (green rectangle) and the {\it Spitzer} LL extraction aperture (yellow rectangle). For galaxies with {\it Spitzer} stare mode spectra, we show a region corresponding to a quarter of the slit length. For scale, the horizontal bar denotes $1^{\prime}$. [{\it See the electronic edition of the Supplement for the complete Figure.}]}
\end{figure}

\clearpage

\begin{figure}[hbt]
\figurenum{\ref{fig:allspec} continued}
\plottwo{f18_053a.ps}{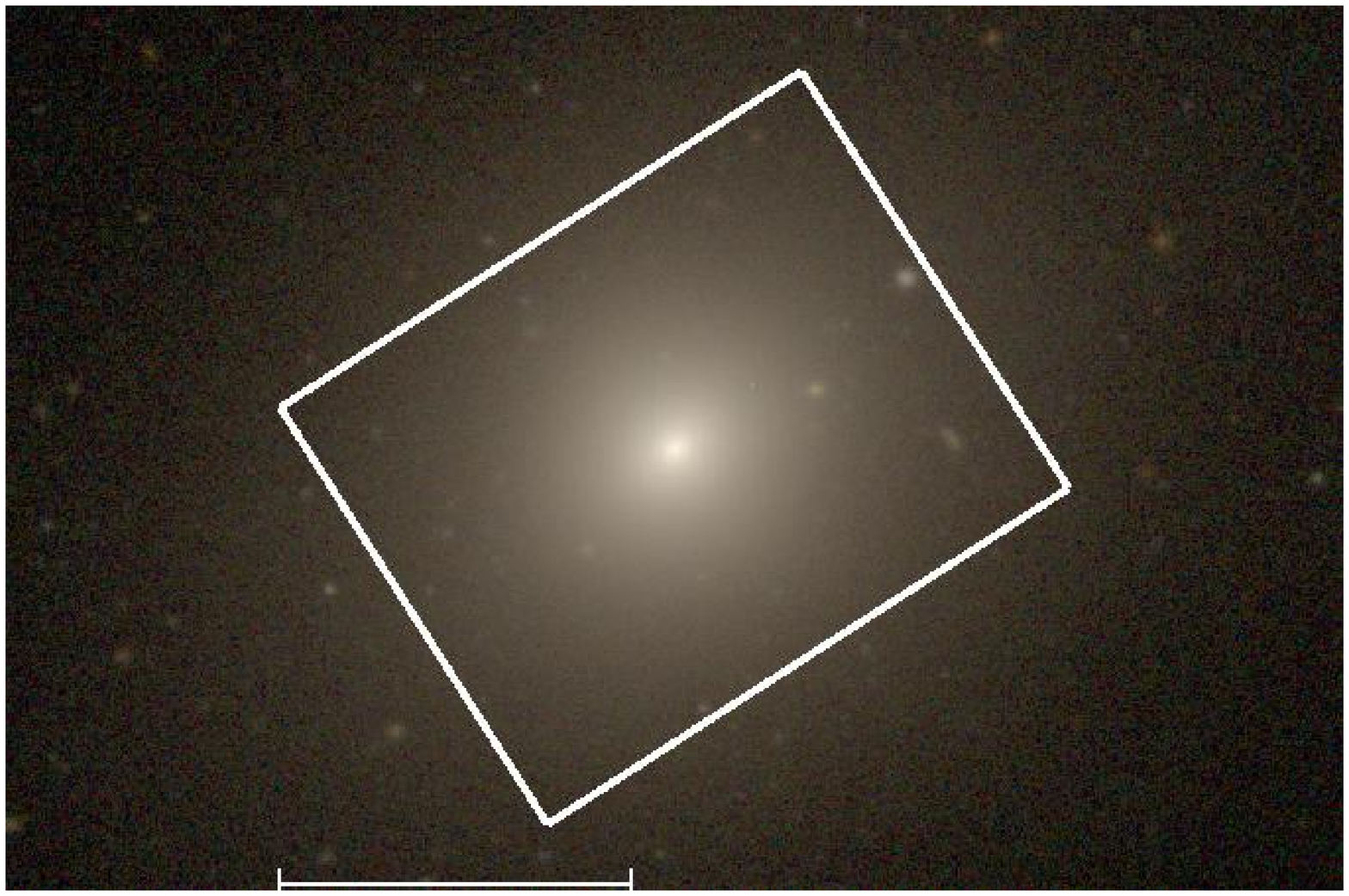}\\
\plottwo{f18_054a.ps}{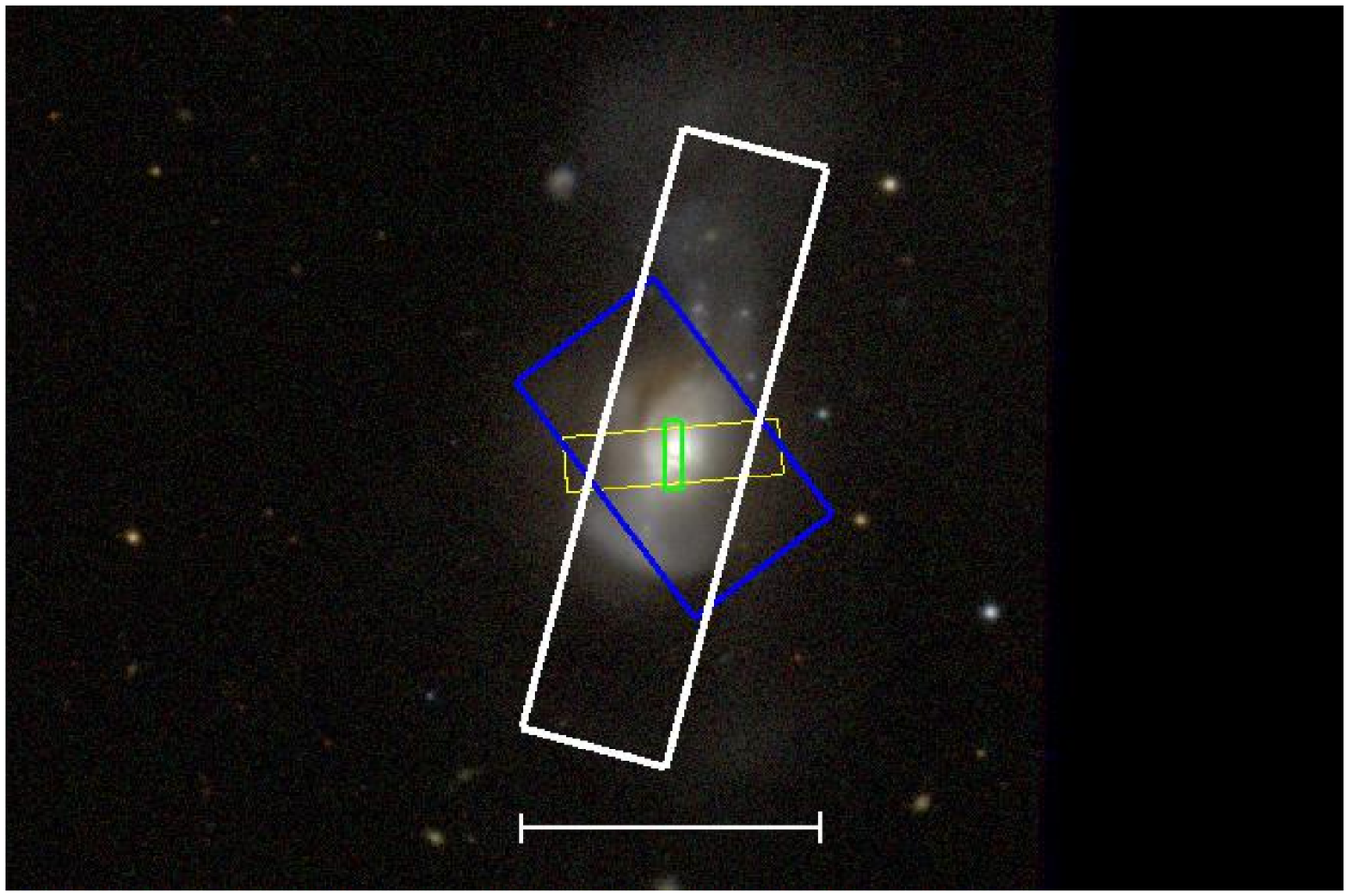}\\
\plottwo{f18_055a.ps}{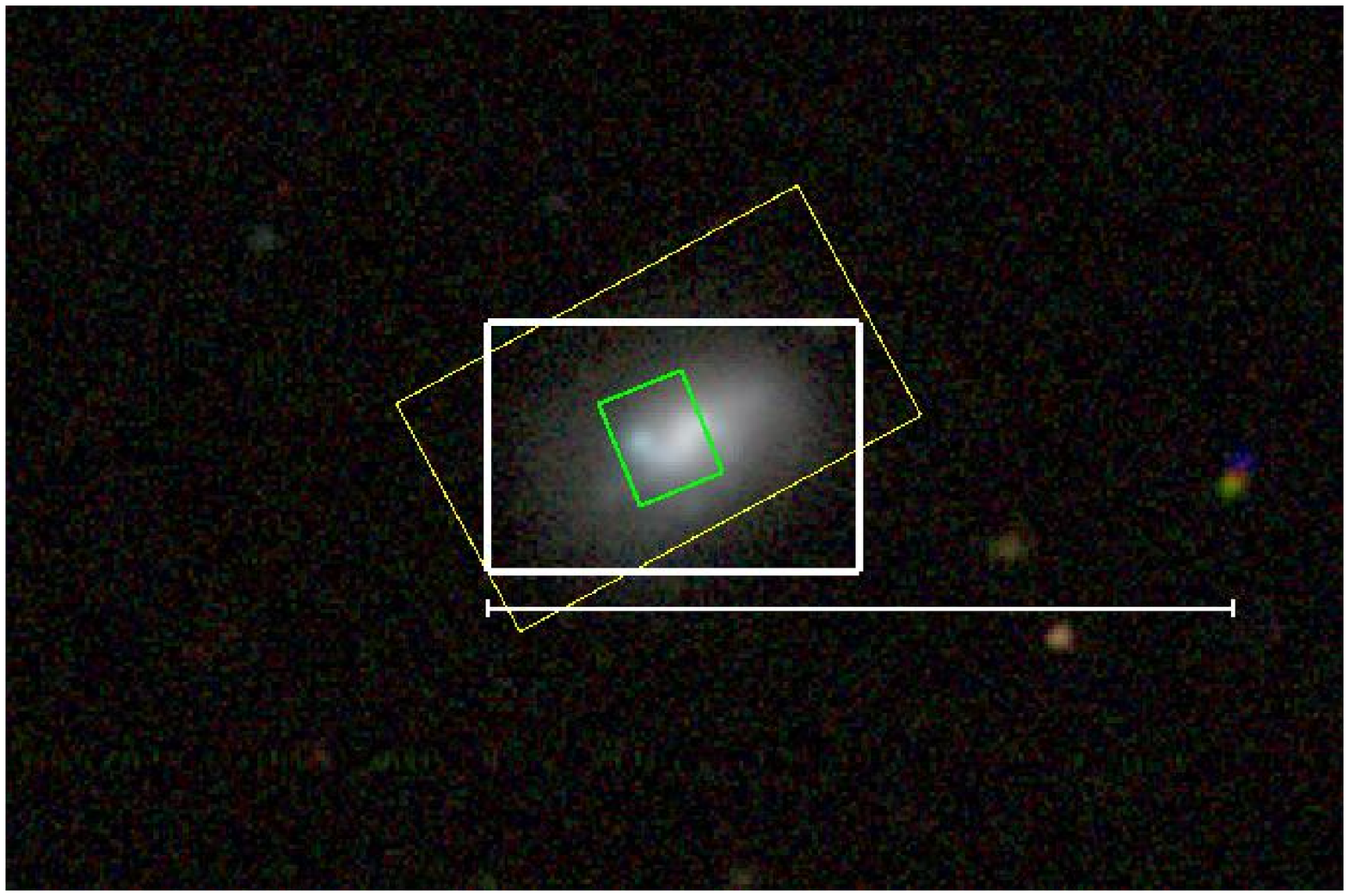}\\
\plottwo{f18_056a.ps}{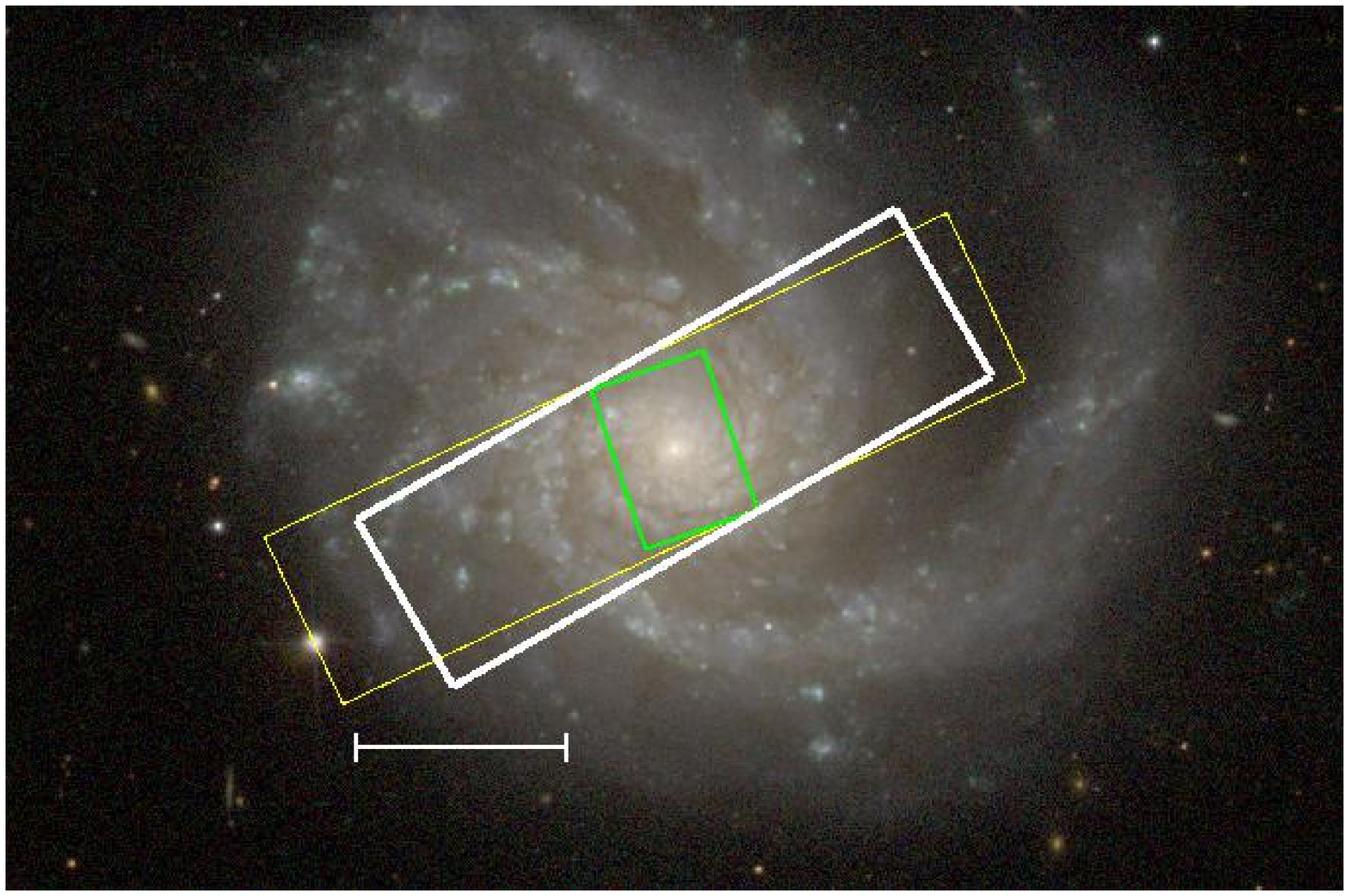}\\
\caption{Galaxy SEDs from the UV to the mid-IR. In the left-panel the observed and model spectra are shown in black and grey respectively, while the photometry used to constrain and verify the spectra is shown with red dots. In the right panel we plot the photometric aperture (thick white rectangle), the {\it Akari} extraction aperture (blue rectangle), the {\it Spitzer} SL extraction aperture (green rectangle) and the {\it Spitzer} LL extraction aperture (yellow rectangle). For galaxies with {\it Spitzer} stare mode spectra, we show a region corresponding to a quarter of the slit length. For scale, the horizontal bar denotes $1^{\prime}$. [{\it See the electronic edition of the Supplement for the complete Figure.}]}
\end{figure}

\clearpage

\begin{figure}[hbt]
\figurenum{\ref{fig:allspec} continued}
\plottwo{f18_057a.ps}{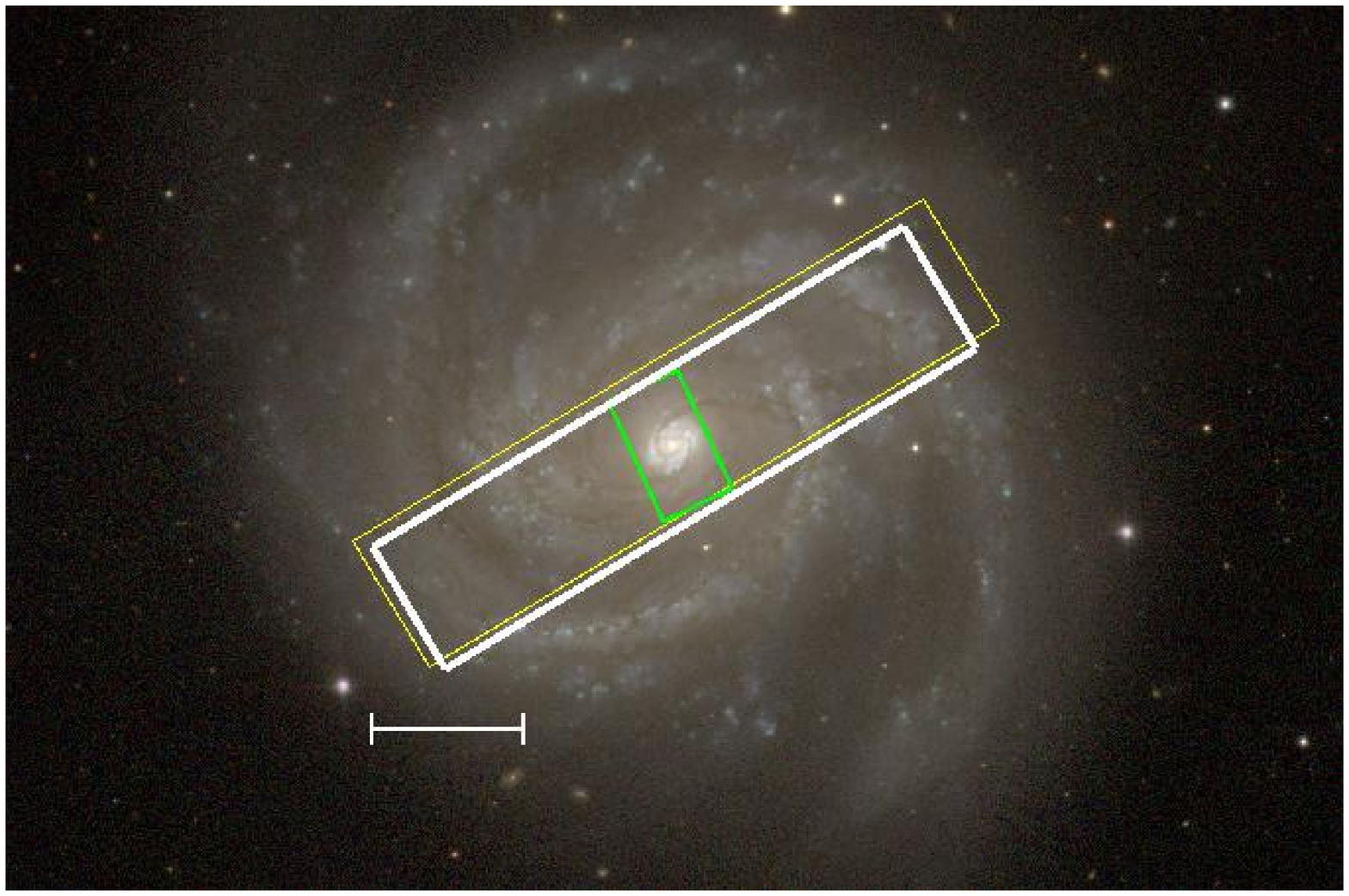}\\
\plottwo{f18_058a.ps}{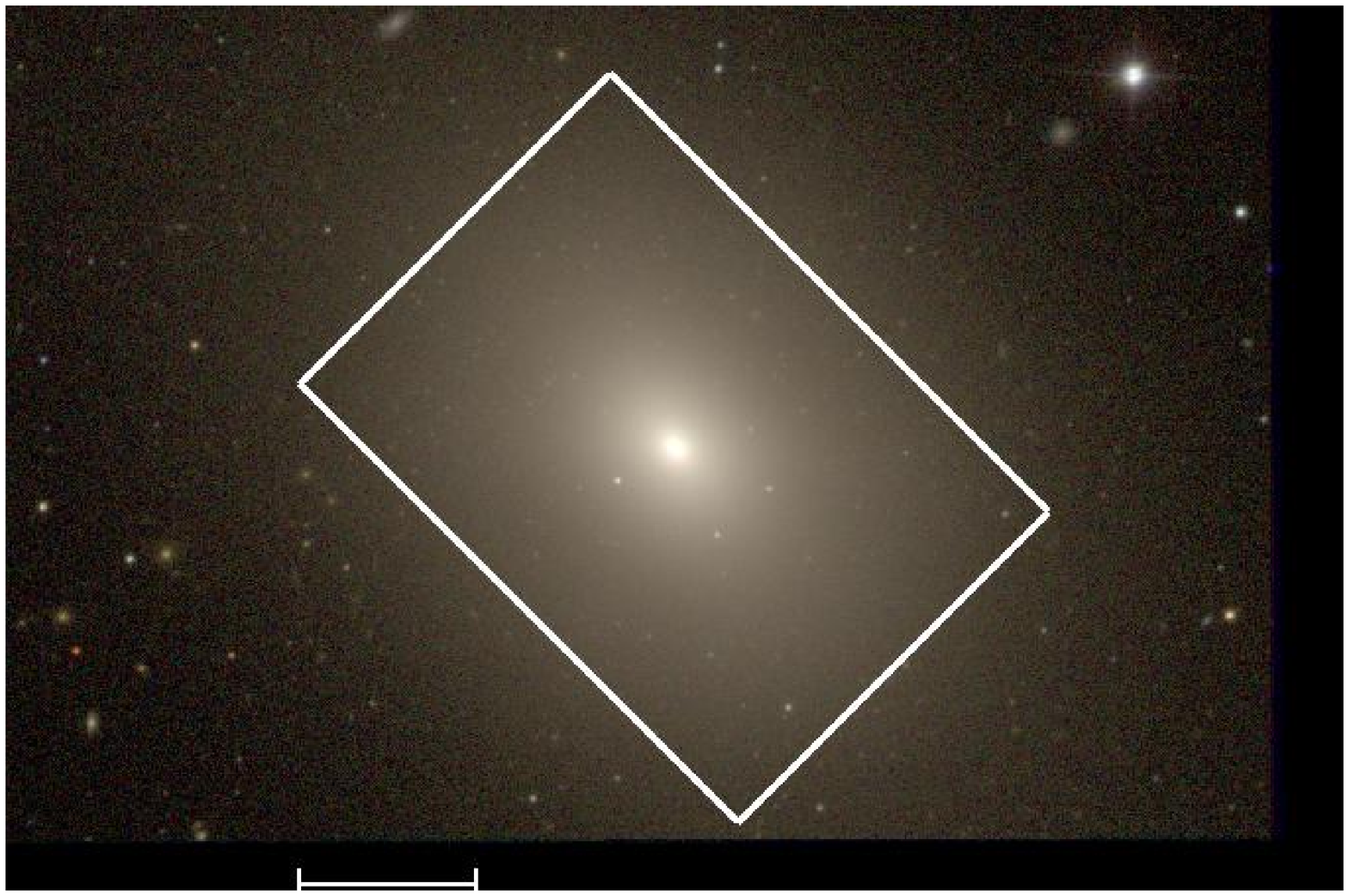}\\
\plottwo{f18_059a.ps}{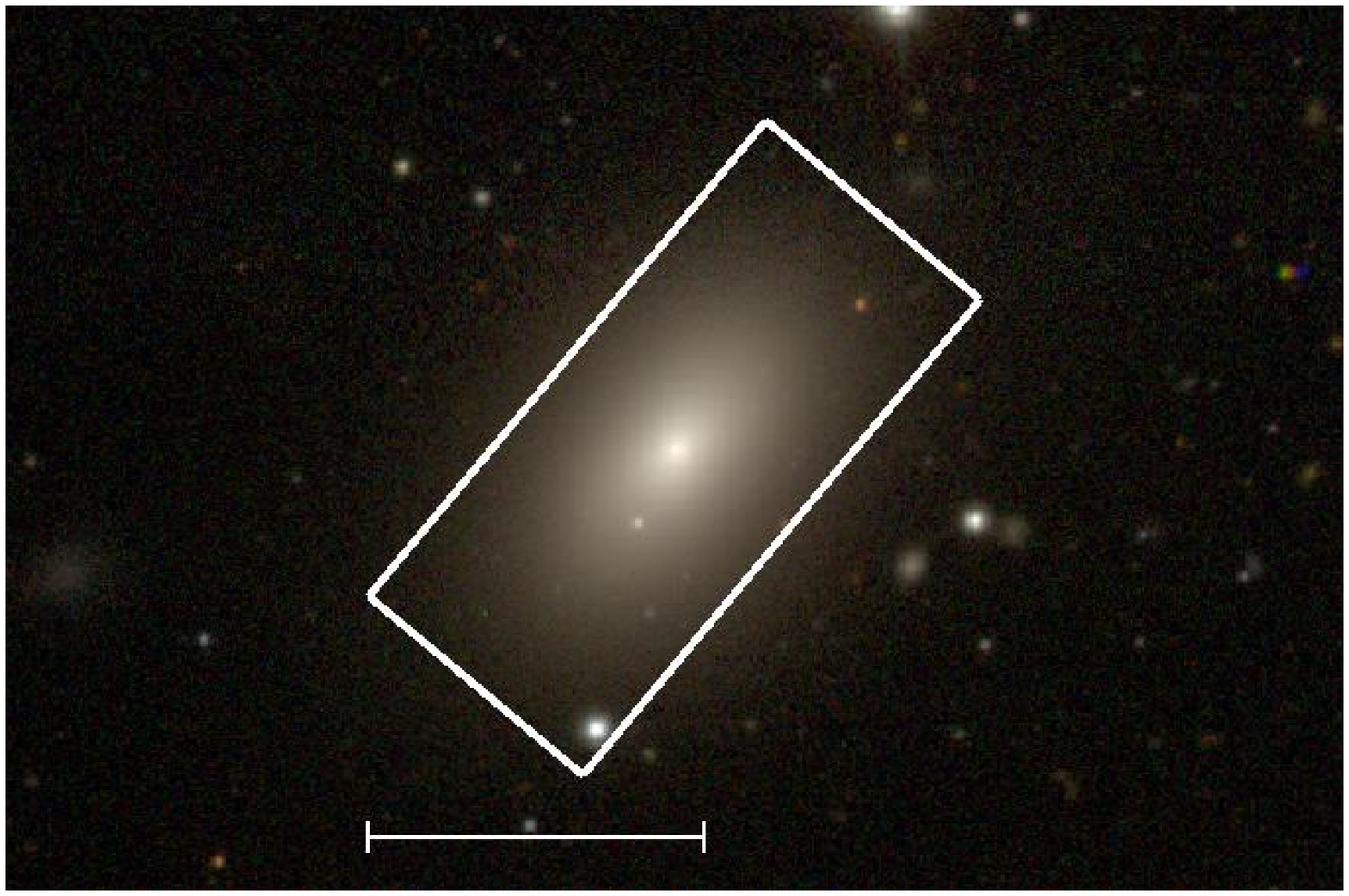}\\
\plottwo{f18_060a.ps}{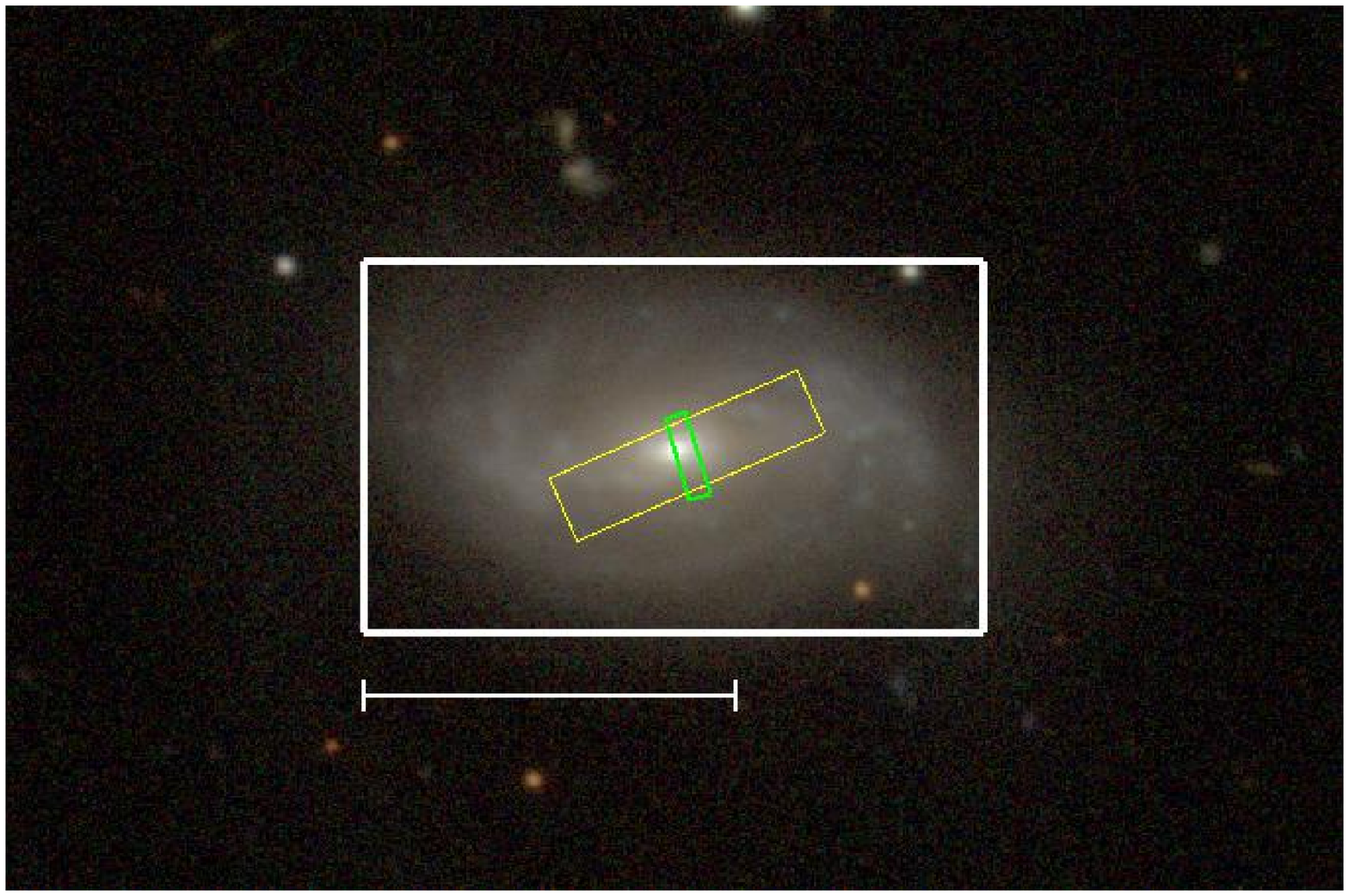}\\
\caption{Galaxy SEDs from the UV to the mid-IR. In the left-panel the observed and model spectra are shown in black and grey respectively, while the photometry used to constrain and verify the spectra is shown with red dots. In the right panel we plot the photometric aperture (thick white rectangle), the {\it Akari} extraction aperture (blue rectangle), the {\it Spitzer} SL extraction aperture (green rectangle) and the {\it Spitzer} LL extraction aperture (yellow rectangle). For galaxies with {\it Spitzer} stare mode spectra, we show a region corresponding to a quarter of the slit length. For scale, the horizontal bar denotes $1^{\prime}$. [{\it See the electronic edition of the Supplement for the complete Figure.}]}
\end{figure}

\clearpage

\begin{figure}[hbt]
\figurenum{\ref{fig:allspec} continued}
\plottwo{f18_061a.ps}{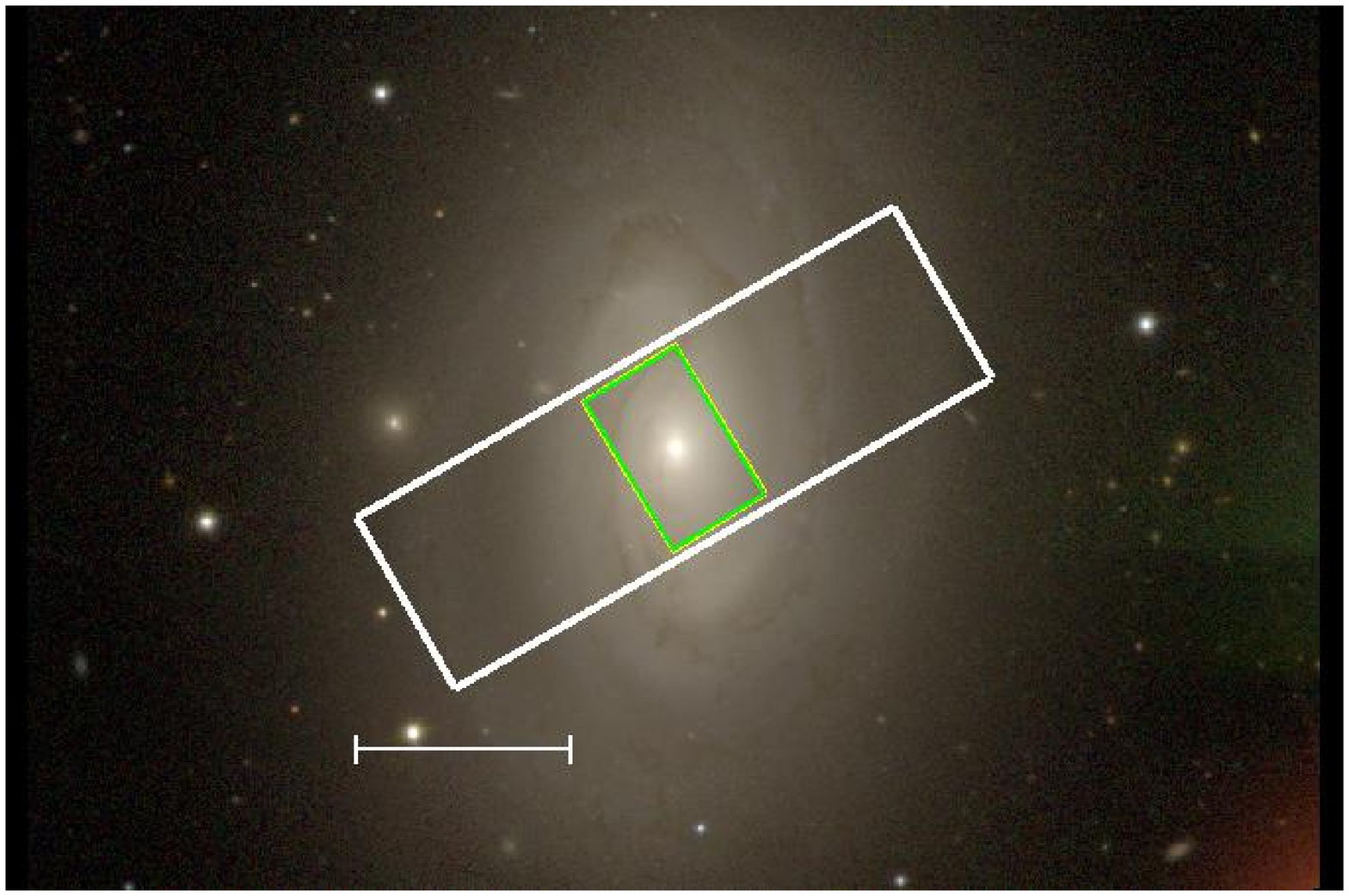}\\
\plottwo{f18_062a.ps}{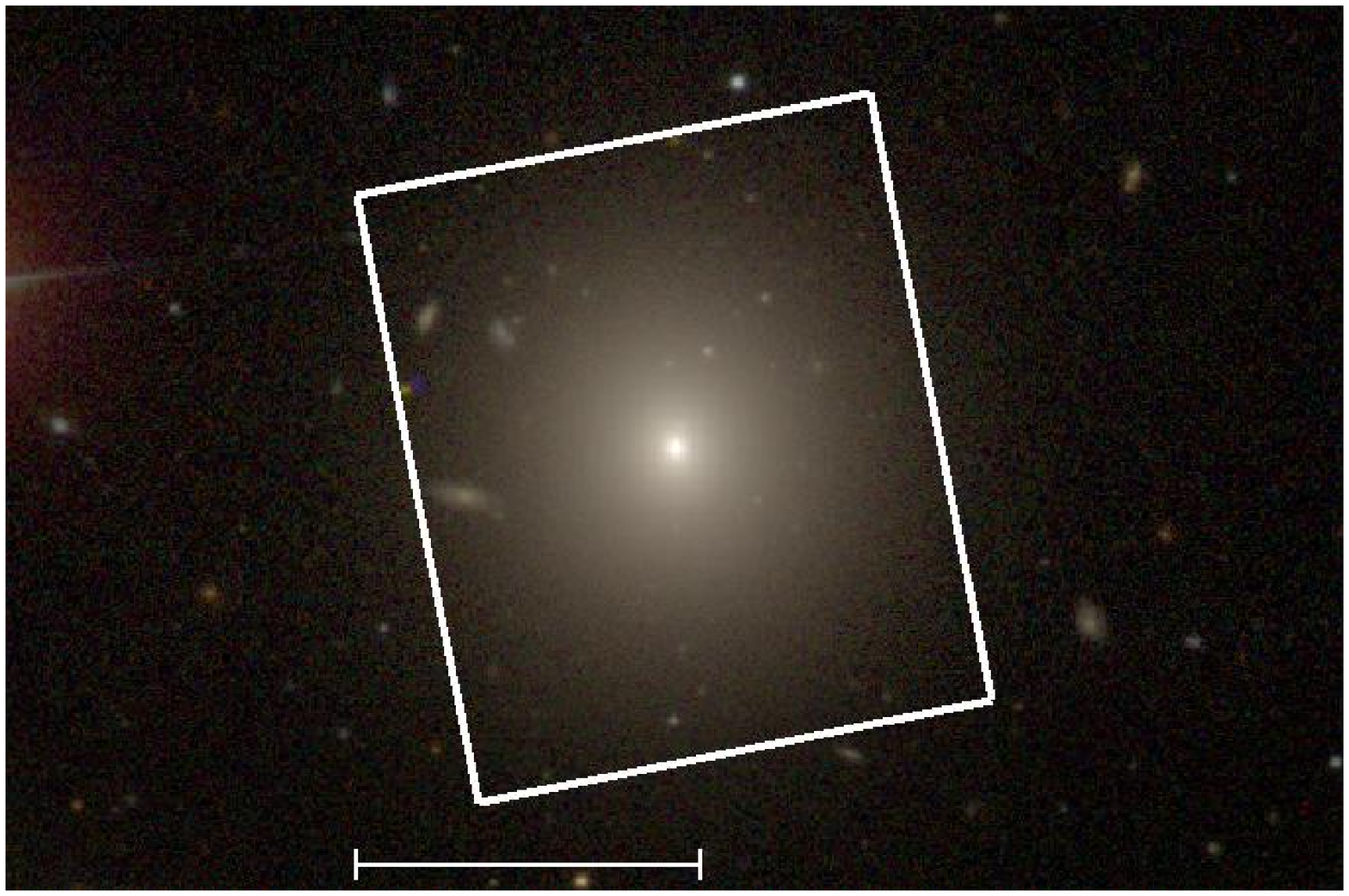}\\
\plottwo{f18_063a.ps}{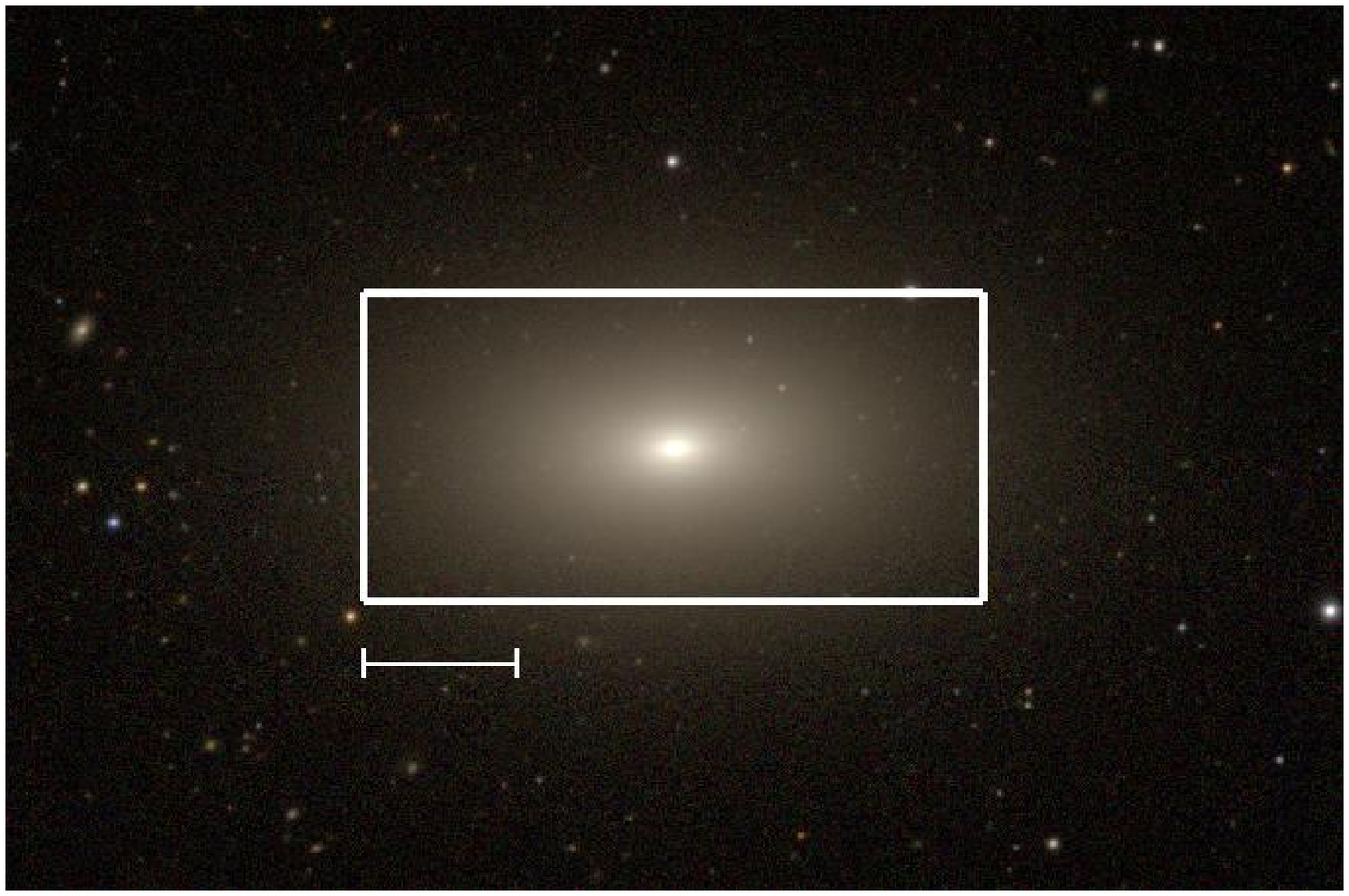}\\
\plottwo{f18_064a.ps}{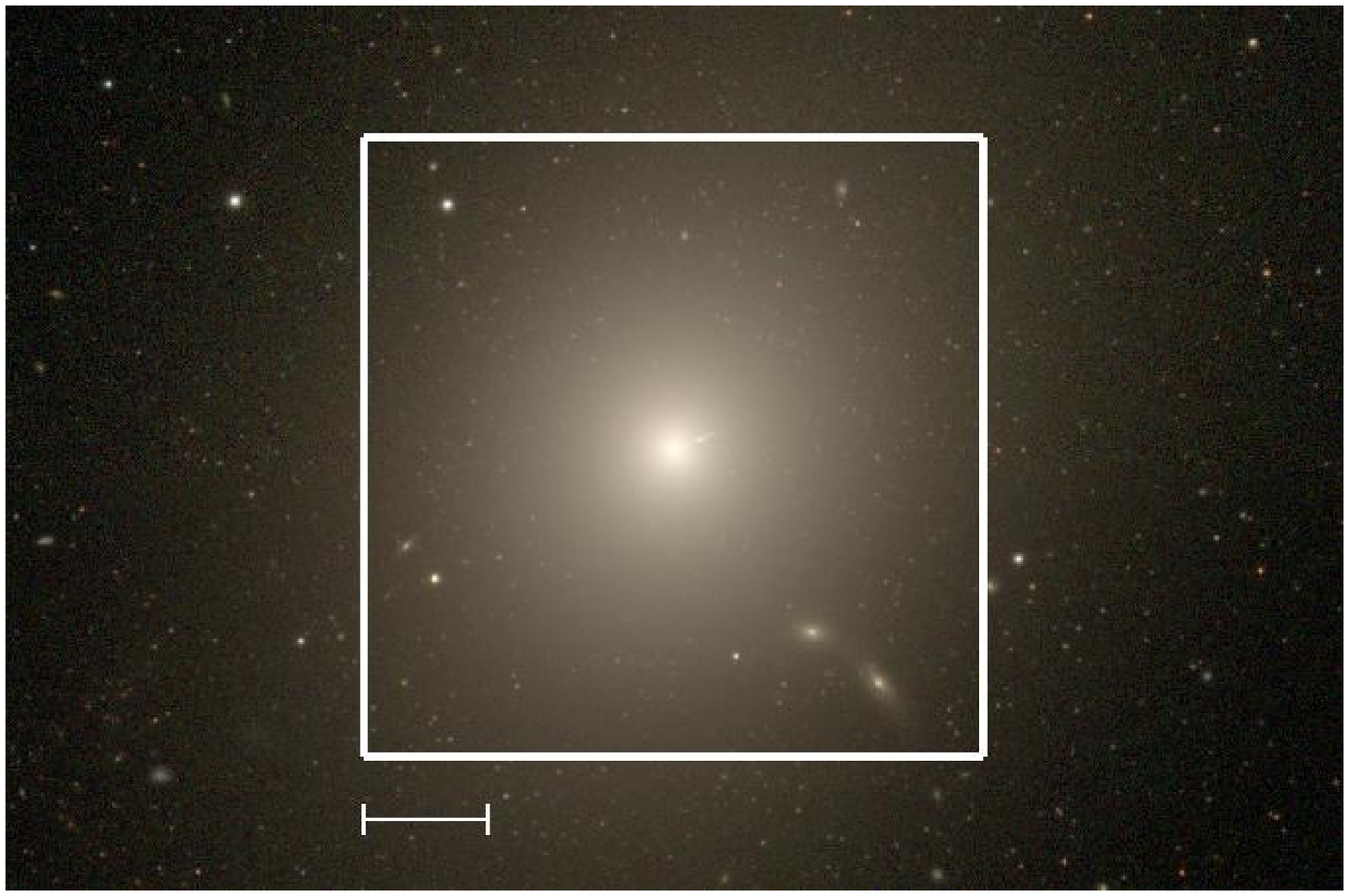}\\
\caption{Galaxy SEDs from the UV to the mid-IR. In the left-panel the observed and model spectra are shown in black and grey respectively, while the photometry used to constrain and verify the spectra is shown with red dots. In the right panel we plot the photometric aperture (thick white rectangle), the {\it Akari} extraction aperture (blue rectangle), the {\it Spitzer} SL extraction aperture (green rectangle) and the {\it Spitzer} LL extraction aperture (yellow rectangle). For galaxies with {\it Spitzer} stare mode spectra, we show a region corresponding to a quarter of the slit length. For scale, the horizontal bar denotes $1^{\prime}$. [{\it See the electronic edition of the Supplement for the complete Figure.}]}
\end{figure}

\clearpage

\begin{figure}[hbt]
\figurenum{\ref{fig:allspec} continued}
\plottwo{f18_065a.ps}{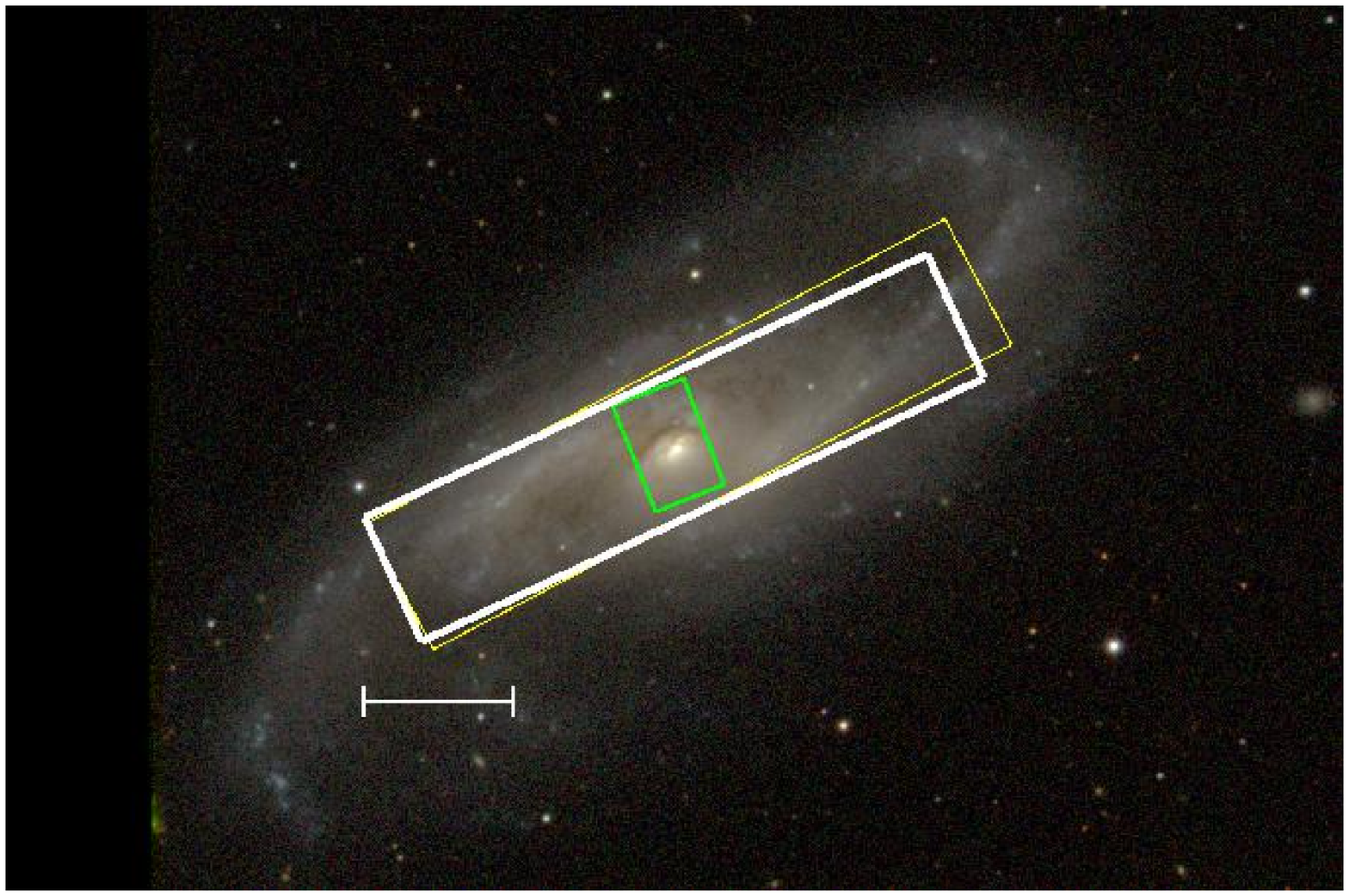}\\
\plottwo{f18_066a.ps}{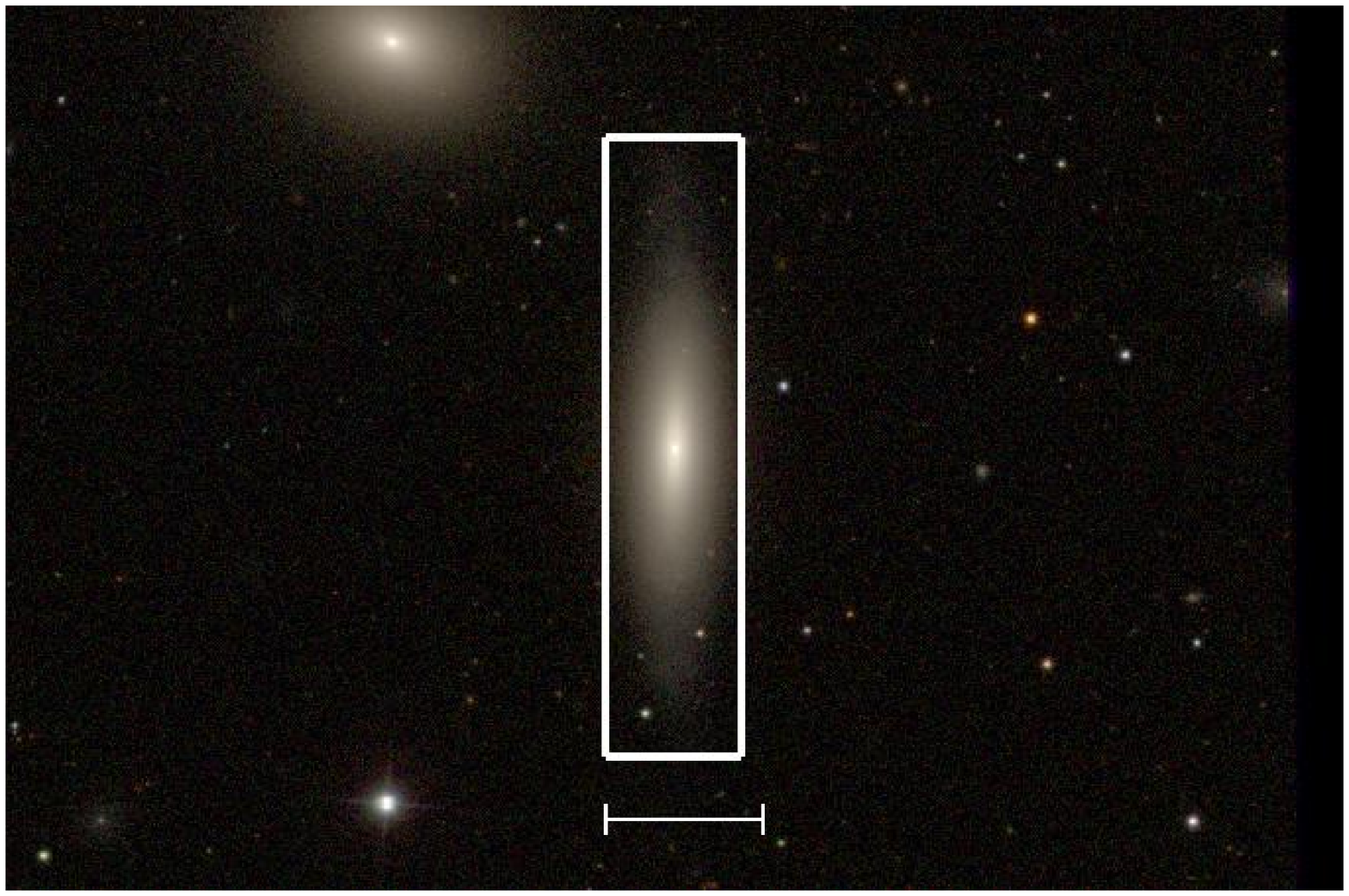}\\
\plottwo{f18_067a.ps}{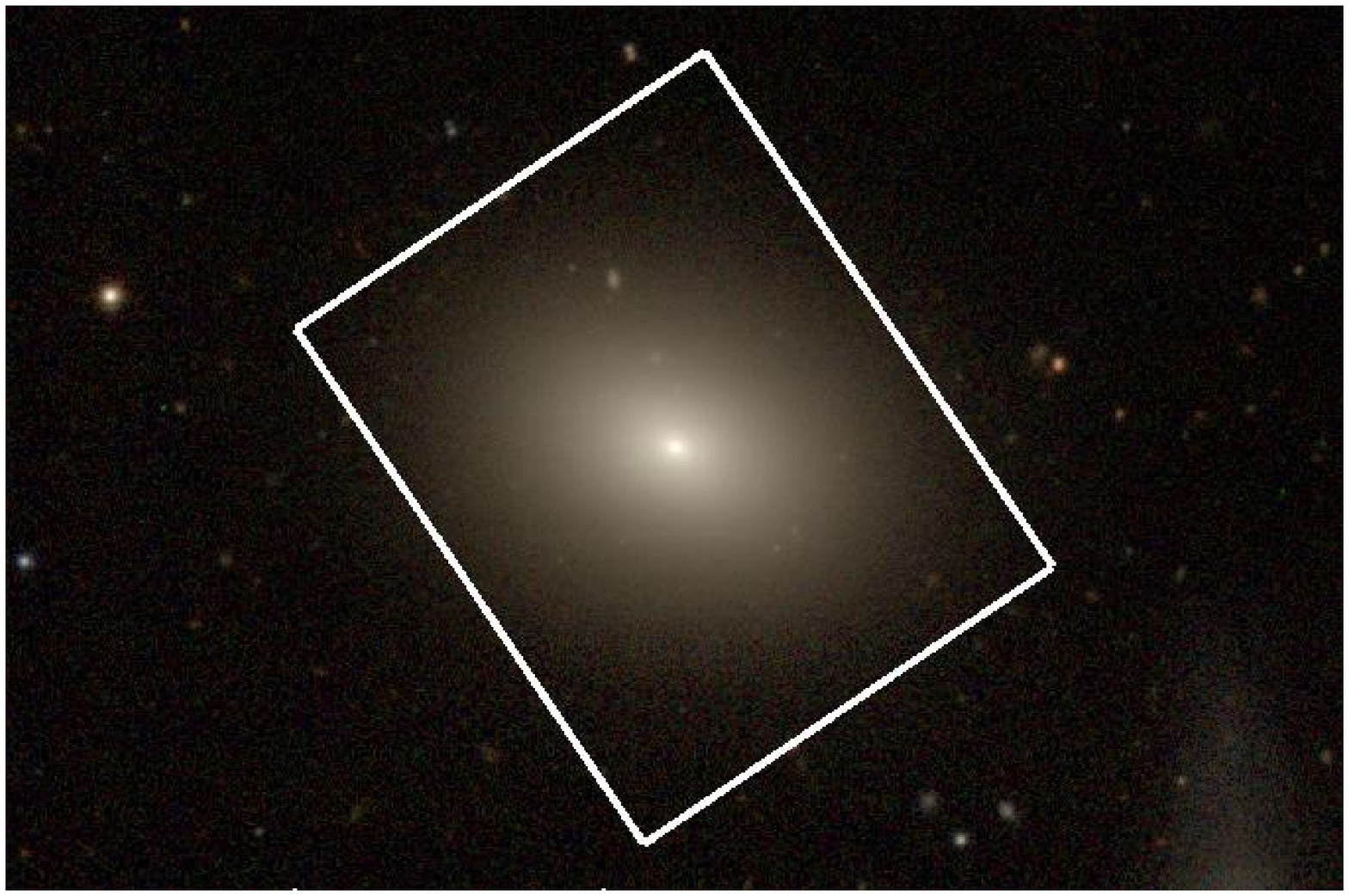}\\
\plottwo{f18_068a.ps}{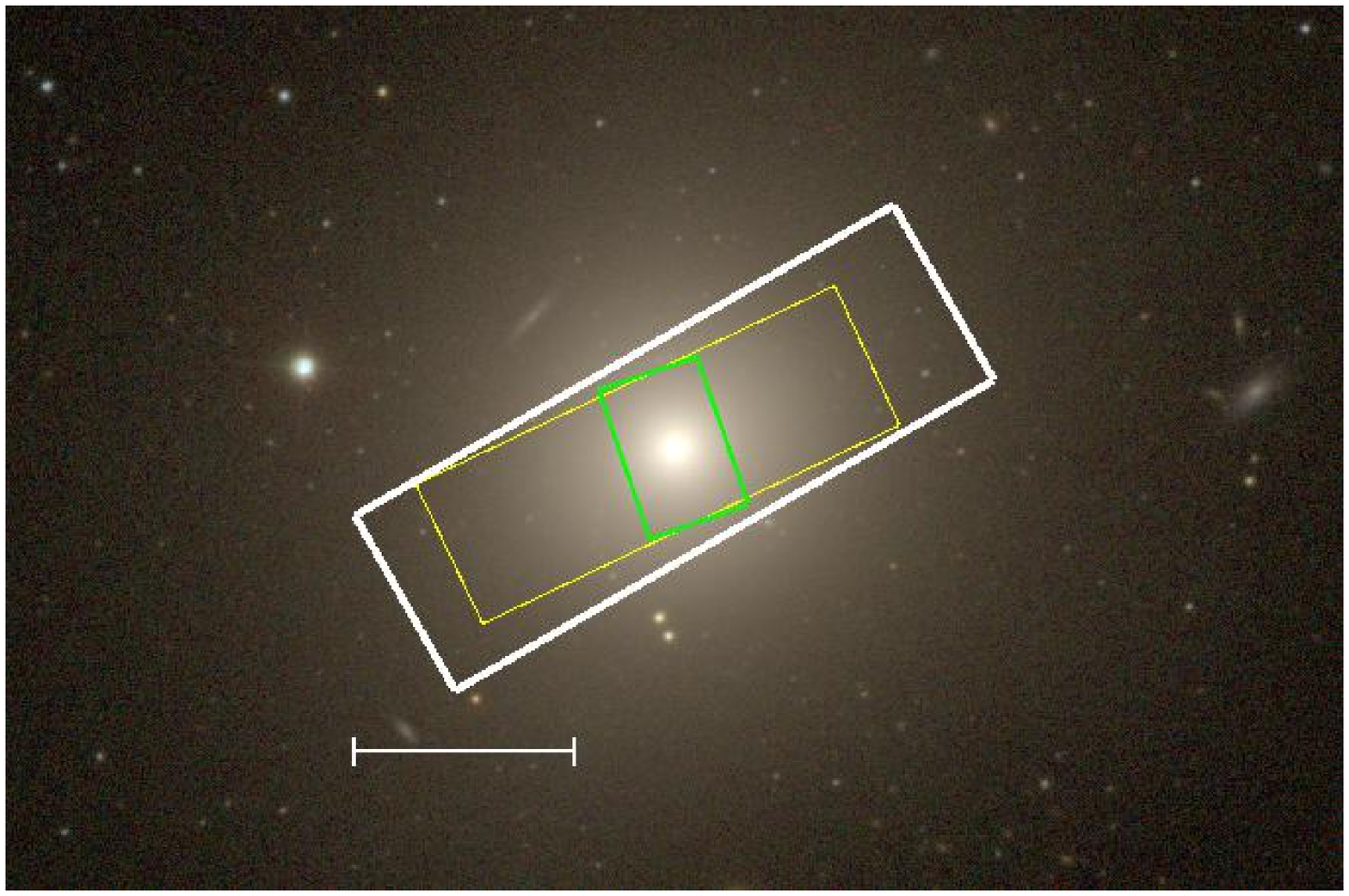}\\
\caption{Galaxy SEDs from the UV to the mid-IR. In the left-panel the observed and model spectra are shown in black and grey respectively, while the photometry used to constrain and verify the spectra is shown with red dots. In the right panel we plot the photometric aperture (thick white rectangle), the {\it Akari} extraction aperture (blue rectangle), the {\it Spitzer} SL extraction aperture (green rectangle) and the {\it Spitzer} LL extraction aperture (yellow rectangle). For galaxies with {\it Spitzer} stare mode spectra, we show a region corresponding to a quarter of the slit length. For scale, the horizontal bar denotes $1^{\prime}$. [{\it See the electronic edition of the Supplement for the complete Figure.}]}
\end{figure}

\clearpage

\begin{figure}[hbt]
\figurenum{\ref{fig:allspec} continued}
\plottwo{f18_069a.ps}{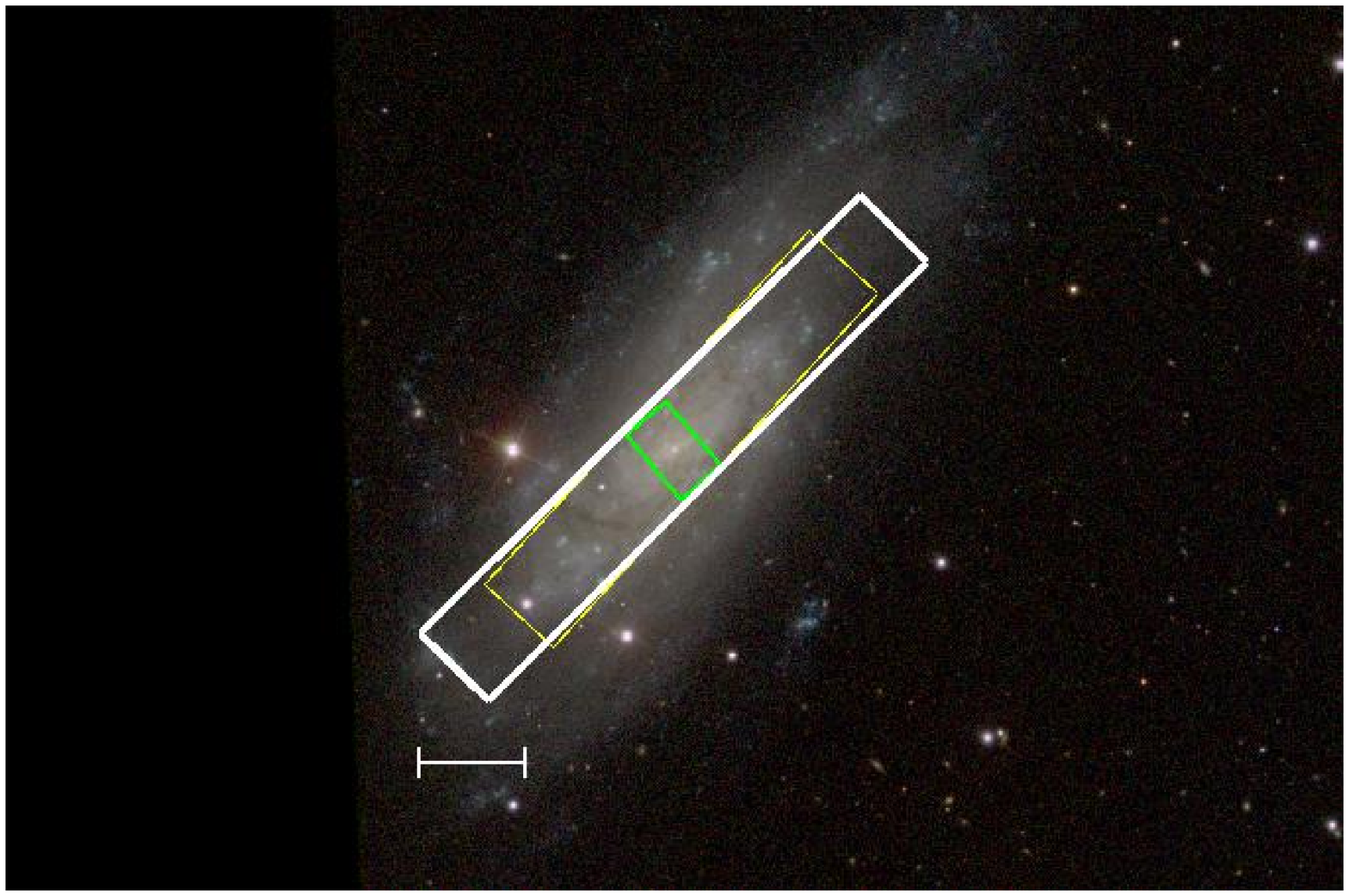}\\
\plottwo{f18_070a.ps}{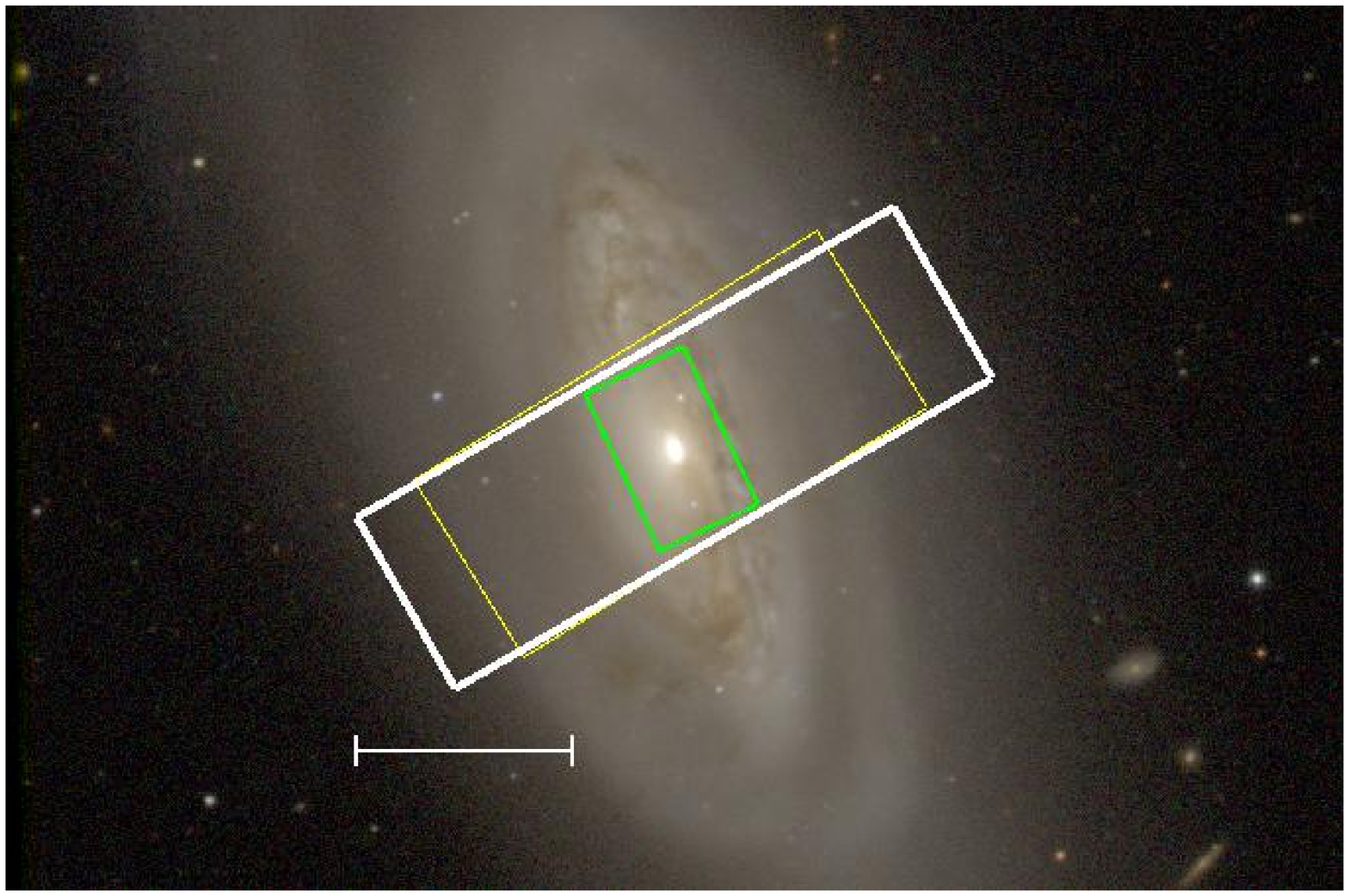}\\
\plottwo{f18_071a.ps}{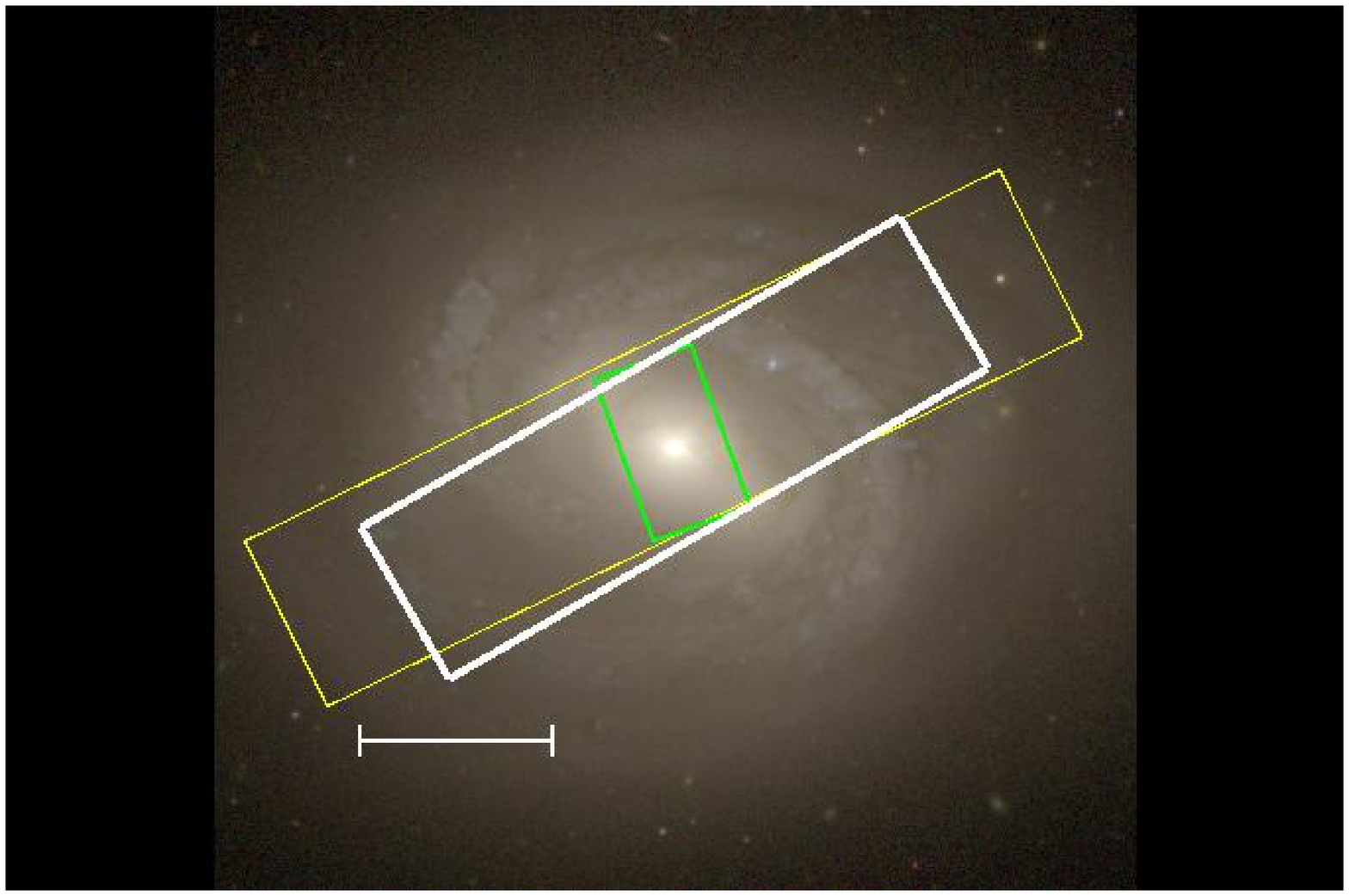}\\
\plottwo{f18_072a.ps}{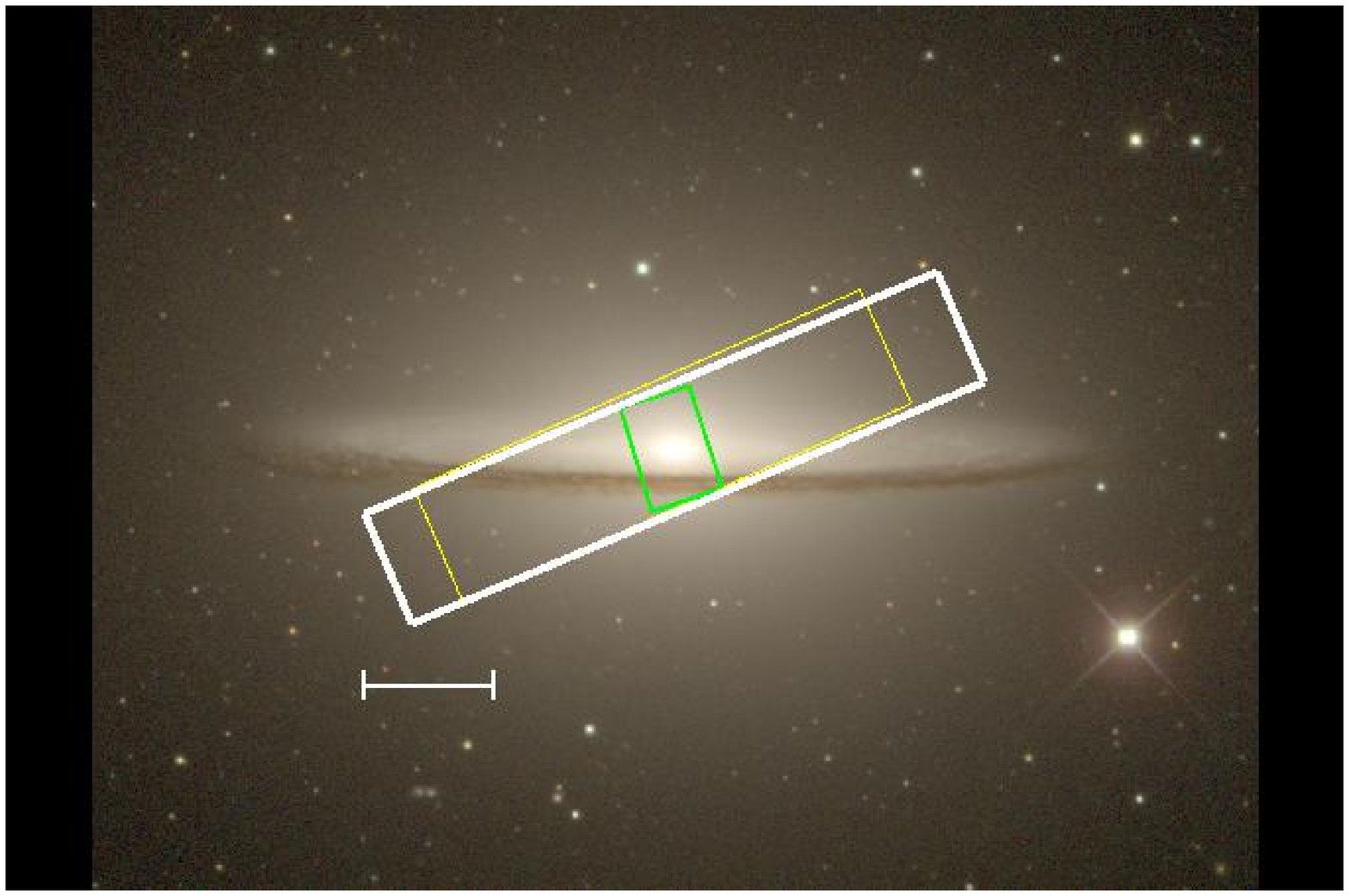}\\
\caption{Galaxy SEDs from the UV to the mid-IR. In the left-panel the observed and model spectra are shown in black and grey respectively, while the photometry used to constrain and verify the spectra is shown with red dots. In the right panel we plot the photometric aperture (thick white rectangle), the {\it Akari} extraction aperture (blue rectangle), the {\it Spitzer} SL extraction aperture (green rectangle) and the {\it Spitzer} LL extraction aperture (yellow rectangle). For galaxies with {\it Spitzer} stare mode spectra, we show a region corresponding to a quarter of the slit length. For scale, the horizontal bar denotes $1^{\prime}$. [{\it See the electronic edition of the Supplement for the complete Figure.}]}
\end{figure}

\clearpage

\begin{figure}[hbt]
\figurenum{\ref{fig:allspec} continued}
\plottwo{f18_073a.ps}{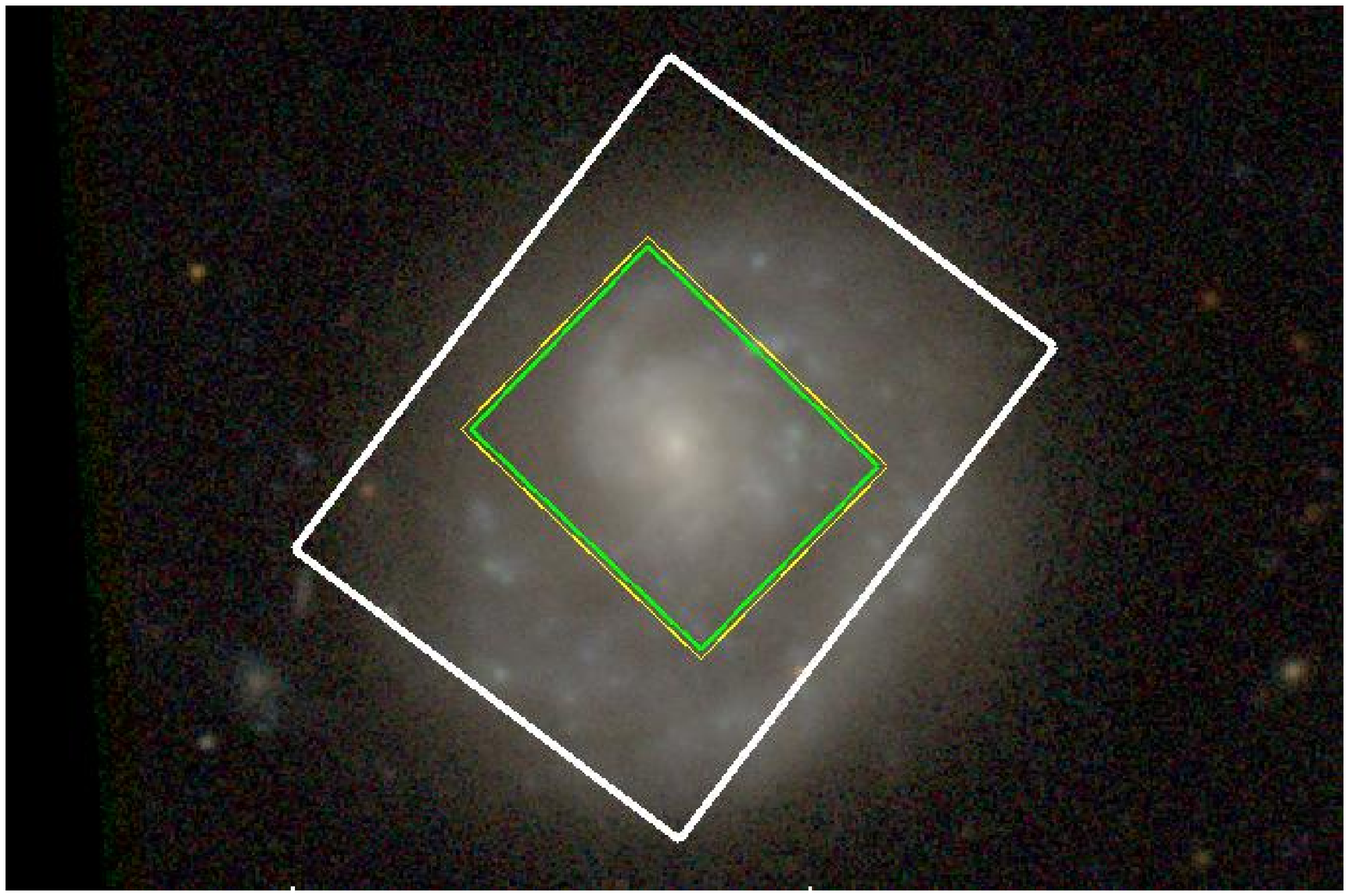}\\
\plottwo{f18_074a.ps}{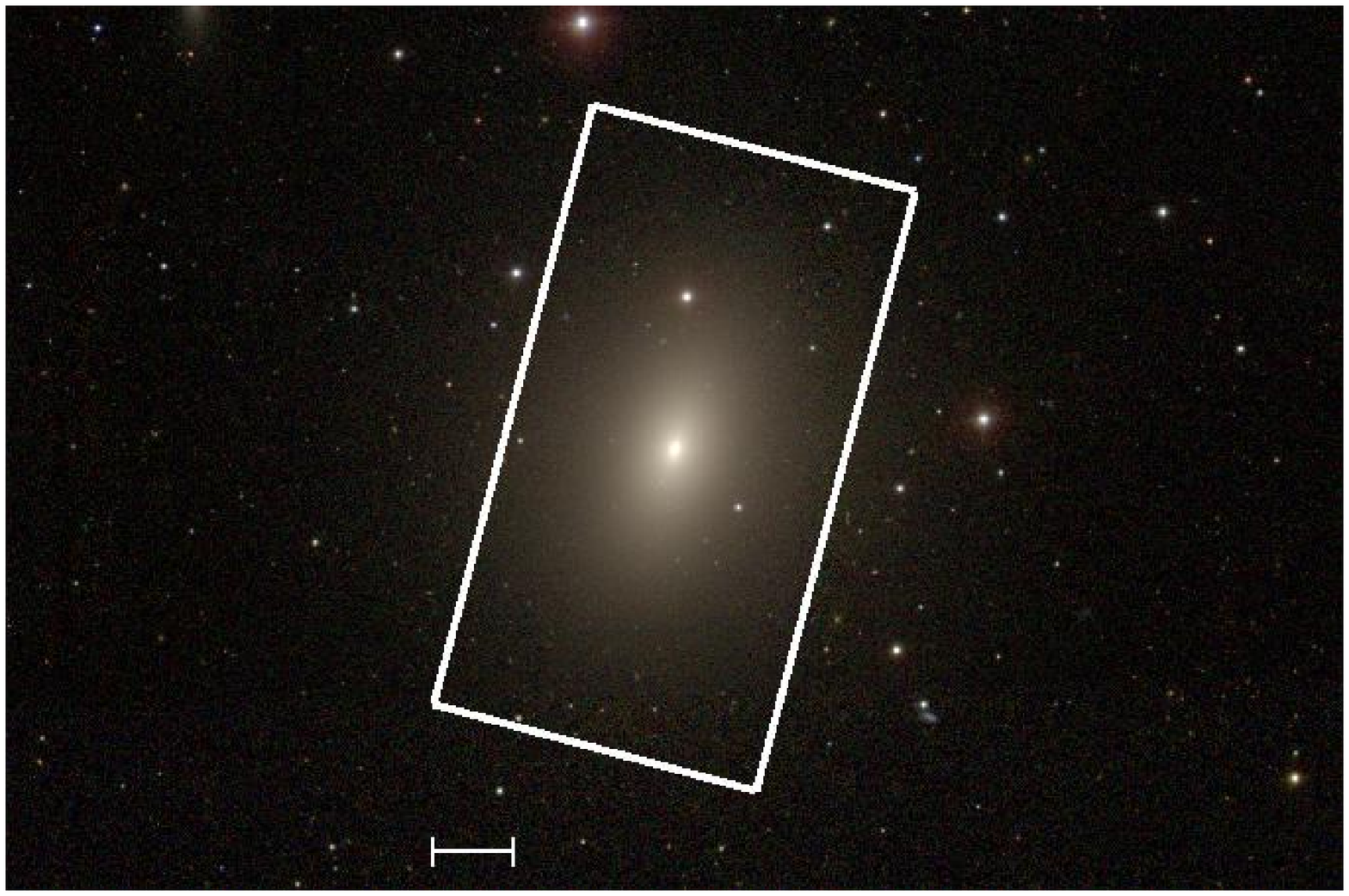}\\
\plottwo{f18_075a.ps}{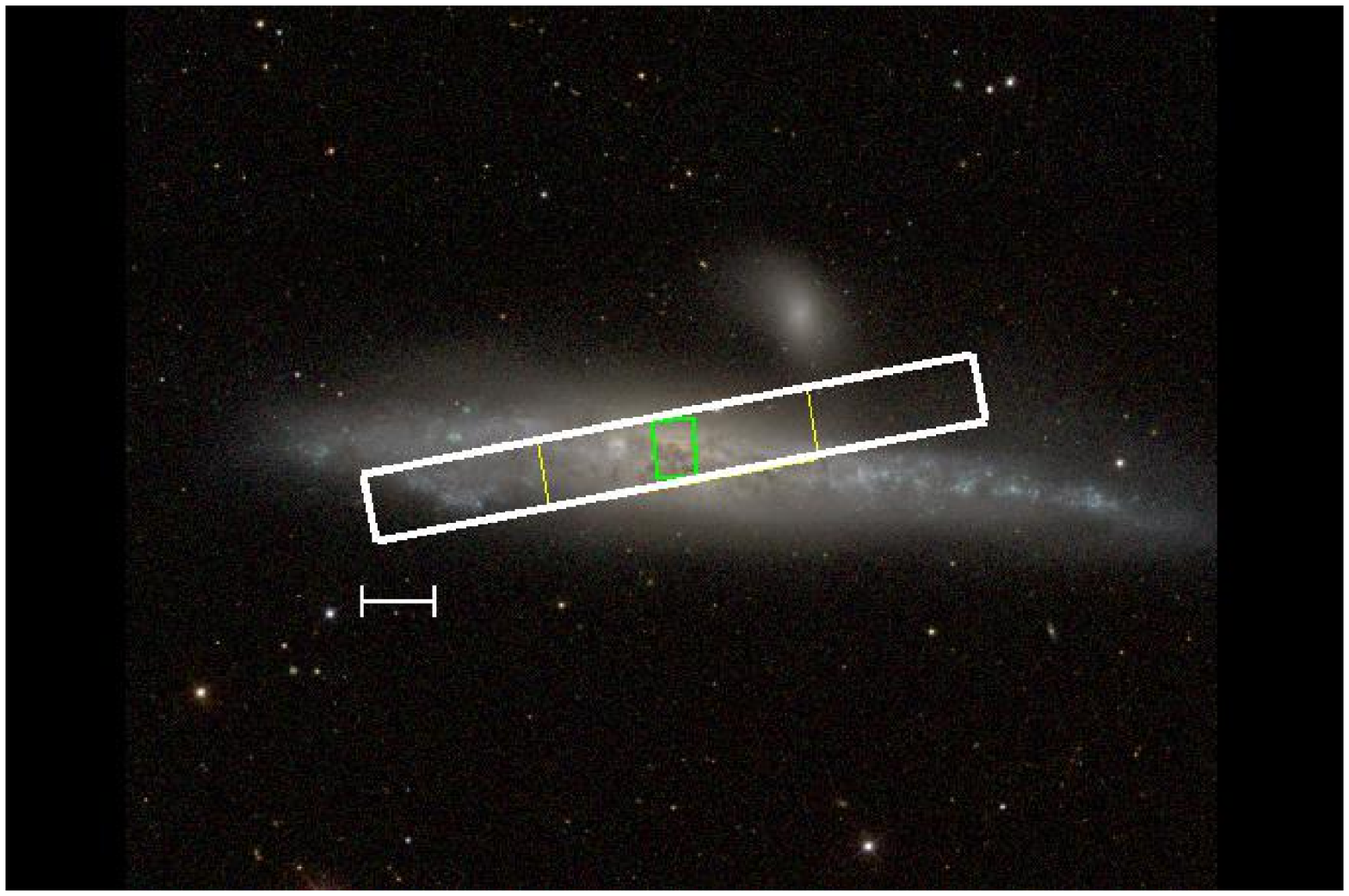}\\
\plottwo{f18_076a.ps}{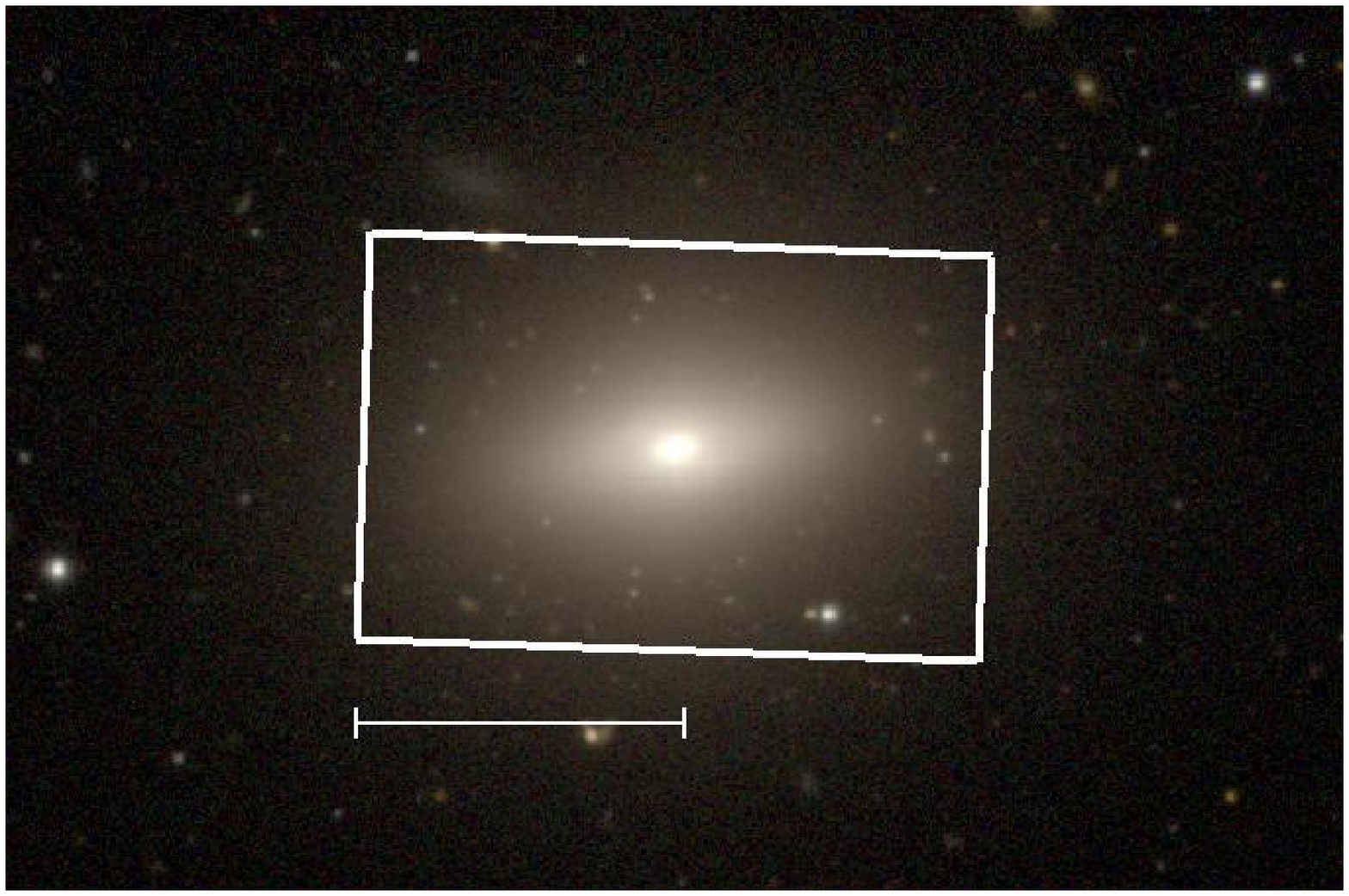}\\
\caption{Galaxy SEDs from the UV to the mid-IR. In the left-panel the observed and model spectra are shown in black and grey respectively, while the photometry used to constrain and verify the spectra is shown with red dots. In the right panel we plot the photometric aperture (thick white rectangle), the {\it Akari} extraction aperture (blue rectangle), the {\it Spitzer} SL extraction aperture (green rectangle) and the {\it Spitzer} LL extraction aperture (yellow rectangle). For galaxies with {\it Spitzer} stare mode spectra, we show a region corresponding to a quarter of the slit length. For scale, the horizontal bar denotes $1^{\prime}$. [{\it See the electronic edition of the Supplement for the complete Figure.}]}
\end{figure}

\clearpage

\begin{figure}[hbt]
\figurenum{\ref{fig:allspec} continued}
\plottwo{f18_077a.ps}{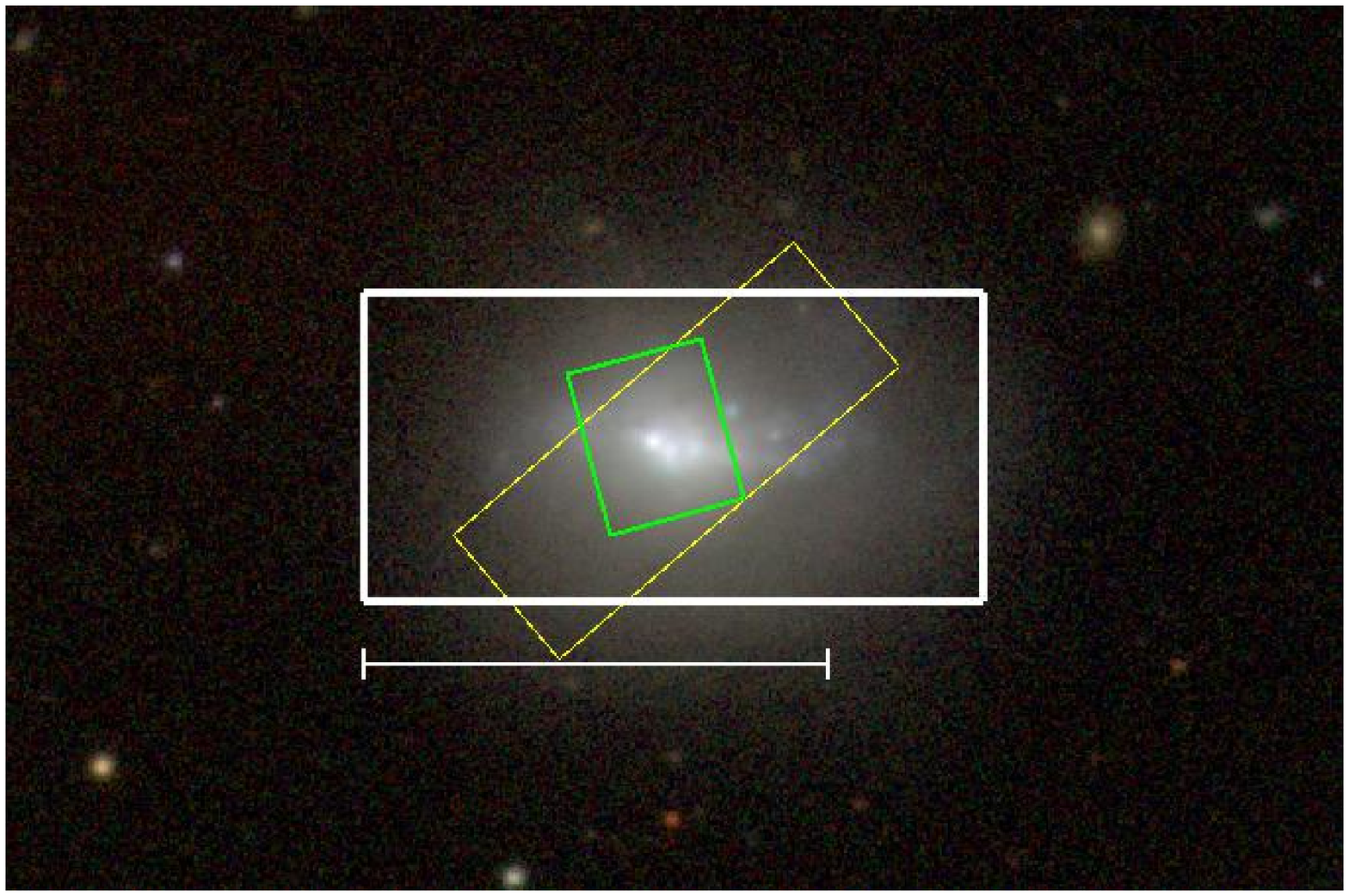}\\
\plottwo{f18_078a.ps}{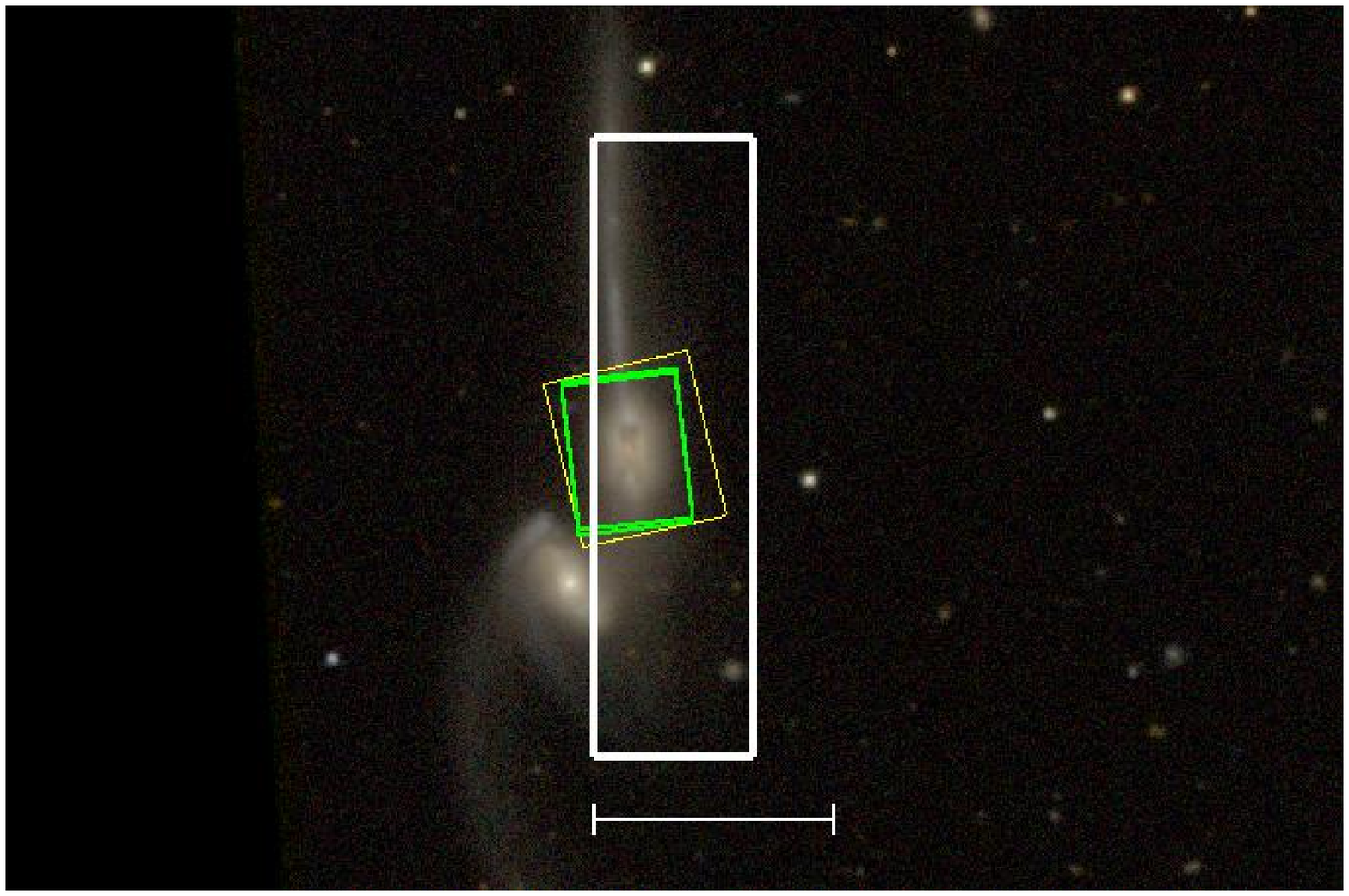}\\
\plottwo{f18_079a.ps}{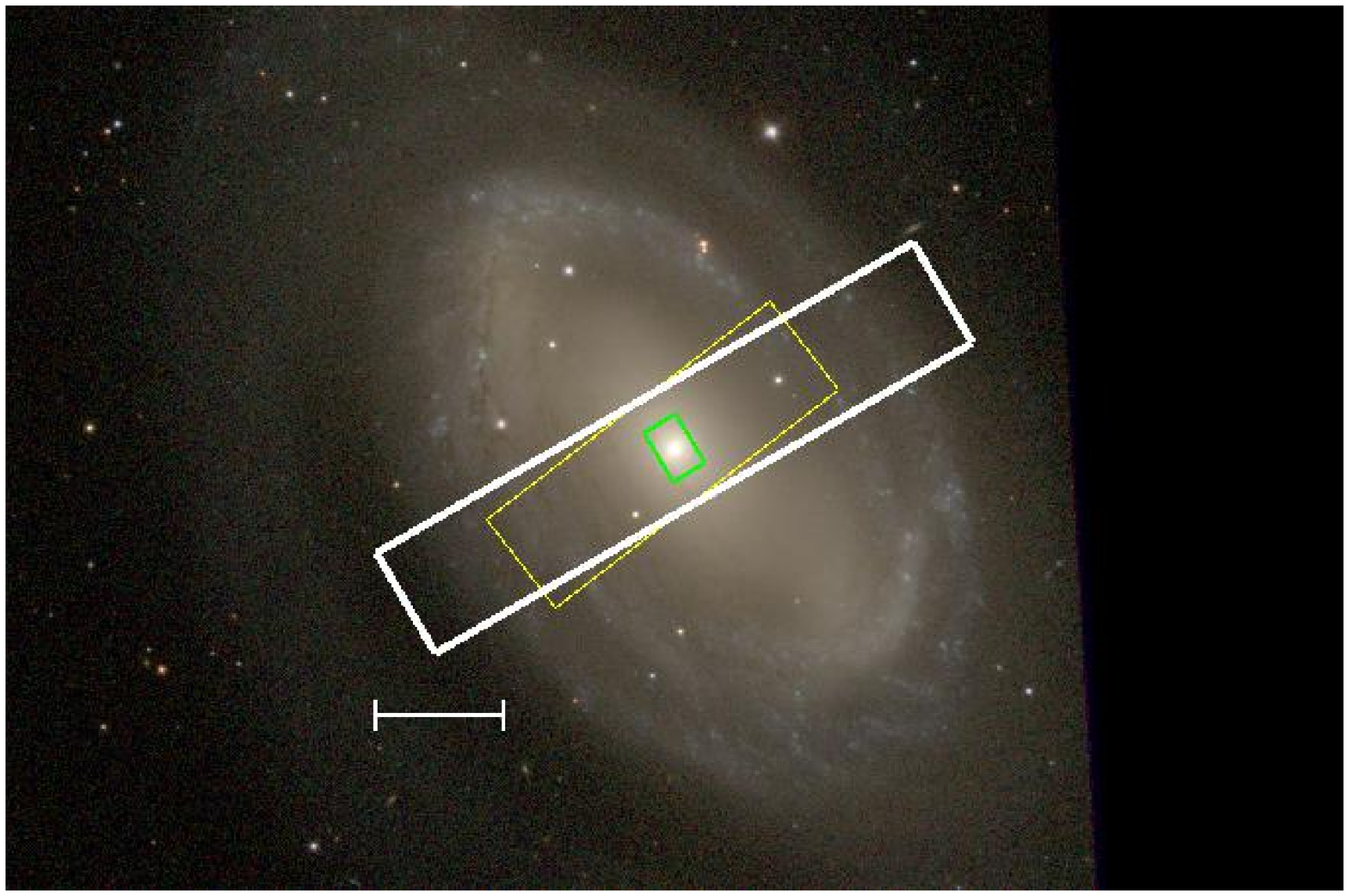}\\
\plottwo{f18_080a.ps}{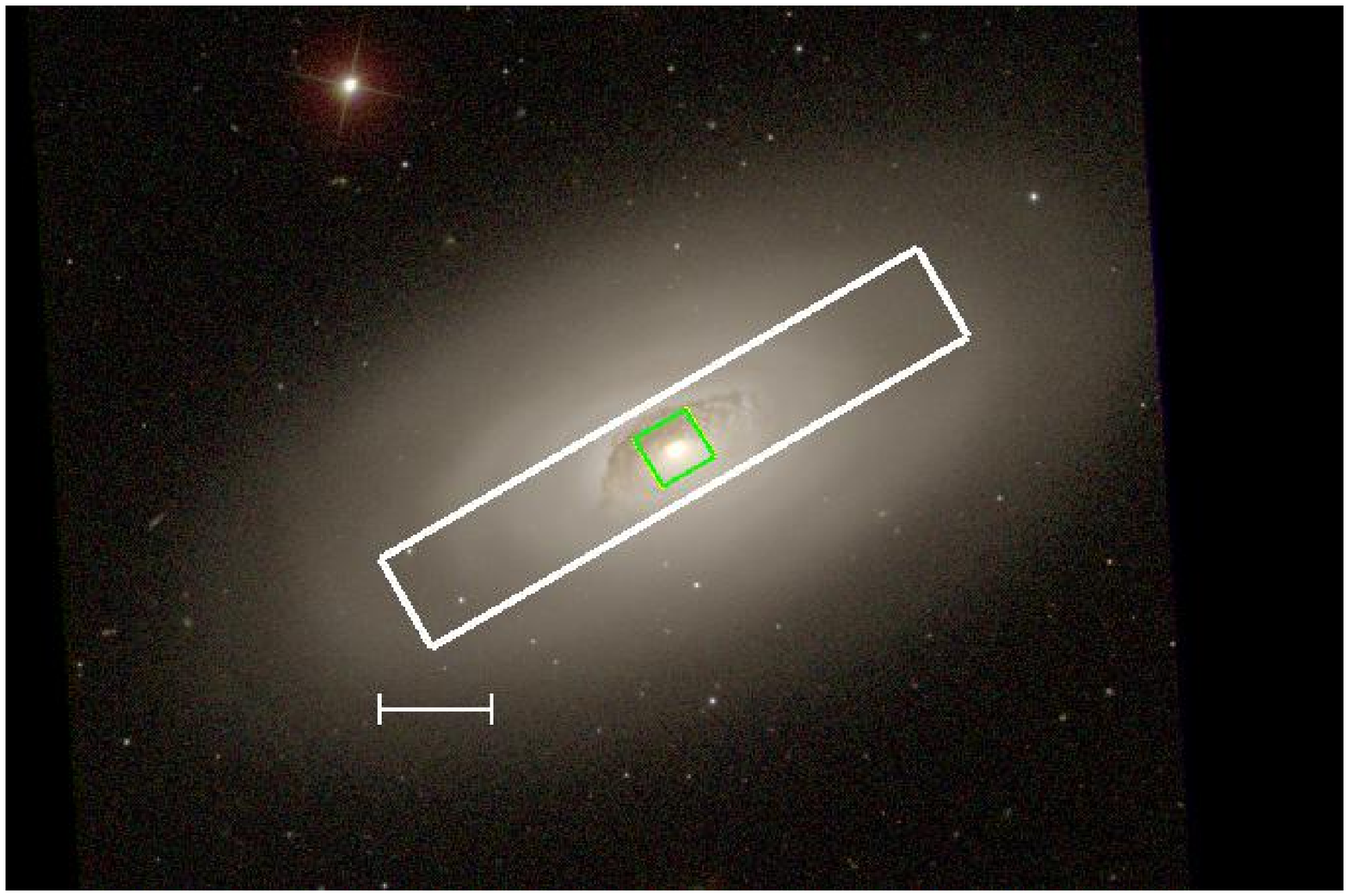}\\
\caption{Galaxy SEDs from the UV to the mid-IR. In the left-panel the observed and model spectra are shown in black and grey respectively, while the photometry used to constrain and verify the spectra is shown with red dots. In the right panel we plot the photometric aperture (thick white rectangle), the {\it Akari} extraction aperture (blue rectangle), the {\it Spitzer} SL extraction aperture (green rectangle) and the {\it Spitzer} LL extraction aperture (yellow rectangle). For galaxies with {\it Spitzer} stare mode spectra, we show a region corresponding to a quarter of the slit length. For scale, the horizontal bar denotes $1^{\prime}$. [{\it See the electronic edition of the Supplement for the complete Figure.}]}
\end{figure}

\clearpage

\begin{figure}[hbt]
\figurenum{\ref{fig:allspec} continued}
\plottwo{f18_081a.ps}{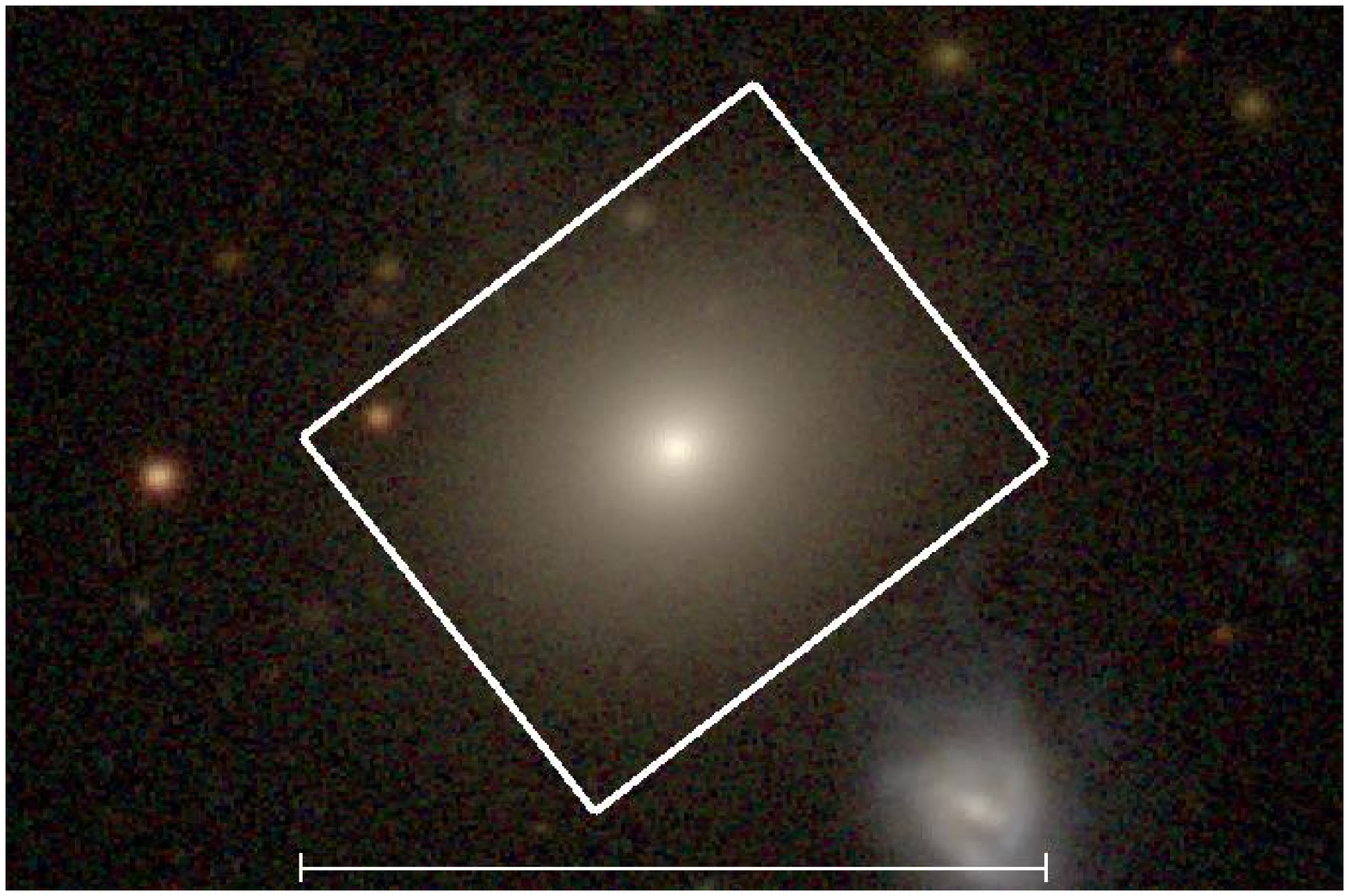}\\
\plottwo{f18_082a.ps}{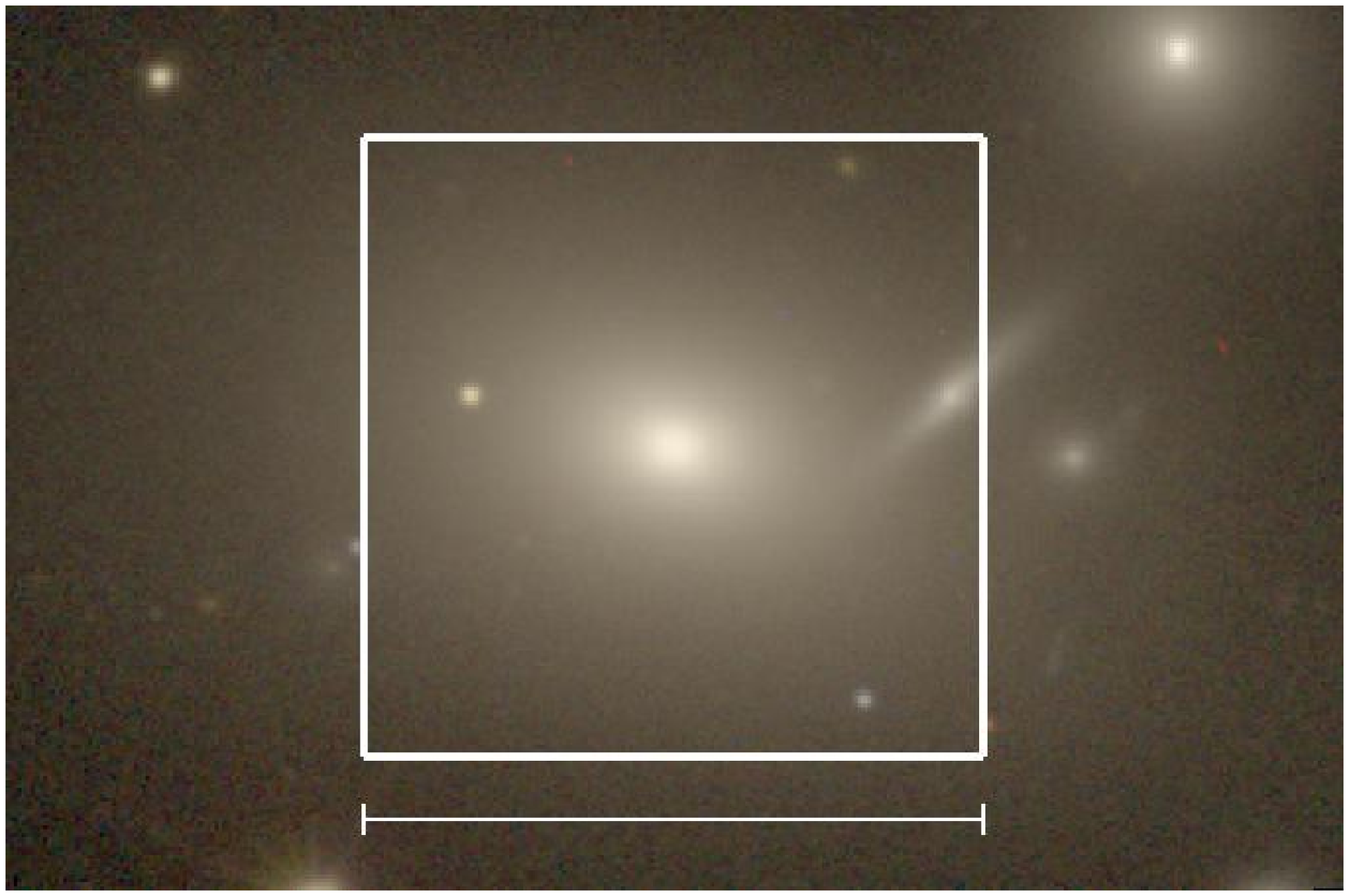}\\
\plottwo{f18_083a.ps}{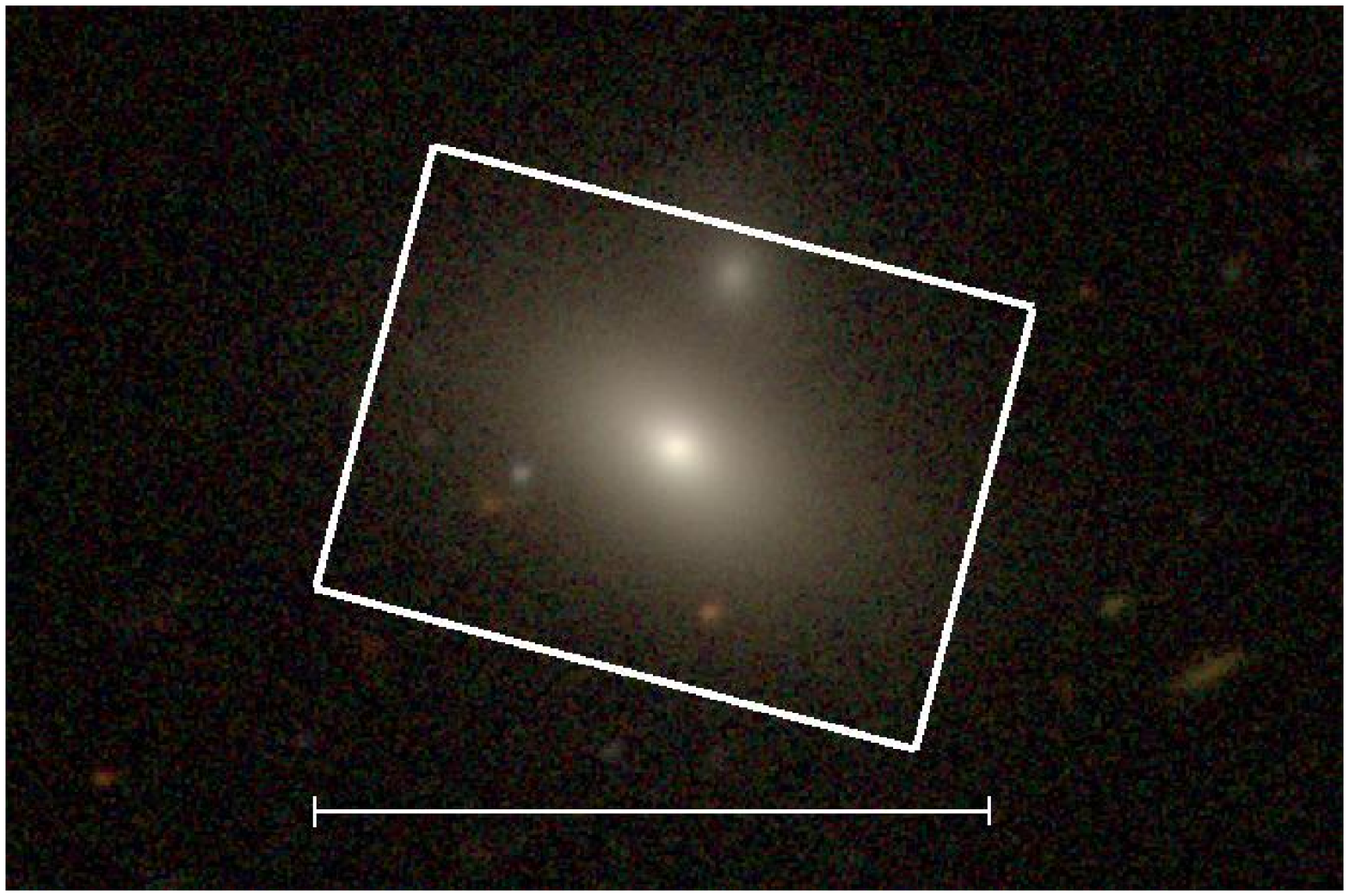}\\
\plottwo{f18_084a.ps}{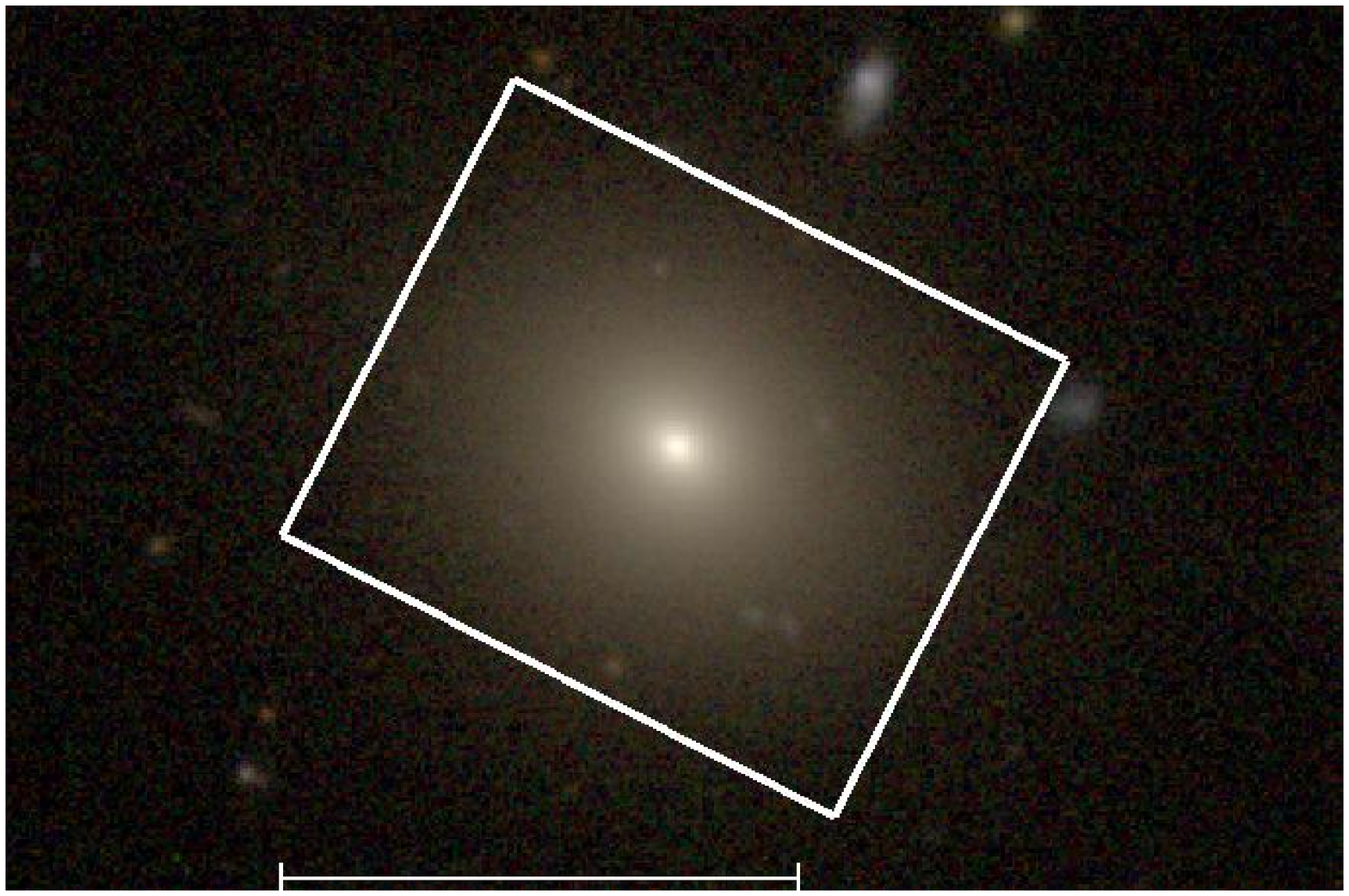}\\
\caption{Galaxy SEDs from the UV to the mid-IR. In the left-panel the observed and model spectra are shown in black and grey respectively, while the photometry used to constrain and verify the spectra is shown with red dots. In the right panel we plot the photometric aperture (thick white rectangle), the {\it Akari} extraction aperture (blue rectangle), the {\it Spitzer} SL extraction aperture (green rectangle) and the {\it Spitzer} LL extraction aperture (yellow rectangle). For galaxies with {\it Spitzer} stare mode spectra, we show a region corresponding to a quarter of the slit length. For scale, the horizontal bar denotes $1^{\prime}$. [{\it See the electronic edition of the Supplement for the complete Figure.}]}
\end{figure}

\clearpage

\begin{figure}[hbt]
\figurenum{\ref{fig:allspec} continued}
\plottwo{f18_085a.ps}{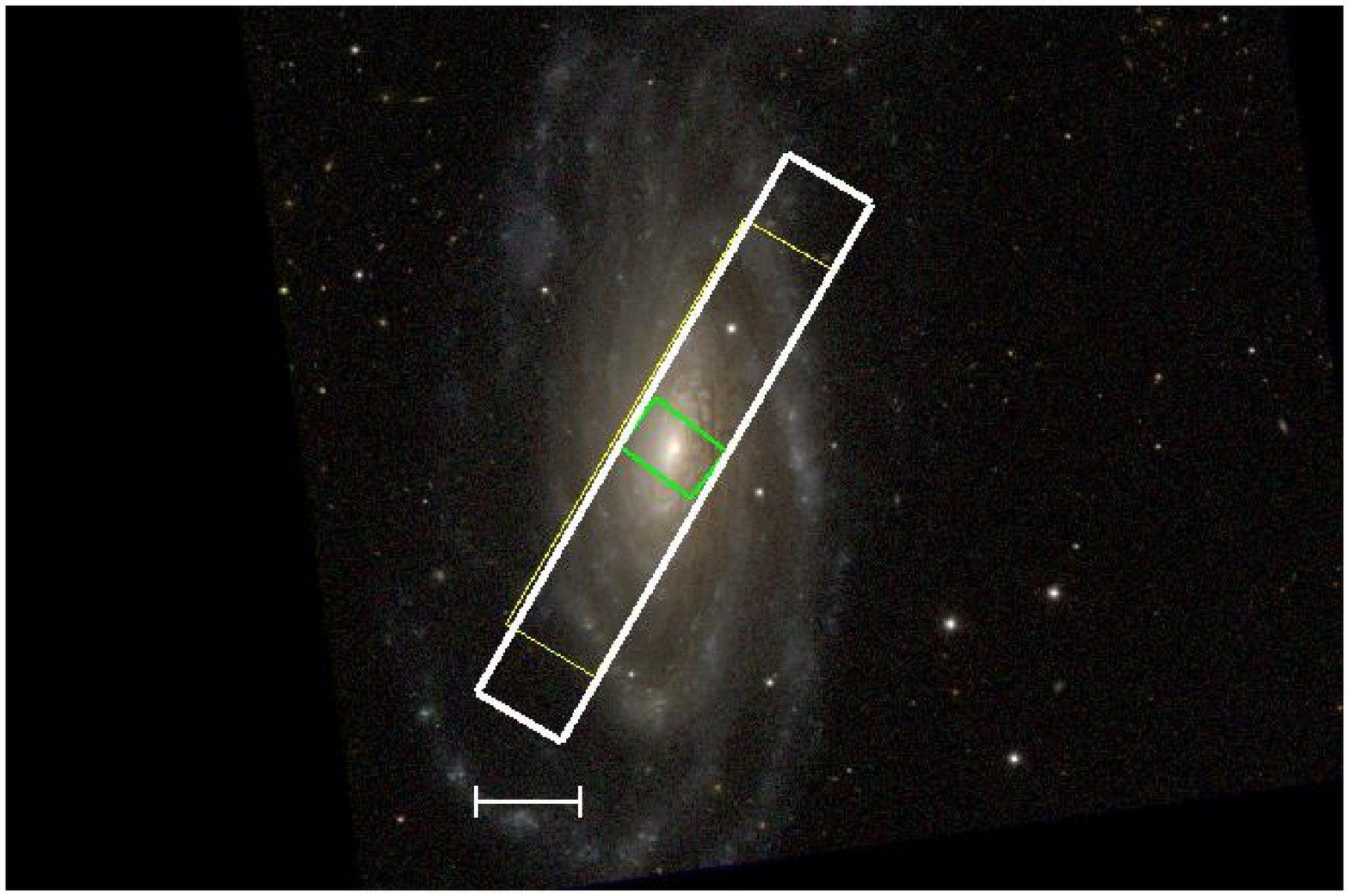}\\
\plottwo{f18_086a.ps}{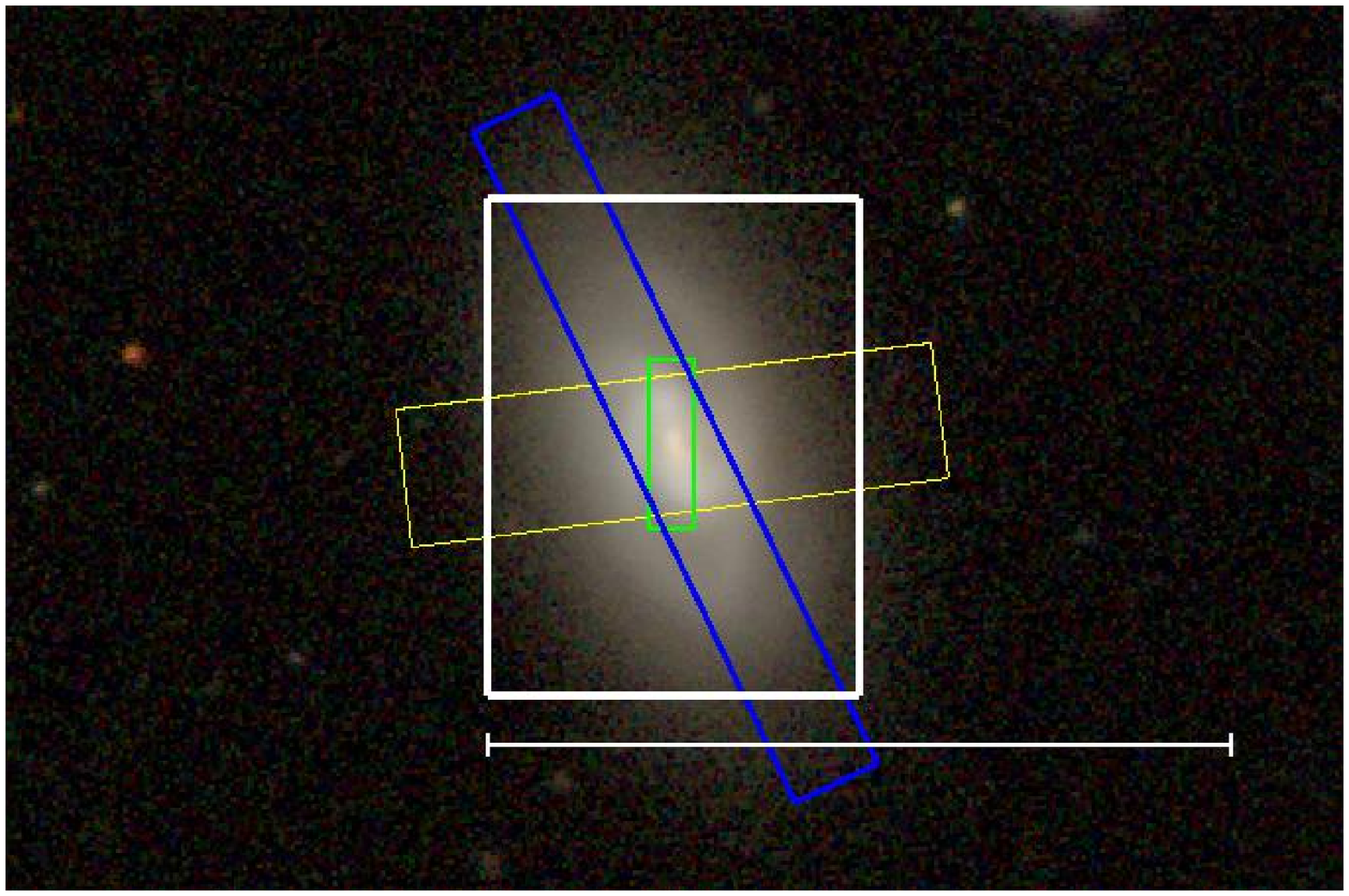}\\
\plottwo{f18_087a.ps}{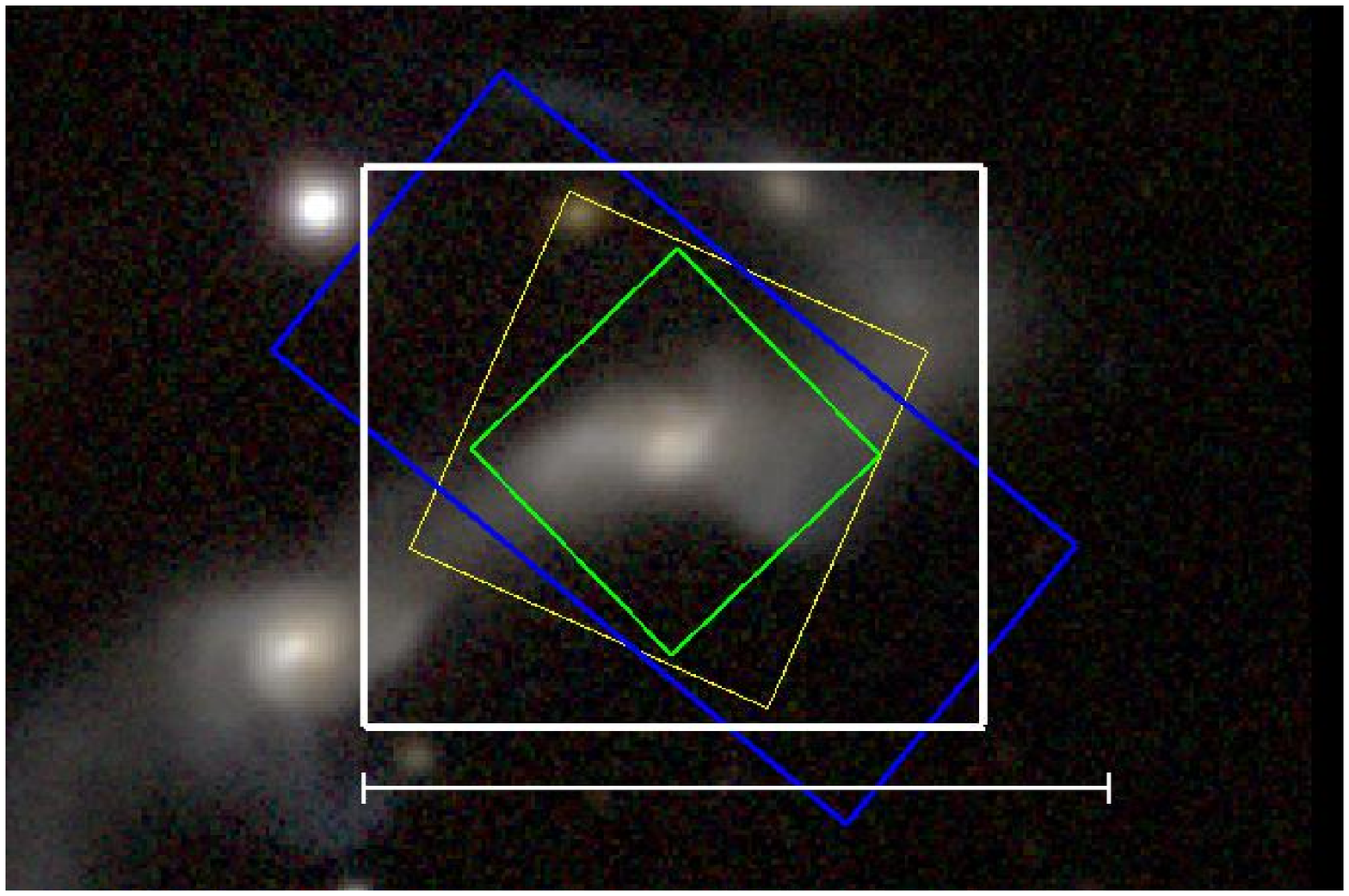}\\
\plottwo{f18_088a.ps}{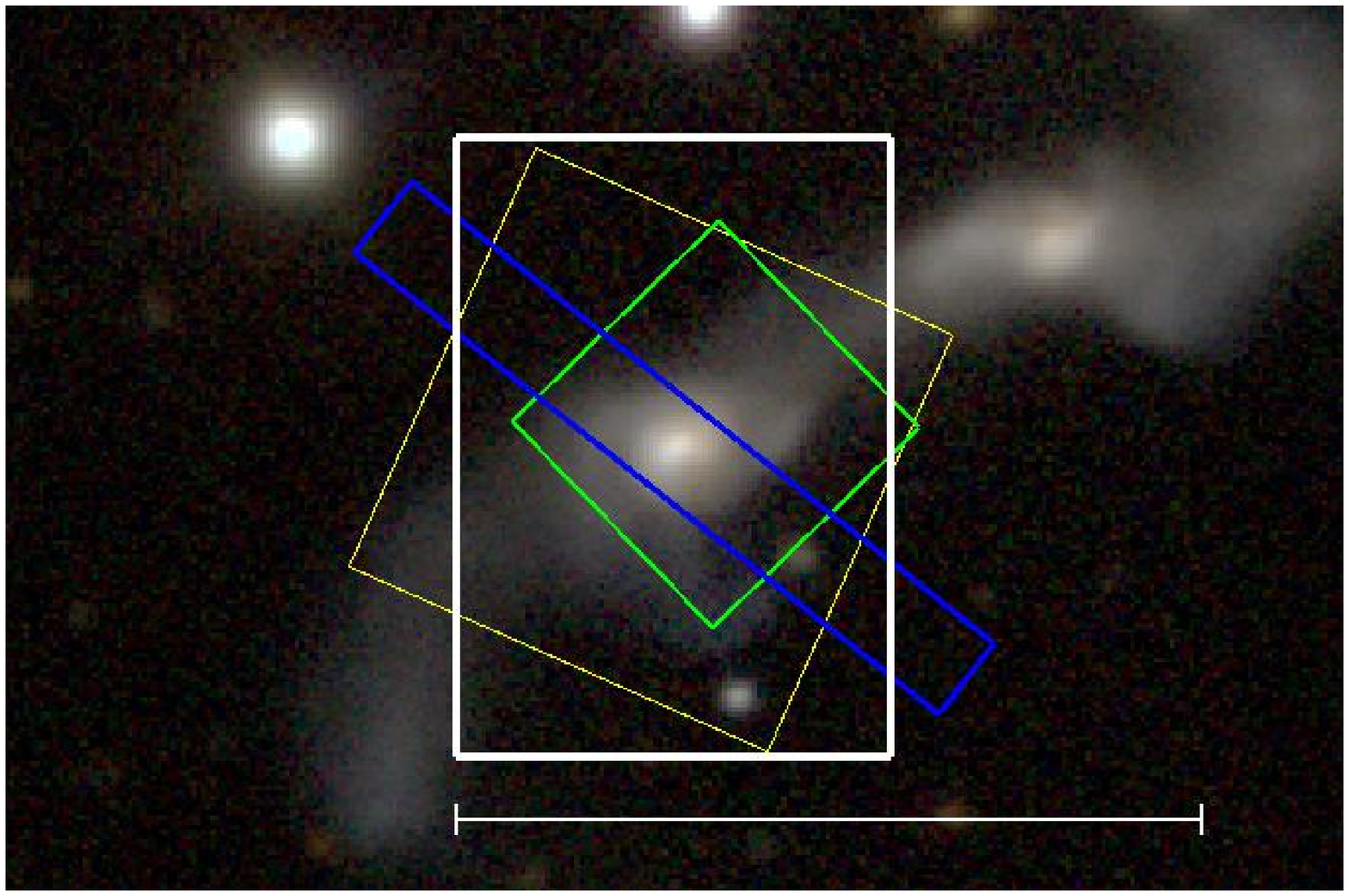}\\
\caption{Galaxy SEDs from the UV to the mid-IR. In the left-panel the observed and model spectra are shown in black and grey respectively, while the photometry used to constrain and verify the spectra is shown with red dots. In the right panel we plot the photometric aperture (thick white rectangle), the {\it Akari} extraction aperture (blue rectangle), the {\it Spitzer} SL extraction aperture (green rectangle) and the {\it Spitzer} LL extraction aperture (yellow rectangle). For galaxies with {\it Spitzer} stare mode spectra, we show a region corresponding to a quarter of the slit length. For scale, the horizontal bar denotes $1^{\prime}$. [{\it See the electronic edition of the Supplement for the complete Figure.}]}
\end{figure}

\clearpage

\begin{figure}[hbt]
\figurenum{\ref{fig:allspec} continued}
\plottwo{f18_089a.ps}{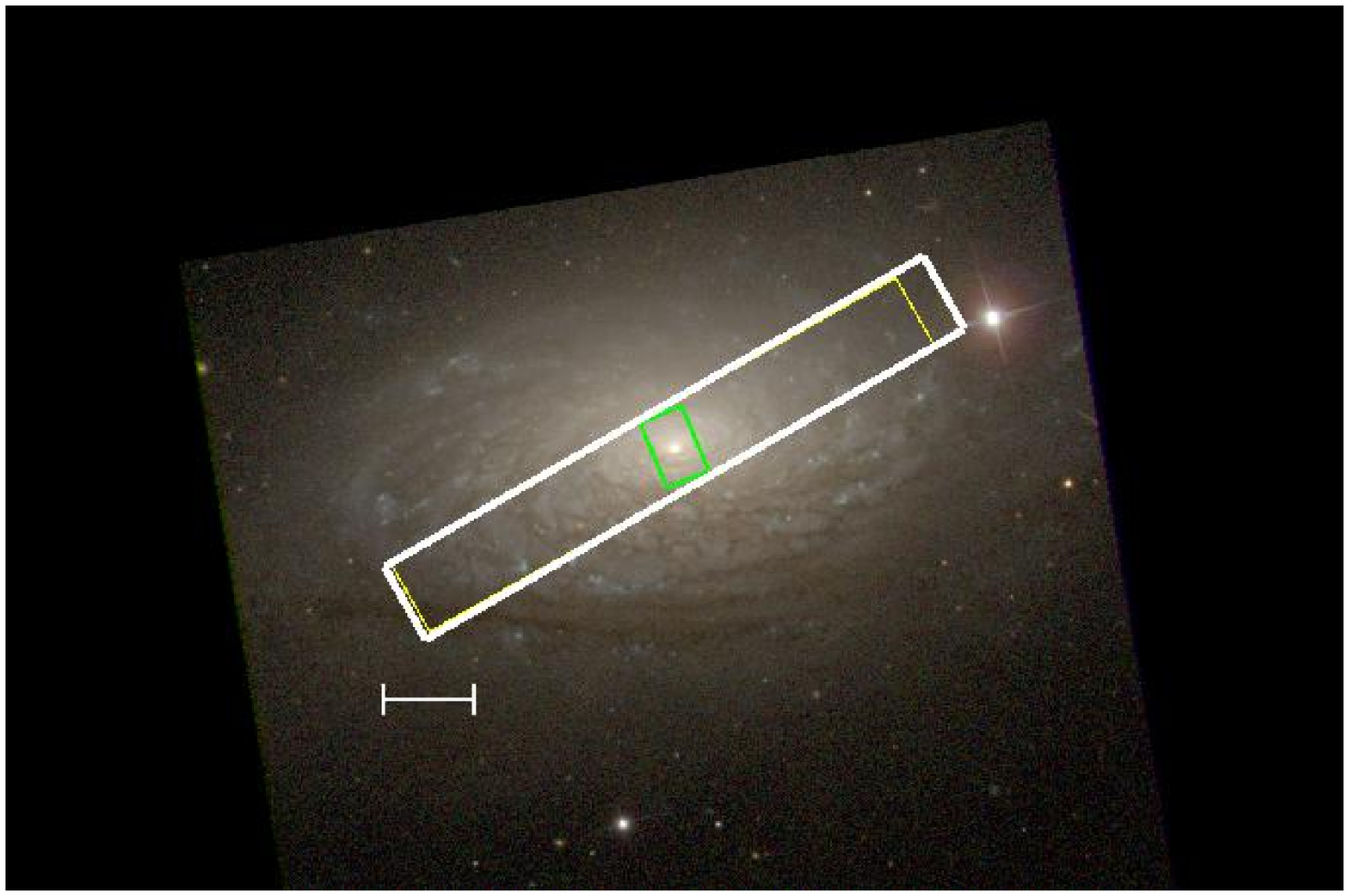}\\
\plottwo{f18_090a.ps}{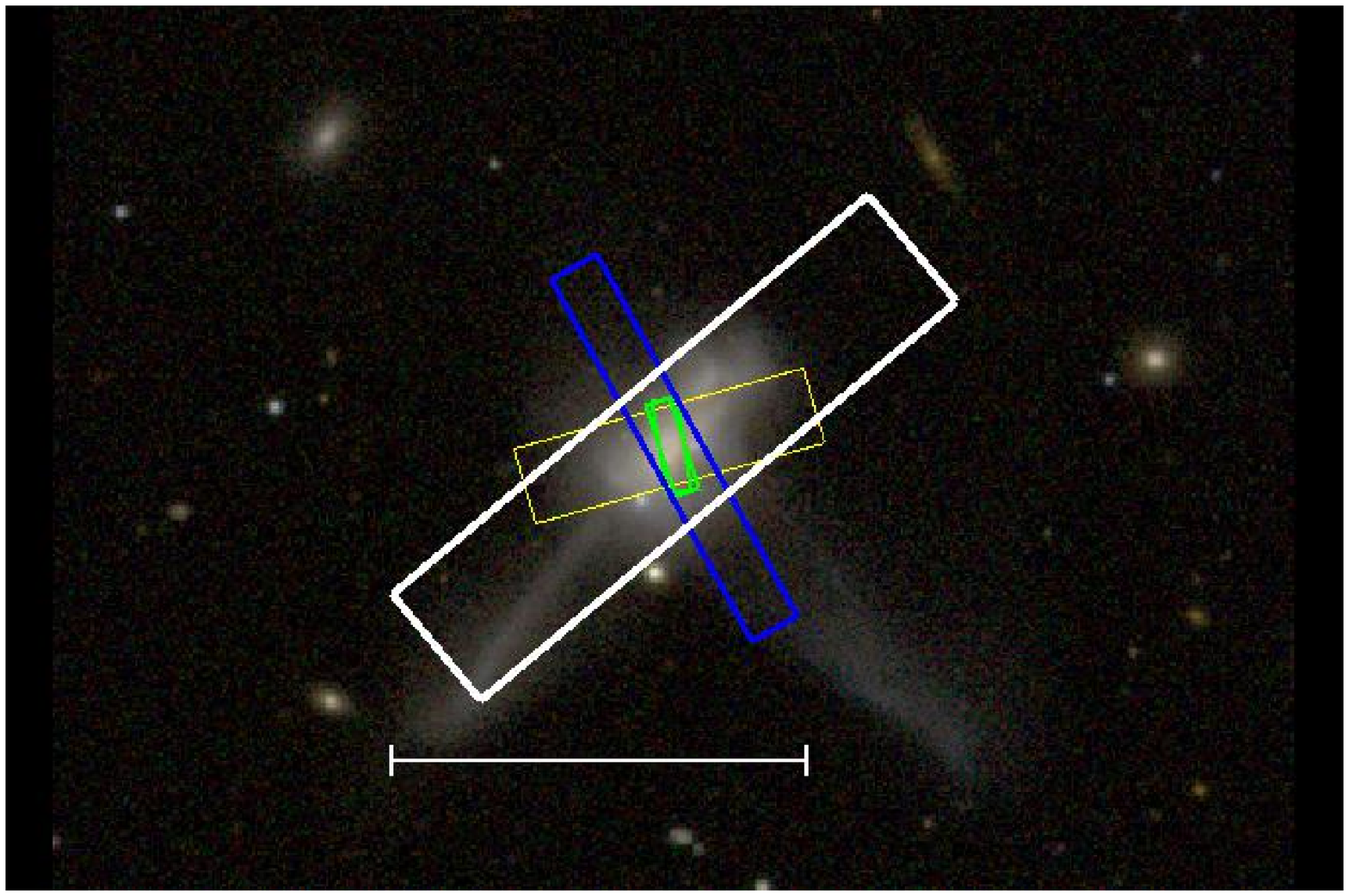}\\
\plottwo{f18_091a.ps}{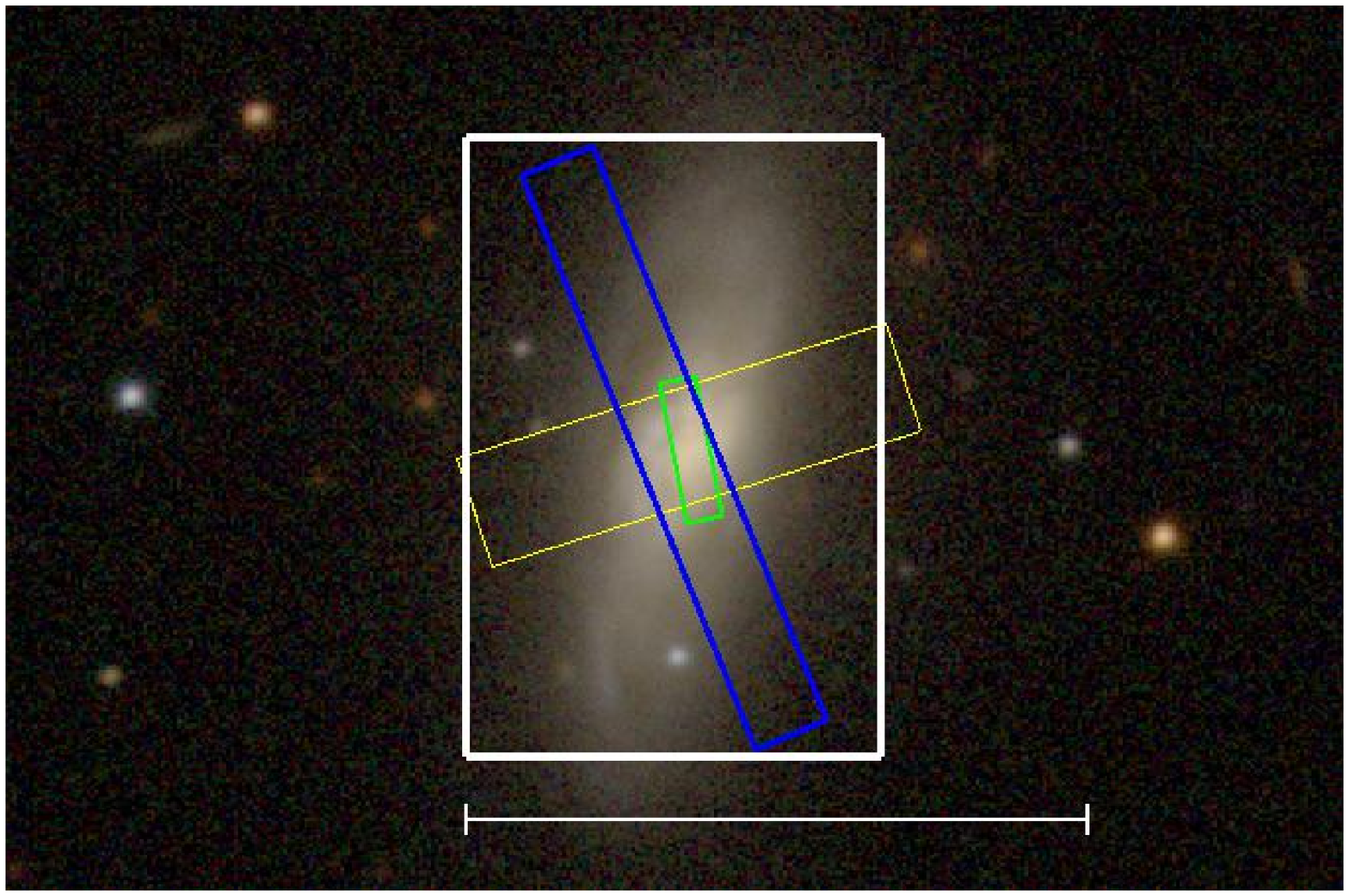}\\
\plottwo{f18_092a.ps}{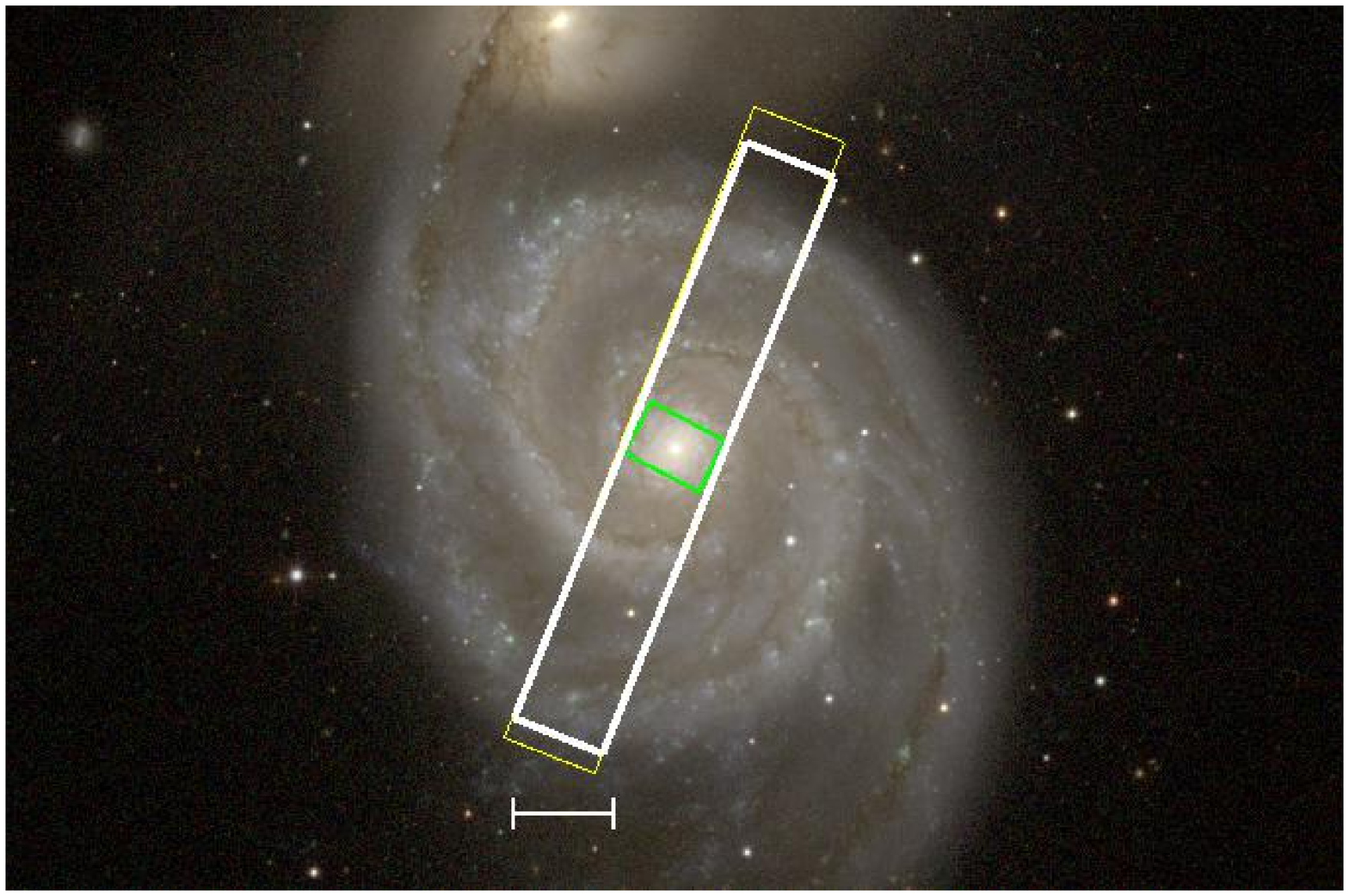}\\
\caption{Galaxy SEDs from the UV to the mid-IR. In the left-panel the observed and model spectra are shown in black and grey respectively, while the photometry used to constrain and verify the spectra is shown with red dots. In the right panel we plot the photometric aperture (thick white rectangle), the {\it Akari} extraction aperture (blue rectangle), the {\it Spitzer} SL extraction aperture (green rectangle) and the {\it Spitzer} LL extraction aperture (yellow rectangle). For galaxies with {\it Spitzer} stare mode spectra, we show a region corresponding to a quarter of the slit length. For scale, the horizontal bar denotes $1^{\prime}$. [{\it See the electronic edition of the Supplement for the complete Figure.}]}
\end{figure}

\clearpage

\begin{figure}[hbt]
\figurenum{\ref{fig:allspec} continued}
\plottwo{f18_093a.ps}{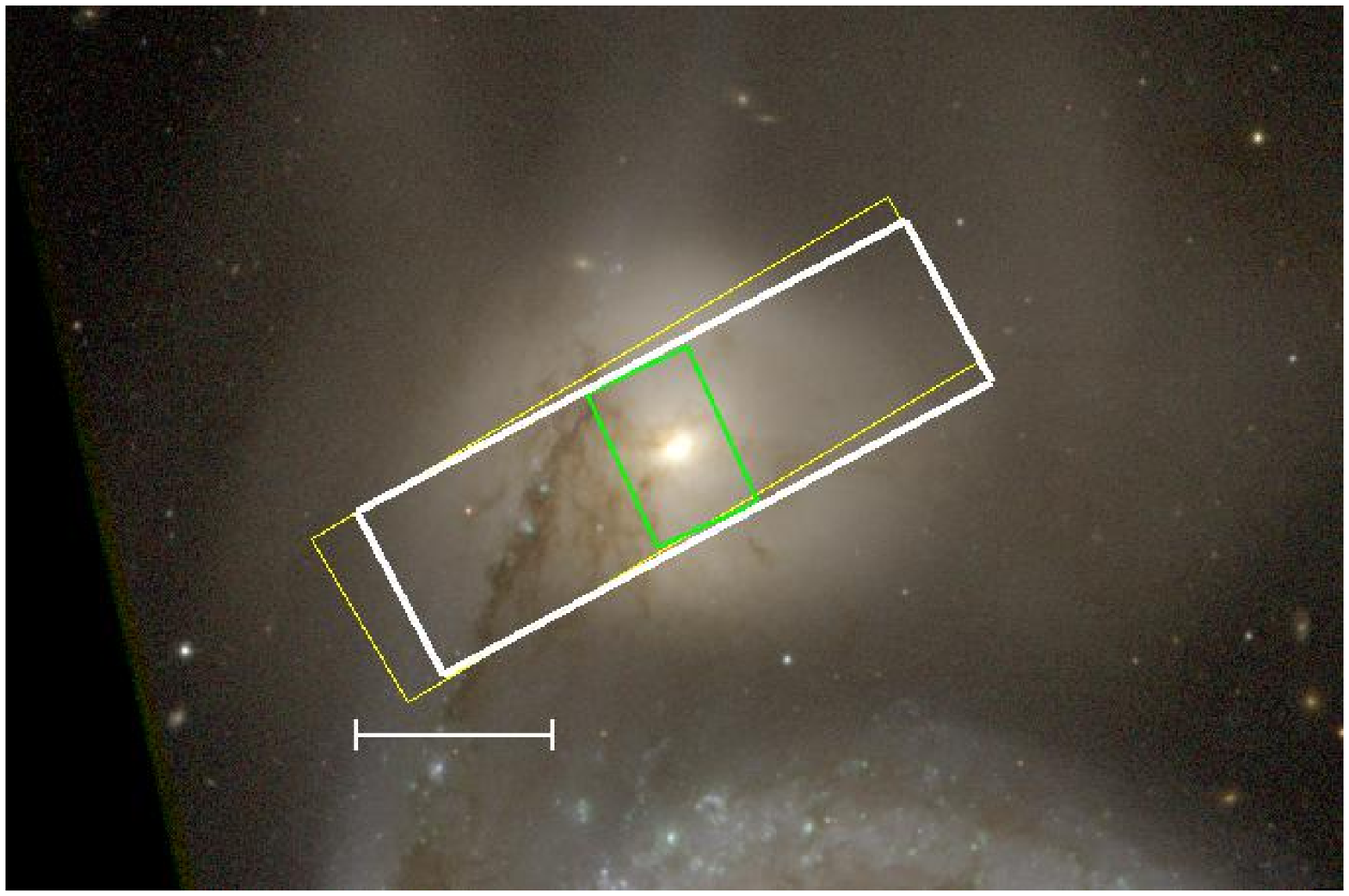}\\
\plottwo{f18_094a.ps}{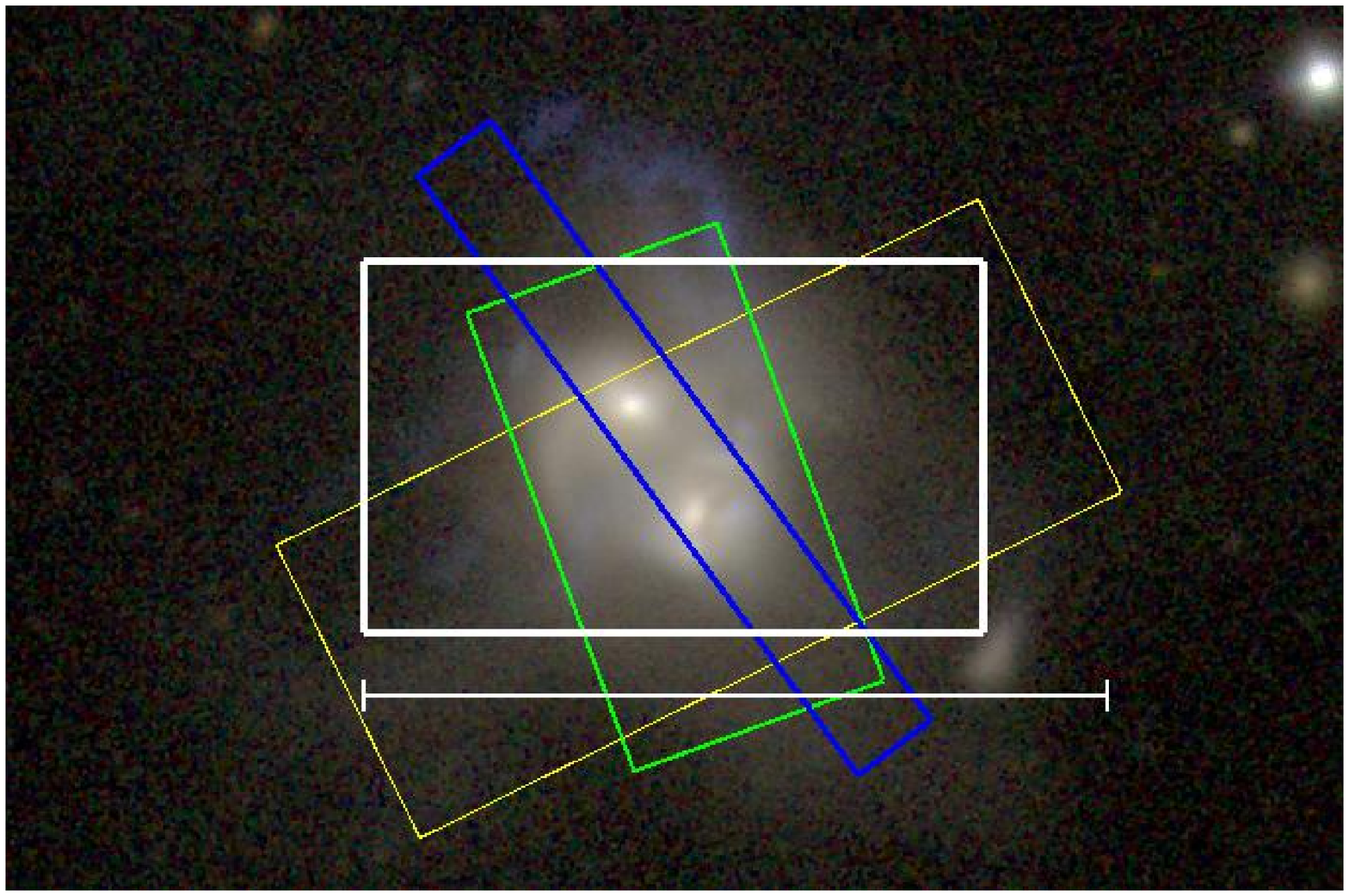}\\
\plottwo{f18_095a.ps}{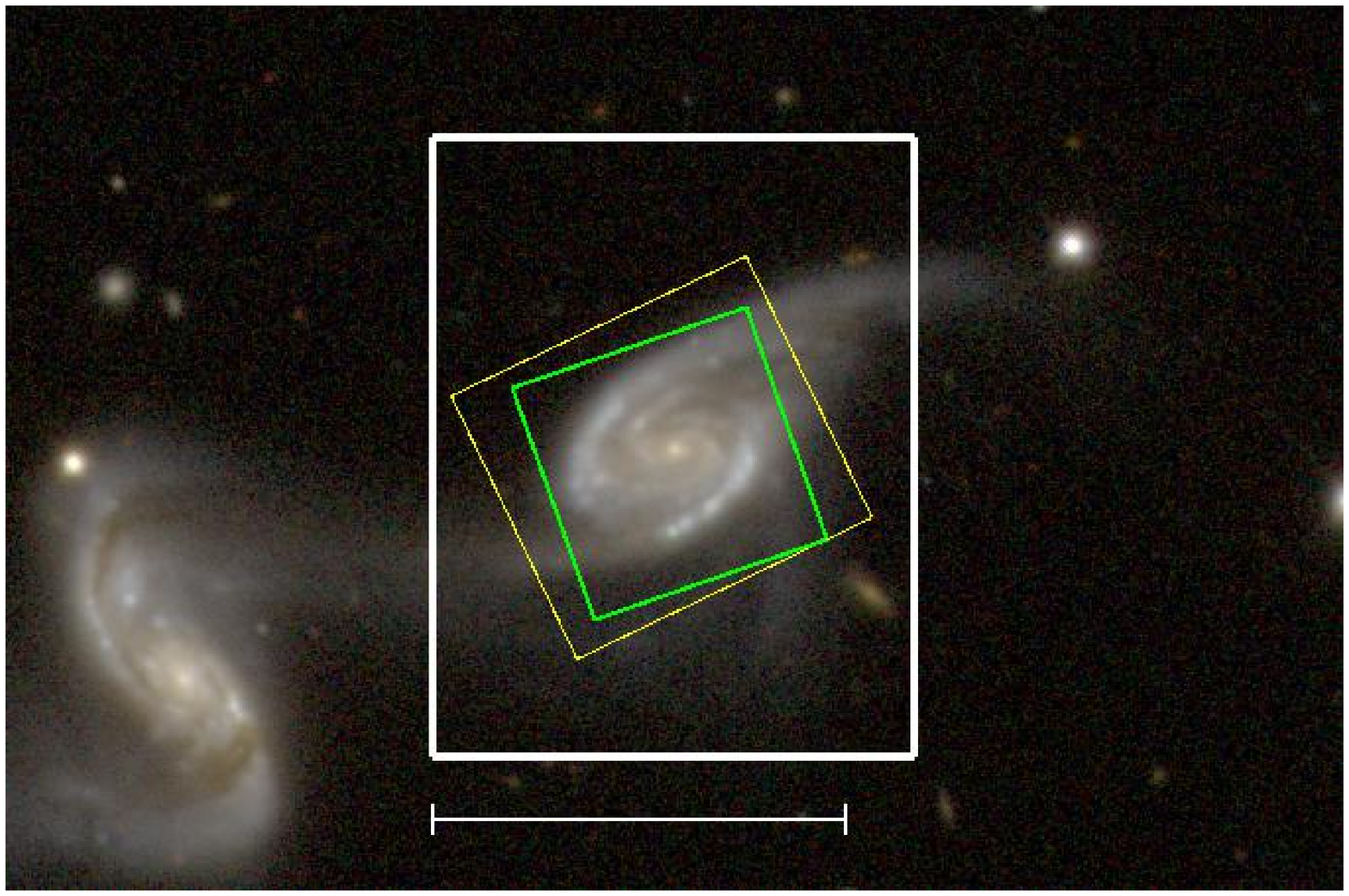}\\
\plottwo{f18_096a.ps}{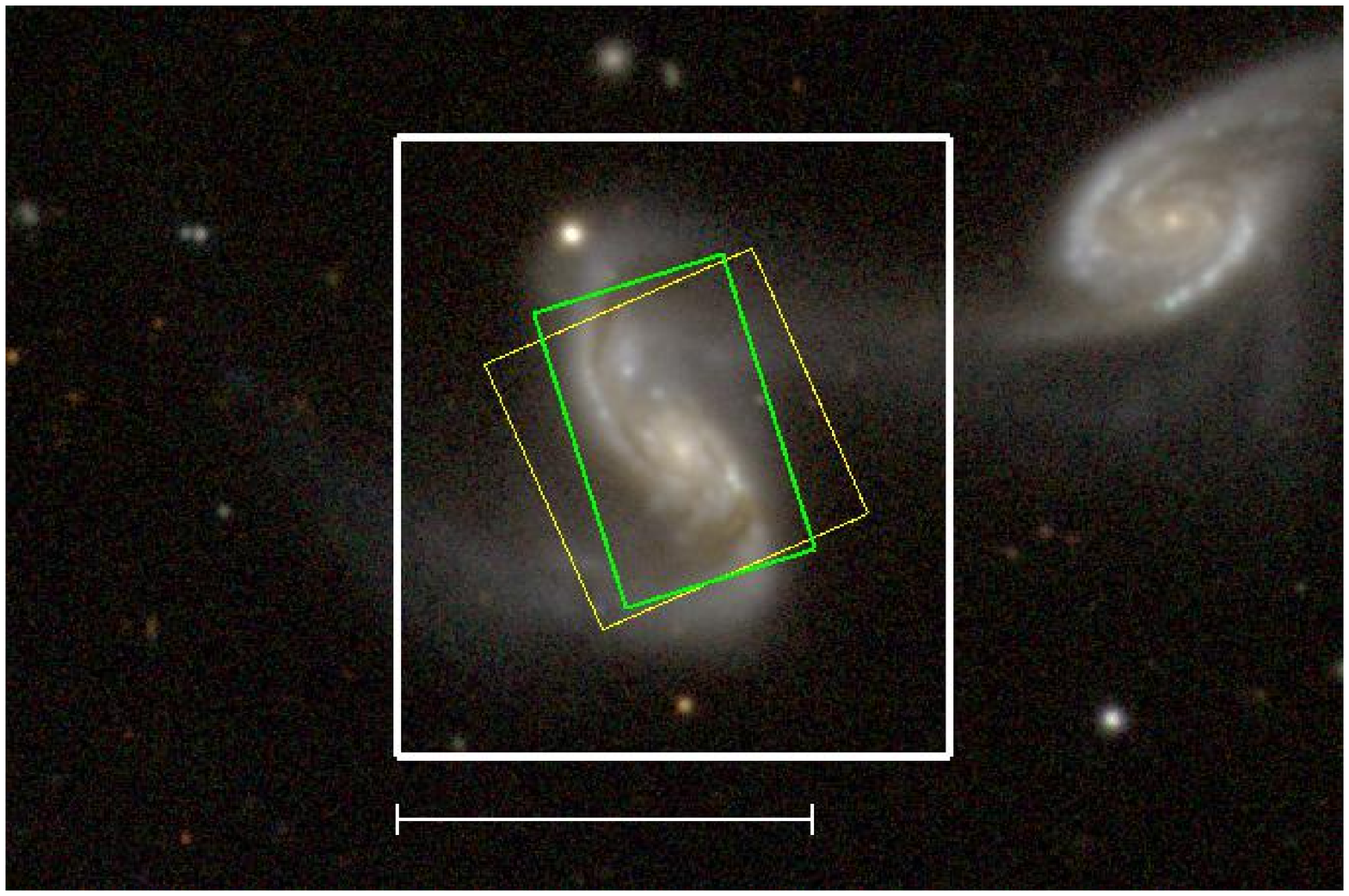}\\
\caption{Galaxy SEDs from the UV to the mid-IR. In the left-panel the observed and model spectra are shown in black and grey respectively, while the photometry used to constrain and verify the spectra is shown with red dots. In the right panel we plot the photometric aperture (thick white rectangle), the {\it Akari} extraction aperture (blue rectangle), the {\it Spitzer} SL extraction aperture (green rectangle) and the {\it Spitzer} LL extraction aperture (yellow rectangle). For galaxies with {\it Spitzer} stare mode spectra, we show a region corresponding to a quarter of the slit length. For scale, the horizontal bar denotes $1^{\prime}$. [{\it See the electronic edition of the Supplement for the complete Figure.}]}
\end{figure}

\clearpage

\begin{figure}[hbt]
\figurenum{\ref{fig:allspec} continued}
\plottwo{f18_097a.ps}{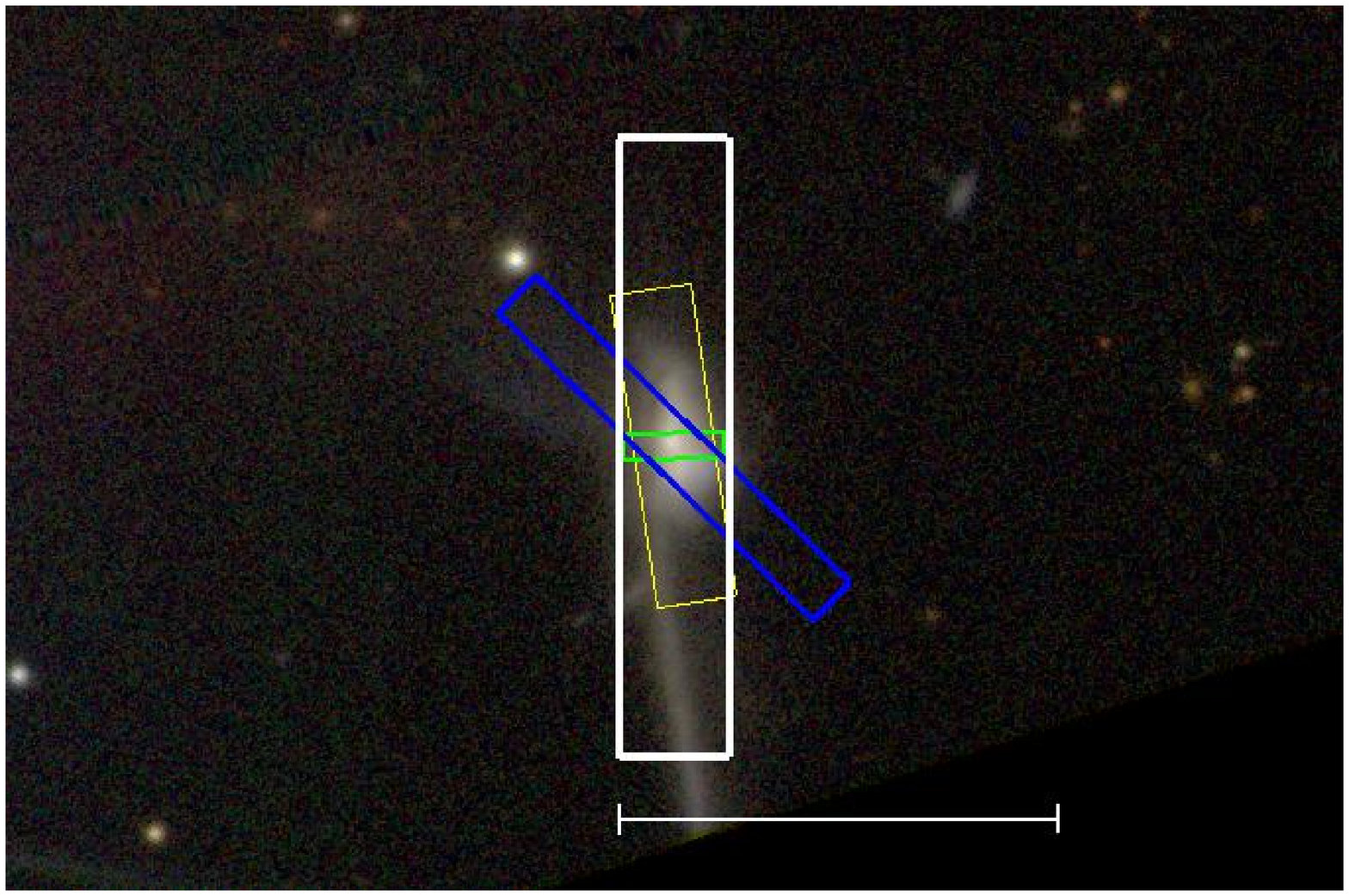}\\
\plottwo{f18_098a.ps}{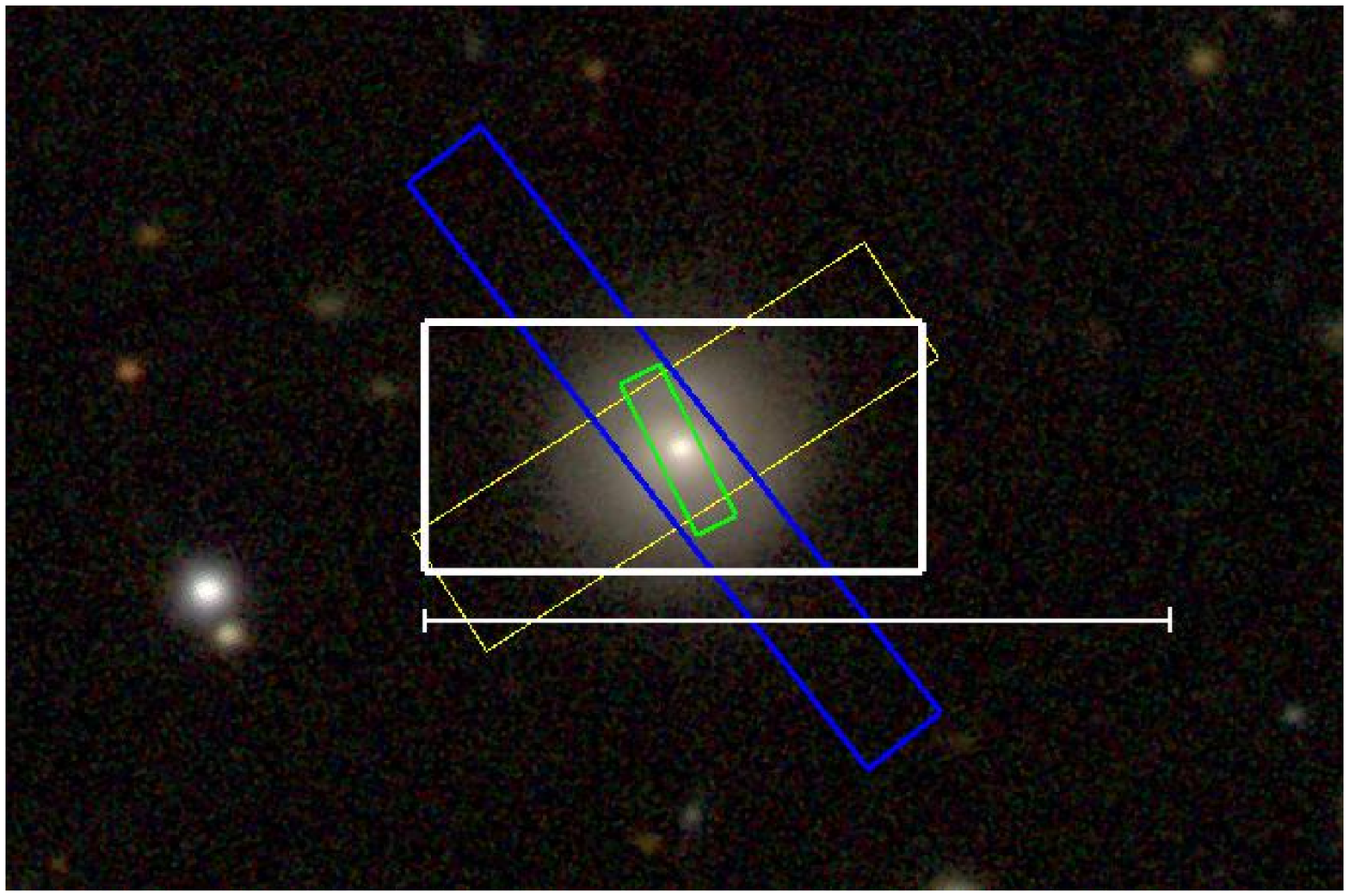}\\
\plottwo{f18_099a.ps}{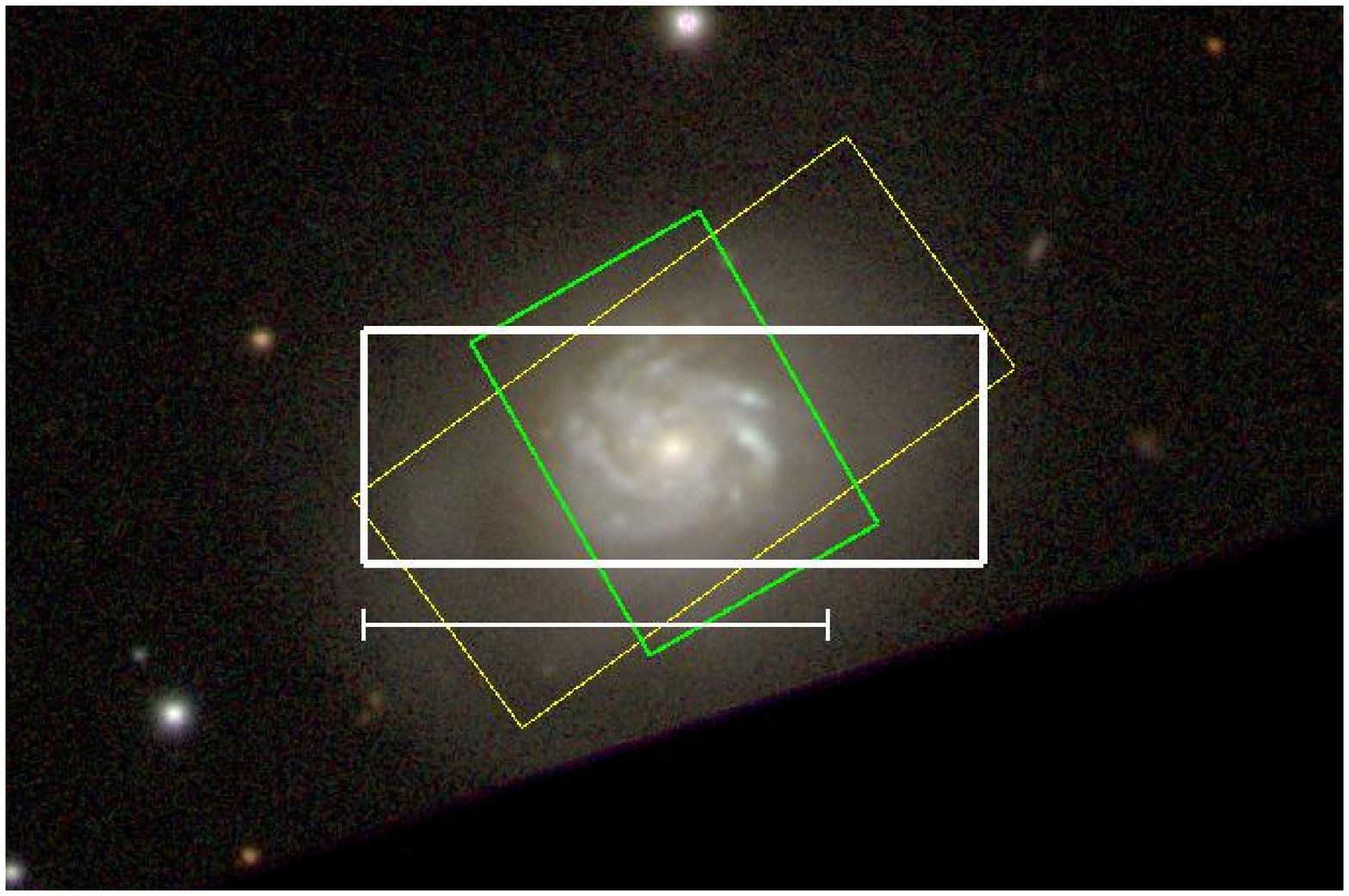}\\
\plottwo{f18_100a.ps}{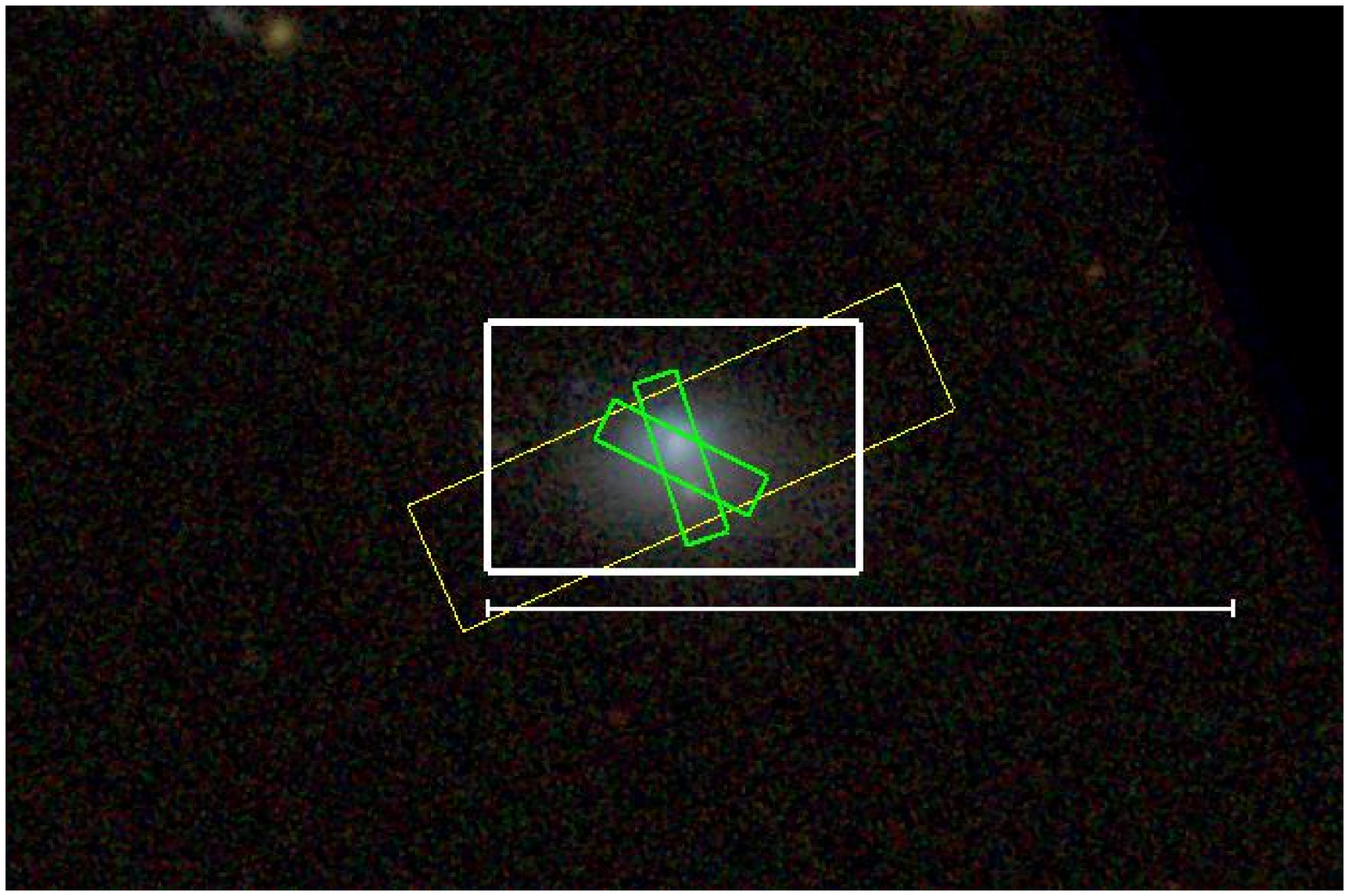}\\
\caption{Galaxy SEDs from the UV to the mid-IR. In the left-panel the observed and model spectra are shown in black and grey respectively, while the photometry used to constrain and verify the spectra is shown with red dots. In the right panel we plot the photometric aperture (thick white rectangle), the {\it Akari} extraction aperture (blue rectangle), the {\it Spitzer} SL extraction aperture (green rectangle) and the {\it Spitzer} LL extraction aperture (yellow rectangle). For galaxies with {\it Spitzer} stare mode spectra, we show a region corresponding to a quarter of the slit length. For scale, the horizontal bar denotes $1^{\prime}$. [{\it See the electronic edition of the Supplement for the complete Figure.}]}
\end{figure}

\clearpage

\begin{figure}[hbt]
\figurenum{\ref{fig:allspec} continued}
\plottwo{f18_101a.ps}{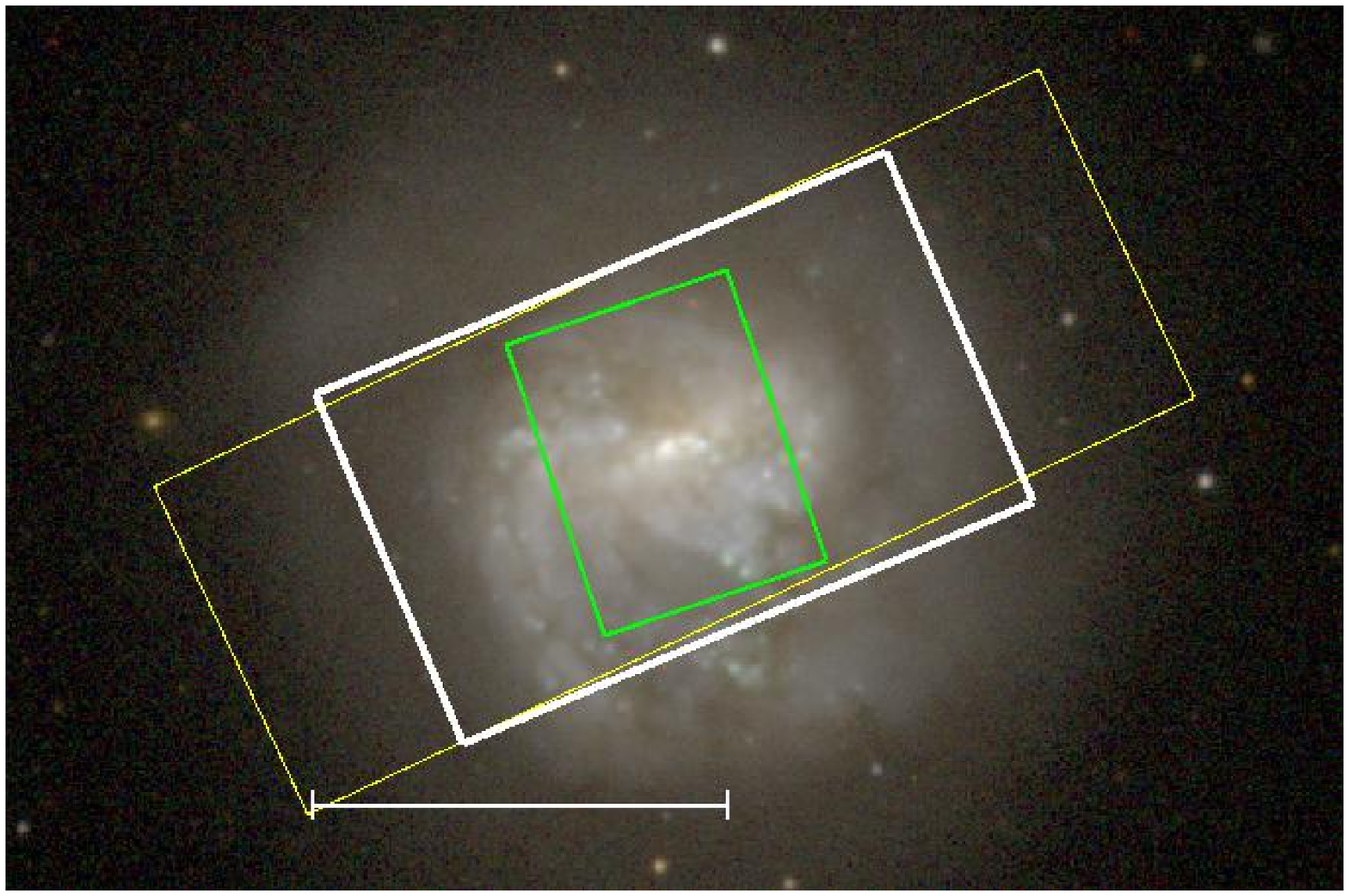}\\
\plottwo{f18_102a.ps}{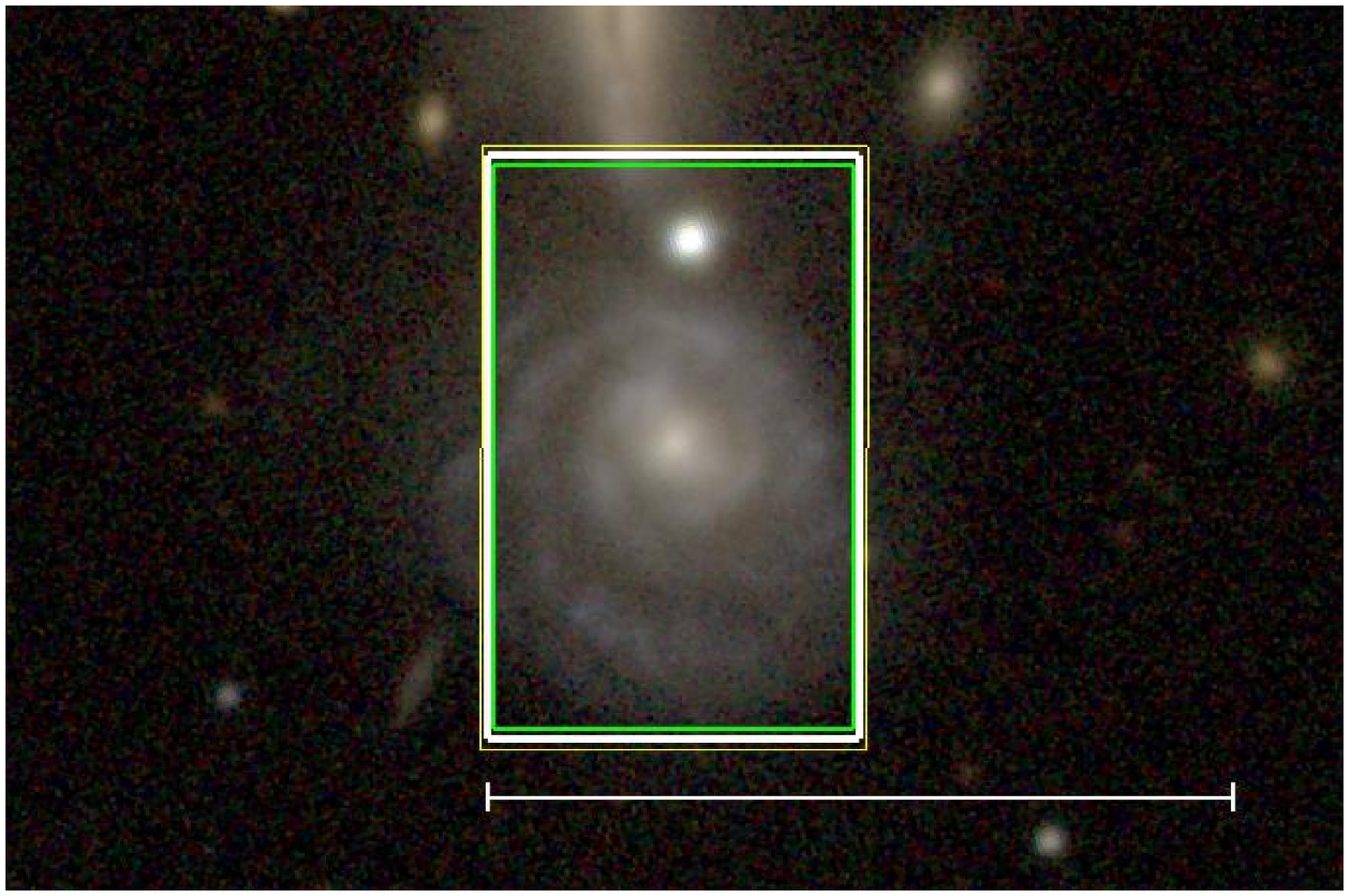}\\
\plottwo{f18_103a.ps}{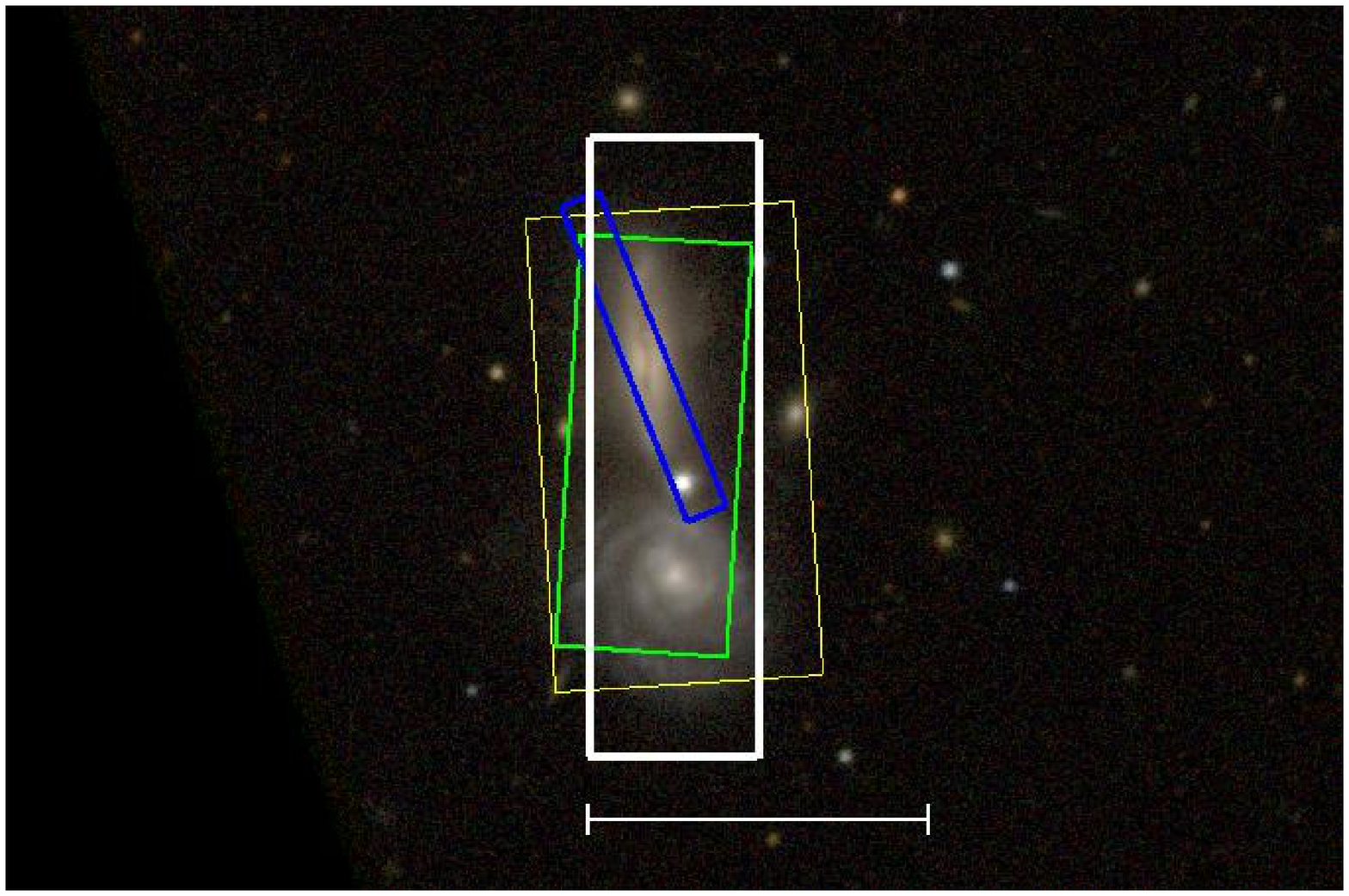}\\
\plottwo{f18_104a.ps}{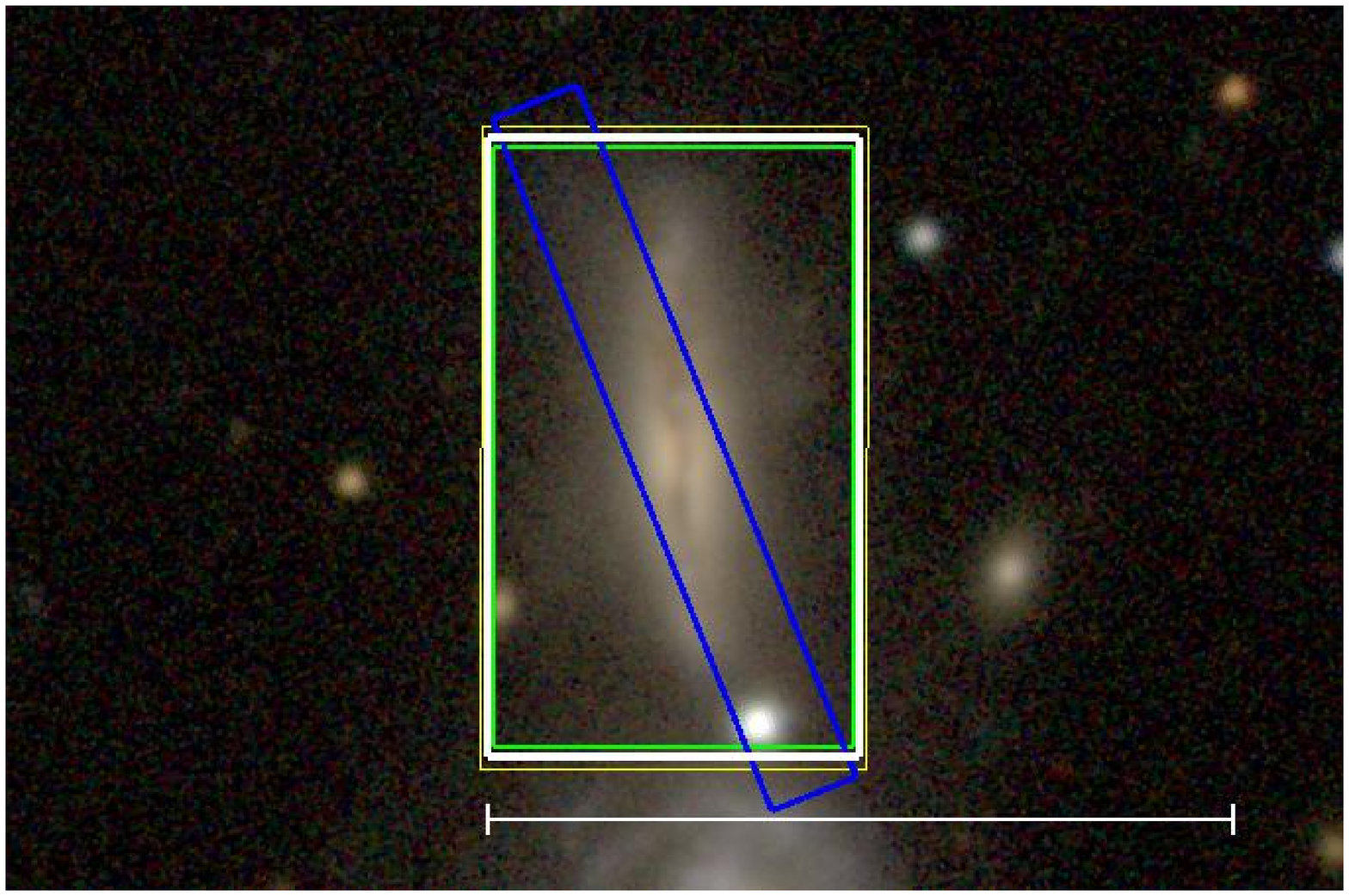}\\
\caption{Galaxy SEDs from the UV to the mid-IR. In the left-panel the observed and model spectra are shown in black and grey respectively, while the photometry used to constrain and verify the spectra is shown with red dots. In the right panel we plot the photometric aperture (thick white rectangle), the {\it Akari} extraction aperture (blue rectangle), the {\it Spitzer} SL extraction aperture (green rectangle) and the {\it Spitzer} LL extraction aperture (yellow rectangle). For galaxies with {\it Spitzer} stare mode spectra, we show a region corresponding to a quarter of the slit length. For scale, the horizontal bar denotes $1^{\prime}$. [{\it See the electronic edition of the Supplement for the complete Figure.}]}
\end{figure}

\clearpage

\begin{figure}[hbt]
\figurenum{\ref{fig:allspec} continued}
\plottwo{f18_105a.ps}{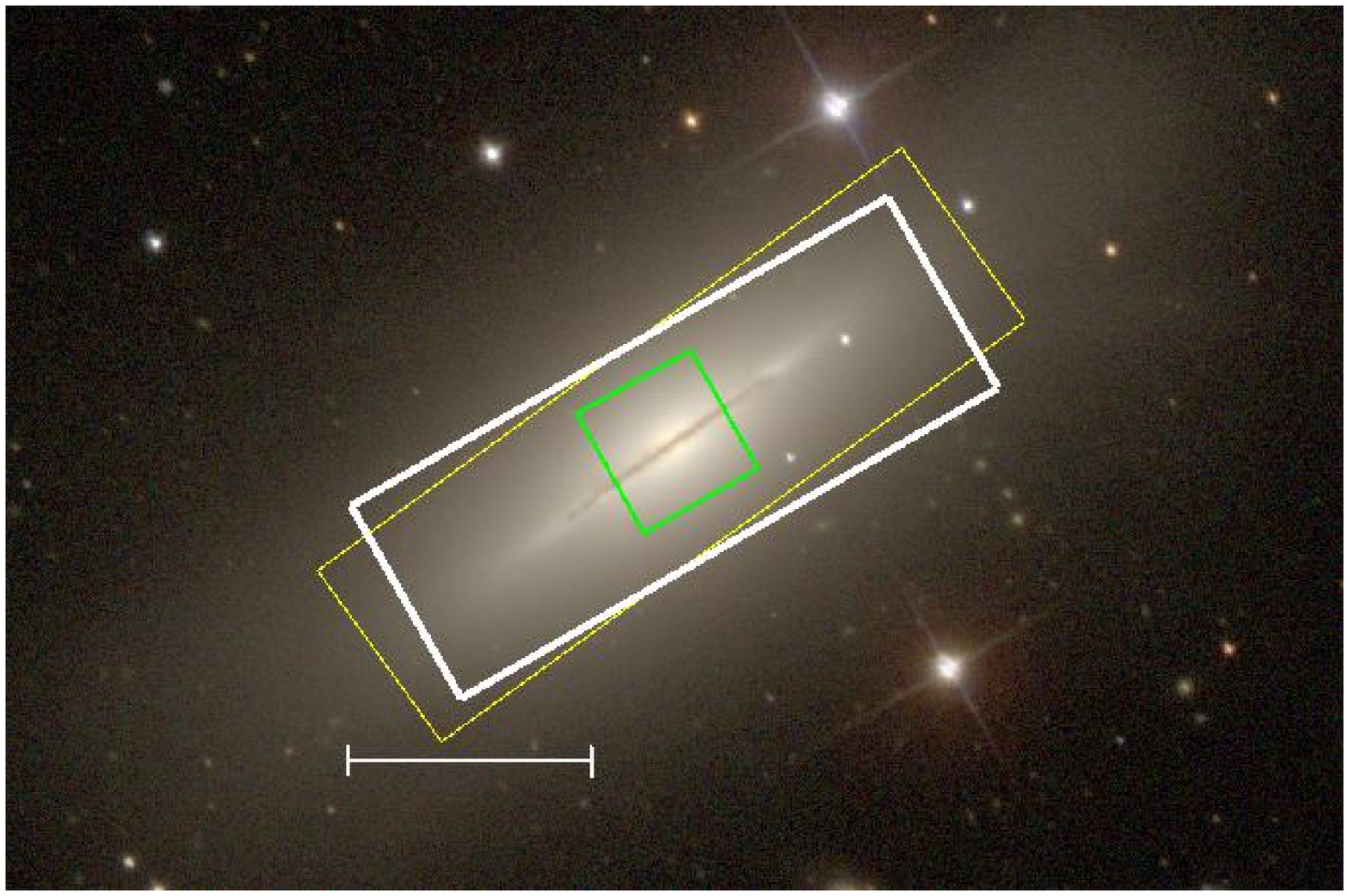}\\
\plottwo{f18_106a.ps}{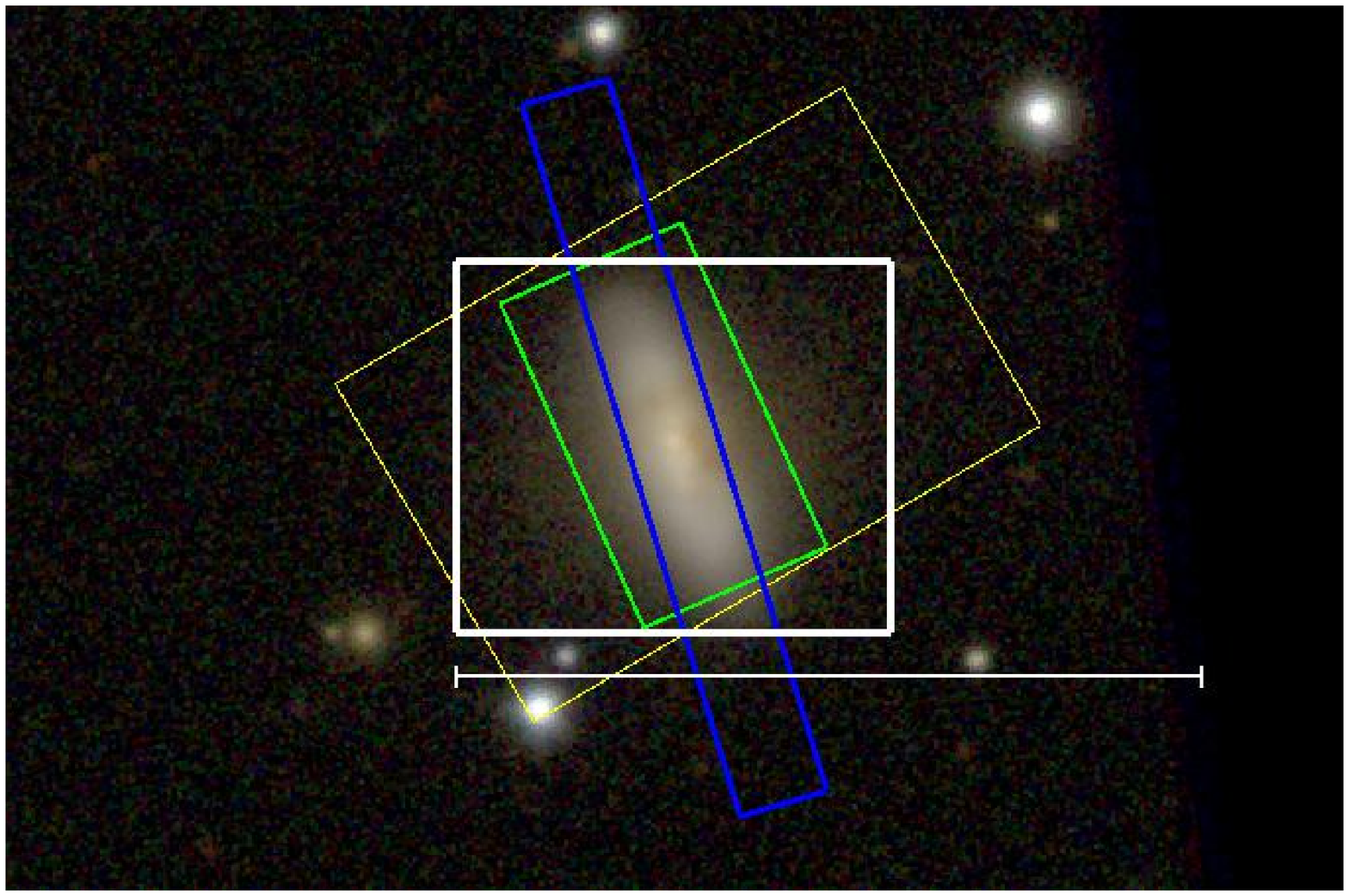}\\
\plottwo{f18_107a.ps}{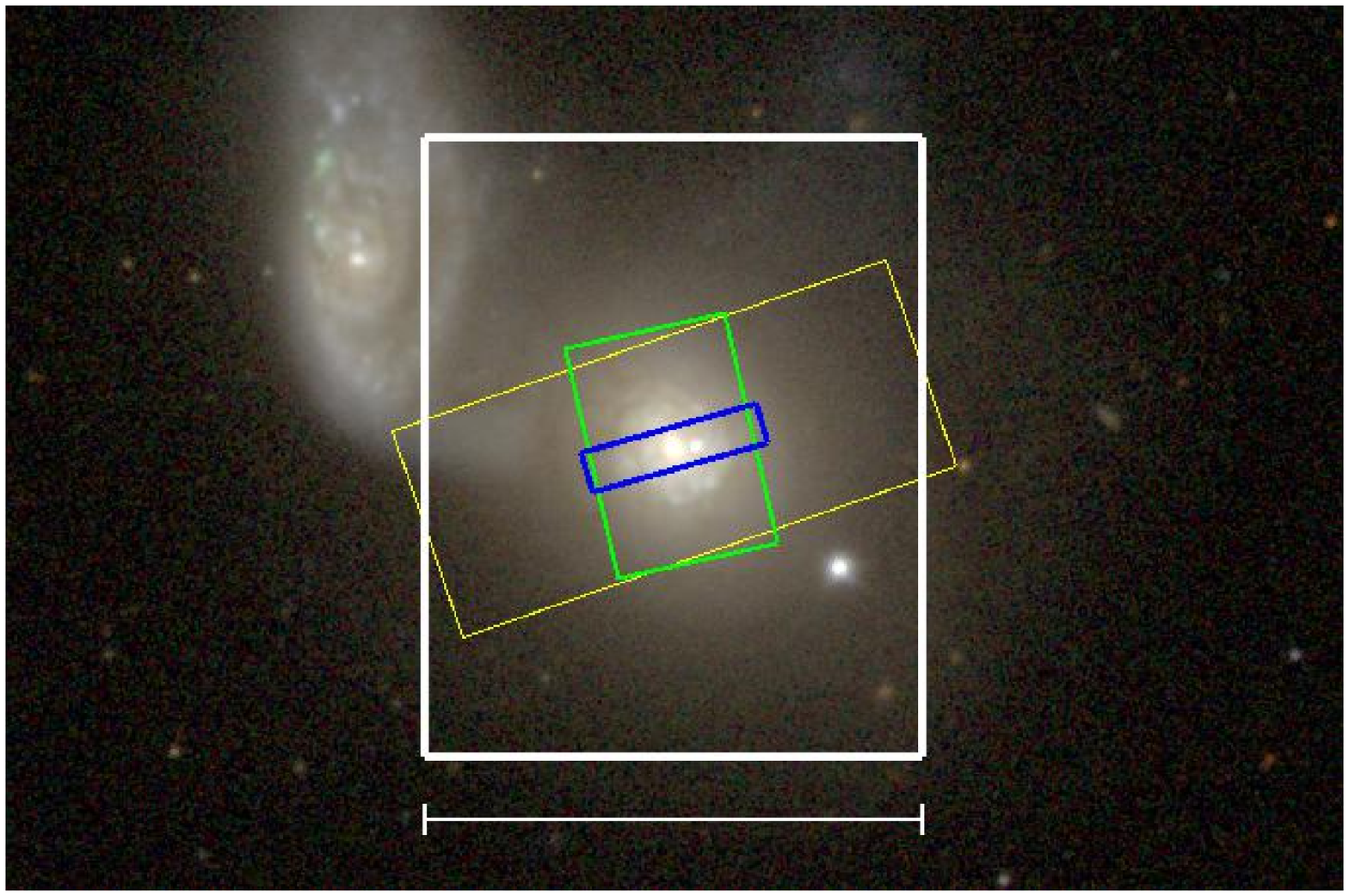}\\
\plottwo{f18_108a.ps}{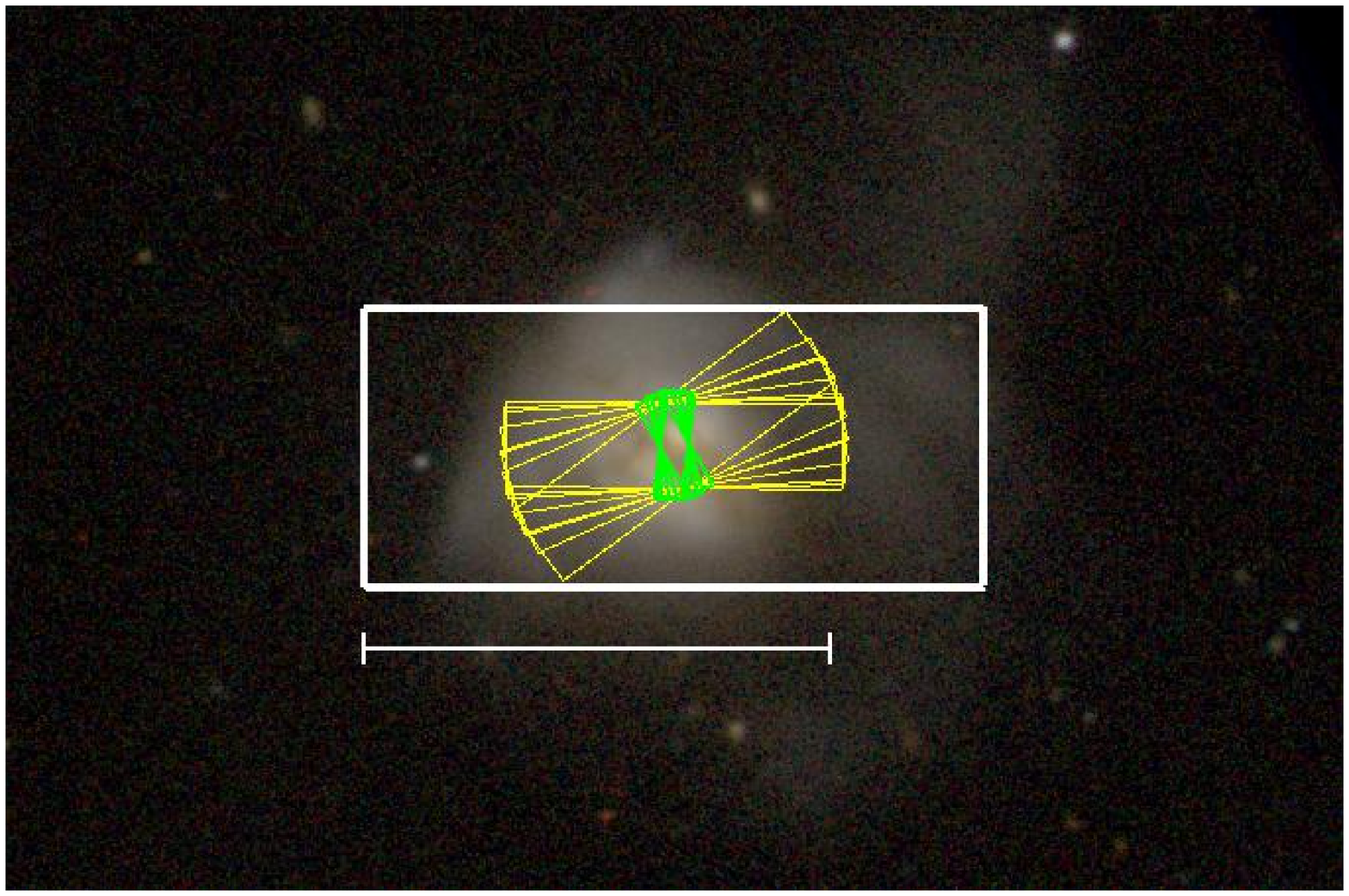}\\
\caption{Galaxy SEDs from the UV to the mid-IR. In the left-panel the observed and model spectra are shown in black and grey respectively, while the photometry used to constrain and verify the spectra is shown with red dots. In the right panel we plot the photometric aperture (thick white rectangle), the {\it Akari} extraction aperture (blue rectangle), the {\it Spitzer} SL extraction aperture (green rectangle) and the {\it Spitzer} LL extraction aperture (yellow rectangle). For galaxies with {\it Spitzer} stare mode spectra, we show a region corresponding to a quarter of the slit length. For scale, the horizontal bar denotes $1^{\prime}$. [{\it See the electronic edition of the Supplement for the complete Figure.}]}
\end{figure}

\clearpage

\begin{figure}[hbt]
\figurenum{\ref{fig:allspec} continued}
\plottwo{f18_109a.ps}{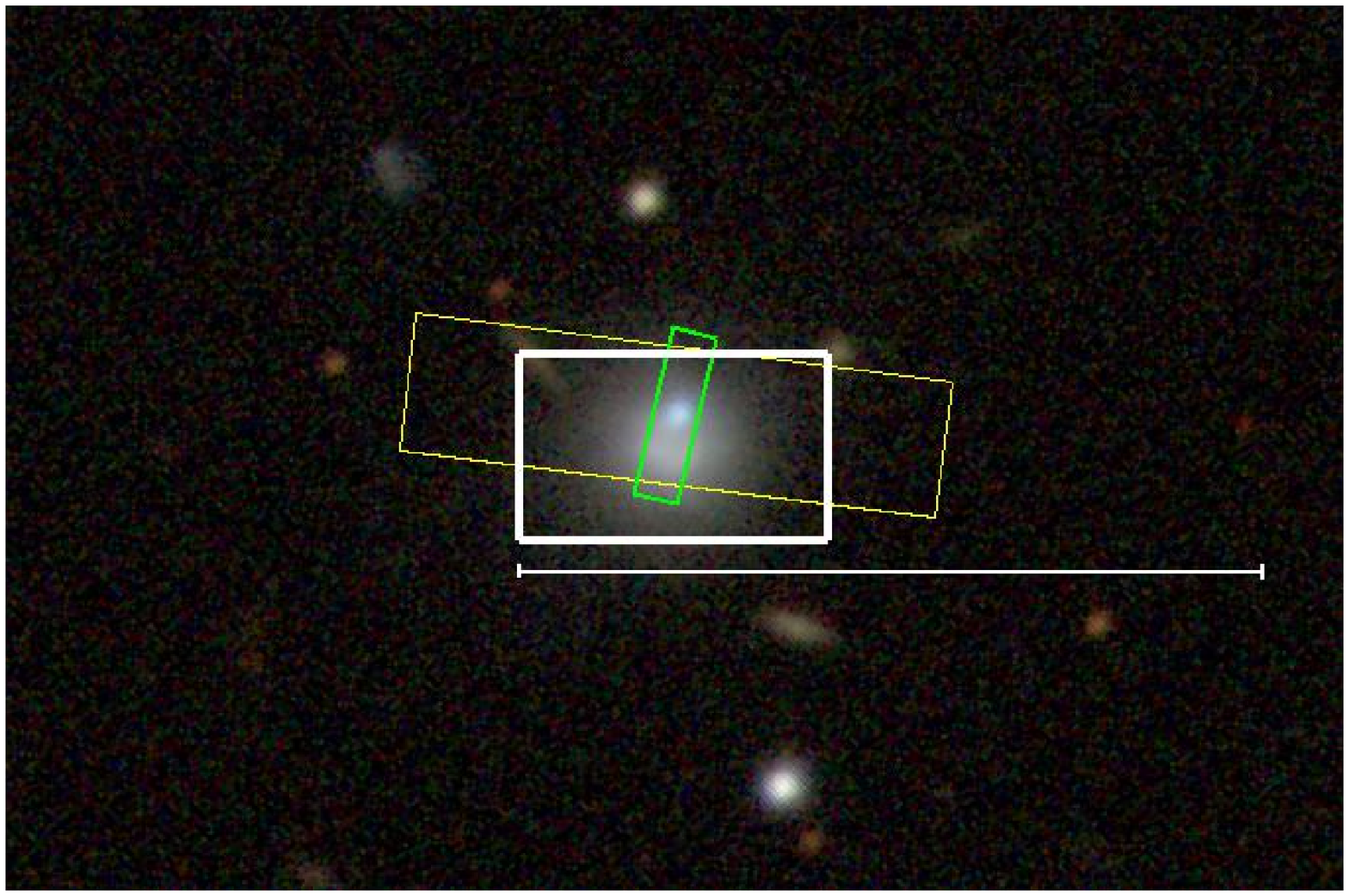}\\
\plottwo{f18_110a.ps}{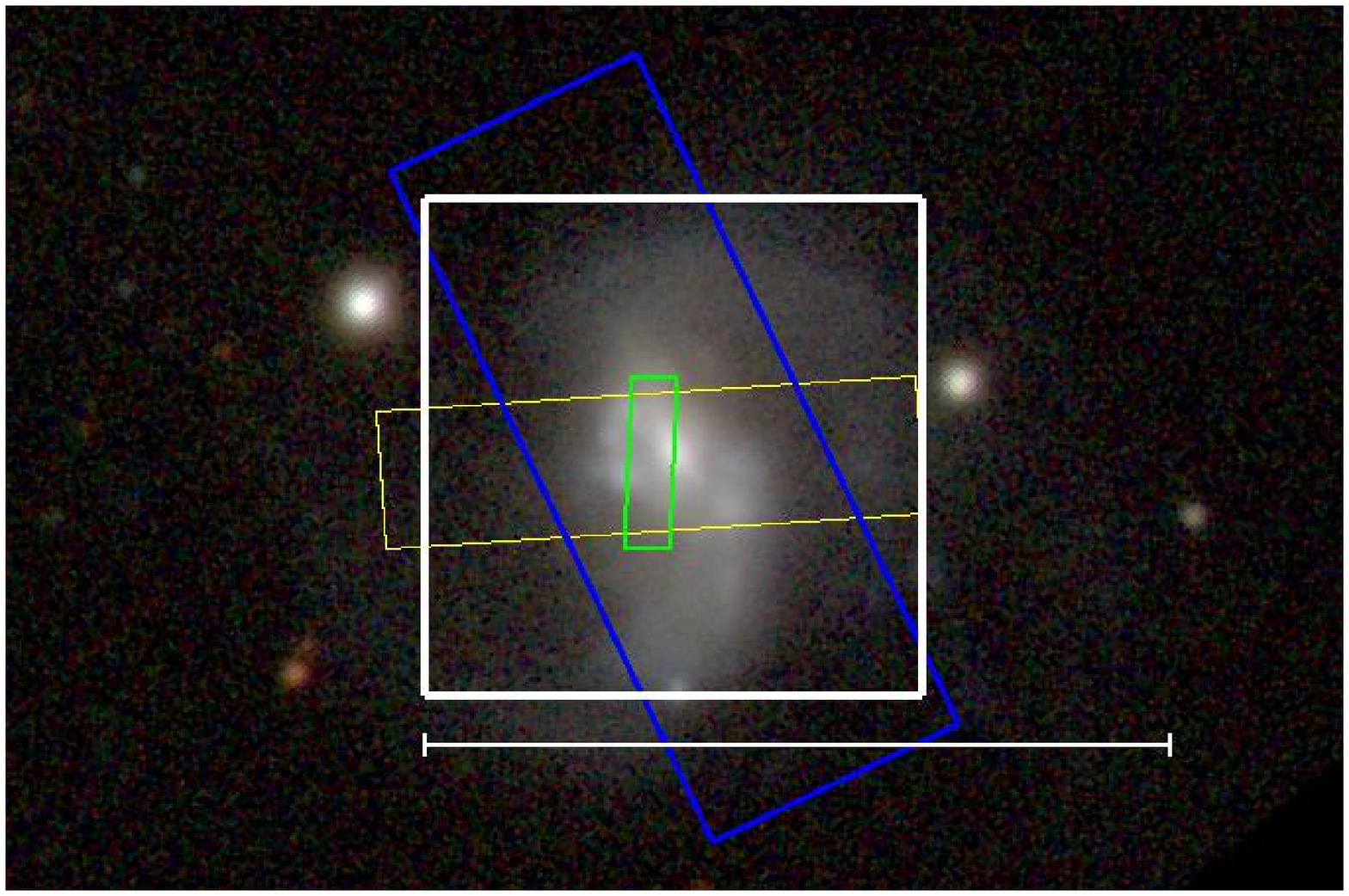}\\
\plottwo{f18_111a.ps}{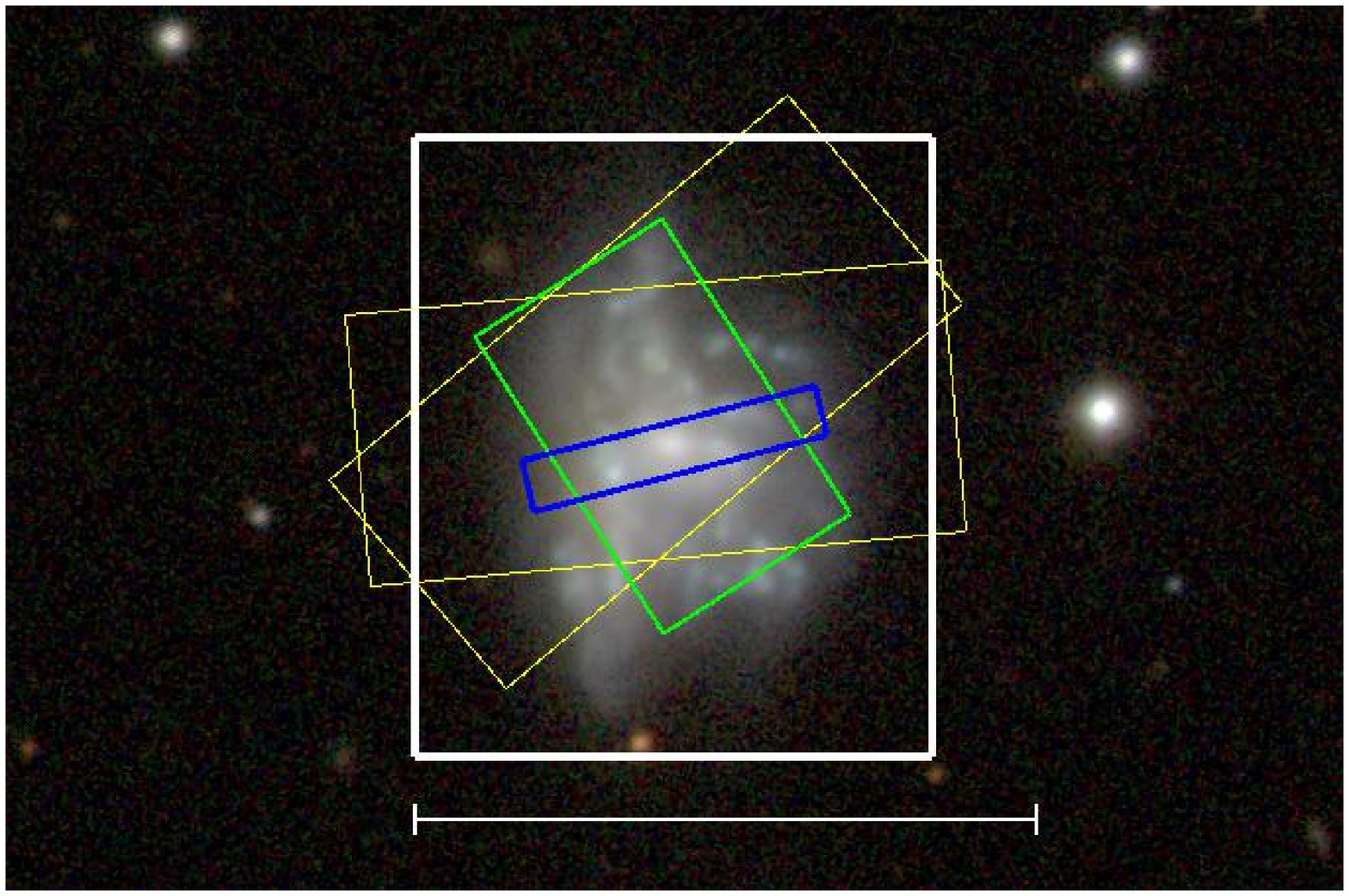}\\
\plottwo{f18_112a.ps}{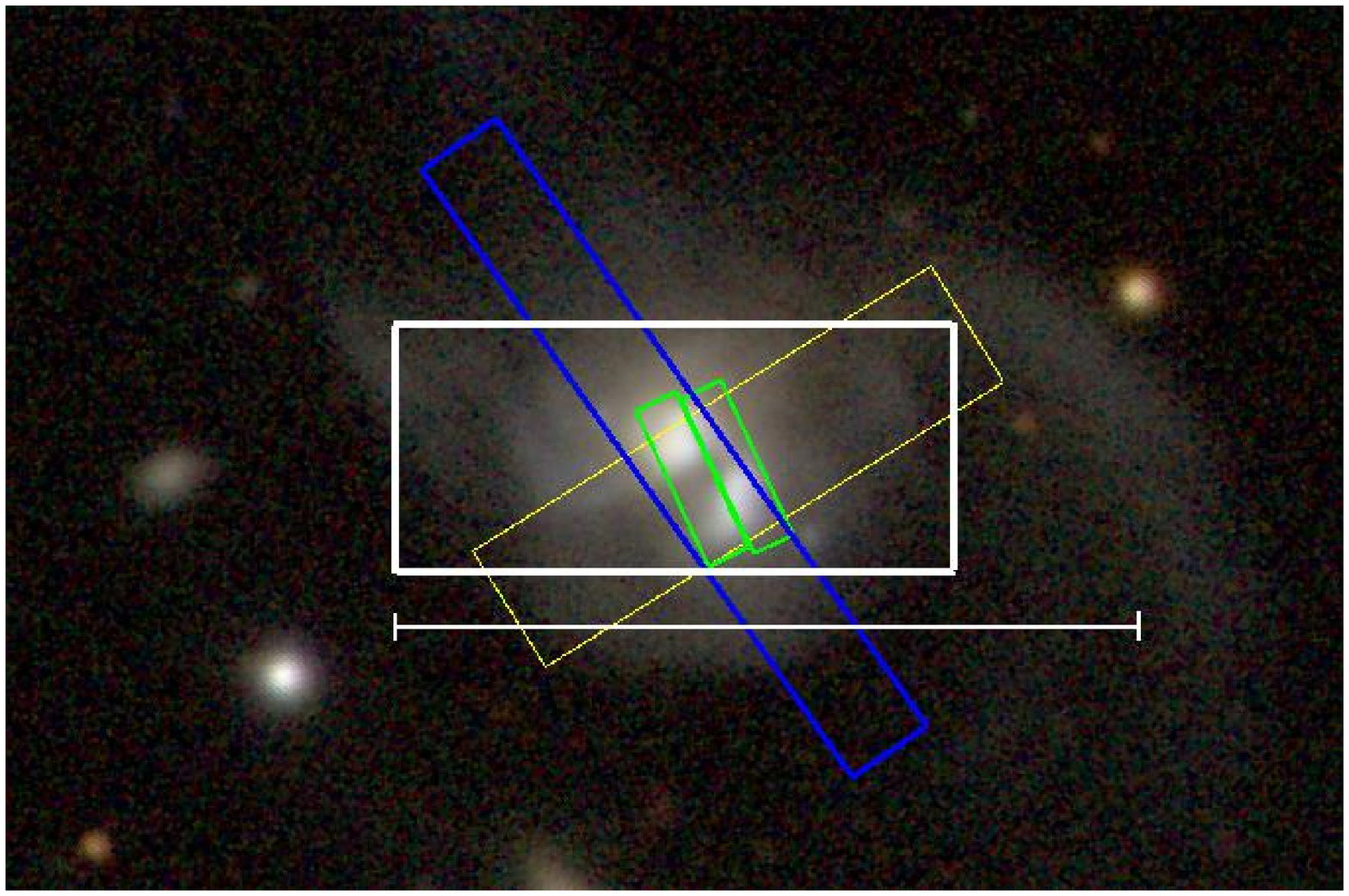}\\
\caption{Galaxy SEDs from the UV to the mid-IR. In the left-panel the observed and model spectra are shown in black and grey respectively, while the photometry used to constrain and verify the spectra is shown with red dots. In the right panel we plot the photometric aperture (thick white rectangle), the {\it Akari} extraction aperture (blue rectangle), the {\it Spitzer} SL extraction aperture (green rectangle) and the {\it Spitzer} LL extraction aperture (yellow rectangle). For galaxies with {\it Spitzer} stare mode spectra, we show a region corresponding to a quarter of the slit length. For scale, the horizontal bar denotes $1^{\prime}$. [{\it See the electronic edition of the Supplement for the complete Figure.}]}
\end{figure}

\clearpage

\begin{figure}[hbt]
\figurenum{\ref{fig:allspec} continued}
\plottwo{f18_113a.ps}{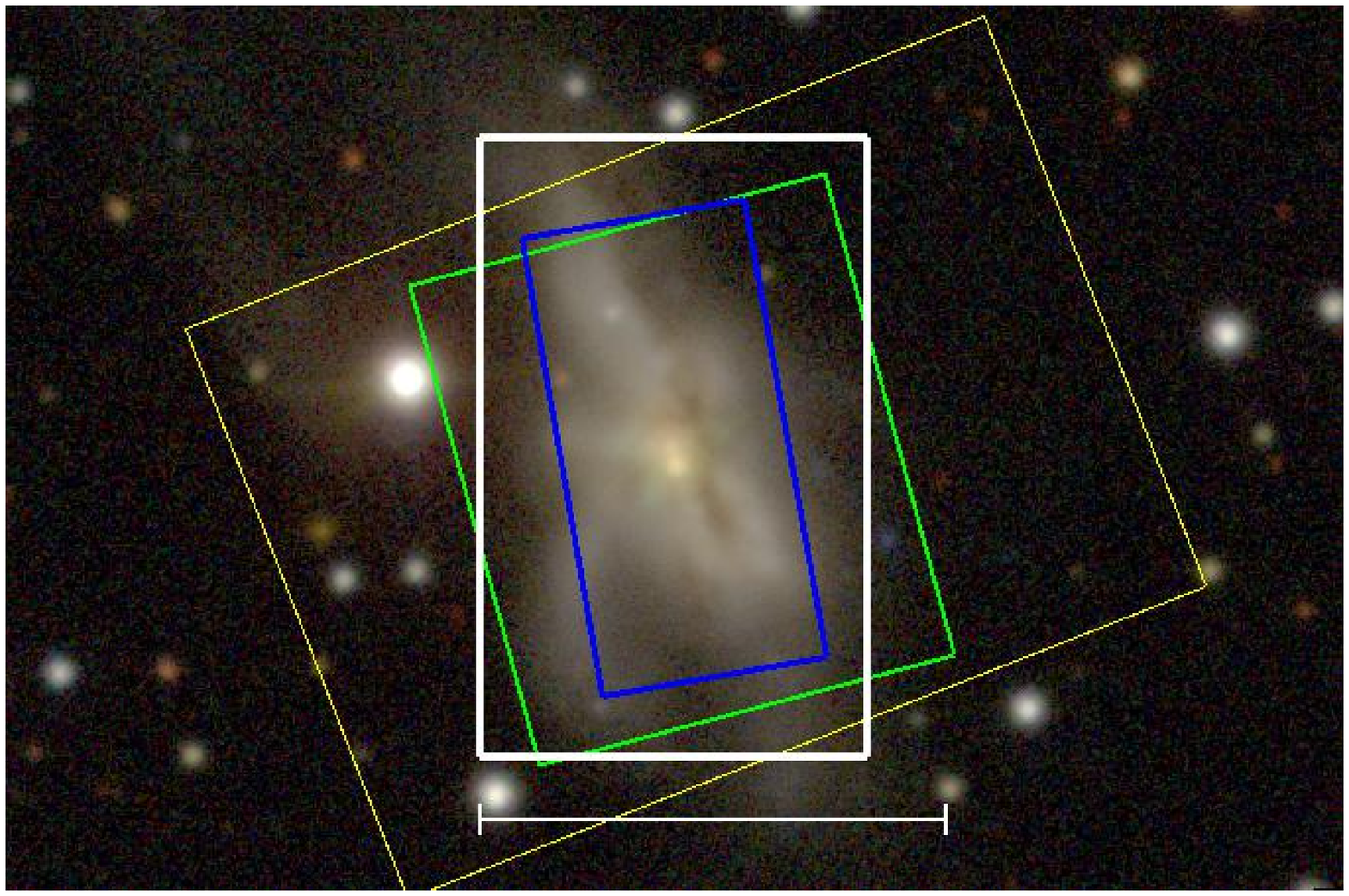}\\
\plottwo{f18_114a.ps}{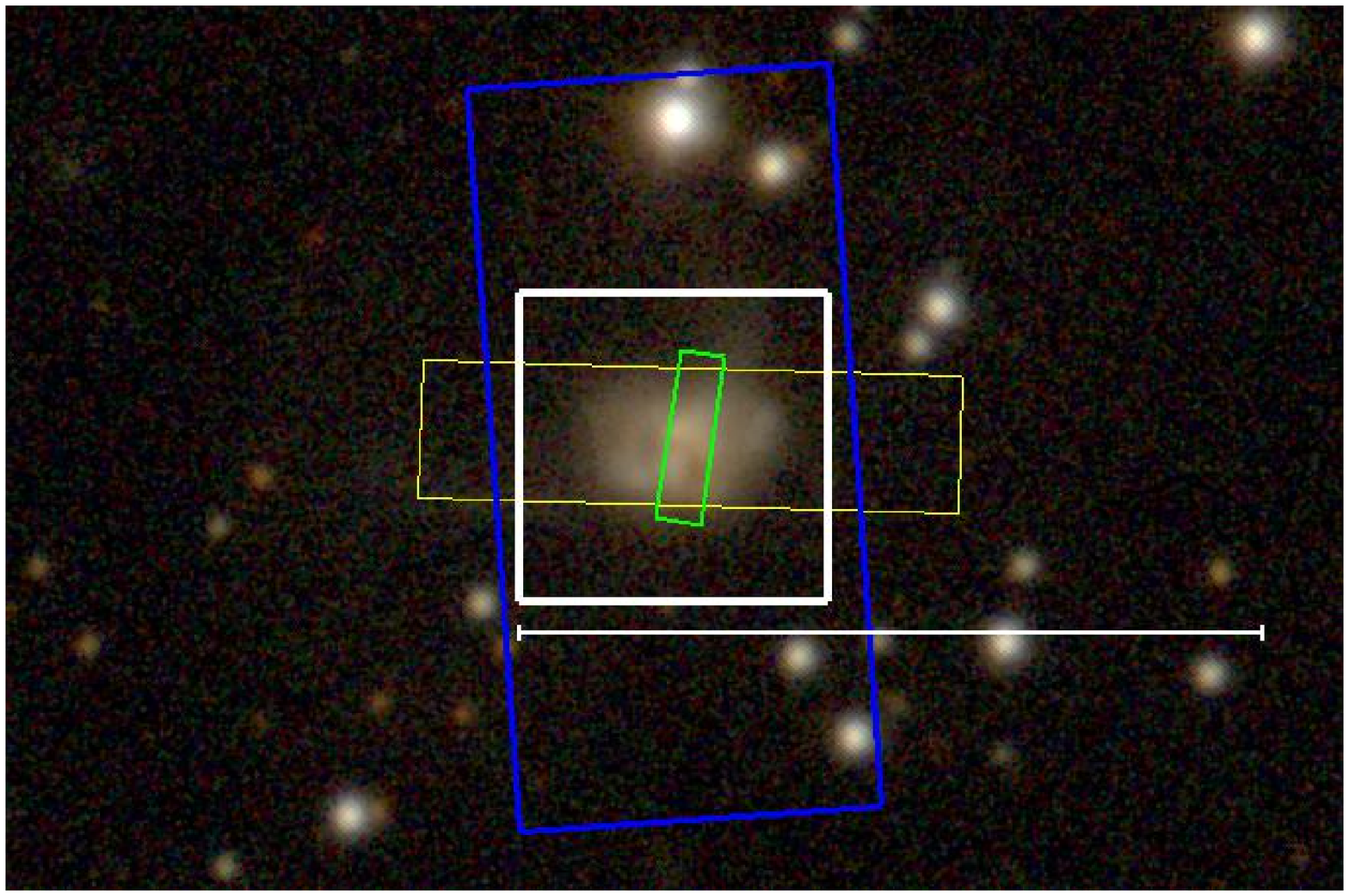}\\
\plottwo{f18_115a.ps}{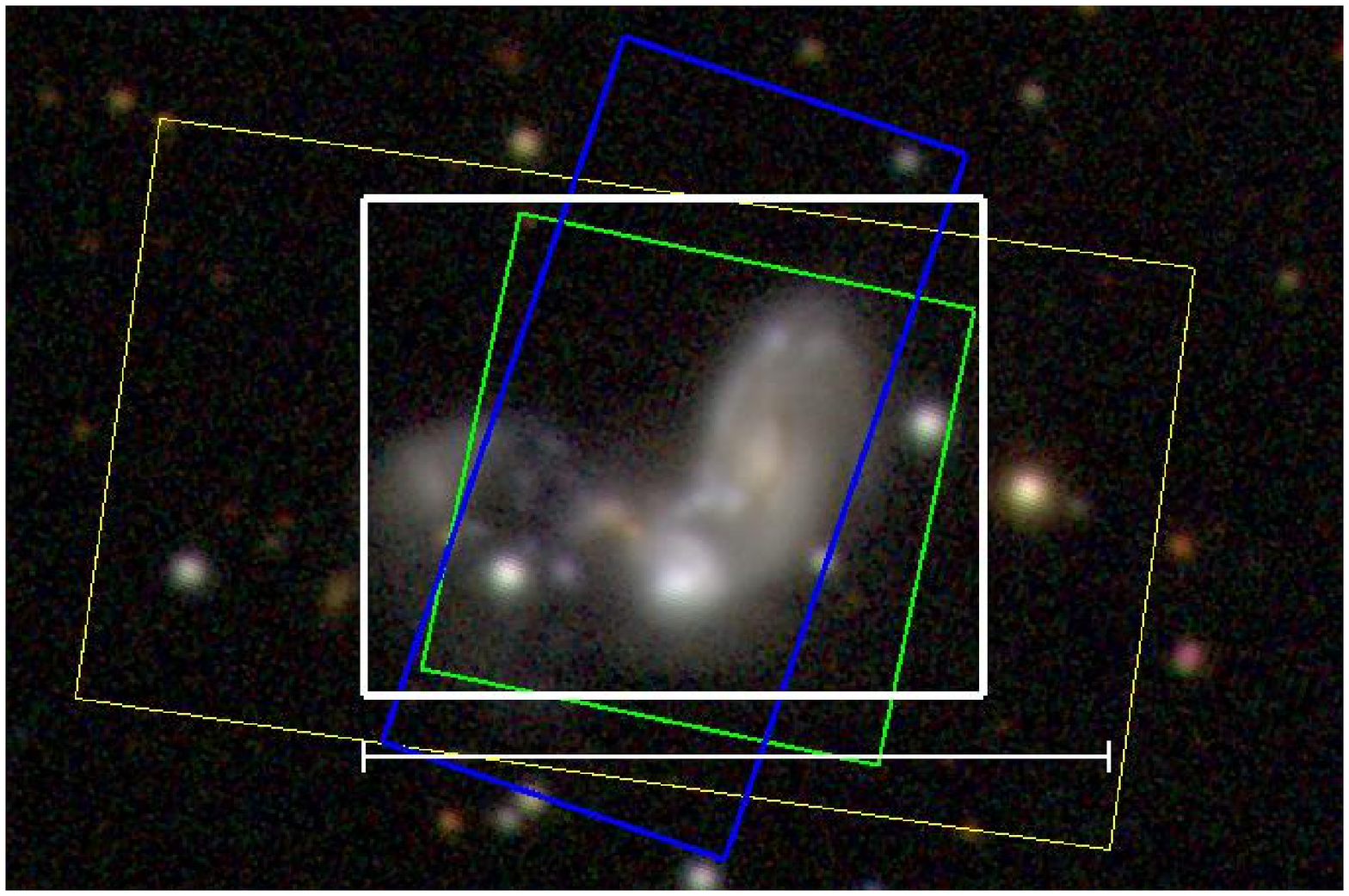}\\
\plottwo{f18_116a.ps}{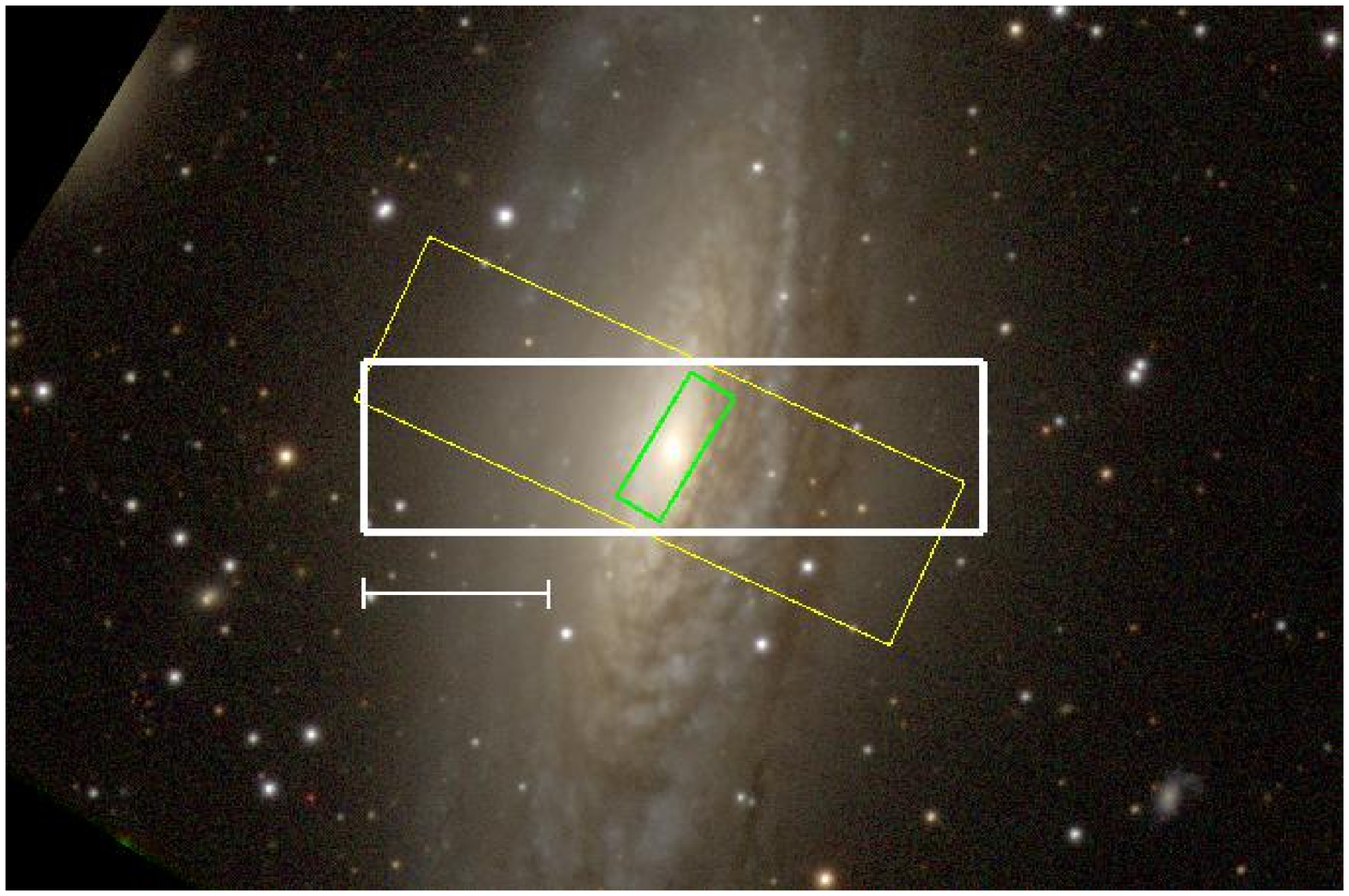}\\
\caption{Galaxy SEDs from the UV to the mid-IR. In the left-panel the observed and model spectra are shown in black and grey respectively, while the photometry used to constrain and verify the spectra is shown with red dots. In the right panel we plot the photometric aperture (thick white rectangle), the {\it Akari} extraction aperture (blue rectangle), the {\it Spitzer} SL extraction aperture (green rectangle) and the {\it Spitzer} LL extraction aperture (yellow rectangle). For galaxies with {\it Spitzer} stare mode spectra, we show a region corresponding to a quarter of the slit length. For scale, the horizontal bar denotes $1^{\prime}$. [{\it See the electronic edition of the Supplement for the complete Figure.}]}
\end{figure}

\clearpage

\begin{figure}[hbt]
\figurenum{\ref{fig:allspec} continued}
\plottwo{f18_117a.ps}{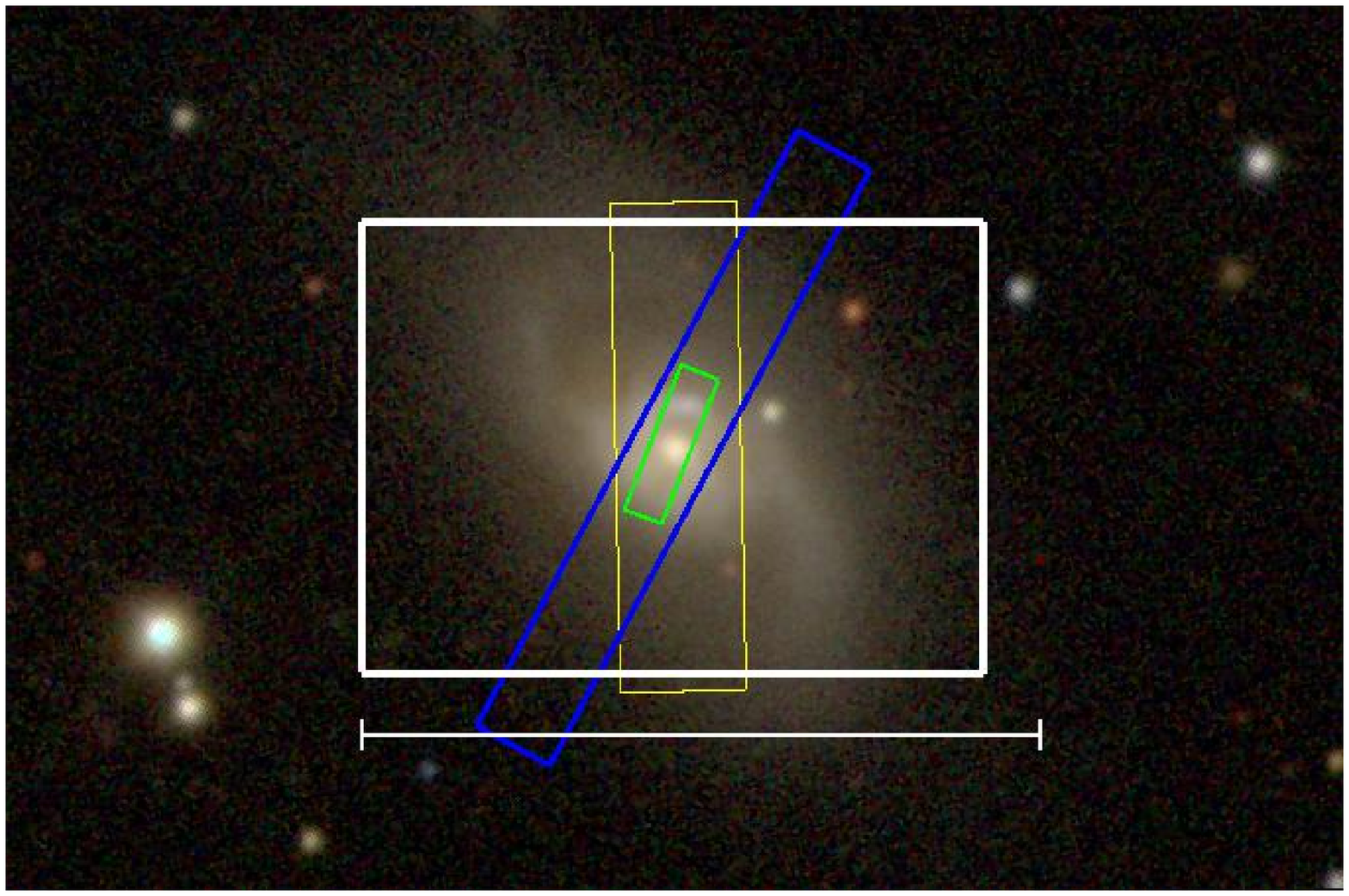}\\
\plottwo{f18_118a.ps}{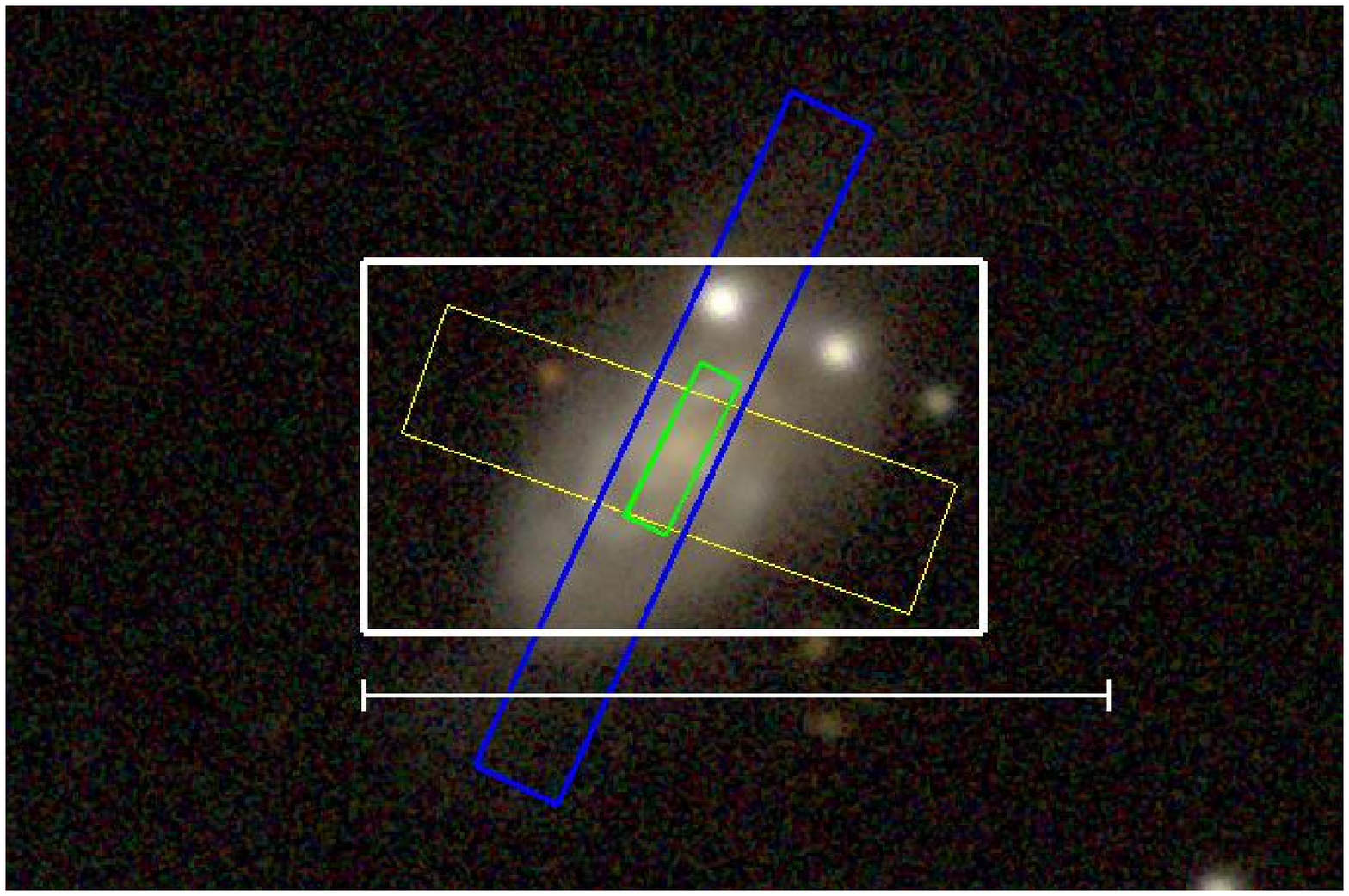}\\
\plottwo{f18_119a.ps}{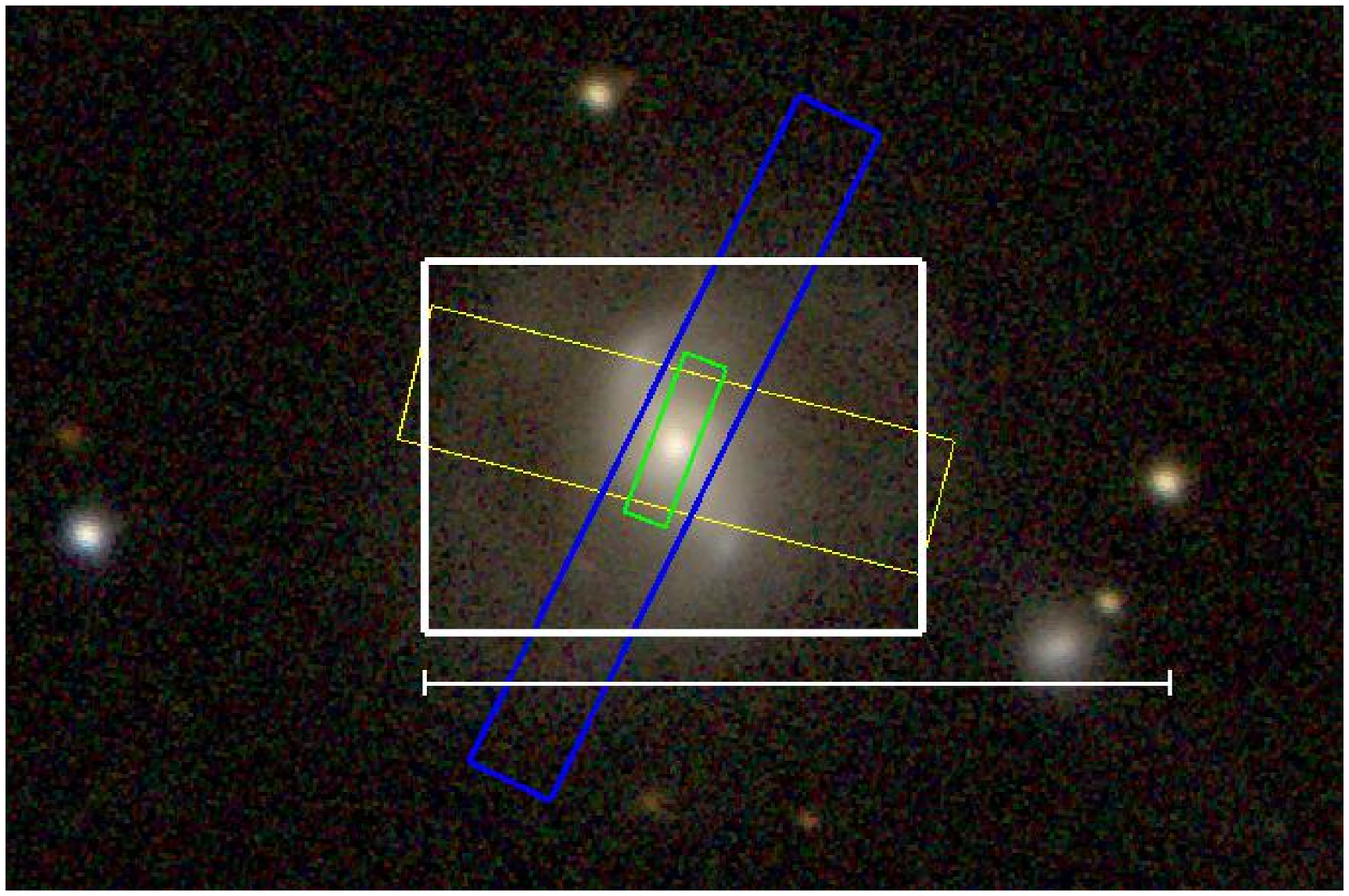}\\
\plottwo{f18_120a.ps}{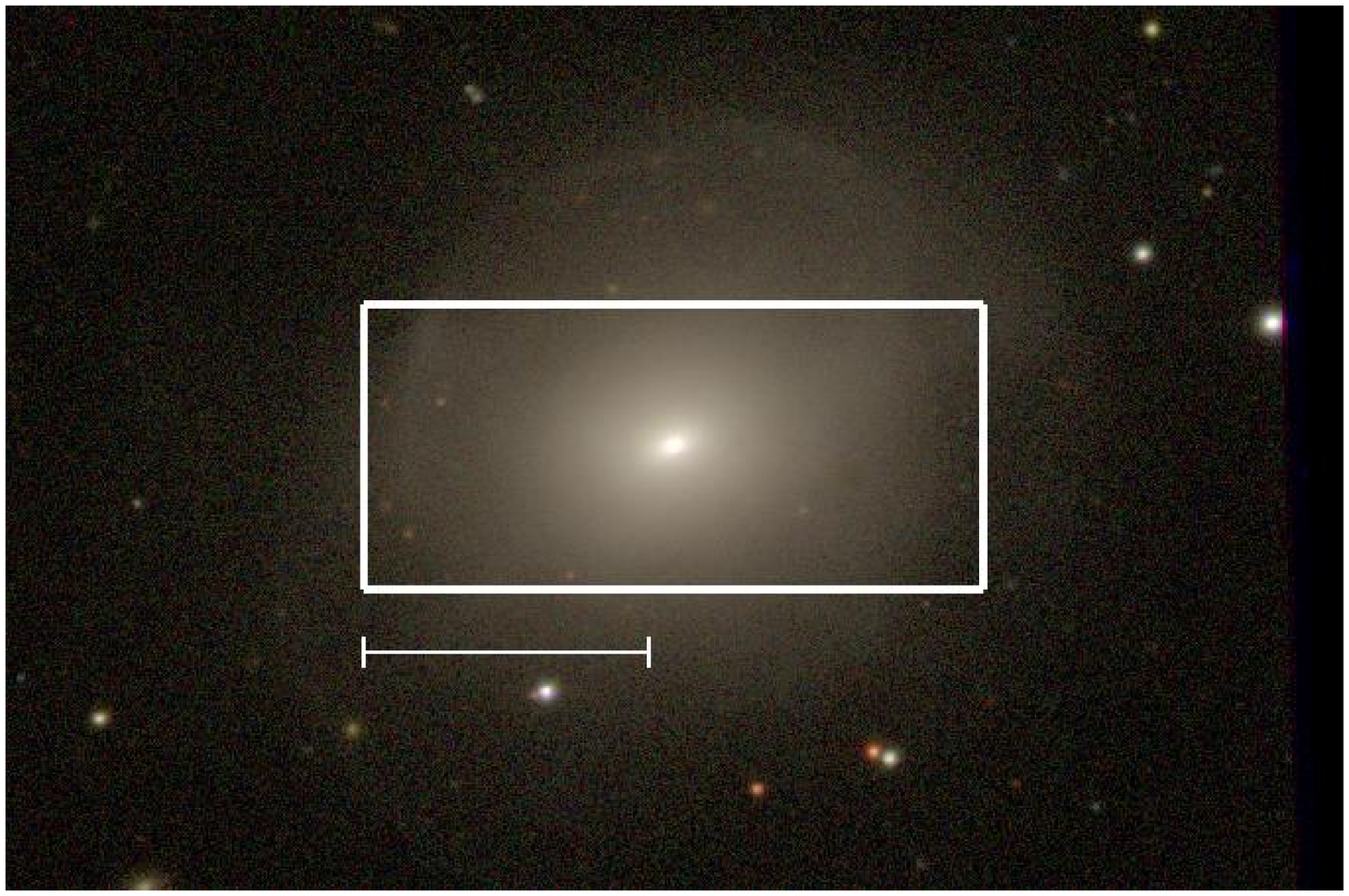}\\
\caption{Galaxy SEDs from the UV to the mid-IR. In the left-panel the observed and model spectra are shown in black and grey respectively, while the photometry used to constrain and verify the spectra is shown with red dots. In the right panel we plot the photometric aperture (thick white rectangle), the {\it Akari} extraction aperture (blue rectangle), the {\it Spitzer} SL extraction aperture (green rectangle) and the {\it Spitzer} LL extraction aperture (yellow rectangle). For galaxies with {\it Spitzer} stare mode spectra, we show a region corresponding to a quarter of the slit length. For scale, the horizontal bar denotes $1^{\prime}$. [{\it See the electronic edition of the Supplement for the complete Figure.}]}
\end{figure}

\clearpage

\begin{figure}[hbt]
\figurenum{\ref{fig:allspec} continued}
\plottwo{f18_121a.ps}{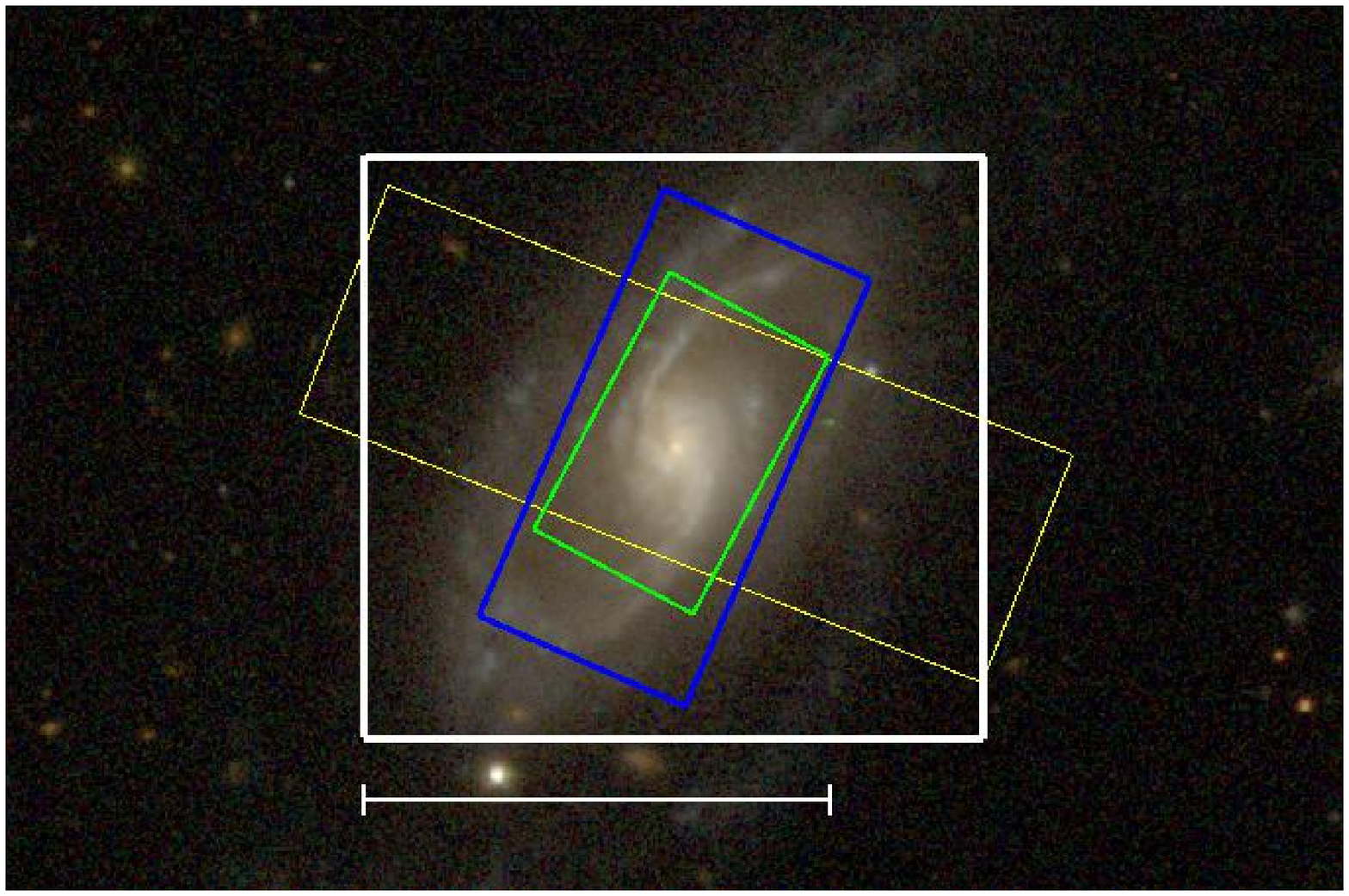}\\
\plottwo{f18_122a.ps}{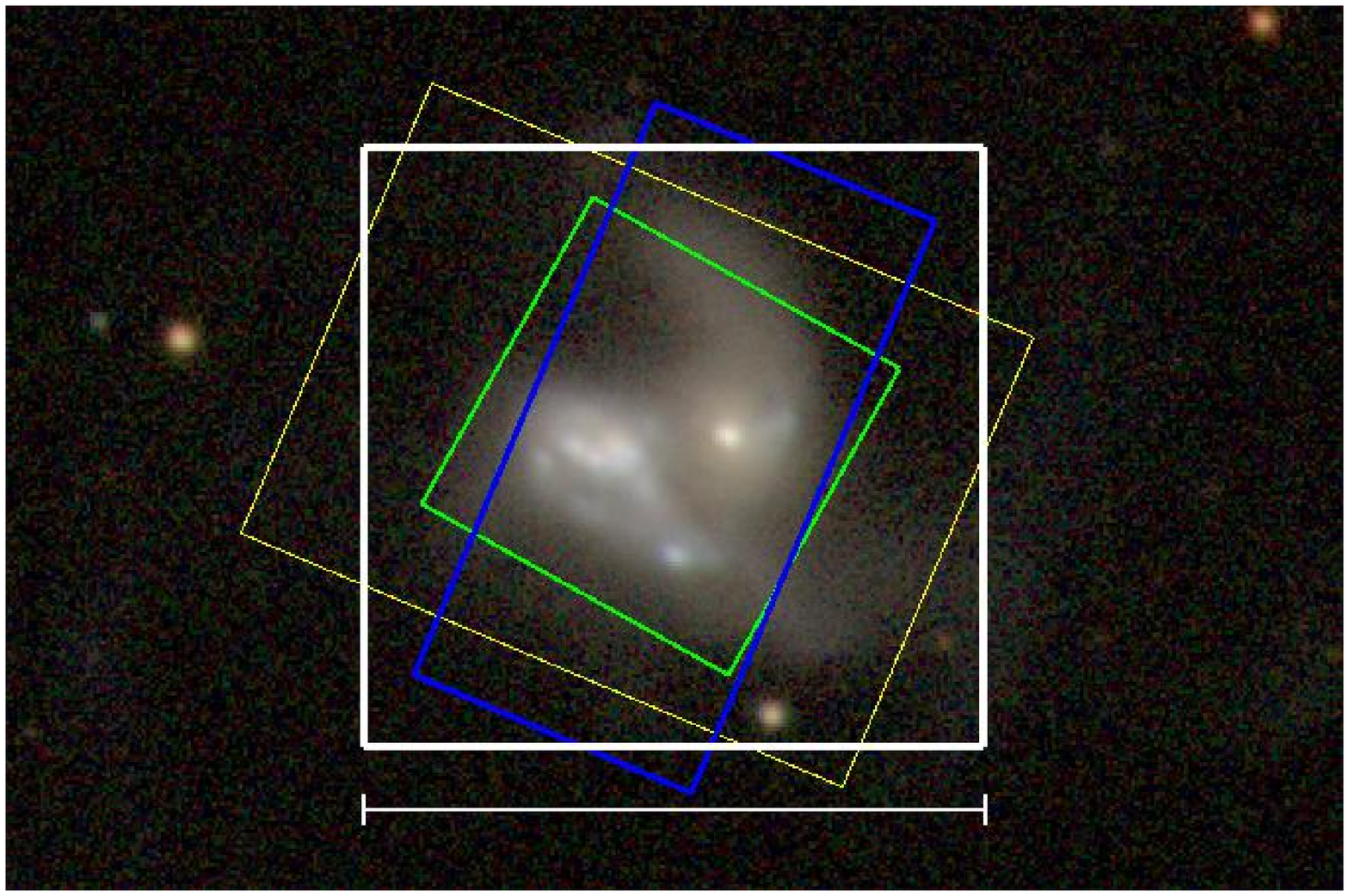}\\
\plottwo{f18_123a.ps}{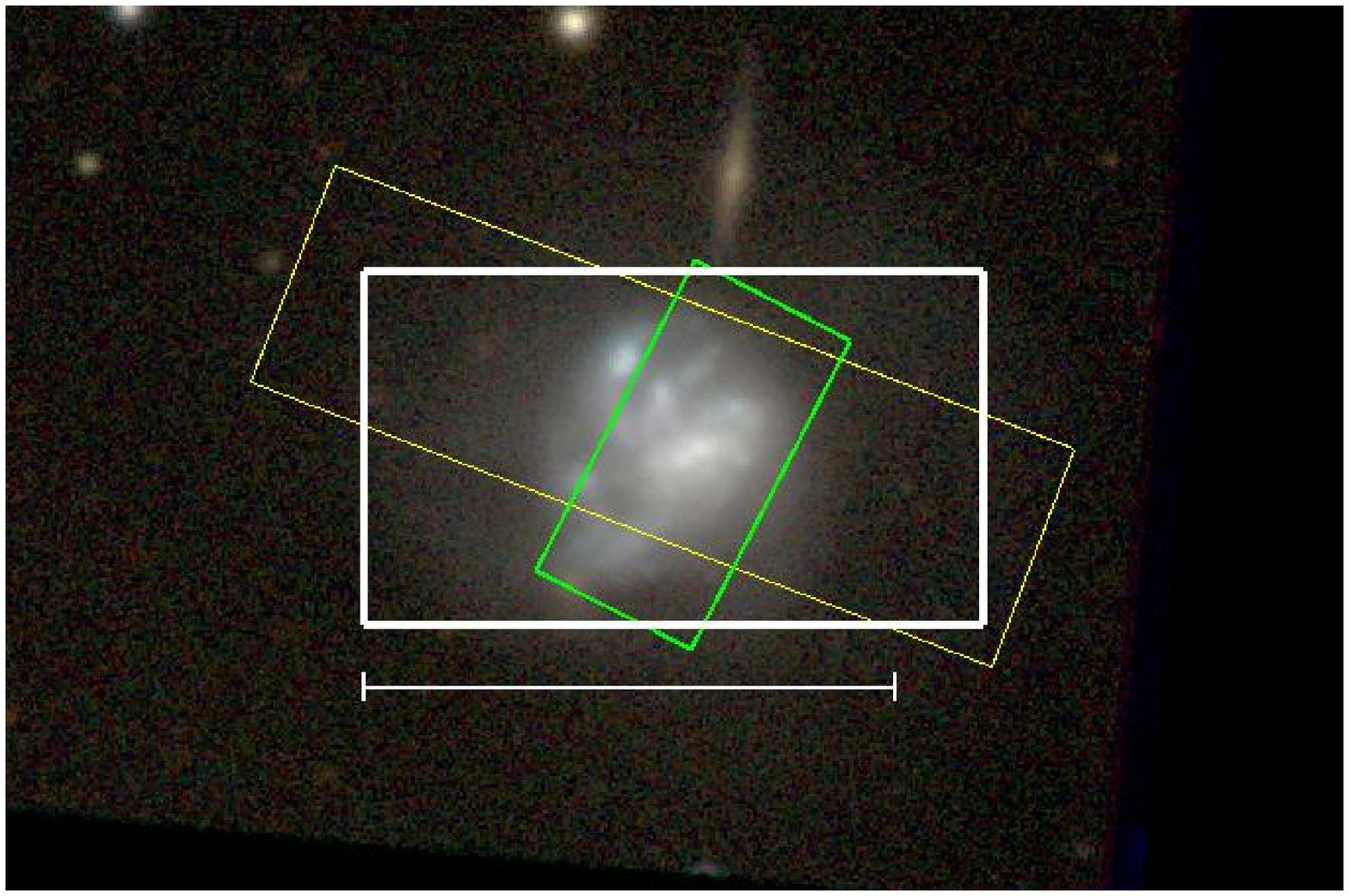}\\
\plottwo{f18_124a.ps}{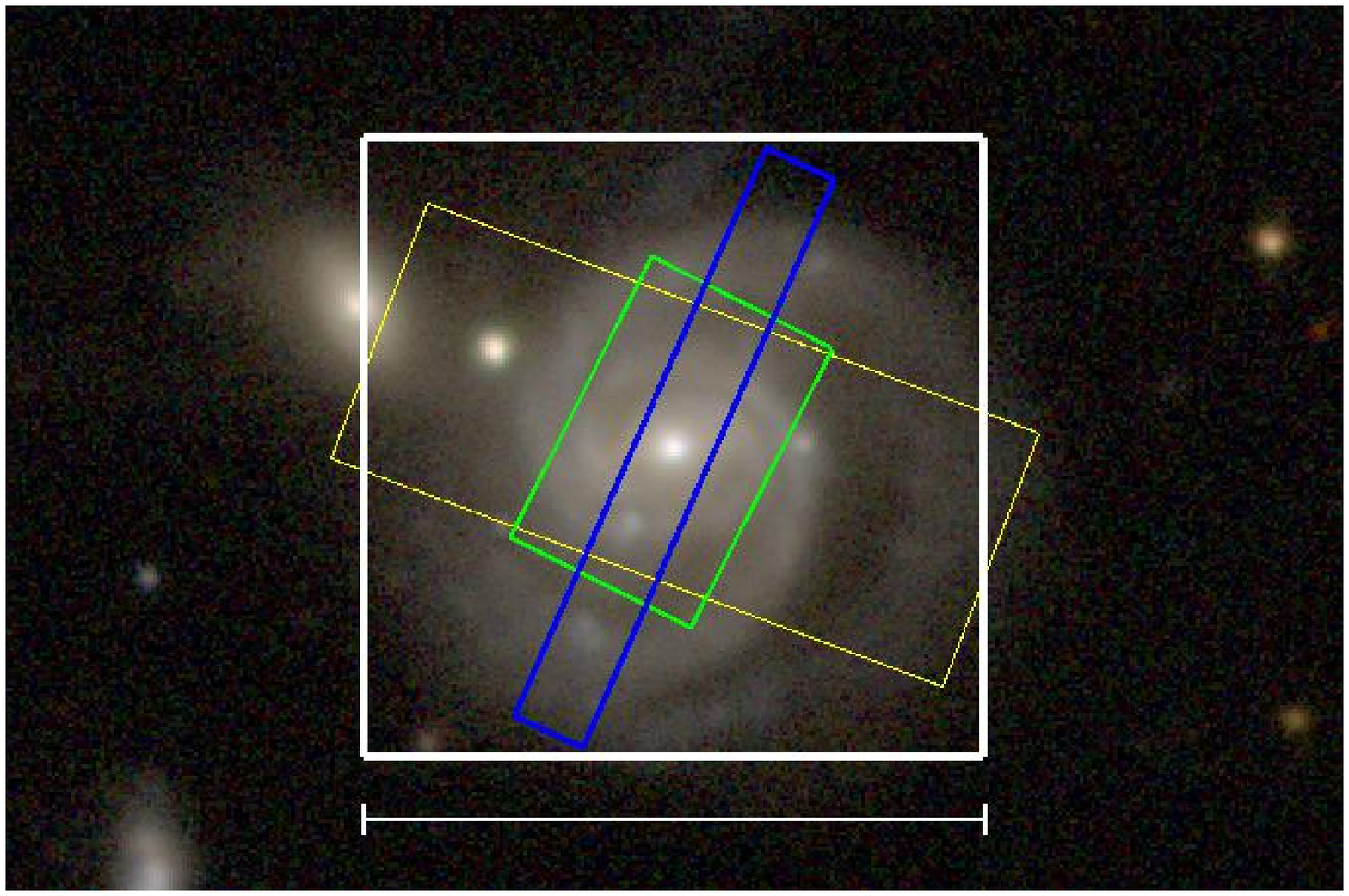}\\
\caption{Galaxy SEDs from the UV to the mid-IR. In the left-panel the observed and model spectra are shown in black and grey respectively, while the photometry used to constrain and verify the spectra is shown with red dots. In the right panel we plot the photometric aperture (thick white rectangle), the {\it Akari} extraction aperture (blue rectangle), the {\it Spitzer} SL extraction aperture (green rectangle) and the {\it Spitzer} LL extraction aperture (yellow rectangle). For galaxies with {\it Spitzer} stare mode spectra, we show a region corresponding to a quarter of the slit length. For scale, the horizontal bar denotes $1^{\prime}$. [{\it See the electronic edition of the Supplement for the complete Figure.}]}
\end{figure}

\clearpage

\begin{figure}[hbt]
\figurenum{\ref{fig:allspec} continued}
\plottwo{f18_125a.ps}{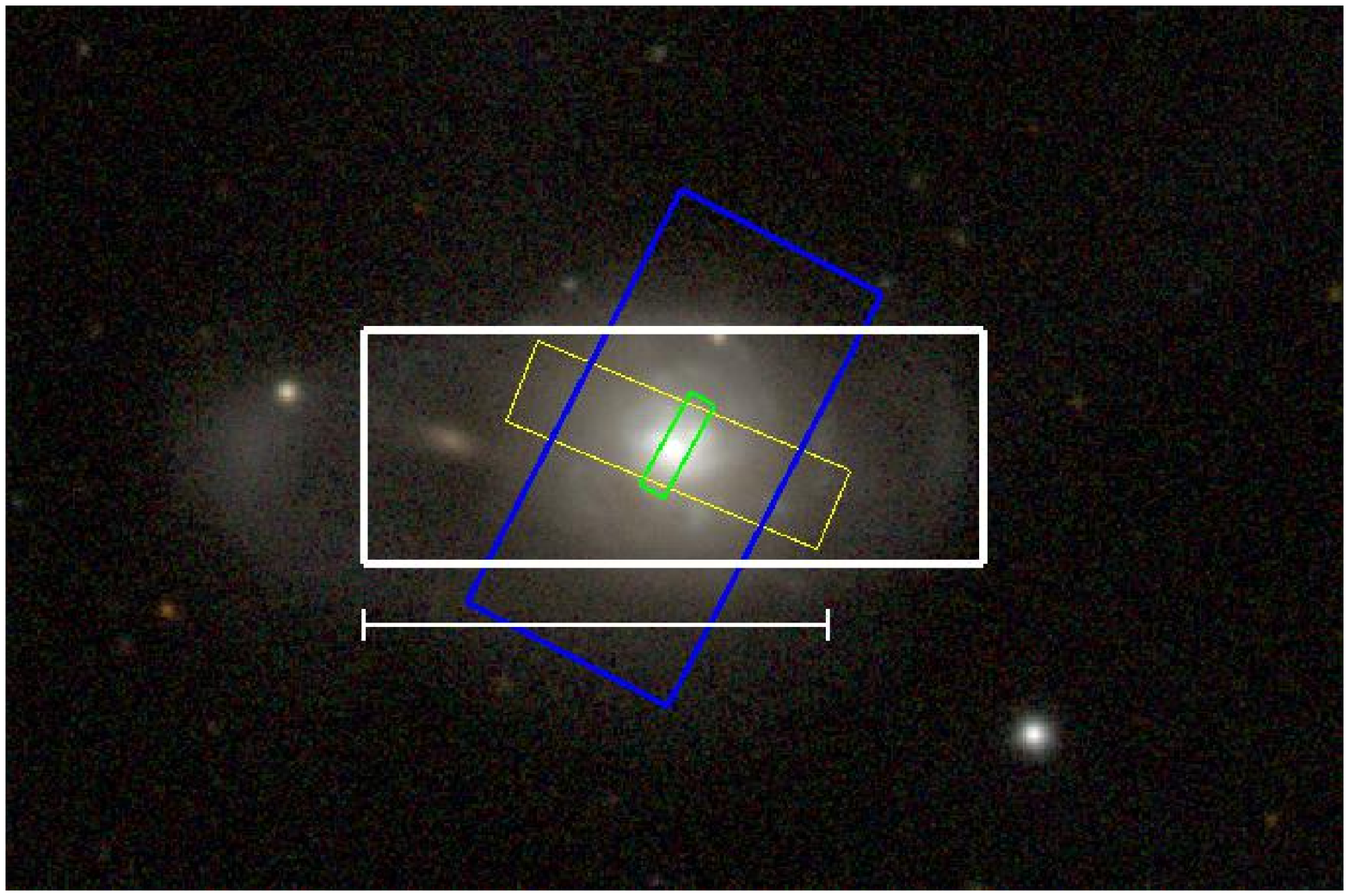}\\
\plottwo{f18_126a.ps}{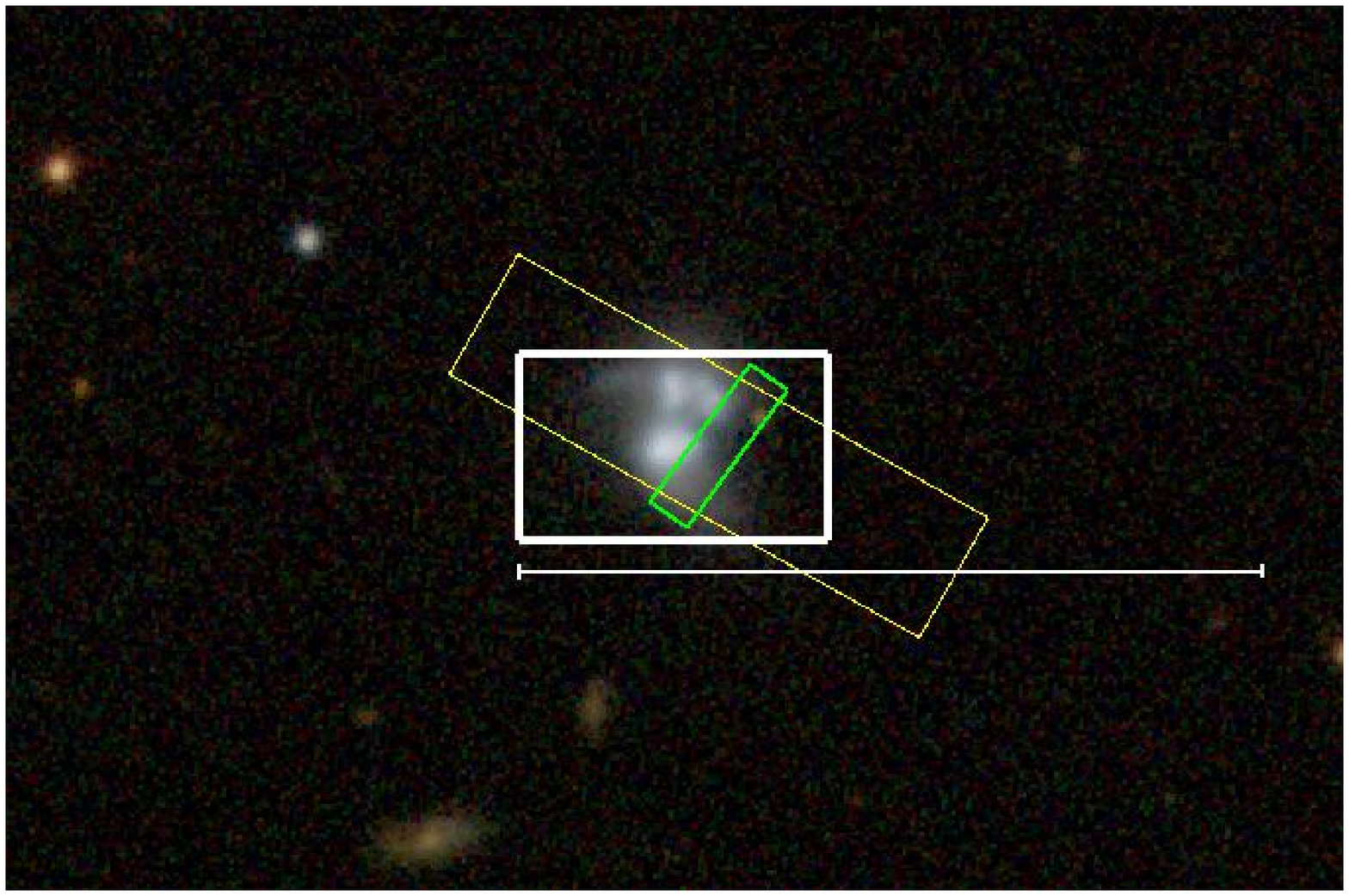}\\
\plottwo{f18_127a.ps}{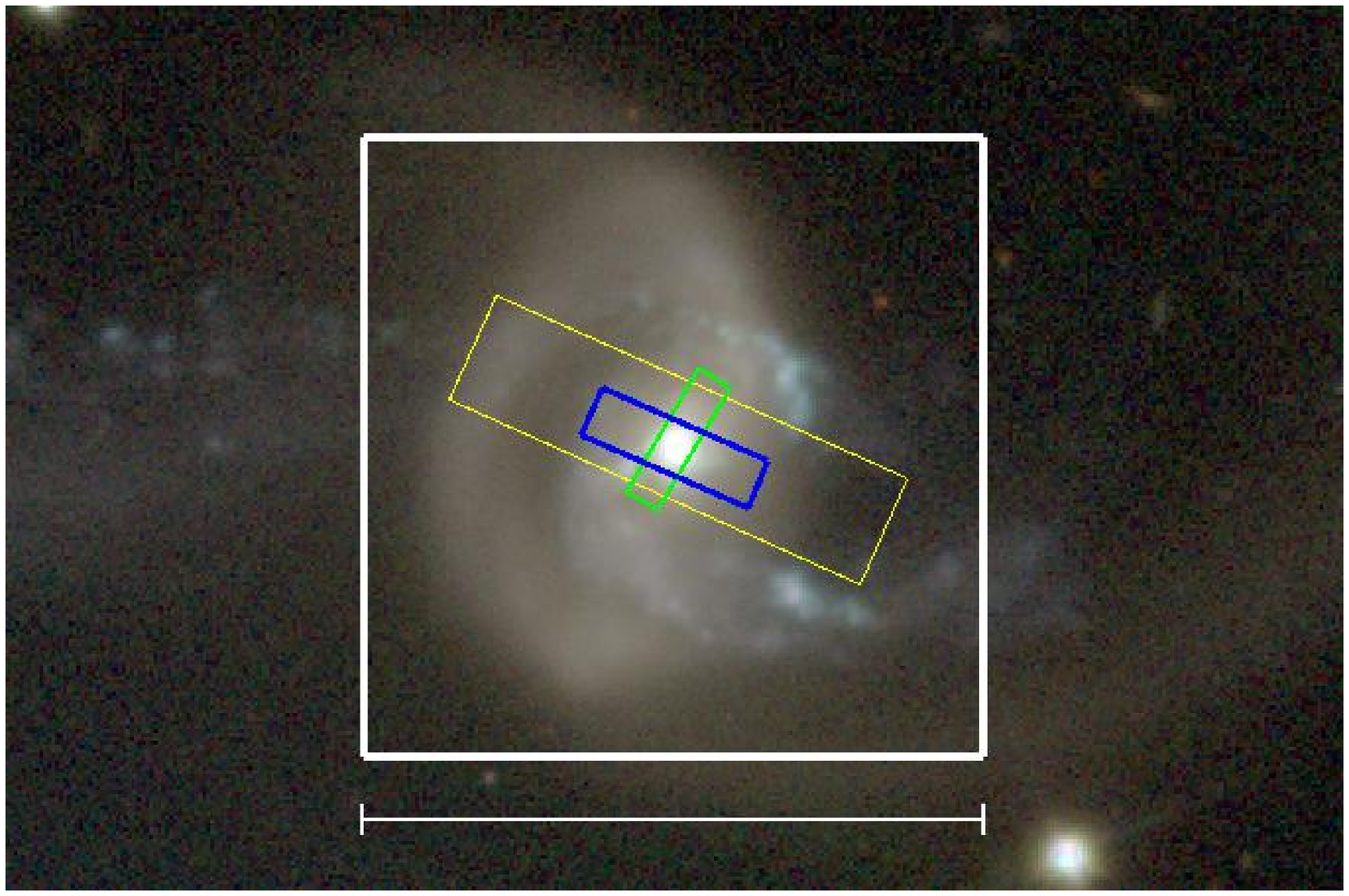}\\
\plottwo{f18_128a.ps}{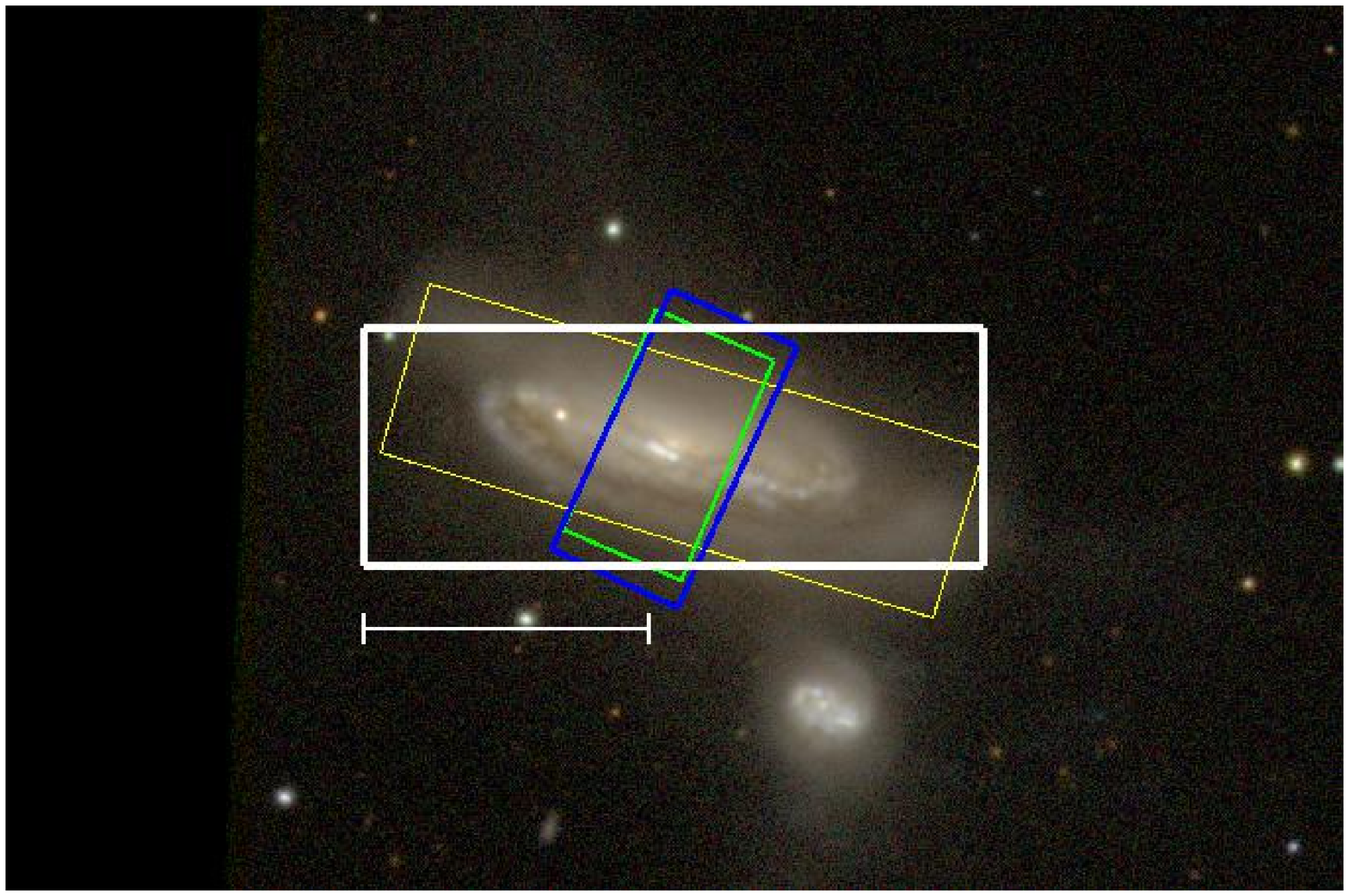}\\
\caption{Galaxy SEDs from the UV to the mid-IR. In the left-panel the observed and model spectra are shown in black and grey respectively, while the photometry used to constrain and verify the spectra is shown with red dots. In the right panel we plot the photometric aperture (thick white rectangle), the {\it Akari} extraction aperture (blue rectangle), the {\it Spitzer} SL extraction aperture (green rectangle) and the {\it Spitzer} LL extraction aperture (yellow rectangle). For galaxies with {\it Spitzer} stare mode spectra, we show a region corresponding to a quarter of the slit length. For scale, the horizontal bar denotes $1^{\prime}$. [{\it See the electronic edition of the Supplement for the complete Figure.}]}
\end{figure}

\clearpage

\begin{figure}[hbt]
\figurenum{\ref{fig:allspec} continued}
\plottwo{f18_129a.ps}{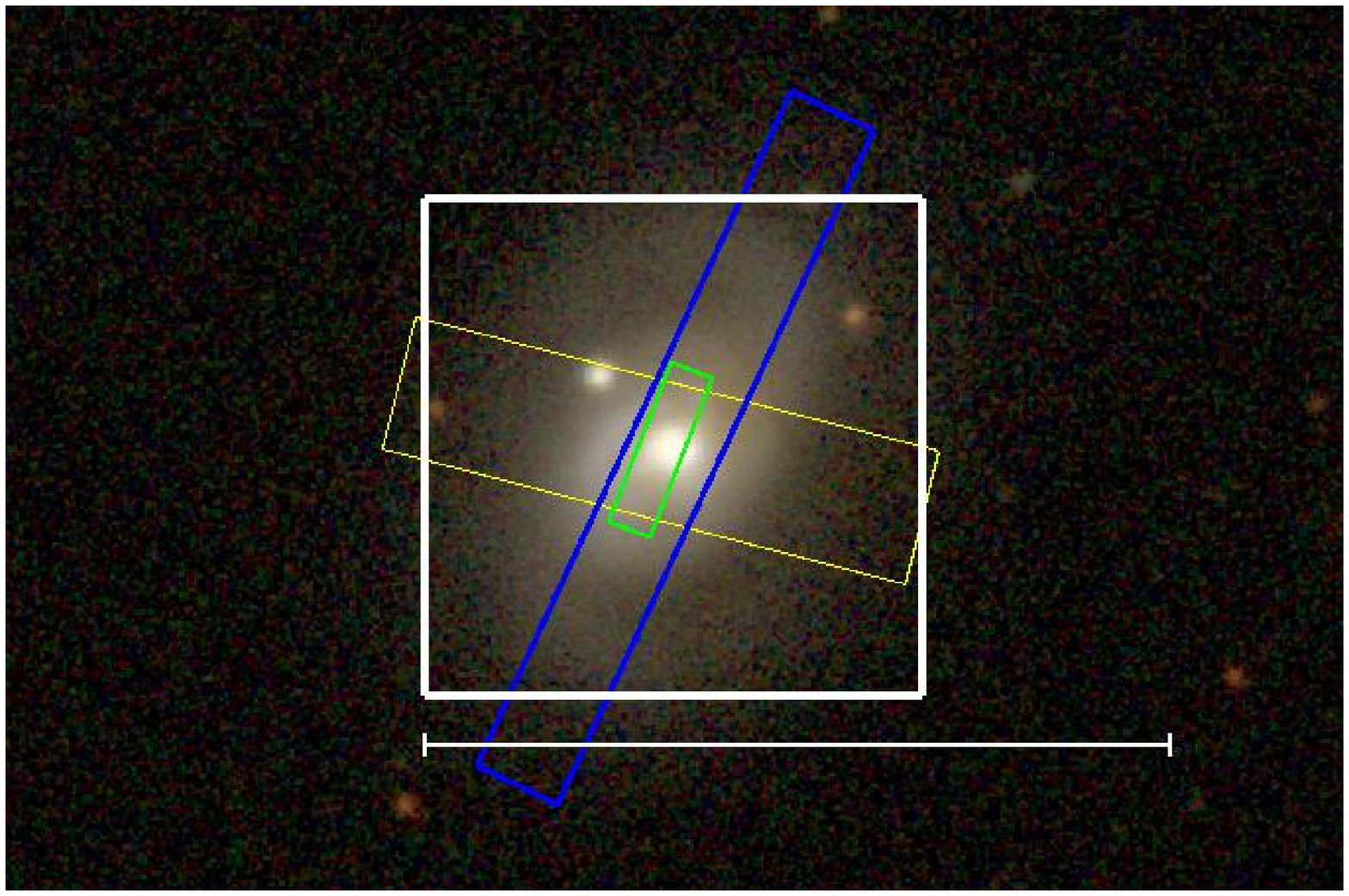}\\
\caption{Galaxy SEDs from the UV to the mid-IR. In the left-panel the observed and model spectra are shown in black and grey respectively, while the photometry used to constrain and verify the spectra is shown with red dots. In the right panel we plot the photometric aperture (thick white rectangle), the {\it Akari} extraction aperture (blue rectangle), the {\it Spitzer} SL extraction aperture (green rectangle) and the {\it Spitzer} LL extraction aperture (yellow rectangle). For galaxies with {\it Spitzer} stare mode spectra, we show a region corresponding to a quarter of the slit length. For scale, the horizontal bar denotes $1^{\prime}$. [{\it See the electronic edition of the Supplement for the complete Figure.}]}
\end{figure}

\clearpage

\end{document}